%% file: main.tex
\input{settings_epfl_template.tex}
\input{settings_custom.tex}  

\begin{document}

\frontmatter
 \input{titlepage.tex}
\include{dedication}

\setcounter{page}{0}
\include{abstracts}

\include{publications}
\include{acknowledgements}

\cleardoublepage
\pdfbookmark{\contentsname}{toc}
\tableofcontents

\cleardoublepage
\phantomsection
\addcontentsline{toc}{chapter}{List of Figures}
\listoffigures

\cleardoublepage
\phantomsection
\addcontentsline{toc}{chapter}{List of Tables}
\listoftables

\cleardoublepage
\phantomsection
\printnoidxglossary[type=acronym, title=List of Acronyms, nogroupskip=true, style=super] %
\printacronyms


\mainmatter

\include{ch1_introduction}

\cleardoublepage
\include{ch2_methods}

\cleardoublepage
\include{ch3_qsim_perspective}

\cleardoublepage
\include{ch4_benchmarking}

\cleardoublepage
\include{ch5_pea_oqs}

\cleardoublepage
\include{ch6_ground_states}

\cleardoublepage
\include{ch7_conclusion}

\addtocontents{toc}{\vspace{\normalbaselineskip}}
\cleardoublepage
\bookmarksetup{startatroot}


\include{appendix}

\backmatter

\include{biblio}

\include{cv}

\end{document}

%% file: settings_epfl_template.tex
\documentclass[a4paper,11pt]{book}

\usepackage[T1]{fontenc}
\usepackage[utf8]{inputenc}
\usepackage{doi}
\usepackage[autostyle]{csquotes}
\usepackage[
    backend=biber,
    style=numeric-comp,
    date=year,
    giveninits=true,
    sorting=none,
    url=false,
    doi=true,
    eprint=true,
    isbn=false
]{biblatex}
\usepackage[french,german,english]{babel}

\usepackage{lmodern} 

\usepackage{fourier} 
\usepackage{setspace} 

\usepackage{graphicx}
\graphicspath{{images/}}
\usepackage[svgnames]{xcolor}

\usepackage{subfig}
\usepackage{booktabs}
\usepackage{lipsum}
\usepackage[final]{microtype}
\usepackage{url}

\usepackage{fancyhdr}

\fancypagestyle{plain}{
	\fancyhf{}

	\fancyfoot[EL,OR]{\thepage}}
\fancypagestyle{addpagenumbersforpdfimports}{
	\fancyhead{}
	
	\fancyfoot{}
	\fancyfoot[RO,LE]{\thepage}
}

\usepackage{listings}
\usepackage{hyperref}
\hypersetup{pdfborder={0 0 0},
	colorlinks=true,
	linkcolor=NavyBlue,
	citecolor=NavyBlue,
	urlcolor=NavyBlue,
	filecolor=black,
	linktocpage=true
}
\ifpdf
\usepackage[final]{pdfpages}
\else
\usepackage{calc}
\usepackage{breakurl}
\usepackage[nlwarning=false]{hypdvips}
\usepackage{backref}
\renewcommand*{\backref}[1]{}
\fi
\usepackage{bookmark}

\makeatletter
\renewcommand\@pnumwidth{20pt}
\makeatother

\makeatletter
\def\cleardoublepage{\clearpage\if@twoside \ifodd\c@page\else
    \hbox{}
    \thispagestyle{empty}
    \newpage
    \if@twocolumn\hbox{}\newpage\fi\fi\fi}
\makeatother \clearpage{\pagestyle{plain}\cleardoublepage}

\usepackage{color}
\usepackage{tikz}
\usepackage[explicit]{titlesec}
\newcommand*\chapterlabel{}
\titleformat{\chapter}[display]  
	{\normalfont\bfseries\Huge} 
	{\gdef\chapterlabel{\thechapter\ }}     
 	{0pt} 
 	  {\begin{tikzpicture}[remember picture,overlay]
    \node[yshift=-8cm] at (current page.north west)
      {\begin{tikzpicture}[remember picture, overlay]
        \draw[fill=black] (0,0) rectangle(35.5mm,15mm);
        \node[anchor=north east,yshift=-7.2cm,xshift=34mm,minimum height=30mm,inner sep=0mm] at (current page.north west)
        {\parbox[top][30mm][t]{15mm}{\raggedleft \rule{0cm}{0.6cm}\color{white}\chapterlabel}};  
        \node[anchor=north west,yshift=-7.2cm,xshift=37mm,text width=\textwidth,minimum height=30mm,inner sep=0mm] at (current page.north west)
              {\parbox[top][30mm][t]{\textwidth}{\rule{0cm}{0.6cm}\color{black}#1}};
       \end{tikzpicture}
      };
   \end{tikzpicture}
   \gdef\chapterlabel{}
  } 
\titlespacing*{name=\chapter,numberless}{-3.7cm}{83.2pt-\parskip}{-3.2pt+\parskip}
\titlespacing*{\chapter}{-3.7cm}{50pt-\parskip-\parskip}{30pt+\parskip+\parskip}
\titlespacing*{\section}{0pt}{13.2pt}{1em-\parskip}  
\titlespacing*{\subsection}{0pt}{13.2pt}{1em-\parskip}
\titlespacing*{\subsubsection}{0pt}{13.2pt}{1em-\parskip}
\titlespacing*{\paragraph}{0pt}{13.2pt}{1em-\parskip}

\newcounter{myparts}
\newcommand*\partlabel{}
\titleformat{\part}[display]  
	{\normalfont\bfseries\Huge} 
	{\gdef\partlabel{\thepart\ }}     
 	{0pt} 
 	  {\ifpdf\setlength{\unitlength}{20mm}\else\setlength{\unitlength}{0mm}\fi
	  \addtocounter{myparts}{1}
	  \begin{tikzpicture}[remember picture,overlay]
    \node[anchor=north west,xshift=-65mm,yshift=-6.9cm-\value{myparts}*20mm] at (current page.north east) 
      {\begin{tikzpicture}[remember picture, overlay]
        \draw[fill=black] (0,0) rectangle(62mm,20mm);   
        \node[anchor=north west,yshift=-6.1cm-\value{myparts}*\unitlength,xshift=-60.5mm,minimum height=30mm,inner sep=0mm] at (current page.north east)
        {\parbox[top][30mm][t]{55mm}{\raggedright \color{white}Part \partlabel \rule{0cm}{0.6cm}}};  
        \node[anchor=north east,yshift=-6.1cm-\value{myparts}*\unitlength,xshift=-63.5mm,text width=\textwidth,minimum height=30mm,inner sep=0mm] at (current page.north east)
              {\parbox[top][30mm][t]{\textwidth}{\raggedleft \rule{0cm}{0.6cm}\color{black}#1}};
       \end{tikzpicture}
      };
   \end{tikzpicture}
   \gdef\partlabel{}
  } 
\titlespacing*{\part}{11.06cm}{26.4pt-\parskip-\parskip}{0pt}

\usepackage{amsmath}
\usepackage{amsfonts}
\usepackage{amssymb}
\usepackage{mathtools}
\makeatletter
\def\resetMathstrut@{%
  \setbox\z@\hbox{%
    \mathchardef\@tempa\mathcode`\(\relax
      \def\@tempb##1"##2##3{\the\textfont"##3\char"}%
      \expandafter\@tempb\meaning\@tempa \relax
  }%
  \ht\Mathstrutbox@1.2\ht\z@ \dp\Mathstrutbox@1.2\dp\z@
}
\makeatother

%% file: settings_custom.tex
\usepackage{amsmath, amssymb, amsthm}
\usepackage{braket}
\usepackage{bm}
\usepackage{IEEEtrantools}
\usepackage{enumitem}

\usepackage[capitalize]{cleveref}
\Crefname{figure}{Fig.}{Figs.}
\Crefname{equation}{Eq.}{Eqs.}

\usepackage{siunitx}
\usepackage[normalem]{ulem}

\usepackage{booktabs}
\newcolumntype{L}[1]{>{\raggedright\let\newline\\\arraybackslash\hspace{0pt}}p{#1}}
\newcolumntype{C}[1]{>{\centering\let\newline\\\arraybackslash\hspace{0pt}}p{#1}}
\newcolumntype{R}[1]{>{\raggedleft\let\newline\\\arraybackslash\hspace{0pt}}p{#1}}

\DeclareCiteCommand{\citenum}
{\usebibmacro{cite:init}%
   \usebibmacro{prenote}}
  {\usebibmacro{citeindex}%
   \usebibmacro{cite:comp}}
  {}
  {\usebibmacro{cite:dump}%
   \usebibmacro{postnote}}

\newcommand{\mycaption}[2]{\caption[#1]{\textbf{#1.} #2}}

\usepackage{indentfirst}

\usepackage[framemethod=TikZ]{mdframed}
\mdfdefinestyle{summarybox}{
    roundcorner=10pt,
    innertopmargin=\topskip,
    frametitle={Summary},
    frametitlerule=true,
    frametitlebackgroundcolor=gray!20,
}

\usepackage[most]{tcolorbox} 
\definecolor{block-gray}{gray}{0.96}

\newtcolorbox{zitat}[2][]{%
    colback=block-gray,
    grow to right by=-7mm,
    grow to left by=-7mm, 
    boxrule=0pt,
    boxsep=0pt,
    breakable,
    enhanced jigsaw,
    borderline west={4pt}{0pt}{gray},
    colbacktitle={block-gray},
    coltitle={black},
    fonttitle={\large\bfseries},
    attach title to upper={},
    #1,
}

\DeclareMathAlphabet{\mathcal}{OMS}{cmsy}{m}{n}

\newcommand{\sherbrooke}[0]{\texttt{ibm\_sherbrooke}}
\newcommand{\torino}[0]{\texttt{ibm\_torino}}
\newcommand{\auckland}[0]{\texttt{ibm\_auckland}}
\newcommand{\kolkata}[0]{\texttt{ibm\_kolkata}}
\newcommand{\cz}[0]{$CZ$}
\newcommand{\qiskit}[0]{\textsc{Qiskit}}
\newcommand{\estimator}[0]{\textsc{Qiskit}'s Estimator}

\def\bea{\begin{IEEEeqnarray}}
\def\eea{\end{IEEEeqnarray}}

\newcommand{\diff}[1]{\mathrm{d} #1 \, }
\newcommand{\sx}{\sigma{\mathstrut}^x}

\newcommand{\sz}{\sigma{\mathstrut}^z}

\def\tf{t_\mathrm{f}}
\def\HP{H_\mathrm{P}}
\def\HM{H_\mathrm{M}}

\def\ndef{n_\mathrm{def}}
\def\Ne{N_\mathrm{e}}
\def\rzz{R_{ZZ}}
\def\cz{CZ}
\def\Eres{E_\mathrm{res}}
\def\tone{\mathrm{T}_1}
\def\ttwo{\mathrm{T}_2}

\def\tr{\mathrm{tr}}
\def\O{\mathcal{O}}

\def\U{\mathcal{U}}
\def\Ut{\tilde{\mathcal{U}}}
\def\Kt{\Tilde{K}}

\newcommand{\qbf}      {{q}}

\newcommand{\Ugs}{U_{\rm GS}}
\newcommand{\Rq}{R_\qbf}
\newcommand{\Rqd}{R_\qbf^\dagger}

\usepackage[acronym,symbols,nogroupskip,nomain,nonumberlist,nopostdot,toc]{glossaries}
\makenoidxglossaries
\glsdisablehyper
\newacronym{vqa}{VQA}{variational quantum algorithm}
\newacronym{vqe}{VQE}{a variational quantum eigensolver}
\newacronym{tdse}{TDSE}{time-dependent Schrödinger equation}
\newacronym{tise}{TISE}{time-independent Schrödinger equation}
\newacronym{qpe}{QPE}{quantum phase estimation}
\newacronym{qmc}{QMC}{Quantum Monte Carlo}
\newacronym{vte}{VTE}{variational time evolution}
\newacronym{pf}{PF}{product formula}
\newacronym{vha}{VHA}{Variational Hamiltonian Ansatz}
\newacronym[firstplural=degrees of freedom (DOFs)]{dof}{DOF}{degree of freedom}
\newacronym{oqs}{OQS}{open quantum systems}
\newacronym{kzm}{KZM}{Kibble-Zurek mechanism}
\newacronym{qkzm}{QKZM}{quantum Kibble-Zurek mechanism}
\newacronym{kz}{KZ}{Kibble-Zurek}
\newacronym{lz}{LZ}{Landau-Zener}
\newacronym{ftlz}{FTLZ}{finite-time Landau-Zener}
\newacronym{tfim}{TFIM}{transverse field Ising model}
\newacronym{cp}{CP}{critical point}
\newacronym{gs}{GS}{ground state}
\newacronym{qa}{QA}{quantum annealing}
\newacronym{sa}{SA}{simulated annealing}
\newacronym{sqa}{SQA}{simulated quantum annealing}
\newacronym{cpt}{CPT}{classical phase transition}
\newacronym{qpt}{QPT}{quantum phase transition}
\newacronym{qv}{QV}{quantum volume}
\newacronym{rb}{RB}{randomized benchmarking}
\newacronym{xeb}{XEB}{cross-entropy benchmarking}
\newacronym{qec}{QEC}{quantum error correction}
\newacronym{em}{EM}{error mitigation}
\newacronym{ems}{EMS}{error mitigation and suppression}
\newacronym{rem}{REM}{readout error mitigation}
\newacronym{trex}{TREX}{twirled readout error mitigation}
\newacronym{m3}{M3}{matrix-free measurement mitigation}
\newacronym{dd}{DD}{dynamical decoupling}
\newacronym{pec}{PEC}{probabilistic error cancellation}
\newacronym{pea}{PEA}{probabilistic error amplification}
\newacronym{zne}{ZNE}{zero noise extrapolation}
\newacronym{ecr}{ECR}{echoed cross resonance}
\newacronym{qaoa}{QAOA}{the quantum approximate optimization algorithm}

\newacronym{hf}{HF}{Hartree-Fock}
\newacronym{se}{SE}{Schrödinger equation}
\newacronym{qft}{QFT}{quantum Fourier transform}
\newacronym{mc}{MC}{Monte Carlo}
\newacronym{jqc}{JQC}{Jastrow quantum circuit}
\newacronym[firstplural=equations of motion (EOMs)]{eom}{EOM}{equation of motion}
\newacronym{lr}{LR}{linear response}
\newacronym{qse}{QSE}{quantum subspace expansion}
\newacronym{qeom}{qEOM}{quantum-EOM}
\newacronym{ps}{PS}{Pauli string}
\newacronym{onv}{ONV}{occupation number vector}
\newacronym{mctdh}{MCTDH}{multiconfigurational time-dependent Hartree}
\newacronym{spsa}{SPSA}{simultaneous perturbation stochastic approximation}
\newacronym{tdvp}{TDVP}{time-dependent variational principle}
\newacronym{povm}{POVM}{positive operator-values measure}
\newacronym{mpf}{MPF}{multiproduct formula}
\newacronym{lcu}{LCU}{linear combination of unitaries}
\newacronym{qw}{QW}{Quantum walk}
\newacronym{qsp}{QSP}{quantum signal processing}
\newacronym{qdrift}{qDRIFT}{quantum stochastic drift}
\newacronym{qsvt}{QSVT}{quantum singular value transformation}
\newacronym{bo}{BO}{Born-Oppenheimer}
\newacronym{mvp}{MVP}{McLachlan variational principle}
\newacronym{dfvp}{DFVP}{Dirac-Frenkel variational principle}
\newacronym{ode}{ODE}{ordinary differential equation}
\newacronym{tn}{TN}{tensor network}
\newacronym{lgt}{LGT}{lattice gauge theory}
\newacronym{qed}{QED}{quantum electrodynamics}
\newacronym{qcd}{QCD}{quantum chromodynamics}
\newacronym{lqed}{LQED}{lattice quantum electrodynamics}
\newacronym{lqcd}{LQCD}{lattice quantum chromodynamics}
\newacronym{ness}{NESS}{non-equilibrium steady states}
\newacronym{tddmrg}{TD-DMRG}{time-dependent density matrix renormalization group}
\newacronym{pes}{PES}{potential energy surface}
\newacronym{dft}{DFT}{density functional theory}
\newacronym{tddft}{TD-DFT}{time-dependent density functional theory}
\newacronym{dmrg}{DMRG}{density matrix renormalization group}
\newacronym[firstplural=matrix product states (MPS)]{mps}{MPS}{matrix product state}
\newacronym{peps}{PEPS}{projected entangled pair states}
\newacronym[firstplural=multiscale entanglement renormalization ansätze (MERA)]{mera}{MERA}{multiscale entanglement renormalization ansatz}
\newacronym{cc}{CC}{coupled cluster}
\newacronym{dmft}{DMFT}{dynamical mean field theory}
\newacronym{cptp}{CPTP}{completely positive and trace preserving}

\newacronym{qcnn}{QCNN}{quantum convolutional neural network}
\newacronym{qnn}{QNN}{quantum neural network}
\newacronym{hva}{HVA}{Hamitlonian variational ansatz}
\newacronym{qdl}{QDL}{quantum data learning}
\newacronym{qml}{QML}{quantum machine learning}
\newacronym{cnn}{CNN}{convolutional neural network}

\newacronym{ed}{ED}{exact diagonalization}
\newacronym{epc}{EPC}{electron-phonon coupling}
\newacronym{ngsed}{NGS-ED}{non-Gaussian exact diagonalization}
\newacronym{ngs}{NGS}{non-Gaussian solver}
\newacronym{heh}{HEH}{Hubbard-extended-Holstein}
\newacronym{cdw}{CDW}{charge density wave}
\newacronym{afm}{AFM}{antiferromagnetic}

%% file: titlepage.tex
\begin{titlepage}

\sffamily

\begin{flushleft}
\parbox{0.3\textwidth}{\includegraphics[width=5cm]{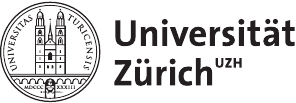}}
\end{flushleft}


\null\vspace{2cm}

\begin{minipage}{4cm}
\end{minipage}
    \hfill
\begin{minipage}{11cm}
{\Large Digital quantum simulation of many-body systems: \\
Making the most of intermediate-scale, \\ noisy quantum computers} \\

\vspace{2cm}

Vorgelegt am 02. Juli 2025 \\[8pt] %
Graduate School for Computational Science\\
Mathematisch-naturwissenschaftliche Fakultät\\
Universität Zürich\\

zur Erlangung der naturwissenschaftlichen Doktorwürde (Dr. sc. nat.)\\[8pt]
von\\ [12pt]
{\Large \textbf{Alexander MIESSEN}}\\[19pt]

Promotionskommission\\[4pt]
%
    Prof. Dr. Guglielmo Mazzola\\
    Dr. Ivano Tavernelli\\
    Prof. Dr. Titus Neupert\\
\end{minipage}
\vspace{2cm}
\begin{flushright}
    2025
\end{flushright}

\end{titlepage}

%% file: dedication.tex
\cleardoublepage
\thispagestyle{empty}

\vspace*{3cm}

\begin{raggedleft}
    bike-strung boy \\
    sun-licked spark of rubber and steel \\
    you soar \\
    down the driveway \\
    arms spread like wings \\
    it is a wonderful thing to fly \\
    and there is no greater occasion \\
    than being alive \\
    --- Lewis Corry \\
\end{raggedleft}

\vspace{4cm}


\vspace{4cm}

\begin{center}
	To my WG, my friends, and my family.
\end{center}

%% file: abstracts.tex

\cleardoublepage
\chapter*{Abstract}
\markboth{Abstract}{Abstract}
\addcontentsline{toc}{chapter}{Abstract (English/Deutsch)}

The landscape of problems that can only be solved computationally is vast and affects many different disciplines.
Quantum mechanical problems are among the most challenging problems to simulate and, in certain cases, cannot be solved even with the most powerful computers.
Quantum computing has emerged as a new technological platform promising to overcome some of these challenges and has seen immense progress in recent years.
Still, current-day quantum machines remain limited in scale and affected by noise.
It is therefore important to identify the overlap between problems that are challenging for conventional computers, of practical relevance, and could be solved with available quantum devices.
Among all applications envisioned to benefit from quantum computing, the real-time simulation of quantum mechanical systems is one of the most anticipated applications to realize such an early, practical quantum advantage.
This thesis is therefore centered around simulating quantum dynamics on quantum devices.

Even within quantum dynamics, one central question is how to best achieve a computational advantage with quantum computation.
This concerns both the specific problem to simulate and the most efficient algorithmic implementation of the time evolution.
The latter is of particular importance since time evolution appears not only in quantum dynamics simulations, but as a subroutine in many other algorithms.
In this regard, we present an overview of the most relevant quantum algorithms to implement quantum dynamics, highlighting their respective advantages and limitations.
This constitutes the first important contribution of this thesis.
Furthermore, we discuss some of the most relevant and most promising problems within quantum dynamics in the natural sciences and identify possible research directions that could benefit from quantum simulation in the near future.

Another important, but difficult, question is how to efficiently gauge the impact that noise has on a simulation outcome and, related to that, how good a result one can expect as the outcome of a simulation.
Recent advances in scale and simulation capabilities of quantum hardware mean that classical emulation becomes increasingly challenging.
As a result, devising new methods to benchmark devices and to provide a scalable and intuitive metric to assess hardware quality has become of key importance.
As a second major contribution of this thesis, we propose such a method for benchmarking hardware and error mitigation algorithms.
Our method is based on well-understood theoretical results, suffers from no scaling issues, and the resulting quality metric is easily interpreted and transferred to other applications.
We successfully implement the scheme on up to 133 qubits, demonstrating coherent evolution up to a two-qubit gate depth of 28, featuring a maximum of 1396 two-qubit gates, before noise becomes prevalent.

While quantum simulation on available hardware might be negatively affected by noise in most cases, harnessing the noise for specialized simulations might prove advantageous in the near term.
Such hybrid digital-analog simulations could help overcome limitations imposed by noise.
As a third contribution of this thesis, we propose a method to implement open quantum dynamics using a variant of probabilistic error amplification specially adapted for this purpose.
Specifically, the method relies on characterizing noise inherent to the hardware and altering it such that it mimics the system-environment interaction under study.
We present a rigorous theoretical analysis of this method, an analytically derived error bound, and numerical tests thereof.
Moreover, we discuss relevant problems that would benefit from such a treatment.

Lastly, the study of time-independent states represents a challenge in the context of and beyond quantum dynamics.
Particularly, ground states of many-body quantum systems are often subject to complex phase structures, the study of which is crucial for many areas of research.
In a fourth contribution, we present two studies related to state preparation and phase classification.
The first study presents a hybrid algorithm to prepare ground states of a realistic condensed matter system, describing the interactions between electrons and the lattice in which they move, and captures several phases across a range of system parameters.
We further investigate how to optimally implement this scheme on available quantum hardware and, in simulation, find it to be robust against realistic levels of hardware noise.
In an independent, but topically related, second study, we present a quantum machine learning-based approach that is successfully employed to distinguish different phases in previously prepared quantum states.

\begin{otherlanguage}{german}
\cleardoublepage
\chapter*{Zusammenfassung}
\markboth{Zusammenfassung}{Zusammenfassung}

Fragestellungen, die nur mit Hilfe von Computern gelöst werden können, sind vielfältig und betreffen verschiedenste Disziplinen.
Zu den herausforderndsten solcher Probleme gehören quantenmechanische Simulationen, die in bestimmten Fällen selbst mit den leistungsfähigsten Computern nicht gelöst werden können.
Quantencomputer stellen eine neue technologische Plattform dar, die verspricht, einige dieser Herausforderungen zu überwinden, und hat in den letzten Jahren immense Fortschritte gemacht.
Dennoch sind die heutigen Quantencomputer nach wie vor durch ihre begrenzte Grösse und hohe Fehlerraten in ihrer Rechenleistung beeinträchtigt.
Es ist daher von zentraler Bedeutung, die Schnittmenge jener Probleme zu identifizieren, die für herkömmliche Computer eine Herausforderung darstellen, von praktischer Relevanz sind und zugleich mit verfügbaren Quantengeräten gelöst werden könnten.
Unter allen Anwendungen, die von Quantencomputern profitieren könnten, zählt die Simulation von zeitabhängigen Prozessen in quantenmechanischen Systemen zu den vielversprechendsten Anwendungen, um einen solchen frühen und praktisch relevanten Vorteil von Quantencomputern gegenüber herkömmlichen Computern zu realisieren.
Diese Arbeit widmet sich daher der Simulation von Quantendynamik mit Hilfe von Quantenprozessoren.

Selbst innerhalb der Quantendynamik stellt sich jedoch die grundlegende Frage, wie sich solch ein rechnerischer Vorteil durch Quantencomputer am besten realisieren lässt.
Dies betrifft sowohl die Auswahl zu simulierender Probleme, als auch die effizienteste algorithmische Implementierung der Zeitentwicklung.
Letztere ist von besonderer Bedeutung, da der Zeitentwicklungsoperator nicht nur im Kontext von Simulationen dynamischer Quantensysteme, sondern auch als Unterroutine vieler anderer Quantenalgorithmen vorkommt.
In diesem Zusammenhang präsentieren wir einen Überblick über die relevantesten Quantenalgorithmen zur Implementierung von Quantendynamik und heben dabei ihre jeweilige Stärken und Schwächen hervor.
Dies stellt den ersten wesentlichen Beitrag dieser Arbeit dar.
Darüber hinaus diskutieren wir einige der relevantesten und vielversprechendsten Problemstellungen innerhalb der Quantendynamik und identifizieren zukünftige Forschungsrichtungen mit hohem Anwendungspotenzial für Quantencomputer in näherer Zukunft.

Eine weitere zentrale, aber schwierige Frage betrifft die effiziente Quantifizierung des Einflusses von Rauschen und Fehlern seitens des Computers auf das Simulationsergebnis und, damit zusammenhängend, die realistisch erreichbare Qualität eines Ergebnisses.
Mit zunehmender Grösse und Leistungsfähigkeit moderner Quantenprozessoren wird eine klassische Emulation immer herausfordernder.
Infolgedessen ist die Entwicklung neuer Methoden zum Benchmarking von Quantenhardware, insbesondere einer skalierbaren und intuitiven Metrik zur Bewertung der Hardwarequalität, von zentraler Bedeutung.
Als zweiten wichtigen Beitrag dieser Arbeit schlagen wir daher eine solche Methode für das Benchmarking von Hardware sowie von Algorithmen zur Fehlerbegrenzung vor.
Unsere Methode basiert auf theoretisch gut verstandenen Konzepten, weist keine Skalierungsprobleme auf und liefert eine leicht interpretierbare sowie auf andere Anwendungen übertragbare Qualitätsmetrik.
Wir implementieren das Verfahren erfolgreich auf bis zu 133 Qubits und zeigen Quantenkohärenz bis zu einer Zwei-Qubit Gate-Tiefe von 28, mit einem Maximum von 1396 Zwei-Qubit-Gates, bevor Rauschen und Fehler das Ergebnis unbrauchbar machen.

Obwohl Rauschen und Fehler auf verfügbaren Quantencomputern Simulationen in den meisten Fällen negativ beeinflussen, könnte sich dessen gezielte Nutzung für spezielle Simulationen in näherer Zukunft als vorteilhaft erweisen.
Solche hybride digital-analoge Simulationsansätze könnten dazu beitragen, die durch Rauschen bedingten Einschränkungen von Hardware zu überwinden.
Als dritten Beitrag zu dieser Arbeit schlagen wir eine Methode zur Simulation der Zeitentwicklung sogenannter offener Quantensysteme vor, die auf einer gezielt angepassten Variante der probabilistischen Fehlerverstärkung (probabilistic error amplification) basiert.
Kernidee dieser Methode ist es, das in der Hardware inhärente Rauschen zu charakterisieren und gezielt zu modifizieren, sodass es die zu untersuchende Wechselwirkung zwischen System und Umgebung nachbildet.
Wir präsentieren eine rigorose theoretische Analyse dieser Methode, eine analytisch hergeleitete Fehlerabschätzung sowie numerische Validierungen.
Zudem diskutieren wir relevante physikalische Probleme, die von derlei Simulationsansätzen profitieren könnten.

Im Rahmen der Quantendynamik und darüber hinaus stellt auch die Untersuchung zeitunabhängiger Zustände eine Herausforderung dar.
Insbesondere die Grundzustände von Vielteilchen-Quantensystemen weisen häufig komplexe Phasenstrukturen auf, deren Verständnis für zahlreiche Forschungsbereiche von entscheidender Bedeutung ist.
Im vierten Beitrag dieser Arbeit präsentieren wir daher zwei Studien zur Präparation von zeitunabhängigen Zuständen und Klassifikation von Phasen.
In der ersten Studie wird ein hybrider Algorithmus zur Präparation von Grundzuständen eines realistischen Festkörpersystems vorgestellt, der die Wechselwirkungen von Elektronen mit dem Gitter, in dem sie sich bewegen, modelliert und mehrere Phasen in Abhängigkeit von verschiedenen Systemparametern abbildet.
Wir untersuchen außerdem, wie dieses Schema optimal auf verfügbarer Quantenhardware implementiert werden kann, und stellen in Simulationen fest, dass es robust gegenüber realistischen Niveaus von Hardware-Rauschen ist.
In einer davon unabhängigen, aber thematisch verwandten zweiten Studie stellen wir einen auf maschinellem Quantenlernen basierenden Ansatz vor, der erfolgreich zur Unterscheidung verschiedener Phasen in zuvor präparierten Quantenzuständen eingesetzt wird.

\end{otherlanguage}

%% file: publications.tex
\chapter*{Publications}
\markboth{Publications}{Publications}
\addcontentsline{toc}{chapter}{Publications}

The following publications are part of and discussed within this thesis.
\begin{itemize}
    \item \textbf{Alexander Miessen}*, P. Ollitrault*, F. Tacchino*, I. Tavernelli (*contributed equally), Quantum algorithms for quantum dynamics, Nature Computational Science \textbf{3} (1), 25 (2023), see Ref.~\citenum{Miessen2023perspective}
    \item \textbf{Alexander Miessen}, D. Egger, I. Tavernelli, G. Mazzola, Benchmarking digital quantum simulations above hundreds of qubits using quantum critical dynamics, PRX Quantum \textbf{5} (4), 040320 (2024), see Ref.~\citenum{Miessen2024benchmarking}
    \item M.M. Denner, \textbf{Alexander Miessen}, H. Yan, I. Tavernelli, T. Neupert, E. Demler, Y. Wang, A hybrid quantum-classical method for electron-phonon systems, Communications Physics \textbf{6} (1), 233 (2023), see Ref.~\citenum{Denner2023phonon}
    \item L. Nagano, \textbf{Alexander Miessen}, T. Onodera, F. Tacchino, I. Tavernelli, K. Terashi, Quantum data learning for quantum field theories, Physical Review Research \textbf{5} (4), 043250 (2023), see Ref.~\citenum{Nagano2023datalearning}
\end{itemize}

Additional publications of mine from before the time of my thesis and not covered in this thesis are the following.
\begin{itemize}
    \item \textbf{Alexander Miessen}, P. Ollitrault, I. Tavernelli, Quantum algorithms for quantum dynamics: a performance study on the spin-boson model, Physical Review Research \textbf{3} (4), 043212 (2021), see Ref.~\citenum{Miessen2021quantum}
    \item P. Ollitrault*, \textbf{Alexander Miessen}*, I. Tavernelli (*contributed equally), Molecular quantum dynamics: A quantum computing perspective, Accounts of Chemical Research \textbf{54} (23), 4229 (2021), see Ref.~\citenum{Ollitrault2021molecular}
    \item P. Ollitrault, S. Jandura, \textbf{Alexander Miessen}, I. Burghardt, R. Martinazzo, F. Tacchino, I. Tavernelli, Quantum algorithms for grid-based variational time evolution, Quantum \textbf{7}, 1139 (2023), see Ref.~\citenum{Ollitrault2023quantumalgorithms}
\end{itemize}



%% file: acknowledgements.tex
\chapter*{Acknowledgements}
\markboth{Acknowledgements}{Acknowledgements}
\addcontentsline{toc}{chapter}{Acknowledgements}

Over the last five and something years of working at IBM Research -- Zürich, and the last three and something years of working on my PhD, I have had the pleasure to work with, be supervised by, meet, and befriend many really great people whose presence in my professional and personal life I very much valued and value.
I want to thank each and every one of them for having made my time as a student and researcher so full of experiences and learnings.

First and foremost, I am immensely grateful to have been supervised by two such kind, experienced, and knowledgeable scientists, Guglielmo Mazzola and Ivano Tavernelli.

Guglielmo, my professor, you have been not only a supervisor and advisor, but a friend.
Your scientific intuition and experience guided me throughout this thesis.
Thank you for never hesitating to encourage me in finding my own path, for always trusting in my abilities, and for your unsolicited and uplifting life advice.
I cannot stress enough how much I value your support and advice, and how important it was for my personal and professional development.

Thank you, Ivano, my supervisor at IBM, for creating an open and friendly atmosphere free of pressure, and for always having an open door.
Thank you for always trying to achieve the best for me and all of your students.
And thank you for having my back when it mattered.

I would also like to thank Prof. Titus Neupert for serving as third committee member, as well as Prof. Giuseppe Carleo for agreeing to review my thesis.

Furthermore, I would like to thank Francesco Tacchino, my informal third supervisor.
It has been so valuable to have you as a colleague and co-supervisor.
Thank you for your ever-open door and the countless advice and support over the years.

Even though not as my supervisor, I would also like to thank you, Stefan, for being a great and fair manager.

As already mentioned above, the last couple of years would not have been the same without my many amazing colleagues at IBM, who became friends over the years.
I could not imagine a better office mate (and dialect teacher) than you, Julian.
Thank you for always being calm and reflected, for sharing the same values, for bringing me sweets that you don't eat, and for all the sweet memories over the years.
Thank you, Anthony, Elena, and Samuele, for the many laughs.
Thank you, Julien and Almu, for introducing me to rock climbing.
Thank you, David, for always geeking out over everything cycling-related with me.
Thanks to all the other quantum kids and especially to Alberto, Anastasia, Christa, Daniel, Laurin, Luciano, Marc, Max, Sabina, and Tim (and Leo, even though not technically a quantum kid).
And thank you, Pauline, for your support back in the day.
It is thanks to all of you that our group is so unique.
You create this warm, welcoming, and friendly atmosphere that I valued so much in the last five years.

I would also like to express my gratitude toward my former high school teachers, Klaus Mulorz and Thorsten Knops.
To this day, I am grateful that you did not give up on me and that later on, you shared your passion for science with me.

I am forever grateful to my mom, dad, and my sister.
Danke für alles, für eure Liebe, eure Geduld, eure Unterstützung, eure vielen Opfer und euren fortwährenden Glauben an mich.

The same holds for my other family, who are my roommates, Anna-Julia, Aline, Jan, Jelena, Luca, Marine, Maxence, and Valerian. You are home to me.

Finally, I would like to thank all of my other dear friends that have made life outside of work so full of love and happy memories on and off the bike, which I would not want to have missed.
Thank you, Etienne, Josephine, Jacob, Jimmy, Nathalie, Felix, Patrick, and Franz.
Lastly, thank you, Christoph, for being my friend and greatest advisor in life, near or far.

\bigskip
 
\hfill Alex

%% file: ch1_introduction.tex
\cleardoublepage

\chapter{Introduction}
\label{chap: intro}

For many years, compute power has approximately doubled every year, mainly owed to an ever-decreasing size of transistors~\cite{Roser2023moore, Campbell2023computer}.
The anticipated end of this trend -- termed Moore's law -- will also limit the computational complexity of what we are able to simulate.
Performing computer simulations of complex processes and solving problems computationally is critical for many industries and areas of research, for example in weather forecasts or materials research.

Some of the most fundamental and challenging problems to simulate are of quantum mechanical nature.
Examples of such problems include, but are not limited to, drug discovery, the search for more efficient batteries, room-temperature superconductors, catalysis, or more energy-efficient fertilizers~\cite{Sliwoski2014computationalDrugDiscovery, Santagati2024drugDesignQC, Reiher2017femoco, Vogiatzis2019catalysis, Bauer2020qcMaterialScienceReview, Miessen2023perspective}.
As a result, a plethora of specialized approximate methods have been developed in the last decades that have been hugely successful in tackling problems with increasing accuracy.
Among the most prominent are \glspl{tn}~\cite{Orus2019natRevPhysTNs, Cirac2021matrix}, \gls{mc} methods~\cite{Foulkes2001quantumMC, Gull2011impurityMC, Becca2017quantumMC}, \gls{dft}~\cite{Burke2012perspectiveDFT, Jones2015revModPhysDFT}, and \gls{dmft}~\cite{Georges1996dmft}.
However, although very powerful in specific cases, these methods cease to be efficient in certain cases of interest~\cite{Troyer2005, Burke2012perspectiveDFT, Jones2015revModPhysDFT, Mazzola2024perspectiveQCMC}.
For example, \glspl{tn} are extremely efficient in describing low-dimensional, low-correlation quantum states, but are difficult to generalize to higher dimensions and become inaccurate in approximating highly entangled states or long-time dynamics~\cite{Schuch2008entropyGrowthTN, Cirac2021matrix} (see also \Cref{chap: perspective}).
And while many problems in quantum mechanics are essentially solved thanks to such highly accurate approximate algorithms, there remains a variety of important problems that are still unsolved or solved only to poor accuracy or for small system sizes~\cite{Manousakis1992heisenbergSpinOneHalf, Norman2016spinLiquidsRMP, Wu2024variationalBenchmarks, Reiher2017femoco, Goings2022cytochrome}.
This is particularly evident for simulating time-dependent phenomena~\cite{Miessen2023perspective}.
Despite further development and the constantly increasing efficiency of the above methods, they cannot fully meet this need for more accuracy.
In fact, simulating quantum mechanics exactly is exponentially expensive and hence infeasible, since the wavefunction, which describes the state of a quantum system, is a $2^N$-dimensional object and $N$ is the number of degrees of freedom in the system.
This calls for an entirely new way to compute.
Richard Feynman first posed the idea of simulating quantum mechanics using a manipulable and itself inherently quantum mechanical system in 1982~\cite{Feynman1982}.

Since then, the idea of a quantum computer has taken shape, with first proof-of-principle experiments in the 1990s, which demonstrated that a quantum computer can indeed be realized~\cite{Chuang1998grover, Weinstein2001qft}.
Today, there is considerable interest -- and even hype -- surrounding quantum computing, with many academics as well as commercial industries hoping it could evolve into a new computational paradigm for highly complex tasks.
This is substantiated by the rapid progress of quantum computing research over recent years in terms of both hardware and quantum algorithmic developments.

In fact, quantum computers have become so capable that a healthy competition between quantum and classical\footnote{
We use the term ``classical'' to refer to anything that is ``non-quantum'', such as conventional, or classical, computers, but also problems that are not of quantum mechanical nature.
Classical algorithms are methods developed for classical computers.
} methods has emerged in recent years.
This competition is created and fueled by the fact that modern quantum computers have reached sizes and error rates that make it challenging for classical methods to reproduce growingly complex quantum simulations.
On one hand, these advances enable increasingly large experiments with simulations that cannot be verified by exact classical algorithms~\cite{Scholl2021rydberg, Ebadi2021quantumPhasesAnalog, Ebadi2022optimizationRydberg, Kim2023utility, Bluvstein2023logicalNeutralAtomsQuera, Google2024dissipativeEngineering, Miessen2024benchmarking, RobledoMoreno2024sqd, Andersen2025analogDigitalKZ, King2025dwaveSupremacy, Haghshenas2025quantinuumDigital, Farrell2025digitalQuantumSimulationScattering}.
On the other hand, approximate classical methods can efficiently reproduce the results of today's largest quantum simulations when tailored to the specifics of the problem at hand~\cite{Tindall20244reproducingUtility, Begusic2024reproducingUtility, Kechedzhi2024reproducingUtility, Patra2024reproducingUtility, Anand2023reproducingUtility, Tindall2025reproducingDwaveSupremacy}.
This boundary between quantum and classical methods will likely continue to blur in the coming years, since classical algorithms will also be further developed and optimized as quantum computers mature~\cite{Daley2022practicalAdvantageAnalog, Miessen2023perspective, Mazzola2024perspectiveQCMC}.
For now, however, even state-of-the-art quantum hardware is still far from a mature product that could be employed to answer unsolved questions of academic or commercial relevance.

As the hardware remains experimental, different viable options for realizing a quantum computer and several technology platforms are being explored and currently competing.
This concerns the specific way of how to engineer a qubit, the fundamental compute unit of a quantum computer, but also the more fundamental question of digital versus analog simulation~\cite{Daley2022practicalAdvantageAnalog}.
While this thesis is concerned solely with digital quantum computation, \Cref{sec: intro hardware} will address the differences between different hardware platforms, including digital and analog platforms, in more detail.
Another major question is which type of application to focus on for achieving a practical quantum advantage.
Since Feynman's original proposal to simulate nature using a quantum computer, the field has evolved to pursue a quantum computational advantage over existing classical methods across a much broader range of computational problems, both of quantum mechanical and classical nature~\cite{Shor1994, Grover1996, Reiher2017femoco, Miessen2023perspective, Abbas2024optimizationWhitePaper}.
We will provide a high-level overview of potential applications of quantum computers in \Cref{sec: intro applications}.
Although not yet demonstrated experimentally, such a quantum advantage has been identified in theory for a few applications.
Those include Shor's algorithm~\cite{Shor1994}, leading to an exponential runtime advantage in prime factorization over the best known classical algorithm, and Grover's search~\cite{Grover1996}, achieving a quadratic speed-up in unstructured search problems\footnote{Note that these runtime advantages are only with respect to known classical algorithms but do not exclude the existence of more efficient, unknown classical algorithms}.
It is further believed that applications within quantum physics and quantum chemistry could generally benefit from a quantum computational treatment due to their natural mapping to a quantum computer and the exponential cost associated with their classical treatment.
However, identifying a specific problem that is well-suited to be simulated with a quantum machine is not trivial.
Moreover, not all areas of application are equally likely to benefit from quantum computing, even in the case of a perfect, noiseless machine.

This thesis aims to address some of these open questions with a particular focus on quantum dynamics within the regime of intermediate scale, noisy quantum computers.
Knowing which problems to solve using quantum computers once they reach a certain level of maturity is hugely important.
The one area of application that is least disputed and seems most promising to benefit from quantum computing remains the simulation of time-dependent phenomena within the natural sciences, as envisioned by Feynman.
For these problems, the terms quantum dynamics, quantum simulation or Hamiltonian simulation are used interchangeably.
While there are many time-dependent problems that could benefit from quantum simulation, it is not entirely clear which one could lead to a quantum advantage first.
Similarly, among the plethora of quantum algorithms for quantum dynamics, it is essential to understand which ones are best suited for simulations on near-term devices.
Moreover, time evolution appears not only in quantum simulation of quantum systems but as a subroutine in many other relevant algorithms~\cite{Shor1994, Harrow2009HHL, Layden2023quantumMCMC}.
An overview of the field of quantum simulation and quantum algorithms to implement time evolution is therefore essential to better assess which combination of algorithms and target problem could most benefit from quantum computing in the near-term.
\Cref{sec: time evo algorithms,chap: perspective} attempt to provide such an overview.

Aside from knowing which problems to simulate with available quantum devices and an adequate choice of algorithm, the ability to compare simulation capabilities across platforms, device upgrades, and different algorithms is just as essential given the fast-paced developments of quantum hardware and algorithms alluded to above~\cite{Amico2023benchmarkingBestPractices}.
Moreover, with processors so large that they can hardly or not at all be verified exactly with classical methods, it becomes more and more challenging to assess how much accuracy can be expected in a given simulation~\cite{Mckay2023layerFidelity}.
Or, in other words, how much error can be accumulated before a simulation outcome becomes too corrupted?
In essence, this means scalable, reliable, and intuitive benchmarks are crucial.
In \Cref{chap: benchmarking}, we present a benchmarking method~\cite{Miessen2024benchmarking} based on quantum dynamics and condensed matter theory, which satisfies these requirements.
Crucially, the method requires no classical verification, meaning it can be scaled to arbitrary system sizes, and provides an intuitive quality metric that can be used as a predictive tool for simulating a range of applications.
We demonstrate the transferability of results from our benchmark to other applications in \Cref{sec: optimization}, specifically to combinatorial optimization, which constitutes another major area of research for applications of quantum computers, the solution of which can be addressed via quantum dynamics.
The scalability of our benchmarking method ensures its practicality beyond existing and near-term quantum devices.
This is particularly desirable in light of increasing efforts to build larger and less noisy devices, indispensable to reach the ultimate goal of fault-tolerance.

In fault-tolerant quantum computation, errors might still occur, but are detected and -- importantly -- corrected in real time~\cite{Folwer2012surfaceCode, Campbell2017reviewFT}.
This necessitates far larger devices and lower error rates than currently available (see \Cref{sec: intro hardware}).
Despite rapid hardware developments, however, the hardware requirements to reach fault-tolerance present tremendous technological hurdles, which is why we will likely not see a fault-tolerant, error-corrected quantum computer before the next decade.
Therefore, it remains crucial to devise methods that work around the noise inherent to current-day and near-term devices.
Typically, this means to mitigate as much of the noise and resulting errors as possible~\cite{Cai2023em}.
For specific applications, however, it could prove advantageous to instead harness the hardware noise and incorporate it in the simulation~\cite{Guimaraes2023partialPEC}.
This is of particular interest for the simulation of open quantum systems~\cite{Breuer2002oqs}, where a system interacts with an environment, which often makes them even more challenging to simulate than closed quantum systems.
In \Cref{chap: pea oqs}, we propose a method to manipulate the device noise in a hybrid digital-analog scheme, such that it models the environment of a given open quantum system.

In addition to benchmarking devices to quantify their simulations capabilities and limitations, as well as devising efficient methods to work with and around the noise inherent to available devices, it is important to remain mindful of limitations resulting from noise.
In fact, many interesting and relevant problems remain out of reach as their resource requirements are too extensive~\cite{Reiher2017femoco}.
Splitting some of the computational workload between a classical and a quantum processor using variational methods could help alleviate some of those limitations~\cite{Cerezo2020variational}.
This is particularly well-suited for problems where a part of the problem requires highly accurate treatment, for example, due to strong correlation, while other parts can be efficiently treated approximately.
We present such a scheme in \Cref{sec: elph} for solving ground states across multiple phases of highly correlated electronic systems interacting with an ionic lattice, together with a feasibility study of a hardware implementation.
In a related work, \Cref{sec: qcnn} presents a framework for classifying phases of previously prepared quantum states.

\section{Quantum computing platforms}
\label{sec: intro hardware}

Fundamentally, a quantum computer is a controllable and manipulable quantum system consisting of many degrees of freedom, which can encode a computational problem.
Typically, these degrees of freedom are qubits, quantum mechanical objects with a two-dimensional state space, though higher-dimensional qudits are in principle conceivable~\cite{Fischer2023quditGates}.
The collective state of its qubits is then manipulated through controlled operations.
Depending on how these unitary transformations are implemented, two kinds of quantum computers can be distinguished -- analog and digital quantum computers~\cite{Daley2022practicalAdvantageAnalog}.
Moreover, each of the two can be realized based on different hardware platforms.
Our main focus in this thesis is digital quantum computation.
Nonetheless, it is important to understand analog simulators and how they compare to digital machines, particularly so when concerned with quantum simulation, i.e., studying time-dependent phenomena, as will become clear in the following.
Here, we will, on a high level, discuss the working principles of both platforms, their main advantages and disadvantages, and their error sources.

Digital quantum computing relies on gate-based computation through a universal gate set that constitutes the machine's most fundamental operations~\cite{NielsenChuang2010}.
Any calculation done on a digital quantum device is carried out through a quantum circuit as a combination of basis gates.
Universality implies that, in principle, any arbitrary operation can be constructed from only basis gates~\cite{NielsenChuang2010}.
In practice, however, doing so is infeasible for arbitrary unitary operations spanning more than a few qubits, since finding the corresponding basis gate decomposition is in most cases highly non-trivial~\cite{Iten2016isometryDecomposition, Ollitrault2020nonadiabatic}.
This means every problem, every practical algorithm, needs to be formulated in terms of a discrete and finite set of operations that can be efficiently translated into a finite set of basis gates.
For example, contrary to analog simulation (see below), complicated unitary operations such as the time evolution operator of a many-body quantum system cannot be implemented exactly.
Instead, they need to be approximated with simpler operations that can be easily implemented using basis gates (see also \Cref{chap: methods} for a more detailed discussion).

One of the greatest strengths of digital platforms is the systematic control over errors.
Discretization errors are generally inevitable due to the mentioned need to decompose complicated operations into basis gates.
In this regard, the digital approach resembles classical algorithms\footnote{Note that modern classical computers are largely digital as well, after widely replacing classical analog computers, although research on classical analog computers is re-emerging~\cite{Schuman2022neuromorphicComputing}.}.
At the same time, however, this enables one to systematically analyze and reduce errors\footnote{Note that heuristic algorithms reduce the possibility for this.}~\cite{Miessen2023perspective}.
Moreover, digital architectures allow for a wide range of error mitigation techniques to be incorporated in the calculation to remove some of the errors affecting a simulation~\cite{Endo2018errorMitigation}.
Many error mitigation methods rely on inserting specific gates into the circuit, for example, to counter, transform, or amplify specific noise.
This is made possible precisely because of the computer's gate-based design and the way that errors occur through gates (see \Cref{sec: error mitigation} for an overview of common error mitigation techniques).
As a result, gate-based computation could imply a more costly computation, as algorithmic errors might need to be compensated with more computational resources.
However, it also implies programmability of the device, meaning the algorithm ultimately determines how a problem is solved, which is in stark contrast to the purpose-driven build of analog devices, as will be discussed below.

A further quintessential benefit of the digital paradigm is the possibility of implementing error correction.
As mentioned previously, to become truly universal, the ultimate goal for digital quantum computing is to reach fault-tolerance~\cite{Campbell2017reviewFT}.
This is only possible through the implementation of error-correction algorithms~\cite{Folwer2012surfaceCode, Gidney2021factoringFT, Google2023surfaceCode, Bravyi2024LDPC, Xu2024neutralAtomsLDPC}, which, in a nutshell, work by encoding information redundantly in many physical qubits to form what is called a logical qubit.
By performing regular checks on the state of all physical qubits during a computation, errors can be detected and corrected on the fly to keep their average state, i.e., the logical qubit state, intact.
Consequentially, error correction implies drastic overheads in both qubit numbers and quantum gates needed to perform a computation~\cite{Folwer2012surfaceCode, Reiher2017femoco, Gidney2021factoringFT, Bravyi2024LDPC}.
Crucially, however, implementing such algorithms is only possible with gate-based computation.
Due to the implied gigantic resource overheads, if and when large-scale error correction, or even full fault-tolerance, might be achieved will depend greatly on future hardware and algorithmic developments.
Nonetheless, the prospect of realizing a fully universal quantum computer remains motivational.

In the meantime, digital quantum computation remains affected by noise~\cite{Miessen2024benchmarking}.
Noise can be broadly separated into errors in the qubit state itself and errors occurring when manipulating the qubit state, e.g., during state preparation, executing gates, and measuring.
Importantly, in a digital computation, errors accumulate with every gate executed.
The analysis of how errors impact digital quantum simulation outcomes is one of the main focuses of this thesis (see also \Cref{chap: benchmarking} and \Cref{sec: benchmarking noisy simulations} in particular, as well as \Cref{chap: pea oqs,sec: elph}).
Qubits are mostly affected by an insufficient isolation from their environment.
Their state gets corrupted over time as they are subject to thermal fluctuations despite cooling the quantum processor and shielding it from external influences as much as possible.
This causes decoherence of the qubit, i.e., loss of superposition, decay of the state, bit and phase flip errors, as well as leakage of the qubit state outside of its computational space.
Another type of error, originating from faulty gates, is often the more severe one.
Sources of noise that affect gate fidelities, particularly those of two-qubit entangling gates, are numerous.
Two of the most dominant sources of error are crosstalk between neighboring qubits, where driving one qubit unintentionally affects the other, and miscalibrations of gates (gates are implemented via physical manipulation of the qubit, such as microwave pulses, that need to be calibrated to implement the correct operation).
Lastly, also state preparation and measurement errors are ever present, i.e., errors in preparing the desired initial qubit state and in reading out the final qubit state.
While they affect a computation independently, they can hardly be separated in noise characterization and benchmarking experiments.
The magnitude of the respective error channel, i.e., which errors dominate in a computation, and their physical origin might differ depending on the physical realization of the qubit.

Digital quantum computers can be realized using different hardware architectures.
Most prominently, these are superconducting qubits~\cite{Kjaergaard2020superconductingQubitsReview}, trapped ions~\cite{Monroe2021trappedIonsRMP}, and neutral (Rydberg) atoms~\cite{Bluvstein2023logicalNeutralAtomsQuera}, each of which have their respective up- and downsides.
Trapped ions and neutral atoms pose few limitations in terms of qubit connectivity due to the ability to physically move qubits, and have large coherence times in the order of seconds or more~\cite{Moses2023quantinuum}.
While trapped ions achieve some of the lowest gate-error rates (in the order of $\qty{0.01}{\%}$)~\cite{Loschnauer2024highFidelityIons}, gate execution times are very slow (in the order of \qty{}{\us} to \qty{}{\ms}) and they are more difficult to scale to larger qubit numbers.
Neutral atoms are relatively easy to scale to large qubit numbers with record systems controlling up to 280 qubits~\cite{Bluvstein2023logicalNeutralAtomsQuera}.
However, they are likewise slow to operate, with gate and reset times in the order of \qty{}{\us} to \qty{}{\ms}, and two-qubit gate errors in the order of $\qty{0.1}{\%}$~\cite{Evered2023highFidelityNeutralAtoms}.

In this thesis, however, we are solely concerned with superconducting qubits.
They offer extremely fast gate operations and readout times, in the order of a few \qty{}{\ns} and \qty{}{\us}, respectively, making them very fast to operate~\cite{IBMQuantumPlatform, Abughanem2025ibmHardware}.
This is a considerable advantage, especially for digital platforms, both in the near- and in the long term, when extensive error mitigation and, eventually, error correction, necessitate large overheads in circuit repetitions and gate counts.
On the downside, superconducting qubits suffer from shorter coherence times (in the order of $\qty{100}{\us}$ to $\qty{300}{\us}$) and relatively high two-qubit gate as well as readout error rates (in the order of $\qty{0.1}{\%}$ and $\qty{1}{\%}$, respectively).
Moreover, they are far more difficult to scale to larger qubit numbers and higher connectivities, as qubits are fixed and physically connected through wires on a chip.
More specific metrics of the respective hardware platform may be found in the references given above.

Analog quantum computers, on the other hand, do not implement unitary operations through discrete quantum gates, but based on continuous changes of the system's qubit interactions.
Concretely, they rely on fine-tuning qubit interactions such that they mimic a given computational problem of interest~\cite{Altmann2021quantumSimulators, Daley2023twentyFiveYearsAnalog, Daley2022practicalAdvantageAnalog}.
In turn, this also means analog simulators are typically designed to simulate only specific classes of problems, whose interactions they can natively implement.
For example, analog quantum annealers of the type built by D-Wave natively implement an Ising-type interaction~\cite{King2022dwave2000IsingChain}, meaning native qubit-interactions of the annealer can be fine-tuned to any Ising-type interaction, but no other kind of interaction (see also \Cref{chap: benchmarking} for details on the Ising model and quantum annealing).

Analog platforms are particularly well-suited for quantum dynamics simulations as the system itself undergoes a time evolution.
In turn, they can simulate the time evolution of specific problems very efficiently and, importantly, without algorithmic errors stemming from a discretized time evolution.
Moreover, depending on the platform, they face fewer operational hurdles than digital devices.
First, certain analog architectures are more easily scalable, reaching over 5000 qubits in analog quantum annealers~\cite{King2023spinGlass} and over 250 qubits in the case of neutral atom simulators\footnote{Note that the scalability advantages of neutral atoms have recently been demonstrated to carry over to digital quantum computing with impressive demonstrations of digital simulation using up to 280 neutral atoms~\cite{Bluvstein2023logicalNeutralAtomsQuera}.}~\cite{Ebadi2022optimizationRydberg}.
This is also due to the fact that they require fewer control operations with high fidelity.
For example, analog quantum annealers only need to implement time evolution under a transverse-field Ising model.
As a result, some of the largest and most impressive experiments of recent years in the realm of quantum simulation have been achieved on analog quantum simulators~\cite{Scholl2021rydberg, Ebadi2021quantumPhasesAnalog, Ebadi2022optimizationRydberg, Daley2023twentyFiveYearsAnalog, King2025dwaveSupremacy}.
Results typically fall in the domain of many-body quantum physics, for example, out-of-equilibrium dynamics~\cite{Choi2016localizationRydberg} and quantum chaos~\cite{Bernien2017probing, Turner2018ergodicityBreaking, Bluvstein2021manyBosyScarsRydberg}, or novel phases of matter~\cite{Semeghini2021spinLiquidRydberg}, routinely challenging classical state-of-the-art simulation methods~\cite{ORourke2023peps}.
Note, however, that many qubits do not automatically imply quantum advantage, as coherent control of these systems long enough to reach regimes that are impossible to simulate using approximate classical methods has not yet been shown~\cite{ORourke2023peps, Tindall2025reproducingDwaveSupremacy}.

Analog devices suffer from similar errors, leading to a corruption of qubit states as digital devices.
However, rather than gate errors, the most dominant source of error in analog devices is calibration errors in tuning up qubit interactions~\cite{Daley2022practicalAdvantageAnalog}.
In order to achieve the level of control over qubit interactions necessary to implement the target problem to good accuracy, all interactions in the system must first be well-understood and, in a second step, calibrated.
Furthermore, high-precision control of the system is difficult, and systematic errors are hard to diagnose.
Just like gate errors in digital devices, these errors are inevitable.
However, unlike in digital platforms, there is far less room for error mitigation in analog simulation~\cite{Amin2023dwaveZNE} (see \Cref{sec: error mitigation}).
Moreover, as digital quantum computers evolve slowly but steadily toward an error-corrected future with the hope of one day achieving full fault-tolerance, there exists no such prospect for analog devices.
Fault-tolerance and error correction require universal computation, which cannot be achieved using common analog devices due to their purpose-targeted design\footnote{Note that theoretical proposals to achieve universal quantum computation through other modes of computation exist, specifically based on adiabatic quantum computation~\cite{Aharonov2008adiabaticQC}.}.
This represents the central disadvantage of analog devices with respect to digital platforms.
Therefore, despite their effectiveness in emulating quantum dynamics, the scope of problems they can solve and algorithms they can implement, as well as their room for development, is naturally limited.

Lastly, hybrid analog-digital computers could offer an alternative to combine the benefits of both platforms~\cite{Bluvstein2022hybridDigitalAnalog}.
Recent experiments have shown impressive results, combining digital and analog control within one simulation~\cite{Andersen2025analogDigitalKZ}.
This is of interest as current and near-term digital devices appear to be somewhere between analog and digital machines, being gate-based but plagued by high levels of \textit{analog} noise.
Beyond the demonstration of Ref.~\citenum{Andersen2025analogDigitalKZ}, employing this noise within the simulation could benefit certain simulations greatly.
We propose such a method for the simulation of open quantum systems in \Cref{chap: pea oqs}.

In summary, current digital platforms might be more affected by noise and, therefore, scaling to larger system sizes and simulation times.
However, in the long term, as hardware improvements will reduce error rates and increase qubit quality, they are likely to surpass analog simulators.
This is due to their programmability, systematic control over errors, and the possibility of realizing error correction.
Quantum computers could follow a similar evolution as conventional computers that evolved into the nowadays purely digital machines.
Nonetheless, the enormous overheads necessary to reach fault-tolerance should not be underestimated and pose a significant hurdle to reaching true universality.
Analog simulators, on the other hand, offer no systematic control over errors and are limited when it comes to the mitigation of errors.
More importantly, they are incompatible with error-correction codes and, consequently, fault-tolerance, naturally disadvantaging them on paper.
Yet, they are appealing for their efficiency in solving specific tasks, such as simulating many-body quantum systems and quantum dynamics in particular.
It is therefore conceivable that they might outperform digital platforms in the near term and could lead to an early -- even if not universal -- quantum advantage.
Furthermore, as we discuss in the next section, we should not forget that, for many potential applications of quantum computation, it remains unclear whether they will actually benefit from quantum computing or whether classical methods will keep the upper hand.
If, for example, it becomes clear that quantum dynamics and related applications could be the only ones where quantum computation could yield a truly practical advantage, analog simulators with their specialized design might be more appropriately suited for the task.
Even more so, considering the large obstacles that need to be overcome to reach full fault-tolerance.

\section{Quantum simulation and other applications of quantum computers}
\label{sec: intro applications}

Ultimately, quantum computing is poised to act as a special-purpose tool, possibly embedded into classical high-performance computing centers~\cite{RobledoMoreno2024sqd, Alexeev2024materialsWhitePaper}, solving very specific problems with exceptionally high computational complexity.
As alluded to before, simulating quantum dynamics was the first use case envisioned for quantum computers.
Certainly, quantum dynamics remains not only the prime and most natural application of quantum computers.
It is also the area of application that is most likely to lead to a practical quantum computational advantage (see also \Cref{chap: perspective}).
Today, however, quantum computing applications research has broadened significantly to include a whole host of computational problems that may or may not benefit from the technology.
These include not only problems of quantum mechanical nature, but also purely classical problems.
Here, we provide a high-level overview of them.

As we describe in detail in \Cref{chap: perspective}, quantum simulation, i.e., simulating quantum dynamics, spans nearly all areas of physics that involve quantum mechanics, such as condensed matter physics, quantum chemistry, and high-energy physics~\cite{Miessen2023perspective}.
Even though the anticipated timelines for gaining an edge over classical methods differ between these research areas, quantum dynamics simulations will almost certainly benefit from quantum computation in every domain.
For more details, we refer the reader to \Cref{chap: perspective}.

Another area of application that seems rather native for quantum computing is the solution of time-\textit{in}dependent quantum mechanical problems.
This includes all of the above fields in which quantum dynamics plays a role.
For example, understanding and mapping out ground state properties of condensed matter or, more generally, many-body quantum systems, in different parameter regimes.
Such systems often exhibit rich phase diagrams, where a ground state could be, for example, superconducting or not, depending on certain system parameters.
Understanding these relationships is crucial in the study of materials~\cite{Bauer2020qcMaterialScienceReview}.
\Cref{sec: elph} introduces a hybrid quantum-classical algorithm to study many-body quantum systems that aim to realistically model interactions of electrons and the lattice they move in in a material~\cite{Denner2023phonon}.
The model has several phases, and our algorithm is able to faithfully reproduce ground state properties across all phases.
Subsequently, \Cref{sec: qcnn} proposes a method based on \gls{qml} to classify phases of states previously generated in quantum simulation~\cite{Nagano2023datalearning}.

As a second example, in quantum chemistry, the knowledge of excited states, in particular energy differences between eigenstates of the system, is central to understanding molecular properties such as binding affinities~\cite{Ryde2016bindingAffinitiesQM}, but also in molecular dynamics~\cite{Ollitrault2021molecular}.
Excited state calculation could present an interesting middle ground application between dynamics and time-independent phenomena, as they can be computed through both time-independent and time-dependent simulation (via dipole perturbation)~\cite{Li2020electronicStructure, Tazhigulov2022correlatedMolecules}.
We explicitly discuss excited state calculation as a suitable application for practical quantum advantage in \Cref{sec: perpective - quantum chemistry}.
Particularly so since, in recent years, focus has slightly shifted away from ground state calculations and more towards quantum dynamics.
Part of the reason might be the realization that a possible quantum advantage in solving ground states might be more difficult to achieve and less obvious to identify than with quantum dynamics~\cite{Lee2023evidenceQuantumAdvantage}.
This is because variational methods~\cite{Cerezo2020variational} come with their own drawbacks and overheads~\cite{Miessen2021quantum, Larocca2025barrenVQA, Scriva2024qaoa} and fault-tolerant methods based on \gls{qpe} necessitate initializing states sufficiently close to the unknown ground state, again inducing computational overheads.
Nonetheless, the study of ground and excited states remains one of the most actively researched applications of quantum hardware~\cite{Tazhigulov2022correlatedMolecules, RobledoMoreno2024sqd, Alexeev2024materialsWhitePaper, Barison2025sqdExcitedStates}.

Beyond quantum mechanical problems, there is a considerable effort to utilize quantum technologies to solve purely classical problems, i.e., problems whose solution is a single basis state of the solution space instead not a superposition of many.
Most prominently, this includes classical, typically combinatorial, optimization problems~\cite{Abbas2024optimizationWhitePaper}.
Although the solution to these problems is a purely classical, single bitstring, the space of possible solutions grows exponentially with system size, just like the dimension of a quantum mechanical problem.
This makes solving many optimization problems challenging for classical algorithms.
The hope in quantum optimization is that quantum computers may exploit quantum effects like superposition, entanglement, and interference in their exponentially large state space to arrive at either better solutions or approximate solutions faster than classical methods.
In practice, however, a speed-up is difficult to formalize~\cite{Abbas2024optimizationWhitePaper}.
On one hand, although in theory quadratic speed-ups based on Grover's search~\cite{Grover1996} can be formalized for specific problems when compared to classical brute-force methods\footnote{Note that classical brute-force methods rely on searching the entire solution space, hence scaling exponentially.
Therefore, this quadratic speed-up still results in exponential scaling, albeit at a reduced rate.}, this speed-up diminishes when compared to state-of-the-art classical algorithms.
On the other hand, many other approaches to quantum optimization are heuristic, for which rigorous performance guarantees are not attainable.
See also Ref.~\citenum{Abbas2024optimizationWhitePaper} for a comprehensive review.
We explore this domain in \Cref{sec: optimization}, where we study a quantum optimization algorithm based on quantum dynamics and discuss how to best use it in conjunction with current noisy hardware.

Lastly, \gls{qml} constitutes another potential application for quantum computers that, for a long time, received considerable interest.
Possibly owed to the general attention surrounding both quantum computing and machine learning, and the immense computational demands of the latter, proposals to combine the two were only consequential.
Early demonstrations of \gls{qml} models to learn and classify data~\cite{Havlivcek2019quantumFeatureMaps} were followed by proofs of quantum speed-up for artificial problems without practical relevance~\cite{Liu2021speedupQML}.
Recently, however, evidence has accumulated that -- at least variational -- \gls{qml} models are in fact efficiently classically simulable~\cite{Thanasilp2023subtletiesTrainability, Thanasilp2024exponentialConcentrationKernal, Cerezo2024barrenAbsenceClassical}.
Moreover, even if theoretical evidence for a practical quantum advantage in \gls{qml} was to be found, it would likely not be realizable for years if not decades, with classical models already featuring hundreds of billions of parameters~\cite{Brown2020gpt3}.
Slightly more promising seem \gls{qml} approaches with quantum mechanical input data.
Note that this does not mean classical data generated in a quantum process, e.g., data measured in physics experiments, but genuine quantum states, for example, generated in quantum simulation.
Here, the hope is that \gls{qml} models might achieve an advantage in processing and recognizing patterns within quantum states due to full consideration of correlations in the quantum state~\cite{Huang2022quantumLearning}.
In \Cref{sec: qcnn}, we propose such a \gls{qdl} framework based on \glspl{qcnn} to classify phases in quantum states prepared in quantum simulation~\cite{Nagano2023datalearning}.

As stated in the previous sections, the progress in hardware and algorithmic developments has been remarkable in recent years.
At the same time, the progress in finding interesting applications for these modern quantum computers has been seemingly slower.
A few holy-grail problems were identified in the last decades that would have an immense impact when solved~\cite{Shor1994, Gidney2021factoringFT, Reiher2017femoco, Goings2022cytochrome, Wu2024variationalBenchmarks}.
However, solving them would certainly require a fully error-corrected, large-scale quantum computer and will therefore not be possible for another decade or so.
Today, experiments demonstrating and pushing the capabilities of available hardware seldom extend beyond simplistic model systems~\cite{RobledoMoreno2024sqd, Farrell2025digitalQuantumSimulationScattering}.
This raises the question, what can be simulated that is of practical relevance but may not require full fault-tolerance?
Although efforts are increasing to find and study such intermediate problems~\cite{Miessen2023perspective, Wu2024variationalBenchmarks, Abbas2024optimizationWhitePaper} (see also \Cref{chap: perspective}), it seems researchers generally struggle to pin down relevant intermediate problems to show the utility of current-day or near-term quantum computers.
This will be a key interest for the community in the coming years.

\section{Outline of the thesis}

This thesis is written for a reader with a background in physics and basic knowledge in quantum mechanics.
It is self-contained to the extent that we introduce the key concepts to follow the more technical parts of the thesis.
These are introduced in \Cref{chap: methods} and expert readers may choose to skip this chapter.
For more fundamental concepts, such as the definition of basis gates, the wavefunction formalism, etc., we refer the interested reader to popular sources such as Ref.~\citenum{NielsenChuang2010,Griffiths2018quantumMechanics}.
All technical contributions to this thesis are contained in  \Cref{chap: perspective,chap: benchmarking,chap: pea oqs,chap: ground states} and briefly outlined below.
Finally, we conclude in \Cref{chap: conclusion}.

\begin{enumerate}[align=parleft, labelsep=1.4cm, leftmargin=*]
	\item[\Cref{chap: methods}]
    	provides a short introduction to quantum dynamics for both closed and open quantum systems.
    	Thereafter, based on Ref.~\citenum{Miessen2023perspective}, we give an overview and our perspective on quantum algorithms for quantum dynamics, discussing their respective advantages and drawbacks, and their feasibility to be implemented in experiments.
		The second part of this chapter introduces error mitigation and suppression techniques relevant to this thesis, many of them employed in the experiments in \Cref{chap: benchmarking}.
		This chapter may be skipped by expert readers or serve as a reference chapter.
	\item[\Cref{chap: perspective}]
		is based on Ref.~\citenum{Miessen2023perspective} as well.
		Here, we discuss potential applications of quantum computers in the domain of quantum dynamics, grouped into many-body quantum systems, open quantum systems, quantum chemistry, and lattice gauge theory.
		Specifically, we give an overview of interesting and possibly classically hard problems.
		Furthermore, we discuss how advanced quantum computing research is in each field and provide our perspective on the timeline of achieving a potential quantum advantage.
	\item[\Cref{chap: benchmarking}] is based on Ref.~\citenum{Miessen2024benchmarking}.
    	We introduce an application-oriented, intuitive benchmarking method to predict how deep a quantum circuit one can expect to execute given a specific hardware and error mitigation.
    	The method is based on digital quantum annealing and well-understood theoretical results from condensed matter physics.
    	We further show that such benchmarking results are transferrable to other applications, such as combinatorial optimization.
	\item[\Cref{chap: pea oqs}] is based on thus far unpublished work.
    	We propose a method for the simulation of open quantum dynamics that aims to utilize hardware noise to model environment interactions.
    	The method prescribes first characterizing device noise to then partially mitigate it.
		We present a rigorous error analysis of the scheme and analytically derive a bound for the inferred random error, which we validate numerically.
    	The working of the method is demonstrated by extrapolating from one to another noise model to simulate the open dynamics of a dissipative Ising model.
	\item[\Cref{chap: ground states}] is based on Ref.~\citenum{Denner2023phonon} (\Cref{sec: elph}) and Ref.~\citenum{Nagano2023datalearning} (\Cref{sec: qcnn}).
    	Although this thesis is focused mainly on quantum dynamics, here we discuss the relevance of solving time-independent problems with quantum computers.
    	Specifically, in the context of phase transitions.
    	\Cref{sec: elph} introduces a variational hybrid quantum-classical algorithm to study ground states of electron-phonon systems across various phases.
    	\Cref{sec: qcnn}, on the other hand, presents a method to classify phases of quantum states using a quantum convolutional neural network (QCNN) in the context of quantum machine learning with quantum states as input data.
	\item[\Cref{chap: conclusion}] concludes this thesis.
    	We highlight its main achievements and provide a condensed outlook on the future of quantum simulation.
\end{enumerate}

%% file: ch2_methods.tex
\chapter{Simulating quantum mechanics}
\label{chap: methods}

\vspace{-20pt}

\begin{zitat}{}
    This chapter is reproduced with permission in parts from Alexander Miessen, Pauline J. Ollitrault, Francesco Tacchino, and Ivano Tavernelli, ``Quantum algorithms for quantum dynamics'', Nature Computational Science \textbf{3}, 25 (2023)~\cite{Miessen2023perspective}.
    It introduces basic concepts of quantum mechanics, how to solve the Schrödinger equation, and quantum algorithms for simulating time-dependent quantum mechanics.
\end{zitat}

\section{Quantum dynamics in a nutshell}
\label{sec: quantum dynamics}

Quantum dynamics studies the behavior of quantum systems as they evolve in time.
This is useful to compute properties like time-dependent correlation functions, relevant, for instance, to reproduce and analyze experimental spectroscopic data.
Such systems can be isolated or embedded in an environment with which they interact.
In the former case, we speak of closed quantum systems, described by pure states, typically represented by the wavefunction $\ket{\Psi(t)}$ at time $t$ (we will use $\ket{\Psi(t)}$ and the short-hand notation $\ket{\Psi}$ interchangeably in the following).
The dynamics of the wavefunction is governed by the \gls{tdse}
\begin{equation}
    i \frac{\partial \ket{\Psi}}{\partial t} = H \ket{\Psi} \ ,
\label{eq: tdse}
\end{equation}
with $\hbar=1$ and $H$ the Hamiltonian.
Solving for $\ket{\Psi}$ gives the evolution of the quantum system as a function of time.
The solution for evolving an initial state $\ket{\Psi(t_0)}$ to time $t$ is most generally given as a unitary map 
\begin{equation}
    \ket{\Psi(t)} = U(t,t_0) \ket{\Psi(t_0)} \ .
\end{equation}
Unitarity ensures the conservation of the total probability of the state and implies reversibility of the process since the inverse $U^{-1} = U^\dagger$ is again unitary.
The unitary time evolution operator for a time-dependent Hamiltonian $H(t)$ is defined as
\begin{equation}
    U(t,t_0) = \mathcal{T} \exp \Biggl( -i \int_{t_0}^{t} \diff{t'} H(t') \Biggr) \ .
\label{eq: time evo op t-dep H}
\end{equation}
The time-ordering $\mathcal{T}$ accounts for any non-commutativity of the Hamiltonian with itself at different times, $[H(t), H(t')] \neq 0$, prescribing a specific time ordering (earlier to later times) to integration when expanding the exponential as a power series.

In the simpler case of a time-independent Hamiltonian $H$, the above expression reduces to
\begin{equation}
    \ket{\Psi(t)} = e^{-iH(t-t_0)} \ket{\Psi(t_0)} \ . 
\label{eq: time evo op t-indep H}
\end{equation}

In the case that the system is embedded in and interacting with an environment, we speak of an open quantum system~\cite{Breuer2002oqs}.
The system state is then appropriately described by a density matrix operator $\rho(t) = \sum_i p_i(t) \ket{\Psi_i(t)} \bra{\Psi_i(t)}$.
This can be interpreted as a probabilistic mixture of pure states $\ket{\Psi_i}$ with probabilities $p_i \geq 0$ and $\sum p_i = 1$.
In fact, to be a valid physical state, the density matrix must be positive semi-definite, $\rho \geq 0$, and have $\tr (\rho) = 1$.
The evolution of the density matrix is governed by the Liouville-von Neumann equation
\begin{equation}
    \frac{\partial \rho}{\partial t} = -i [ H, \rho ] + \mathcal{L}[\rho] \ .
\label{eq: von neumann}
\end{equation}
Here, $[\cdot,\cdot]$ is the commutator.
The interaction between the system and an environment is included through the super-operator $\mathcal{L}$, meaning it acts on an operator to produce another operator.
It is often assumed that the system-environment interaction is independent of previous states of the environment, or memoryless, a feature termed Markovian.
With this and a few additional assumptions~\cite{Breuer2002oqs}, the most general form of the super-operator $\mathcal{L}$ is given by the Lindbladian
\begin{equation}
    \mathcal{L}[\cdot] = \sum_k\gamma_k V_k \cdot V_k^\dagger - \frac12 \bigl\{ V_k^\dagger V_k, \cdot \bigr\} \ .
\label{eq: lindbladian}
\end{equation}
Here, $\{\cdot,\cdot\}$ is the anti-commutator, and the so-called jump operators $V_k$ mediate the system-environment interaction, turning \Cref{eq: von neumann} into the Gorini–Kossakowski–Sudarshan–Lindblad, often simply Lindblad, master equation.

Formally, the solution to \Cref{eq: von neumann} is obtained by writing the entire right-hand side as a super-operator, $\partial \rho / \partial t = \mathcal{A} [\rho]$, and integrating it,
\begin{equation}
	\rho(t) = e^{\mathcal{A}[\cdot] (t-t_0)} \rho(t_0) \ .
\label{eq: time evo op open systems}
\end{equation}
Importantly, this highlights that open quantum dynamics is non-unitary.
In general, operations $\Lambda$ that map one density matrix to another must be \gls{cptp}, which means the resulting state is again a density matrix, i.e., Hermitian, positive semi-definite, and of the same trace as the input state.
Contrary to the unitary evolution of closed systems, \gls{cptp} maps can describe irreversible processes since the inverse $\Lambda^{-1}$ is not necessarily \gls{cptp} itself.
The exponential of a super-operator in \Cref{eq: time evo op open systems} and its action are best understood when expanding the exponential as a power series.

\section{Time evolution algorithms}
\label{sec: time evo algorithms}

\begin{figure}
    \centering
    \includegraphics[width=\textwidth]{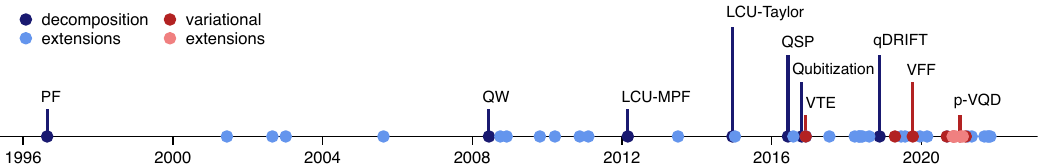}
    \mycaption{Development timeline of quantum algorithms for quantum dynamics}{
        Non-exhaustive timeline visualizing the development of decomposition and variational quantum algorithms for quantum dynamics and respective further developments.
        Each point represents the date of first appearance, typically as \texttt{arXiv} preprint.
        All corresponding references are summarized in \cref{tab: decomposition scalings}.
    }
    \label{fig: timeline algorithms}
\end{figure}

While \Cref{eq: time evo op t-indep H,eq: time evo op t-dep H} look deceivingly concise,
in most cases, they cannot be implemented directly using quantum gates native to digital quantum computers.
This is because time evolution operators span the entire system and finding a decomposition into single- and two-qubit gates is highly non-trivial, as mentioned in \Cref{sec: intro hardware}.
Therefore, the time evolution must be approximated using a suitable time evolution algorithm.
Since Lloyd's first proposal of a universal quantum algorithm for quantum dynamics~\cite{Lloyd1996}, the field of digital quantum simulation has seen immense progress.
Increasingly sophisticated protocols have been devised, achieving asymptotically optimal complexities.
In this section, we will summarize the state-of-the-art in quantum algorithms for dynamical simulations.
The notation used herein and nomenclature specific to the quantum algorithms literature are defined in the following, where the important notions of decomposition and variational methods are introduced, and detailed scaling laws for the most important decomposition methods are collected.
Our goal is not to give a technical review of all methods, which can be found in various other sources~\cite{Motta2021emergingWIRES,Cerezo2020variational,Childs2018toward}, but rather to put into perspective the most recent advancements.
The following sections provide a high-level overview of the most important methods.
\Cref{fig: timeline algorithms} highlights the current surge in attention for all classes of quantum algorithms for quantum dynamics.
Furthermore, \Cref{fig: rating algorithms and applications} provides a qualitative perspective on different methods in terms of their applicability and resource requirements.

While we focus only on quantum dynamics here, it is also important to recall that, for a realistic description of physical processes (as would be monitored in an experiment), careful preparation of a suitable initial state is required, e.g., a ground state or a thermal Gibbs state.
In general, however, this can be a hard task even for quantum computers~\cite{Schuch2009computational,Ogorman2022intractability}.
The design of quantum algorithms addressing this issue is a very active field of research and will be touched upon in \Cref{chap: ground states}.

\subsection{Nomenclature}

When discussing quantum algorithms, one commonly finds the following nomenclature in the literature;
\textit{Ancilla qubits} are auxiliary qubits necessary to perform a certain operation.
\textit{Query complexity} refers to the number of times a certain operation (the decomposed exponential in \glspl{pf}, the walk operator of quantum walks, etc.) needs to be queried.
This can also refer to the number of calls to an unspecified \textit{black-box oracle}, i.e., an unknown unitary that implements a desired operation.
A \textit{block-encoding} encodes a (not necessarily unitary) operator into a unitary operator defined on a higher-dimensional space using additional ancilla qubits.

The literature commonly distinguishes between long- and near-term quantum algorithms.
Methods of the former type require circuit depths and potential qubit-overheads (ancilla qubits) that are far too large to be implemented on noisy near-term quantum devices and will likely only become feasible once fault-tolerant quantum computers are available.
On the other hand, near-term algorithms admit circuit depths small enough to be executed on noisy near-term hardware with little or no additional ancilla qubits.

Here, we instead focus on the distinction between decomposition and variational methods.
As the name indicates, the aim of the former is to derive a quantum circuit, $U$, implementing an approximate decomposition of the unitary time evolution operator itself to given accuracy\footnote{Note that we do not specify a norm or measure of accuracy since we are only interested in asymptotic scaling laws instead of absolute accuracies. This is customary in the literature~\cite{Childs2018toward}.}, $\lVert e^{-iHt} - U \rVert \leq \epsilon$~\cite{Motta2021emergingWIRES,ChildsPHD2000}.
Importantly, the way in which these algorithms are constructed yields rigorous asymptotic scaling laws for the required resources (gate-counts and ancilla qubits) as a function of system size and target accuracy.
Vice versa, one can infer scaling laws for the corresponding approximation errors and improve their accuracy simply by incorporating more resources.
The rigor inherent to decomposition methods is hugely beneficial for the systematic study of quantum dynamics in an exploratory field such as quantum computing.
In particular, it allows for reasonable predictions of the resources required for future simulations.
Seminal decomposition methods include PFs~\cite{Lloyd1996, Childs2018toward}, qDRIFT~\cite{Campbell2019qdrift}, LCU~\cite{Berry2015lcu}, QSP~\cite{Low2017qsp}, and Qubitization~\cite{Low2019qubitization}, for which the scaling laws can be found in the accompanying table.

Variational methods, on the other hand, directly approximate the time evolution of the state, without necessarily requiring an implementation of the time evolution operator itself~\cite{Cerezo2020variational}.
The state is parameterized time-dependently and is evolved variationally.
Although these techniques have been shown to be versatile and efficient in certain cases~\cite{Li2017efficientVariational, Miessen2021quantum, Ollitrault2023quantumalgorithms}, their variational nature makes them inherently heuristic and subject to large degrees of unpredictability.
This is because, in contrast to decomposition methods, they do not allow for general estimates of their accuracy and resource requirements (gate counts and variational parameters).
In fact, accuracy and cost of variational methods are highly dependent on the system of interest, as well as on the selection of a suited variational ansatz, which is a highly non-trivial task in itself.

In the following, we will discuss the main advantages and drawbacks of the most important decomposition and variational methods for time evolution.
\glspl{pf} will be discussed in more detail since they are the most-employed class of methods for time evolution in practice, and because we use \glspl{pf} in \Cref{chap: benchmarking}.
All other methods will be discussed only on a high level.

Throughout, we use the notation specified in the following.
We assume a target accuracy $\epsilon$ as defined above and Hamiltonians given either as a weighted sum of unitary operators,
\begin{equation}
    H = \sum_{l=1}^L a_l H_l
\label{eq: hamiltonian lcu}
\end{equation}
or as a $d$-sparse matrix accessed via oracles to $m$-bit precision, acting on an $N$-qubit space.
Further, we take $\tau = \lVert H \rVert_\mathrm{max} t$ and $\tilde{\tau} = t \sum_l a_l$ as a rescaled simulation time $t$, where $\lVert H \rVert_\mathrm{max}$ is the largest element of $H$ in absolute value, such that $t$ cannot be made arbitrarily small.

\renewcommand{\arraystretch}{1.2}
\begin{table}[htb]
\begin{center}
        \begin{tabular}{L{4cm}C{4cm}C{4cm}}
            \hline\hline
            {Algorithm}
                    & {Gate complexity}
                        & {Ancillas} \\
            \\[-12pt] \hline \\[-12pt]
            {PF order 1}~\cite{Lloyd1996}
                    & $\mathcal{O} ( L^3 \tau^2 / \epsilon )$
                        & none \\
            {PF order $2k$}~\cite{Childs2018toward}
                    & $\mathcal{O} (5^{2k} \frac{L(L \tau)^{1+1/2k}}{\epsilon^{1/2k}})$
                        & none \\
            {qDRIFT}~\cite{Campbell2019qdrift}
                    & $\mathcal{O}(\tilde{\tau}^2 / \epsilon)$
                        & none \\
            {LCU-Taylor}~\cite{Berry2015lcu}
                    & $\mathcal{O}(\tilde{\tau} \frac{\log(\tilde{\tau}/\epsilon)}{\log\log(\tilde{\tau}/\epsilon)})$
                        & $\mathcal{O}(\frac{\log(L)\log(\tilde{\tau}/\epsilon)}{\log\log(\tilde{\tau}/\epsilon)})$ \\
            {QSP}~\cite{Low2017qsp}
                    & $\mathcal{O}(\tau + \frac{\log(\epsilon^{-1})}{\log\log(\epsilon^{-1})})$
                        & $N + \mathcal{O}(m)$ \\
            {Qubitization}~\cite{Low2019qubitization}
                    & $\mathcal{O}(\tilde{\tau} + \frac{\log(\epsilon^{-1})}{\log\log(\epsilon^{-1})})$
                        & $\lceil \log_2 L \rceil + \mathcal{O}(1)$ \\[4pt]
            \hline\hline
        \end{tabular}
    \caption[Asymptotic scaling laws of decomposition methods]{Asymptotic scaling laws of the most common decomposition methods.}
    \label{tab: decomposition scalings}
\end{center}
\end{table}

\subsection{Product formulas}
\label{sec: product formulas}

\Glspl{pf} offer a uniquely simple and straightforward approach to quantum simulation by approximating the complicated unitary time evolution operator with a product of simpler-to-implement exponentials~\cite{Lloyd1996}, requiring no additional ancilla qubits, but dividing $t$ into $n$ time steps $\Delta t = t/n$. 

The simplest digitized implementation of the time evolution operator with time-independent Hamiltonian, \Cref{eq: time evo op t-indep H}, is via a first-order \gls{pf}, also known as Lie-Trotter expansion,
\begin{equation}
    U \approx \Biggl( \prod_{l=1}^L e^{ -i a_l H_l \Delta t} \Biggr)^n \ .
\label{eq: PF-1 time-indep}
\end{equation}
For $H_l$ being $N$-qubit Pauli operators, each of these exponentials can be efficiently decomposed into hardware native basis gates of the digital machine.
Following the notation defined above, the error scales as
\begin{equation}
    \mathcal{O} \Bigl( \sum^L_{i>j} \lVert [H_i, H_j] \rVert \tau^2 / n \Bigr) \ .
\end{equation}
Most generally, this amounts to
\begin{equation}
    \mathcal{O} \bigl( L^2 \tau^2/n \bigr) \ .
\end{equation}

This error may be reduced at the cost of an increased gate-count (see \Cref{tab: decomposition scalings}), either by increasing $n$, i.e., choosing a finer time step, or by employing higher-order \glspl{pf} of order $2k$ ($k \geq 1)$.
The error of higher-order \glspl{pf}, also known as Trotter-Suzuki expansions, scales as
\begin{equation}
    \mathcal{O} \bigl( (L \tau)^{2k+1} / n^{2k} \bigr)
\end{equation}
but with a gate-count scaling exponentially with $k$, which complicates the choice for an optimal \gls{pf}~\cite{Wiebe2010higher,Childs2018toward}.

It is worth highlighting that, for most physical problems, the dependence of the error on the Hamiltonian's commutator structure is rather powerful.
In practice, this can significantly reduce the scaling in $L$~\cite{Childs2021trotterError}.
This has been underpinned by practical simulations, showing that \glspl{pf} perform significantly better than expected from formal bounds~\cite{Childs2021trotterError}. 
In \Cref{chap: perspective}, we highlight applications relying employing \glspl{pf}.
Furthermore, in \Cref{fig: rating algorithms and applications}, we stress the influence of the commutator structure in practical applications by improving the grade given to \glspl{pf} compared to the one expected when considering only asymptotic scaling.

Importantly, \glspl{pf} can also be used to approximate time evolution operators of time-dependent Hamiltonians, $H(t) = \sum_{l=1}^L a_l(t) H_l$~\cite{Poulin2011quantum, Wiebe2011simulating}.
A first order \gls{pf} approximating the integral in \Cref{eq: time evo op t-dep H} as a Riemann sum, $\int_0^{t_f} \diff{t'} H(t') = \lim_{n\rightarrow \infty} \sum_{m=1}^n H(m \Delta t) \Delta t$ with $\Delta t = t_f/n$.
The time ordering in the Trotterized $U(t_f,t_0)$ is enforced with a ``right-to-left ordering'', resulting in
\begin{equation}
    U \approx \prod_{m=n}^1 e^{ -i \Delta t \sum_l a_l(m \Delta t) H_l} \ .
\label{eq: first step PF-1 time-dep}
\end{equation}
Therefore, in contrast to the case of a time-independent Hamiltonian, splitting the exponent into discrete time steps introduces a first approximation.
The Hamiltonian at time step $m \Delta t$ is further decomposed using a first order \gls{pf}, or any higher-order \gls{pf},
\begin{equation}
    U \approx \prod_{m=n}^1 \prod_{l=1}^L e^{-i a_l(m \Delta t) H_l \Delta t} \ .
\label{eq: full PF-1 time-dep}
\end{equation}

Due to this broad applicability and their simplicity, more recent work focused on improving the error-scaling of \glspl{pf}, e.g., by randomizing the order of Hamiltonian terms~\cite{Childs2019fasterquantum}. 
Moreover, the \gls{qdrift} protocol~\cite{Campbell2019qdrift,Chen2021qdriftConcentration} constructs a time evolution operator by randomly sampling individual Hamiltonian terms biased by their weights $a_l$, and achieves asymptotic gate-counts independent of $L$ without additional ancillas or complicated oracle implementations.
On the downside, the method exhibits a worse scaling with simulation time compared to higher-order \glspl{pf} and is oblivious to the potentially powerful commutator structure of the Hamiltonian.
Later generalizations include continuous \gls{qdrift}~\cite{Berry2020timedependent} to simulate time-dependent Hamiltonians and the higher-order method qSWIFT~\cite{Nakaji2023qswift}.

In order to mitigate algorithmic errors of \glspl{pf}, \glspl{mpf}~\cite{Chin2010multiproduct}, i.e., linear combinations of \glspl{pf}, were employed in quantum algorithms.
The linear combination can be handled in place (as an integral part of the quantum circuit)~\cite{Low2019wellconditioned} or by classical post-processing~\cite{Endo2019multiproduct, Vazquez2023wellconditioned}.
However, \glspl{mpf} are often ill-conditioned and care must be taken to avoid an exponential reduction in the success probability of the in-place linear combination~\cite{Low2019wellconditioned}, or an explosion of alternative errors (e.g., sampling errors) when combining the formulas in classical post-processing~\cite{Vazquez2023wellconditioned}.

\subsection{Linear combinations of unitaries}
\label{sec: lcu}

One of the main approaches traditionally proposed as an alternative to \glspl{pf} is the \gls{lcu}~\cite{Berry2015lcu}. 
Given a Hamiltonian that can be written as an \gls{lcu} \Cref{eq: hamiltonian lcu}, the exponential of the time evolution operator is Taylor-expanded to a given order that is determined by the desired accuracy.
The decomposition of the Hamiltonian as a sum of unitaries \Cref{eq: hamiltonian lcu} is then used to write the time evolution operator itself as a sum of unitaries.
Compared to \glspl{pf}, \gls{lcu} reduces the complexity of quantum simulation to a product in $\tilde{\tau}$ and $\epsilon$ (see \Cref{tab: decomposition scalings}).
However, this comes at the cost of introducing many ancilla qubits (scaling linearly with the Taylor truncation order and $\log L$), and an exponentially decreasing (with the number of ancillas) success probability of the algorithm.
Concerning \Cref{fig: rating algorithms and applications}, \gls{lcu} ranks poorly in terms of ancillas needed and sub-optimally in terms of measurements (due to the decreasing success probability).
One of the method's major advantages is its versatility to be extended to simulations of time-dependent Hamiltonians~\cite{Kieferova2019lcuDysonSeries,ChenYi2021lcuDysonSeries} or in the interaction picture~\cite{low2019hamiltonian}.

\subsection{Quantum walks, quantum signal processing, and qubitization}
\label{sec: qubitization}

\glspl{qw} emerged as an analog of the widely applicable classical random walks~\cite{Childs2009quantumWalk, Childs2009contiDiscreteQW}.
If $H$ has eigenvalues $\lambda$, Hamiltonian simulation by \gls{qw} constructs a walk operator $W$ with eigenvalues $\mu = \pm e^{\pm i \arcsin \lambda}$.
First conceptualizations~\cite{Berry2012quantumWalk} decreased the query complexity to linear in $\tau$ but a sub-optimal $\epsilon^{-1}$.

\Gls{qsp}~\cite{Low2017qsp} alleviates this sub-optimality by approximating any operator $W$ with eigenvalues $e^{i\theta_\lambda}$ with a unitary that has transformed eigenphases $e^{-i t \sin(\theta_\lambda)}$.
For the \gls{qw} operator with $\theta_\lambda = \arcsin\lambda$, this amounts to implementing time evolution.
\gls{qsp} achieves optimal, additive query complexity (see \Cref{tab: decomposition scalings}) at the cost of doubling the size of the qubit register.
Importantly, however, \gls{qsp} still requires element-wise access to the Hamiltonian through black-box oracles. 
This can be impractical and expensive to realize, particularly in applications where the Hamiltonian is given as a sum of Pauli operators.

Qubitization~\cite{Low2019qubitization,Dong2021qubitization} leverages a block-encoding of the Hamiltonian combined with \gls{qsp} and concepts of \gls{lcu} to concretely simulate Hamiltonians, maintaining optimal query complexity (see \Cref{tab: decomposition scalings}) and without necessitating oracle-access to the Hamiltonian matrix.
At the same time, it reduces the ancilla overhead to a constant, independent of $t$ and $\epsilon$.
Moreover, for Hamiltonians with specific structures, the computational cost of qubitization can be further reduced~\cite{Berry2019qubitization, Lee2021qubitization}.
Because of its optimal query complexity, we give qubitization the best grade in terms of circuit depth among the decomposition methods in \Cref{fig: rating algorithms and applications}.
However, empirical studies of the constant scaling factors~\cite{Elfving2020qubitization} are needed to refine this ranking.
Note that, on the downside, qubitization in its current formulation does not allow for the simulation of time-dependent Hamiltonians, excluding a broad range of complex problems from its scope of application.

Recent work has focused on improving and generalizing qubitization.
The \gls{qsvt}~\cite{Gilyen2019qsvt,Martyn2021qsvt} provides a unified algorithmic framework based on \gls{qsp} and qubitization, encompassing primitives for several fundamental quantum algorithms, such as Hamiltonian simulation, quantum search, or phase estimation.
Moreover, a hybridized method for Hamiltonian simulation in the interaction picture employs continuous \gls{qdrift} for time-dependent Hamiltonians in combination with qubitization (or \glspl{pf}) to achieve improved scalings~\cite{Rajput2022qubitizationHybridized}.

\subsection{Variational methods}
\label{subsec: variational methods}

As summarized in \Cref{tab: decomposition scalings}, the decomposition methods summarized above tend to require many resources.
As a result, in recent years, several variational approaches have been proposed to overcome these resource requirements of the decomposition methods for applications with noisy near-term quantum hardware.
They may be loosely categorized into algorithms that involve solving either an \gls{eom} derived from a variational principle or an optimization problem.

Leveraging well-established variational principles~\cite{Yuan2019variational}, the idea of constructing a time-dependent variational wavefunction (a parameterized quantum circuit) and propagating the parameters in time by solving an \gls{eom} was first proposed by Li \textit{et al.}~\cite{Li2017efficientVariational}.
Algorithms of this kind differ mainly in how they construct the wavefunction ansatz~\cite{Yao2021adaptiveVariational, Bharti2021variational}. 
However, they require solving a linear system of equations and are thus sensitive to noise.
Moreover, due to their iterative nature and the need to evaluate the corresponding \gls{eom} matrix elements at each time step, these algorithms require a large number of measurements, limiting their applicability to larger systems~\cite{Miessen2021quantum}.

Recasting the problem as an optimization problem might alleviate the numerical issue of solving a linear system of equations and partially reduce the measurement costs at the expense of deeper circuits or additional ancillas.
Proposed methods include, for instance, variationally approximating the Hamiltonian or the time evolution operator in diagonal form~\cite{Cirstoiu2020fastForwarding, Commeau2020variational}.
Another approach is, at each step, to optimize the fidelity between a suitably small time step with a \gls{pf} and a variational state~\cite{Barison2021efficientQuantum, Benedetti2021variational, Zhang2023adaptivePF}, including promising proposals of ansätze based on \glspl{tn}~\cite{Lin2021tensorNetworkTimeEvo,Barratt2020tensorNetworkTimeEvo}. 
Overall, as emphasized in \Cref{fig: rating algorithms and applications}, both \gls{eom}- and optimization-based algorithms remain measurement-intensive and inherently heuristic, making it impossible to predict accuracy and cost prior to simulation~\cite{Miessen2021quantum,Zhoufal2023errorBounds}.

\subsection{Methods for open quantum systems and finite temperature dynamics} 
\label{subsec: algorithms open systems}

Many interesting problems extend to solving the dynamics of open quantum systems, which is, in general, non-unitary, as stated in \Cref{sec: quantum dynamics}.
This task can still be addressed with unitary quantum simulators, achieving advantages similar to the case of closed systems~\cite{Lloyd1996,Kliesch2011dissipativeChurchTuring}.
Conceived methods include the time evolution using \glspl{pf}~\cite{Kliesch2011dissipativeChurchTuring,Wang2011openSystems,Han2021openSystems} and \glspl{lcu}~\cite{Schlimgen2021openSystems}, as well as real and imaginary time approaches~\cite{Endo2020variational,Kamakari2022openSystems}. 
In some cases, open dynamics is reproduced through an explicit extension of the system using additional \glspl{dof} accounting for the presence of the environment.
These effective models can be constructed based on the target microscopic system-environment interaction~\cite{Wang2011openSystems}, possibly in the form of collisional processes~\cite{Cattaneo2021collisionalModels}.

At an algorithmic level, the use of additional qubit resources is also leveraged to obtain suitable representations of quantum dynamical maps and master equations.
This can be achieved via dilation theorems, such as Stinespring's and its variations~\cite{Wang2013solovay,Hu2020openSystems,HeadMardsen2021nonMarkovian}, or through vectorization of the density matrix~\cite{Kamakari2022openSystems,Ramusat2021openSystems}, which naturally leads to the appearance of (anti-)Hermitian effective Hamiltonians.

Moreover, the study of finite temperature dynamics necessitates additional efforts, such as explicit equilibration with a thermal bath or Gibbs thermal averaging, e.g., using imaginary time evolution~\cite{Motta2020thermalStates,Sun2021finiteTemperature}.

\section{Error mitigation and suppression}
\label{sec: error mitigation}

Current-day quantum devices are affected by various kinds of noise, and errors occur at all stages of a computation.
The strength and kind of noise encountered depend strongly on the application of interest, i.e., the quantum circuits and observables, and the quantum device at hand.
We summarized their most common error sources in \Cref{sec: intro hardware}.
Here, we will describe \gls{ems} techniques to circumvent or overcome some of these errors, which is central when executing simulations on current-day hardware~\cite{Cai2023em, Zimboras2025mythsQuantumComputating}.

Error suppression refers to methods that reduce errors at the circuit level.
In general, this encompasses modifying or adding operations in the circuit such that the circuit becomes less susceptible to certain noise, without changing the overall unitary the circuit is meant to implement.
Error mitigation, on the other hand, refers to methods that invert or remove errors in post-processing. 
We employ some of the most established techniques for hardware experiments in \Cref{chap: benchmarking}, and propose a method to leverage them for simulation of open quantum systems in \Cref{chap: pea oqs}.
Here, we provide brief introductions to each of them.
More detailed descriptions can be found in the respective literature.
Further, we assume standard definitions of quantum gates~\cite{NielsenChuang2010, IBMQuantumPlatform}.

\subsection{Dynamical decoupling}
\label{sec: dyanmical decoupling}

\begin{figure}[t!]
	\centering
    \includegraphics[]{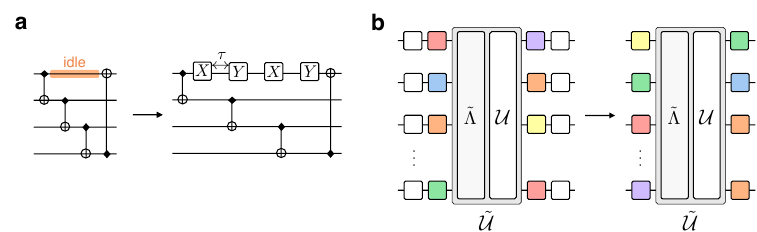}
    \mycaption{Dynamical decoupling and Pauli twirling}{
        \textbf{a} Dynamical decoupling sequences can be entered during idle times in circuits, typically in ladders of two-qubit gates as they have the longest execution times.
        Here, an XY4 sequence with spacing $\tau$ is entered to keep the top qubit from idling and decohering.
        \textbf{b} Pauli twirling randomly samples Paulis (indicated by different colors) around noisy gate layers, typically two-qubit gate layers, to simplify the noisy channel to Pauli noise upon averaging over many samples.
        This does not increase circuit depth since multiple single-qubit gates can be contracted to form different single-qubit gates.
    }
    \label{fig: dd twirling}
\end{figure}

As outlined in \Cref{sec: intro hardware}, some of the most common and severe sources of noise in a quantum computer are unwanted qubit-environment interactions.
\Gls{dd} is an error suppression technique to systematically remove such interactions.
This is achieved by inserting specific quantum gate sequences during qubit idle times to decouple the qubit from its environment (see \Cref{fig: dd twirling}~\textbf{a}).
Importantly, these sequences implement an identity and so do not alter the overall operation implemented by the circuit.
Moreover, \gls{dd} does not increase circuit execution times since idle times are simply filled.
Here, we will briefly outline \gls{dd}, closely following Ref.~\citenum{Ezzell2023dd}.
The joint Hamiltonian of the qubit interacting with its environment may be written as
\begin{equation}
H = H_\mathrm{Q} + H_\mathrm{E} + H_\mathrm{err} \ ,
\end{equation}
where $H_\mathrm{Q}$ is the qubit Hamiltonian, $H_\mathrm{E}$ is the environment Hamiltonian, and $H_\mathrm{err}$ contains the qubit-environment interaction and potential other error terms.
We wish to isolate the qubit and remove all unwanted interactions, i.e., cancel $H_\mathrm{err}$.
\Gls{dd} adds a time-dependent control-Hamiltonian to the system to achieve this cancellation,
\begin{equation}
\tilde{H}(t) = H_\mathrm{Q} + H_\mathrm{E} + H_\mathrm{err} + H_\mathrm{c}(t) \ .
\end{equation}
In an ideal setting, the control Hamiltonian consists of error-free, instantaneous pulses, 
\begin{equation}
    H_\mathrm{c}(t) \propto \sum_k \delta(t - t_k) H_{P_k} \ .
\end{equation}
Here, pulses $H_{P_k}$ are applied instantaneously at times $t_k$.
As shown in Ref.~\citenum{Ezzell2023dd}, the total time evolution for time $T = \sum_{k=1}^n \tau_k$ and $\tau_k = t_k - t_{k-1}$ is then given by a sequence of evolution operators
\begin{equation}
\tilde{U}(T) = U_{\tau_n} C_n \ldots U_{\tau_1} C_1 \ ,
\end{equation}
where $U_{\Delta t_k} = \exp(-i \tau_k H)$ and $C_k = \exp(-i \pi H_{P_k} / 2)$.
We can decompose the total evolution $\tilde{U}(T) = U_\mathrm{Q}(T) U_\mathrm{E}(T)$ into the desired qubit-evolution $U_\mathrm{Q}(T) = \exp(-i T H_\mathrm{Q}) \otimes \mathbb{1}_\mathrm{E}$ and a rest $U_\mathrm{E}$ that includes the evolution of the environment and all unwanted errors.
In the ideal, error-free case, the latter acts only on the environment, $U_\mathrm{E}(T) =\mathbb{1}_\mathrm{Q} \otimes \exp(-i T H_\mathrm{E})$.
Therefore, the goal of \gls{dd} is to sequentially cancel unwanted errors and interactions such that, in the resulting total evolution, qubit and environment are decoupled.
In practice, with non-ideal pulses, the resulting evolution will only be approximately decoupled and still carry an error, $U_\mathrm{E}(T) =\mathbb{1}_\mathrm{Q} \otimes \exp(-i T H_\mathrm{E}) + \mathrm{err}$.

As an example~\cite{Ezzell2023dd}, consider a generic single-qubit environment interaction $H_\mathrm{E} + H_\mathrm{err} = \sum_\alpha \gamma_\alpha \sigma_\alpha \otimes B_\alpha$, with $B_\alpha$ environment operators coupling to the qubit through Pauli matrices $\sigma_\alpha$, and $\gamma_\alpha$ coefficients.
Because Pauli operators anticommute with each other, the \gls{dd} sequence
\begin{equation}
\mathrm{PX} = X - U_{\tau} - X - U_{\tau}
\end{equation}
cancels $Y \otimes B_y$ and $Z \otimes B_z$ interactions.
The resulting effective error Hamiltonian after a duration $2 \tau$ and for ideal pulses reads $F^\mathrm{eff}_\mathrm{PX} = \gamma_x X \otimes B_x + \mathbb{1}_\mathrm{Q} \otimes H_\mathrm{E} + \mathcal{O}(\tau^2)$.
This means the $\mathrm{PX}$ sequence is not universal as it does not cancel all single-qubit interactions.
By adding a second $Y$ term to the sequence
\begin{equation}
\mathrm{XY4} = Y - U_{\tau} - X - U_{\tau} - Y - U_{\tau} - X - U_{\tau} \ ,
\end{equation}
the remaining $X \otimes B_x$ can be canceled.
The $\mathrm{XY4}$ sequence yields universal decoupling at first order after duration $4 \tau$ (see \Cref{fig: dd twirling}~\textbf{a}).

Unwanted interactions are not restricted to single-qubit interactions.
In fact, one of the most prominent sources of noise for transmon qubits is crosstalk, which can be described as an effective $ZZ$-coupling.
While simultaneous $\mathrm{PX}$ or $\mathrm{XY4}$ sequences on both qubits remove single-qubit interactions, $ZZ$-interactions remain due to the commutativity relations of Pauli operators.
To solve this issue, the \gls{dd} sequences on both qubits can be staggered, i.e., shifted relative to each other.
A staggered $\mathrm{PX}$ sequence takes the form
\begin{equation}
X_0 - U_{\tau} - X_1 - U_{\tau} - X_0 - U_{\tau} - X_1 - U_{\tau}
\end{equation}
and removes interactions $Z_0 \otimes Z_1$ as well as single-qubit $Z$- and $Y$- interactions.
Similarly, staggering the universal $\mathrm{XY4}$ sequence~\cite{Barron2024qaoa}, removes $ZZ$-interactions additional to all single-qubit interactions,
\begin{equation}
Y_0 - U_{\tau} - Y_1 - U_{\tau} - X_0 - U_{\tau} - X_1 - U_{\tau} - Y_0 - U_{\tau} - Y_1 - U_{\tau} - X_0 - U_{\tau} - X_1 - U_{\tau} \ .
\end{equation}

There exists a plethora of different \gls{dd} sequences, each canceling different errors or canceling errors to different orders~\cite{Ezzell2023dd}.
In the following, unless otherwise specified, we will always use staggered the XY4 sequence.

\subsection{Pulse-efficient transpilation}
\label{sec: pulse efficient}

Pulse-efficient transpilation is an error suppression technique at the level of transpiling gates to pulse sequences.
It can effectively reduce overall circuit execution times when transpiling parameterized two-qubit gates.
Instead of transpiling these gates to standard entangling gates such as CNOT gates, they are transpiled to shorter, more hardware-native, and, ideally, equivalent entangling pulses.
However, it is important to note that this technique is not generally applicable but is subject to device constraints.
First, it requires pulse-level control of the device, i.e., being able to send instructions to the device as a set of custom pulse sequences instead of a set of basis gates, as is usually the case.
Second, the coupler-structure of the hardware must allow direct implementation of entangling pulses with continuously varying entanglement.
For these reasons, the specific implementation of this technique outlined below is only applicable on a subset of IBM Quantum processors, namely the Eagle device family that has by now been retired from public access.
Although similar techniques could remain useful, it is desirable in the mid- and long-term to step away from techniques that require low-level access (e.g., pulse-level control) to a quantum device.
Instead, quantum computers are evolving to a more out-of-the-box tool that does not require experimental expertise.

The transpilation method used in parts of this work targets parameterized two-qubit gates, such as $\rzz (\theta)$, where $\theta \in \mathbb{R}$ is a continuous gate angle.
One way to implement these gates\footnote{Note that, in general, the specific implementation of a two-qubit gate can vary strongly from one quantum device to another.} is to transpile them to a parameterized single-qubit $R_Z(\theta)$ gate, sandwiched between two maximally entangling gates, such as the CNOT gate.
This is the case on certain IBM Quantum hardware\footnote{Note that, on IBM hardware, newer \qiskit{} capabilities allow for more flexibility in compiling parameterized two-qubit gates to non-maximally entangling two-qubit gates.}.
Backends of the IBM device family Eagle~\cite{IBMQuantumPlatform} have a CNOT gate constructed from $R_\mathrm{ZX}(\pi/2)$ rotations as their basis two-qubit gate~\cite{Earnest2023pulse}.
Those gates are implemented through \gls{ecr} pulses that are exposed to the user.
Due to the freedom, $R_{ZX}(\theta)$ gates with angles $\theta$ different from $\pi/2$ can be implemented by scaling the area of the \gls{ecr} pulse.
Other parameterized two-qubit gates, such as $\rzz(\theta)$, can be related to $R_{ZX}$ through a simple basis rotation.
This allows for transpiling any parameterized two-qubit gate to a single \gls{ecr} gate with scaled pulse area instead of the standard double-CNOT implementation (where each CNOT is an \gls{ecr} gate with maximum area).
A more detailed description can be found in Ref.~\cite{Earnest2023pulse}.

Care is to be taken when the gate angles become so small that, to scale the pulse area, not only the width of the flat-top pulse needs to be scaled but also its amplitude (see Ref.~\citenum{Earnest2023pulse}).
For such small angles, the rotation implemented by the pulse depends non-linearly on the angle $\theta$.
These calibration errors when scaling the pulse area can lead to a loss of equivalence between the target operation and the pulse-efficiently transpiled circuit.
Additional fine-amplitude calibrations can mitigate this issue~\cite{Vazquez2023wellconditioned}.

In general, this leads to an overall reduction of the circuit duration.
For example, in the application discussed in \Cref{chap: benchmarking}, two-qubit gate angles start off being small at the beginning of our time evolution circuit and grow towards the end of the circuit.
For these circuits, pulse-efficient transpilation on average yields a reduction in circuit duration of $\sim 40\%$.

\subsection{Pauli twirling}
\label{sec: twirling}

Pauli twirling, or randomized compiling, is a technique to transform the arbitrary noise of a quantum gate into more structured noise~\cite{Wallman2016twirling, Berg2023pec}.
It suppresses coherent contributions in an error channel and transforms the error into a Pauli error channel.
This can already yield an improvement compared to untransformed noise.
In addition, transforming arbitrary noise to Pauli noise is a necessary prerequisite for error mitigation techniques such as \gls{pec} and \gls{pea} (see \Cref{sec: methods pea}).

More concretely, following Ref.~\citenum{Berg2023pec}, assume a noisy $N$-qubit operation $\tilde{\mathcal{U}} = \mathcal{U} \circ \tilde{\Lambda}$, where $\mathcal{U}$ is the ideal operation and $\Lambda$ is a noise channel.
We can twirl this operation by dressing it with randomly sampled $N$-qubit Paulis $P, P' \in \{ I, X, Y, Z \}^{\otimes N}$,
\begin{equation}
    \tilde{\mathcal{U}}' = P \tilde{\mathcal{U}} P' \ , 
\end{equation}
such that the overall operation remains unchanged.
A schematic representation of the procedure is shown in \Cref{fig: dd twirling}~\textbf{b}.
The simplification of the noise channel is achieved by generating and averaging over many independent realizations of circuits with randomly sampled Paulis $P_i, P'_i$,
\begin{equation}
\Lambda(\cdot) = \mathbb{E}_i \bigl[ P'_i \tilde{\Lambda} (P_i \cdot P_i) P'_i \bigr] \ .
\end{equation}
Averaging over many samples, the resulting error channels will become Pauli error channels of the form
\begin{equation}
    \Lambda(\cdot) = \sum_i c_i P_i \cdot P_i^\dagger \ .
\label{eq: pauli noise model}
\end{equation}
Again, for an $N$-qubit channel, $P_i \in \{ I, X, Y, Z \}^{\otimes N}$ are Pauli strings consisting of $N$ Pauli operators.
Note that the basis $P_i$ for expanding $\Lambda$ here is exponentially large in the number of qubits, with $4^N$ Pauli strings of length $N$.

Finally, Pauli twirling comes with no overhead in terms of circuit depth since single-qubit gates such as Pauli gates used for twirling can be combined with each other to form new single-qubit gates.
However, it can require an increased number of measurements to guarantee sufficiently large numbers of circuit samples when averaging.

\subsection{Readout error mitigation}
\label{sec: readout mitigation}

Readout, or measurement, errors constitute one of the most common sources of errors in quantum computers, as mentioned in \Cref{sec: intro hardware}.
Qubit states after a computation are measured in terms of computational basis states $\ket{0}$ and $\ket{1}$, eigenstates of the $Z$ operator.
The measurement of an $N$-qubit state is fully described by a probability vector $\bm{p} \in \mathbb{R}^{2^N}$ for all $2^N$ computational basis states.
Readout errors occur as wrongly assigned qubit states, e.g., falsely reporting a qubit in state $\ket{1}$ that is actually in state $\ket{0}$.
This means readout errors result in a wrong, noisy probability distribution $\bm{p}_\mathrm{noisy}$.
In general, the mapping of ideal to noisy probabilities can be described by a $2^N \times 2^N$ assignment matrix $A$~\cite{Nation2021m3},
\begin{equation}
    \bm{p}_\mathrm{noisy} = A \bm{p}_\mathrm{ideal}  \ .
\label{eq: readout error assignment matrix}
\end{equation}
Each entry $A_{ij}$ gives the probability of measuring basis state $\ket{i}$ instead of $\ket{j}$, corresponding to probabilities $p_i$ and $p_j$, respectively.
If $A$ is known exactly, \Cref{eq: readout error assignment matrix} can be inverted to obtain $\bm{p}_\mathrm{ideal}$.
In principle, the entries $A_{ij}$ can be learnt by preparing all possible $N$-qubit basis states and measuring the outcome sufficiently many times.
In practice, it is noteworthy that inverting $A$ may result in a quasi-probability distribution~\cite{Maciejewski2020readoutMitigation, Julien2024thesis}, i.e., a vector with entries that sum to $1$ but may contain negative entries,
\begin{equation}
    \bm{q}_\mathrm{mitig} = A^{-1} \bm{p}_\mathrm{noisy} \ .
\label{eq: readout error assignment matrix inverted}    
\end{equation}

More importantly, however, fully constructing $A$ is infeasible for system sizes of interest as $A$ grows exponentially in $N$.
That is why, in practice, $A$ is often approximated to a reduced form~\cite{Bravyi2021readoutMitigation}.
For example, assuming readout errors between qubits are uncorrelated, the assignment matrix can be written as a tensor product of $N$ single-qubit assignment matrices,
\begin{equation}
    A = A_1 \otimes \ldots \otimes A_N \ ,
\label{eq: readout error assigment matrix tensored}
\end{equation}
reducing the number of degrees of freedom to learn to $4N$.
However, correlations between errors are typically present and need to be taken into account when approximating $A$~\cite{Bravyi2021readoutMitigation}.
Therefore, more sophisticated techniques to approximate $A$ have been devised.

In this work, we employ two methods to mitigate readout errors, which we will briefly summarize below.
When expectation values are evaluated directly, i.e., through \qiskit{}'s Estimator primitive without evaluating single-measurement bitstrings, \gls{trex} is used~\cite{Berg2022trex}.
Instead, when sampling and processing individual bit strings (using \qiskit{}'s Sampler primitive), for example, when solving optimization problems, we use \gls{m3}~\cite{Nation2021m3} for readout error mitigation instead of \gls{trex}.

\subsubsection{Twirled readout error mitigation (TREX)}

\gls{trex}~\cite{Berg2022trex} twirls the readout error channels to obtain a simplified error model through mechanisms similar to those described in \Cref{sec: twirling}.
Instead of sampling Paulis from the Pauli group, twirling the readout channel consists of uniformly sampling $P \in \{ I, X \}$ before the measurement.
By averaging expectation values over many realizations of circuits with twirled readouts, the assignment matrix $A$ approximately diagonalizes, with a single multiplicative factor per basis state.
First, calibration experiments are conducted to learn these multiplicative factors on the diagonal of the assignment matrix through readout-twirling experiments on empty circuits.
In a second step, every measurement is twirled when executing the actual circuits that one aims to mitigate.
Readout errors in the final measurement outcome are then mitigated by dividing out the diagonal elements learned in the calibration experiments.
A detailed derivation of this protocol and proof of the effective diagonalization of the assignment matrix can be found in Ref.~\citenum{Berg2022trex}.
Importantly, since twirling relies on averaging over expectation values of different circuit realizations, \gls{trex} can only mitigate readout errors in expectation values but not on individual bit strings.

\subsubsection{Matrix-free readout error mititgation (M3)}

Instead, \gls{m3}~\cite{Nation2021m3} aims to mitigate readout errors in individual bit strings. 
It works on the assumption that only a subset of all possible basis states contribute to the result of a computation.
In this case, it suffices to construct only a reduced assignment matrix $\tilde{A}$ on a subspace of the full computational space of all basis states instead of the full $A$.
Concretely, $\tilde{A}$ is constructed based on noisy bit strings observed in the result $\bm{p}_\mathrm{noisy}$.
Each entry $\tilde{A}_{ij}$ is computed from a tensored construction as in \Cref{eq: readout error assigment matrix tensored}.
Furthermore, the sparsity of $\tilde{A}$ can be adjusted by considering only entries $\tilde{A}_{ij}$ whose corresponding bit strings $\ket{i}$ and $\ket{j}$ are less than a maximum Hamming-distance apart.
Lastly, the reduced matrix needs to be renormalized to ensure left-stochasticity.

\Gls{m3} relies on being calibrated on the output strings of the computation itself.
For large system sizes and in the presence of stronger noise, however, the number of unique, noisy bit strings to mitigate might be equivalent to the number of measurements taken, i.e., every bit string is only measured once.
In this case, even $\tilde{A}$ can become prohibitively large.
With the ability to compute individual elements, matrix-free methods to compute $\tilde{A}^{-1} \bm{p}_\mathrm{noisy}$ can be employed~\cite{Nation2021m3} to circumvent this issue.
Nonetheless, the method becomes more and more approximate with larger system sizes.

\subsection{Probabilistic error amplification}
\label{sec: methods pea}

\begin{figure}[t!]
	\centering
    \includegraphics[width=\textwidth]{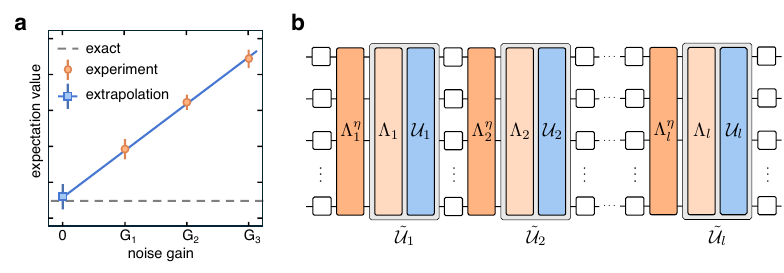}
    \mycaption{Zero noise extrapolation and probabilistic error amplification}{
        \textbf{a} Schematic representation of zero noise extrapolation.
        Note that this does not represent any real data and only serves to illustrate the concept.
        An expectation value is measured at several noise gains $G_i$.
        Those values are fitted and the exact expectation value at zero noise is approximated via extrapolation of the fit to $G=0$.
        \textbf{b} In probabilistic error amplification assumes every real gate layer $\tilde{\mathcal{U}}_l$ to be composed of a noise channel $\Lambda_l$ and the actual, noise-free gate $\mathcal{U}_l$.
        If this noise channel can be learned, it can be amplified by sampling more of the same noise into the circuit with amplitude $\eta$, resulting in a noise gain $G = \eta + 1$.
    }
    \label{fig: pea}
\end{figure}

\Gls{pea} is a type of \gls{zne} that can be used to mitigate noise in expectation values~\cite{Li2017efficientVariational, Temme2017zne}.
Although we do not employ \gls{pea} for active error mitigation in hardware experiments (e.g., in \Cref{chap: benchmarking}), we introduce a time evolution algorithm based on \gls{pea} in \Cref{chap: pea oqs}.
Moreover, a more detailed and technical description of the method, as well as approximate error bounds, can be found there.
Here, we briefly outline its working principles.

Rather than working to reduce the noise in a quantum circuit through error suppression or post-processing of the measurement results, the idea of \gls{zne} is to purposefully amplify the noise to several strengths, or noise gains, without changing the target operation of the ideal, i.e., noiseless, circuit.
Expectation values are then measured under this amplified noise, a curve is fitted through these expectation values as a function of different noise gains, and extrapolated back to zero noise (see \Cref{fig: pea}).
\Gls{zne} merely prescribes to amplify the noise and extrapolate to zero noise, but does not specify how to do so.
There exist different approaches to implement the amplification.
The simplest one is digital \gls{zne}~\cite{Dumitrescu2018digitalZNE, He2020digitalZNE}, where the most error-prone gates, usually entangling gates such as the CNOT gate, are repeated $M$ times in every occurrence of the gate, such that the amplified circuit remains unchanged in the absence of noise.
For example, two CNOTs yield an identity.
Therefore, appending two, four, and so on CNOTs after every CNOT in the original circuit amplifies the noise induced by the gate by factors $3, 5, \ldots$.
Another technique for amplification is to stretch the pulses that implement entangling gates~\cite{Kandala2019zne, Kim2021scalable, Vazquez2023wellconditioned}.
However, these techniques each have their respective drawbacks.
Digital \gls{zne} only offers large noise gains, potentially introducing so much noise already with a noise gain of $3$ that the signal is lost entirely, and no signal for extrapolation remains.
Pulse \gls{zne}, on the other hand, shares the complications of pulse-efficient transpilation (see \Cref{sec: pulse efficient}), i.e., requiring pulse-level access to the device, and potentially additional calibrations.
In addition, those techniques are indifferent to the structure of the noise channels, making detailed control of the noise amplification impossible.

\Gls{pea} on the other hand, alleviates these issues~\cite{Kim2023utility}.
In brief, the method relies on four steps;
(1) Transforming the noise to Pauli noise via Pauli twirling (see \Cref{sec: twirling}).
(2) Learning the $N$-qubit Pauli noise channels up to a fixed order $N$.
(3) Amplifying the learned noise channels by sampling additional noise in the form of Paulis into the circuit, where the rate of sampling is set by the noise gain.
And (4) performing the \gls{zne}.

Again, we assume noise in a quantum circuit to occur primarily through two-qubit gates.
A two-qubit gate layer is again modeled as $\tilde{\mathcal{U}}_l = \mathcal{U}_l \circ \tilde{\Lambda}_l$, with $\mathcal{U}_l$ the ideal operation and $\Lambda_l$ a noise channel.
Pauli twirling is applied, approximately transforming $\Lambda_l$ to a Pauli noise model \Cref{eq: pauli noise model}.
In principle, an $N$-qubit Pauli noise model includes $4^N$ Pauli channels.
For the subsequent step of characterizing the noise model, it is therefore customary to truncate the model at second order (often termed sparse Pauli noise model), i.e., learn only up to two-qubit Pauli channels and neglect higher orders, $P_i \in \{IX, XI, IY, \ldots, XX, XY, XZ, YY, \ldots \}$.
Once learned, the noise model can be amplified by sampling more of the same noise into the circuit, as shown in \Cref{fig: pea}.
The rates at which samples are drawn are set by the wanted noise amplification.
Specifically, the noise model can be written as $\Lambda = \exp(\mathcal{L})$, generated by the Lindbladian with Pauli rates $\lambda_k$,
\begin{equation}
	\mathcal{L} [\cdot] = \sum_k \lambda_k (P_k \cdot P_k - \cdot) \ .
\end{equation}
It is shown in the supplementary of Ref.~\citenum{Berg2023pec} that the exponential $\Lambda$ applied to a state $\rho$ can then be written as a composition in the form of
\begin{equation}
	\Lambda [\rho] = \prod_k \Bigl( w_k \cdot + (1 - w_k) P_k \cdot P_k^\dagger \Bigr) \rho \ .
\end{equation}
Note the slight misuse of notation here, as the product actually denotes a sequential composition of terms.
The probability of an error $P_k$ occurring is $1-w_k$, where
\begin{equation}
	w_k = \frac{1 + e^{-2 \lambda_k}}{2} \ .
\label{eq: methods pea probability}
\end{equation}
As a result, since we can write
\begin{equation}
    \mathcal{U}_l \circ {\Lambda}_l \circ {\Lambda}_l^\eta = \mathcal{U}_l \circ {\Lambda}_l^{\eta + 1} \ ,
\end{equation}
amplifying the noisy channel from $1 \rightarrow \eta + 1$ for some real number $\eta > 0$ amounts to appending the additional noise channel $\Lambda_l^\eta$ with scaled Pauli rates $\eta \lambda_k$ to each noise gate layer $l$.
The probabilities $w_k$ for the scaled noise channel are calculated from the scaled Pauli rates.
Each Pauli error $P_k$ in the learned noise model is then applied with probability $1-w_k$.
Note that each sample is realized through one circuit execution.
Expectation values are measured for each realization and averaged over to obtain the expectation value at amplified noise $\braket{\mathcal{O}}^{(G)}$.
Finally, this is repeated for several noise gains $G = 1 + \eta$, fitted against $G$, and extrapolated to $G=0$ as described above (see \Cref{fig: pea}).

It should be highlighted that the accuracy of this approach crucially depends on three components.
First and most importantly, the ability to accurately model and, in a second step, learn the noise occurring in the circuit accurately.
Both steps are non-trivial and the subject of ongoing research~\cite{Layden2024nonCliffordLearning, Fischer2024tem, Chen2025pauliLearning}.
Second, the accuracy of the amplification step is set by the number of samples and is controllable.
Third, the accuracy of the extrapolated expectation value is determined by the chosen model and the accuracy of the fit.
As shown in the supplementary of Ref.~\citenum{Kim2023utility} and \citenum{Filippov2024peaTheory}, the functional form of the expectation value as a function of $G$ is most generally a linear combination of exponentials.
Typically, this is approximated through a single or double-exponential, introducing more inaccuracies.
More details are provided in \Cref{chap: pea oqs}.
Lastly, note that, for $\eta < 0$, the above procedure results in cancellation of the noise, with \gls{pec}~\cite{Berg2023pec} being the limit case of fully cancelling the noise.
In this case, however, the sampling probabilities are drawn from a quasi-probability distribution.
This implies significant sampling overheads exponential in the strength of the cancelled noise.

%% file: ch3_qsim_perspective.tex
\chapter{Quantum simulation -- where we're headed}
\label{chap: perspective}

\begin{zitat}{}
    This chapter is reproduced with permission in parts from Alexander Miessen, Pauline J. Ollitrault, Francesco Tacchino, and Ivano Tavernelli, ``Quantum algorithms for quantum dynamics'', Nature Computational Science \textbf{3}, 25 (2023)~\cite{Miessen2023perspective}.
    We will outline some key topics in the natural sciences, which, in our opinion, will greatly benefit from the successful implementation of quantum simulation protocols.
    A visual summary of the perspective given in the following is shown in \Cref{fig: rating algorithms and applications}.
\end{zitat}

To determine the impact of quantum computing in specific areas of application, it is crucial to understand
challenges faced by classical algorithms, which are similar across different domains.
For example, \gls{tn} methods have become increasingly popular for quantum dynamics simulations, experiencing fast-paced developments due to their broad applicability to many-body physics~\cite{Cirac2021matrix}, quantum chemistry~\cite{Glaser2022tensor}, and \gls{lgt}~\cite{Magnifico2021tensorLGT}.
However, their efficiency is rooted in reproducing low-entanglement states, while their cost rapidly increases with more than one spatial dimension and, more importantly, as entanglement grows over time or in out-of-equilibrium states~\cite{Banuls2021tensor, Giudice2022temporal}.
As another example, methods based on \gls{mc}~\cite{Becca2017quantumMC} sampling allow for the treatment of large system sizes but fail in certain physical regimes due to severe sign problems~\cite{Werner2009, Troyer2005, Cohen2015}.

Complexity considerations of both classical and quantum simulations are also affected by the specific properties targeted.
In the context of quantum simulation, it was recently shown that the purity of an observable $A$, defined as $\mathrm{Tr}(\rho_A^2)$, where $\rho_A=A/ \mathrm{Tr}(A)$, plays an important role in the error of the computed expectation value~\cite{Poggi2020quantifying}.
Experimental evidence further indicates that some quantities of interest, such as low-weight or local operators and oscillation frequencies, could be more resilient to noise~\cite{Chiesa2019quantum, Neill2021accurately, Kim2021scalable}.

\section{Quantum many-body dynamics} 
\label{sec: perpective - many body}

The availability of effective quantum simulation platforms will significantly impact the field of complex many-body and highly entangled quantum systems.
Indeed, problems in this area can often be naturally mapped onto quantum hardware and, at least in some interesting regimes, pose significant challenges to classical methods already at relatively small system sizes. 
Here, quantum computers could quickly reach -- likely sooner than in other classes of applications -- the level of maturity at which they become useful resources to tackle relevant open questions.
Following the general discussion made at the beginning of this chapter, highly entangled systems, as well as long-time and multi-dimensional simulations, represent the primary terrain of competition with the most advanced classical techniques such as \glspl{tn}, including \glspl{mps}~\cite{Cirac2021matrix}, \gls{dmrg}, \gls{mera}~\cite{Rizzi2008simulation}, \gls{peps}~\cite{ORourke2023peps}, and neural network quantum states~\cite{Sharir2022neural}, or time-dependent mean field methods~\cite{Murakami2017tdmfm}.

\begin{figure}[t!]
    \centering
    \includegraphics[width=\textwidth]{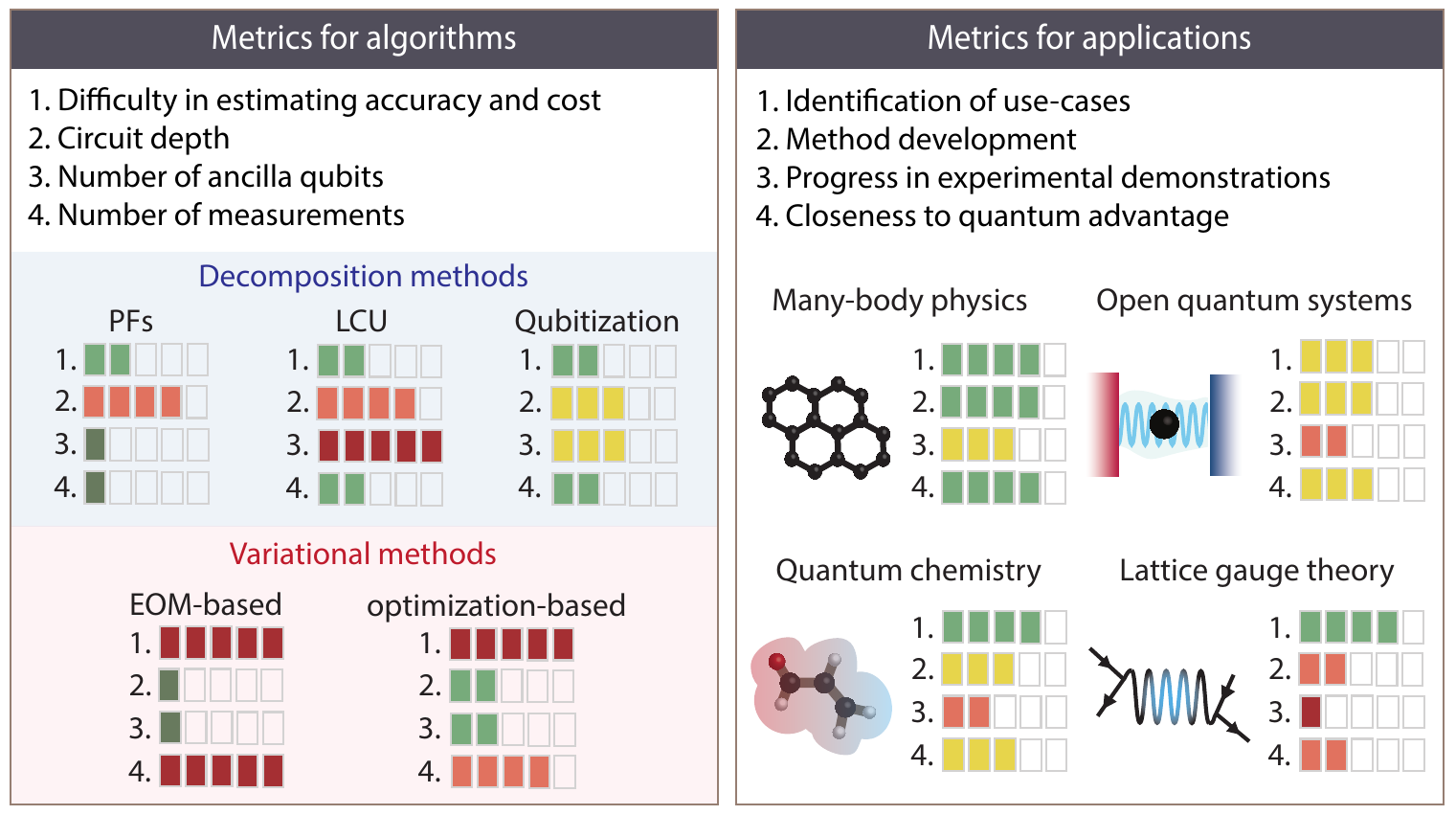}
    \mycaption{Rating of time evolution algorithms and applications}{Assessment of the practicability of quantum algorithms for quantum dynamics and progress of applications grouped into sub-areas.
    For each metric (row), we provide a color-coded rating, where dark green, green, yellow, orange, and red indicate best to worst, respectively.
    For clarity, the same information is conveyed by the filling of the bars, where less is best in the left panel and more is best in the right panel.
    We classify algorithms by how difficult it is to estimate their cost and accuracy prior to simulation, as well as their resource requirements.
    For example, PFs rank worse in 1. than LCU and qubitization since their performance strongly depends on the Hamiltonian's commutator structure, making predictions based on scaling laws less tight.
    Applications are evaluated based on the level of advancement of the research in each sub-field, i.e., how well quantum computing use-cases are characterized.
    Importantly, these rankings are relative to the hardness of the corresponding classically tractable simulation.
    The fourth metric proposed for applications can be understood as a summary of the other three and indicates, based on the current state-of-the-art, concrete potential for achieving quantum advantage with noisy quantum processors.
    }
    \label{fig: rating algorithms and applications}
\end{figure}

As our rating in \Cref{fig: rating algorithms and applications} reflects, several target models have already been identified, 
and experimental proof-of-principle demonstrations are among the most advanced ones across all fields.
So far, these mostly rely on \glspl{pf}, covering for example paradigmatic spin models in one and two dimensions~\cite{Miessen2024benchmarking, Chiesa2019quantum, Smith2019simulating, Crippa2021simulating, Berthusen2021quantum, Kim2021scalable, Kim2023utility, Farrell2024SchwingerDynamics, Farrell2025digitalQuantumSimulationScattering}, and prototypical quantum materials~\cite{Google2020fermiHubbardDynamics, Neill2021accurately}.

A possible key to quantum advantage is the opportunity offered by quantum dynamics to move beyond low-energy/-entanglement problems, as for instance in the simulation of out-of-equilibrium and quenched systems.
So far, progress in this domain has been driven mostly by analog simulators~\cite{Smith2016many, Roushan2017spectroscopic, Browaeys2020rydbergReview}, with system sizes close to the limits of exact diagonalization approaches ($\sim 50$ spins)~\cite{Bernien2017probing, Zhang2017observation, Leseluc2018accurate}.
More recently, control over a record number ($196$-$256$) of Rydberg atoms in 2D arrays was demonstrated, challenging \glspl{mps} methods~\cite{Ebadi2021quantumPhasesAnalog, Scholl2021rydberg}.
Conducting experiments of comparable scale on digital platforms could provide full control over general forward and time-reversed dynamics and unlock otherwise inaccessible regimes.
This, in turn, gives access to a plethora of descriptors, from dynamical correlations and Green's functions~\cite{Chiesa2019quantum,Endo2020greensFunction,Motta2020thermalStates,Sun2021finiteTemperature,Hongxiang2021variational,Baker2021lanczos,Rizzo2022greensFunction,Libbi2022greensFunction}, to the more demanding out-of-time-order correlators (OTOCs)~\cite{Garttner2017otoc,Gonzalez2019otoc,Manoj2020tocScrambling,Braumuller2022otoc}. 
Quantum computers could therefore shed new light on phenomena such as information scrambling and transport~\cite{Pappalardi2018scrambling,Google2021scrambling,Geller2022scrambling,Richter2021hydro,Richerme2014nonlocal}, ergodicity breaking and many body scar states~\cite{Turner2018ergodicityBreaking,Ho2019scars,Desaules2022scars}, as well as general equilibration and thermalization dynamics~\cite{Eisert2015nonEquilibrium}, eventually touching upon open questions concerning many-body localization and the eigenstate thermalization hypothesis~\cite{Nandkishore2015localization}.
In the context of dynamical phase transitions and quantum-classical chaos, the simulation of dynamically localizing models~\cite{Benenti2001complexDynamics,Benenti2003localization,Pizzamiglio2021localization} could also serve as a strict benchmark for near-term quantum processors thanks to the high sensitivity to quantum coherence.

On a parallel track, a more exploratory but nevertheless fascinating perspective application of quantum simulation algorithms lies in the field of exotic materials~\cite{Smith2022topologicalPT,feng2020topological,Malz2021topological,Rodriguez2022realtime,Harle2023topological,Stenger2021majorana}.
Here, quantum hardware platforms could be used as true experimental test beds for detecting elusive phases of matter, such as time crystals~\cite{Ippoliti2021timeCrystal,Google2022timeCrystal,Randall2021timeCrystal,Frey2022timeCrystal}, blurring the boundary between simulation and actual engineered physical phenomena.

As they experimentally lead the way, many-body closed systems applications can also conveniently be used to highlight unifying trends and recurring themes.
An example is the focus on error mitigation strategies~\cite{Temme2017errorMitigation, Endo2018errorMitigation, Kim2023utility, Fischer2024tem}, which will crucially support quantum advantage before the era of full error correction.
While incoherent noise and systematic coherent errors can become particularly detrimental in the context of digital quantum simulations, many specifically designed countermeasures~\cite{Chiesa2019quantum,Google2020fermiHubbardDynamics,Kim2021scalable} 
have already demonstrated great potential in extending the reach of current quantum processors.
Remarkably, very accurate information can be recovered even from noisy dynamics via Fourier analysis~\cite{Roushan2017spectroscopic,Chiesa2019quantum,Neill2021accurately} -- an observation that, in our opinion, makes spectroscopy-inspired
experiments particularly well-suited to achieve robust results (cf. also \Cref{sec: perpective - quantum chemistry}).
At the same time, \gls{pec} techniques are emerging as promising tools to systematically obtain noise-free dynamical observables on near-term digital quantum processors~\cite{Berg2023pec}.

\section{Open quantum systems}
\label{sec: perpective - open systems}

Besides unitary dynamics, the study of open quantum systems has recently gained momentum, with several paradigmatic experiments reported on near-term quantum processors~\cite{Barreiro2011openSystems, GarciaPerez2020openSystems, DelRe2020drivenDissipative, Kamakari2022openSystems}. 
Despite being a relatively little-explored field in quantum simulations, open quantum dynamics could represent a lively playground for early intermediate-scale demonstrations of quantum advantage.
Intuitively, the inherent open-system nature of current noisy quantum computers, together with new hardware solutions such as dynamic circuits~\cite{Corcoles2021dynamicCircuits}, makes them potentially well-suited for the study of engineered dissipative processes~\cite{potocnik2018lightHarveting}. 
However, the identification of appropriate use cases, namely good candidate problems with well-characterized classical benchmarks, is perhaps more delicate compared to closed dynamics.
On the one hand, a direct approach to open systems requires a full density matrix formulation, which generally limits the sizes and time scales of what can reliably be achieved classically~\cite{Hendrik2021openSystemsRMP}.
On the other hand, strong dissipation can, among other effects, limit the growth of entanglement~\cite{Kshetrimayum2017steadyStates}, simplifying the classical description.
This falls in line with the common wisdom that sufficiently noisy quantum processors can be efficiently simulated~\cite{Zhou2020simulatingQC}.
As a consequence, similarly to the case of closed systems, while multi-dimensional (2D, 3D) and highly entangled settings remain among the most challenging regimes, some peculiar features arising from dissipation must be taken into account when trying to identify promising research directions.

Typical tasks in open quantum dynamics concern both real-time evolution and the calculation of \gls{ness}.
The former is generally hard to accomplish on classical computers and, hence, represents the most natural application of quantum simulation methods~\cite{Kliesch2011dissipativeChurchTuring, Barthel2012markovian}.
Here, dissipative models featuring native long-range interactions, such as those realized by platforms based on Rydberg atoms~\cite{Helmrich2018openSystems}, could represent a natural target towards quantum advantage.
In fact, modelling these systems is currently regarded among the most challenging and actively investigated problems in \glspl{tn} research for open quantum systems~\cite{Hendrik2021openSystemsRMP, ORourke2020simplified}.

Other open questions relate instead to non-equilibrium phase diagrams and dissipative phase transitions~\cite{Kshetrimayum2017steadyStates, Helmrich2018openSystems, Fink2017photonBlockade}.
Although \gls{ness} problems are not by themselves of dynamical nature, some of the conventional approaches employed for their solution can be traced back to real-time simulation techniques.
In fact, a straightforward approach to obtain \gls{ness} makes use of the direct integration of the system dynamics over long times.
In this context, quantum simulation may become specifically useful when dealing with many-body systems~\cite{Jin2016steadyState, Hendrik2021openSystemsRMP} and for exploring transient behavior -- e.g.,  equilibration and response dynamics to external perturbations.
This would offer the chance to study, for example, unconventional many-body relaxation effects, such as the ones dominating the transient dynamics of dissipative quantum glasses~\cite{Olmos2021dissipativeQuantumGlasses}.
Moreover the potential of quantum simulation to treat high entanglement and long-time dynamics could alleviate the severe limitations of classical methods whenever the intermediate states become highly entangled.
This can happen even if the \gls{ness} itself is not highly entangled, or in the presence of critical slowing down~\cite{Vicentini2018dissipativeBoseHubbard} (when the time required to reach the \gls{ness} becomes prohibitive).

Classical methods for direct steady state solution, bypassing the real-time evolution~\cite{Mascarenhas2015steadyStateMPO, Kshetrimayum2017steadyStates, Finazzi2015dissipative, Hartmann2019dissipative, Nagy2019openSystems}, are particularly effective for little to moderately correlated systems, but often have high costs or limited accuracy otherwise.
Quantum alternatives have been proposed, based on variational techniques~\cite{Yoshioka2020variationalSteadyState} and \gls{qpe}~\cite{Ramusat2021openSystems}.
The latter method, although quite challenging to be implemented in the near term, achieves a provable advantage over classical counterparts by leveraging, as a subroutine of \gls{qpe}, quantum circuits for Hamiltonian simulation applied to an effective dynamics.

Intriguing use-cases for quantum simulators in the domain of open quantum dynamics include paradigmatic driven-dissipative models (e.g., Ising, Heisenberg, or Bose-Hubbard ) with their rich phase diagrams~\cite{Hendrik2021openSystemsRMP}.
Moreover, fundamental tests on the emergence of dissipative Markovian and non-Markovian dynamics from unitary interactions with large systems~\cite{GarciaPerez2020darwinism} as well as bottom-up algorithmic approaches to quantum thermodynamics experiments~\cite{Solfanelli2021experimentalVerification, Melo2022quantumHeatEngine} could be of interest.
Finally, in a more application-oriented context, dynamical methods for open quantum dynamics could be employed for the study of topologically ordered materials~\cite{Iemini2016dissipativeTopological}, to model collective dissipative effects~\cite{Cattaneo2023dissipative}, and to investigate charge and energy transport in physics and chemistry~\cite{Somoza2019dissipation, Hu2022openSystems}.

\section{Quantum chemistry}
\label{sec: perpective - quantum chemistry}

The potential impact of quantum methods introduced for many-body quantum dynamics naturally extends to the field of quantum chemistry~\cite{Tazhigulov2022correlatedMolecules}.
However, due to the limited number of identified use-cases and a less natural representation of physical systems on hardware, this area of application remains rather unexplored.
Indeed, only few experimental demonstrations and concrete proposals are currently available, as reflected in \Cref{fig: rating algorithms and applications}.
Here, we discuss two challenging computational problems where we foresee a potential quantum advantage: the study of dynamical electronic properties and the coupled time evolution of electrons and nuclei (non-adiabatic dynamics)~\cite{Ollitrault2021molecular}.

The accurate calculation of electronic excited state properties is a formidable task, for which recent progress in classical algorithms has been significant~\cite{Dreuw2005chemrev,Lischka2018chemrev,Westernayr2021chemrev}.
On the other hand, simulating spectroscopic experiments involving dense or high-energy transitions, e.g., near or above the ionization potential, or nonlinear responses, remains challenging.
In these cases, explicitly solving the \gls{tdse} appears as the most natural route~\cite{Goings2018electronicStructure,Li2020electronicStructure}, as is also the case for the study of out-of-equilibrium processes such as molecular conductance, electron transport phenomena, solvent-solute interactions, etc.
In this context, a recent review on time-dependent electronic structure approaches~\cite{Goings2018electronicStructure} still underlines as a pressing question: ``How can we go beyond \gls{dft} and \gls{hf} to properly account for electron correlation in real time?''.
Recent attempts in this direction rely on the \gls{tddmrg}~\cite{Chan2011dmrg,Baiardi2020dmrg}.
Promising results have been reported for the simulation of charge dynamics after ionization and in the presence of a weak electromagnetic perturbation in larger systems.
Nonetheless, \gls{dmrg} is not expected to perform well for the dynamics of highly entangled states as stated previously.
Likewise, a recent time-dependent \gls{cc} implementation~\cite{Wang2022realTimeCC} enabled simulations of the linear absorption spectrum of a water tetramer interacting with an electric field.
However, the underlying scaling of \gls{cc} methods continues to limit their access to larger system sizes.
Moving to more complex molecules, quantum computing techniques appear as valid alternatives for electronic dynamics, offering memory savings and potentially shorter run times.
For instance, electronic dynamics can be performed with any decomposition method as long as valid basis functions can be found to express the Hamiltonian in second quantization.

Concerning the quantum dynamics of the nuclear part of the molecular wavefunction, recent developments in classical approaches mainly include different extensions of \gls{mctdh}~\cite{Meyer2009mctdh} or of \gls{tddmrg}, allowing to accurately simulate molecules like pyrazine~\cite{Baiardi2019mps}.

One of the main shortcomings of these methods is the need to pre-compute multi-dimensional \glspl{pes}.
Classically, one can resort to direct dynamics approaches~\cite{Persico2014nonadiabatic, Worth2008nonadiabatic, Ben2000abInitio, Lasorne2007gaussianWave, Richings2018mctdh}, calculating the \glspl{pes} on-the-fly during the propagation.
Another option are pre-\gls{bo} methods, which treat electronic and nuclear degrees of freedom on the same footing~\cite{Abedi2010exact, Matyus2012molecular, Bubin2013born, Pavosevic2020multicomponent, Yang2019density, Muolo2020nuclear}.
In both cases, however, accurately describing nuclear wavefunction dynamics remains too costly to be extended to large systems.
Having to pre-sample such high-dimensional \glspl{pes} remains a problem even if quantum computers could accelerate the point-by-point solution of the time-independent Schrödinger equation.
Therefore, in analogy with the classical case, quantum advantage should be sought in a direct dynamics or pre-\gls{bo} fashion~\cite{Ollitrault2021molecular}.

This poses the question of how to digitally encode and manipulate fermionic and bosonic (molecular vibrations) degrees of freedom simultaneously.
In particular, the latter requires truncation of the occupation space, leading to mapping strategies similar to those employed for generic spin $S\geq 1$ operators~\cite{Chiesa2015spinPhoton,Tacchino2018electromechanical,Macridin2018electronPhonon,McArdle2019molecularVibrations,Sawaya2020vibrational,Ollitrault2020vibrational,Tacchino2021spinQudits}.
Alternatively, it could also be advantageous to express molecular quantum dynamics by means of a grid-based approach~\cite{Chan2023gridBased, Ollitrault2020nonadiabatic,Ollitrault2023quantumalgorithms}.
The problem is then formulated in first quantization, i.e., the particle exchange statistics are accounted for directly in the wavefunction.
A spatial grid is employed to express the wavefunction or the density matrix, and to evaluate the Hamiltonian (described by position and momentum operators) without the need to introduce further approximations other than the choice of the grid spacing.
However, in this case, the implementation of decomposition methods can become cumbersome~\cite{Mitarai2019analogDigitalConversion, Woerner2019quantumRiskAnalysis, Haner2018arithmetic, Ollitrault2020nonadiabatic} and variational approaches may be better suited~\cite{Ollitrault2023quantumalgorithms}.

\section{Lattice gauge theory}
\label{sec: perpective - lgt}

Quantum computational approaches to \gls{lgt} range from 
simulations and hardware calculations for \gls{qed}, particularly for the Schwinger model~\cite{Martinez2016fewQubitLGT,DeJong2022schwinger}, to foundational work and feasibility assessments for more general non-abelian \glspl{lgt}~\cite{Lamm2019lgt,Kan2021lgt,Gonzalez2022nonAbelianGaugeTheory}, aiming to identify potentially scalable solutions~\cite{Mathis2020scalableLGT}. 

Classical simulations of \gls{lgt} are dominated by methods based on \gls{mc} sampling, which have enabled the simulation of large system sizes~\cite{Tanabashi2018reviewParticlePhysics} but fail in regimes of finite-matter density and real-time evolution due to sign problems.
\Glspl{tn} have gained significant attention in recent years as an alternative approach~\cite{Magnifico2021tensorLGT}, but the associated challenges in scaling to larger dimensions and long-time dynamics severely limit their current reach.
Within this context, quantum computers could offer the possibility to perform \gls{lgt} simulations free of sign-problems and without being restricted to low entanglement and low dimensions.
Similarly to \glspl{tn}, however, they face the practical challenge of dynamically treating fermionic (matter) and bosonic (gauge fields) degrees of freedom, requiring large amounts of resources and, possibly, novel information-encoding strategies~\cite{Gonzalez2022nonAbelianGaugeTheory}. 
Combined with the system sizes required to produce results complementary to state-of-the-art \gls{mc} simulations for lattice \gls{qcd}, this will necessitate millions of (error-corrected) qubits.
We therefore conclude that, while fully fault-tolerant quantum computing would undoubtedly have a tremendous impact in this research field~\cite{Bauer2023highEnergyPhysics}, little to no advantage might be accessible in the near term.

\section{Outlook}
\label{sec:conclusion}

The development of quantum algorithms for the solution of quantum dynamics is a very active and promising field of research.
It holds the potential to solve quantum dynamical problems which cannot be efficiently treated with classical methods, and, maybe, to one day provide the long-sought quantum advantage.
Considering the preceding discussion, we foresee crossover points to first occur in the out-of-equilibrium simulation of lattice spin Hamiltonians with a few hundred sites or in the simulation of electron dynamics in molecules with order fifty to one hundred electrons.
As quantum advantage requires simultaneous achievements in hardware, theory, and software development, concrete timelines are very difficult to predict.
Even more so, as classical algorithms evolve constantly in a head-to-head race with the most advanced quantum algorithms and hardware experiments.
However, the system sizes mentioned above suggest that selected demonstrations could be within reach of noisy near-term technology.
Through a combination of error mitigation techniques~\cite{Berg2023pec}, high-speed classical control~\cite{Wack2021qualitySpeedScale}, and, possibly, basic error correction tools, quantum dynamics could hence fuel a continuous transition towards the fault-tolerant regime~\cite{Google2023surfaceCode, Bravyi2024LDPC, Xu2024neutralAtomsLDPC}.
Furthermore, quantum computers can be of intermediate value before achieving a clear, practical quantum advantage through so-called quantum utility~\cite{Kim2023utility}.
The aforementioned competition between quantum simulations and approximate classical methods (see also \Cref{chap: intro}) can benefit both paradigms by serving a yet another way to cross-check simulation results.
In this regime, quantum simulation constitutes one of many computational methods to reach approximate solutions to complex problems, adding one more independent data point to an approximate solution space.

In fact,  we expect quantum advantage to occur in a regime that precedes the strict applicability of the asymptotic scaling laws presented in \cref{tab: decomposition scalings}.
That is why one of the main challenges for these intermediate-size problems is to assess which algorithm is optimal to simulate a given application.
We therefore argue that the most relevant distinction to make today is not the common one between long- vs.~near-term methods, but rather between decomposition vs.~variational approaches.
On the one hand, variational methods can be used to reduce the circuit depth and provide a suitable framework for the solution of, e.g., grid-based dynamics~\cite{Ollitrault2023quantumalgorithms}.
However, the associated measurement costs could render them impractical for relevant system sizes~\cite{Miessen2021quantum}, and their heuristic nature does not allow for a precise control over errors and convergence.
In addition, variational time evolution algorithms often incur numerical instabilities~\cite{Ollitrault2023quantumalgorithms}, and would require fast quantum-classical communication capabilities to operate at scale.
On the other hand, decomposition methods provide a systematic way to control errors and trade resources for accuracy.
Therefore, we consider them the best candidates for early demonstrations of quantum advantage.
\Glspl{pf} in particular offer a versatile and simple framework for quantum dynamics with practical applications often requiring less deep circuits than suggested by scaling laws.
Today, precisely for these reasons, \glspl{pf} appear to be the most effective and widely adopted family of algorithms for implementing quantum dynamics.
They will likely play a central role in outperforming classical simulations using quantum approaches.

Concerning applications, the study of many-body dynamics currently represents the most advanced and promising candidate for quantum advantage, see \Cref{fig: rating algorithms and applications}.
Here, classically intractable systems appear rather quickly as quantum correlations and entanglement grow.
Looking forward, the same conditions are also encountered in other relatively less explored fields like open quantum systems and quantum chemistry, e.g., when addressing problems such as out-of-equilibrium evolution.
In parallel, hybrid approaches -- where quantum computers are employed to solve a specific sub-problem embedded in a larger classical calculation -- could be leveraged to extend the reach of intermediate-size quantum resources toward, e.g., complex materials simulations~\cite{Bauer2016hybrid}.
Real-time evolution also appears as a sub-routine in a variety of different quantum algorithms that could themselves lead to practical quantum speed-ups~\cite{Layden2023quantumMCMC}.

Lastly, it is worth mentioning that analog quantum simulation platforms represent an interesting alternative to achieve impactful results in the near future~\cite{Scholl2021rydberg, Ebadi2021quantumPhasesAnalog,Flannigan2022errorPropagationAnalog,Daley2022practicalAdvantageAnalog}.
However, achieving such a demonstration with digital architectures would provide a far stronger result, as it would pave the way toward universal quantum computation.

%% file: ch4_benchmarking.tex
\chapter[Benchmarking digital quantum simulations using quantum critical dynamics]{Benchmarking digital quantum\\simulations using quantum critical\\dynamics}
\label{chap: benchmarking}

\vspace{20pt}

\begin{zitat}{}
    This chapter is reproduced with permission in parts from Alexander Miessen, Daniel J. Egger, Ivano Tavernelli, and Guglielmo Mazzola, ``Benchmarking digital quantum simulations above hundreds of qubits using quantum critical dynamics'', PRX Quantum \textbf{5}, (4), 040320 (2024)~\cite{Miessen2024benchmarking}.
    \Cref{sec: quantum annealing} provides the necessary technical background to \gls{qa} and the \gls{qkzm}.
    In \Cref{sec: benchmarking}, we introduce and implement our qualitative application-oriented benchmark of quantum simulation against universal behavior.
    Moreover, \Cref{sec: benchmarking} presents the main results of this chapter -- a comparison of different levels of \gls{ems} on two quantum processors with up to 133 qubits using our \gls{qkzm}-based benchmarking.
    Lastly, we study combinatorial optimization in \Cref{sec: optimization} and show that our benchmarking results are transferrable to this application.
    Alexander Miessen carried out all calculations, simulations, and experiments, implemented the code, conducted the data analysis and visualization, and wrote the manuscript.
\end{zitat}

Quantum critical dynamics occur when a quantum system reaches a 
\gls{cp}, characterized by a non-analytic change in the system's ground state energy as a function of a Hamiltonian parameter.
The system's correlation length and relaxation time are maximal at the \gls{cp} and diverge in the thermodynamic limit.
Crossing it, the system undergoes a \gls{qpt}.
As a result, system details no longer affect macroscopic quantities, causing the emergence of universal behavior, a key property of critical phenomena~\cite{Stanley1971phaseTransitions,Sachdev2011book,Sondhi1997continuousQPT}.

\Glspl{qpt} are experimentally realized by changing the control parameters of the system Hamiltonian over time.
However, understanding the real-time dynamics of many-body quantum systems close to the critical point is a formidable task.
Many believe that only quantum simulators~\cite{Feynman1982}, i.e., controllable quantum systems that can emulate others~\cite{Georgescu2014quantumSimulationRMP}, can tackle this problem at scale~\cite{Altmann2021quantumSimulators, Monroe2021trappedIonsRMP}.
Quantum simulators were first realized with ultracold dilute gas in an optical lattice~\cite{Jaksch1998coldAtoms} and have since been implemented on a variety of platforms~\cite{Altmann2021quantumSimulators, Monroe2021trappedIonsRMP,Lewenstein2012ultracold,Bernien2017probing,Scholl2021rydberg}.
Most of these platforms are \textit{analog} quantum simulators, which are subject to calibration errors and decoherence, which we described in \Cref{sec: intro hardware}.

As also detailed in \Cref{sec: intro hardware}, a parallel approach is to perform the simulation on \textit{digital} quantum computers using suitable algorithms compiled to the native basis gate set of the hardware~\cite{Miessen2023perspective, Abrams1999exponentialEigenvalues, Kim2023utility, Google2022timeCrystal, Keenan2023KPZscaling, Miessen2021quantum}.
However, current digital machines are prone to errors, such as calibration errors, crosstalk, and decoherence.
In the future, quantum error correction could potentially enable fault-tolerant simulations of many-body quantum systems on digital machines~\cite{Bravyi2024LDPC}.
Ref.~\citenum{Daley2022practicalAdvantageAnalog} gives a perspective on the relative strengths and weaknesses of analog and digital platforms.

Crucially, quantum critical dynamics have implications beyond condensed matter and statistical physics.
For example, \Glspl{qpt} occur ubiquitously in quantum optimization, where a quantum algorithm helps solve a classical optimization problem.
Such applications are among the most anticipated and economically impactful use cases for quantum computers~\cite{Abbas2024optimizationWhitePaper}.
\Gls{qa} is an algorithm to find ground states that can be used to solve combinatorial optimization problems~\cite{Albash2018annealingRMP}.
It evolves an easy-to-prepare ground state of one Hamiltonian to the unknown ground state of another problem Hamiltonian, which corresponds to the classical optimization problem to solve.
If the evolution is adiabatically slow, the system follows the instantaneous ground state of the time-dependent Hamiltonian and ends in the solution of the optimization problem.
Particularly for \gls{qa}, \glspl{qpt} create algorithmic bottlenecks~\cite{Kadowaki1998quantumAnnealing,Knysh2016spinGlass,Albash2018annealingRMP}, as they imply an energy gap between the ground and the first excited state that closes in the thermodynamic limit.
For finite annealing times $\tf$, the evolution may therefore not be adiabatic due to a diverging relaxation time at the \gls{cp}.
This mechanism produces defects that carry through to the final state~\cite{Zurek2005dynamicsQPT}.

In a noise-free, closed system, the density of defects is a non-increasing function of $\tf$, and several scaling regimes can be identified as $\tf$ increases~\cite{Zurek2005dynamicsQPT,Zeng2023kzmFastQuenches}.
The \gls{qkzm}~\cite{Zurek2005dynamicsQPT,Kibble1980cosmologicalPT,Zurek1985cosmologicalHelium} quantifies the relationship between $\tf$ and the number of defects produced during the annealing run in a regime of sufficiently slow, yet finite, $\tf$.
It predicts a universal scaling, a power-law decay, of the density of defects in the final solution as a function of $\tf$ and the system's critical exponents~\cite{Bando2020dwave}, i.e., a set of numbers characterizing the system's behavior near its \gls{cp}.
It has been the subject of a range of largely analog experimental studies~\cite{Li2023probingLongRangeKZM, Keesling2019rydbergKZM, King2022dwave2000IsingChain, King2023spinGlass,Ebadi2021quantumPhasesAnalog}.

We first present an application-oriented benchmarking method that utilizes the predictability of known universal scaling laws, such as the \gls{kz} scaling.
Those attributes make it ideal for an intuitive and easily scalable metric to assess the quality of large-scale quantum simulations.
This is increasingly important as digital quantum computing devices and algorithms have left the infantile stage of a few tens of qubits with error rates prohibiting more than a handful of two-qubit gates~\cite{Bravyi2022future, Vazquez2024circuitCutting}.
Today's quantum devices can exceed 100 qubits at error rates that, combined with \gls{ems} techniques~\cite{Cai2023em}, allow a coherent simulation of thousands of two-qubit gates~\cite{Kim2023utility, Shtanko2023localIntegrability}.

\begin{figure}[t!]
    \centering
    \includegraphics[width=\textwidth]{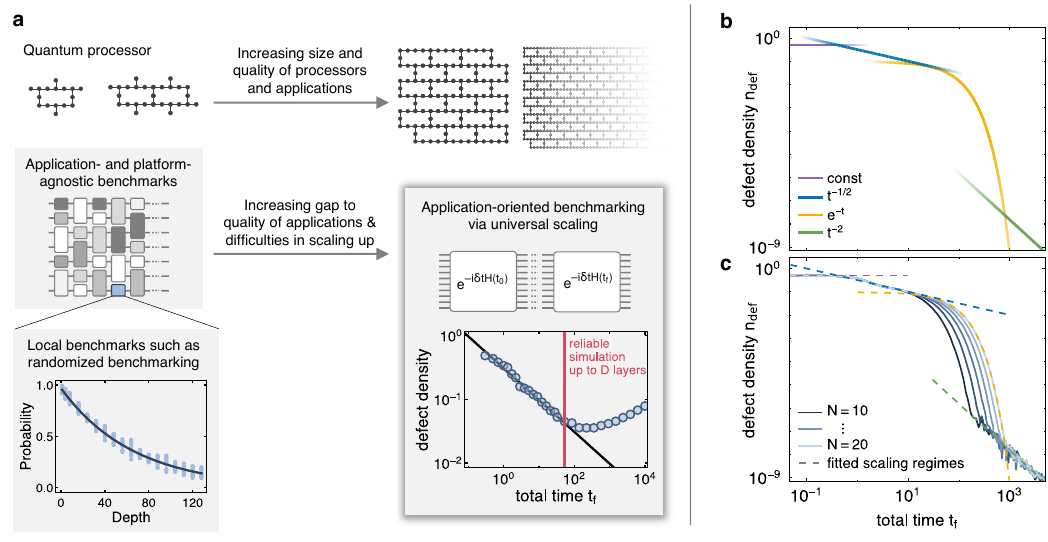}
    \mycaption{Application-oriented benchmarking of quantum simulations}
    {\textbf{a} Benchmarking experiments give detailed insight into device characteristics such as error rates and decoherence times.
    As devices scale, however, it is desirable to have a benchmarking method that resembles the applications that are executed.
    By simulating critical dynamics and measuring an observable that follows a known universal scaling, we track for how many time steps, which correspond directly to the number of circuit layers, we can reproduce the expected universal scaling.
    Here, we measure the density of defects produced in quantum annealing, predicted by the Kibble-Zurek scaling.
    This provides a direct and concrete metric predicting how many layers of two-qubit gates can be reliably simulated given, e.g., a certain device, qubit subset, and error mitigation technique.
    Note that \textbf{a} contains only schematic plots and shows no actual data.
    \textbf{b} For finite system sizes away from the thermodynamic limit, the density of defects exhibits two scaling regimes beyond the Kibble-Zurek scaling that are theoretically known (see main text).
    These can benchmark deep circuits.
    \textbf{c} The departure from Kibble-Zurek scaling is determined by the system size and all three scaling regimes can be identified in statevector simulations with $N=10, \ldots, 20$ spins (solid lines).
    Shown here are results from Trotterized simulations with $\Delta t = 0.01$ for $\tf \leq 0.1$ and $\Delta t = 0.1$ otherwise.
    }
    \label{fig: schematic and statevector KZM}
\end{figure}

Many benchmarking methods employ randomized circuits.
This makes them more comparable, objective, and device-agnostic while accounting for, e.g., qubit connectivity and basis gate set~\cite{Amico2023benchmarkingBestPractices}.
Moreover, randomness helps collect many of the error sources into the same benchmarking process.
Examples of such methods span an entire field of research, including variants of randomized benchmarking~\cite{Magesan2012rb, Proctor2022mirrorRB, Hines2024scalableRB}, cross-entropy benchmarking~\cite{Boixo2018xeb}, and quantum volume~\cite{Cross2019qv}.
These methods are crucial to assess gate fidelities and characterize different kinds of noise.
However, they do not appropriately represent the capabilities of current quantum hardware to run a specific application.
Indeed, applications for noisy quantum devices have circuits with highly structured repeated layers that are often tailored to the connectivity and the basis gate set of the hardware.
It is therefore of little surprise that such simulations achieve much higher gate counts and qubit numbers than one would predict from generic benchmarks.
Even though there are very promising new benchmarking methods designed to overcome these issues~\cite{Mckay2023layerFidelity}, they still only provide a fidelity without relating to applications.
Pure application benchmarks are the other extreme.
Either by testing against a set of well-understood problems and, possibly, corresponding solutions~\cite{Lubinski2023benchmarksApplicationOriented, Wu2024variationalBenchmarks}, or based on specific applications such as quantum optimization~\cite{Santra2024quantumOptimization} or discrete time crystals~\cite{Zhang2023benchmarkingTimeCrystals}.

We go a step further and propose a concrete, application-oriented benchmarking scheme to assess how many two-qubit gate layers of a given application circuit can be reliably simulated given a specific device and, importantly, \gls{ems} method.
A schematic of the principle is shown in \Cref{fig: schematic and statevector KZM}\textbf{a}.
The result is not an abstract fidelity but simply the number of circuit layers that can be reliably simulated, which can be directly transferred to other applications.
Here, we focus on the \gls{kz} scaling, though other universal scalings, e.g., at short times, can also be used~\cite{Zeng2023kzmFastQuenches}.

This benchmark is transferrable to other applications involving time evolution, for example, dynamics of spin systems, quantum optimization, \gls{qml}, and variational algorithms.
To showcase this, we apply the benchmark to combinatorial optimization specifically as
the underlying circuits directly implement digitized \gls{qa} or \gls{qaoa}.
QAOA is an annealing-inspired variational ansatz, with variational layers representing time steps in digitized \gls{qa}, used in conjunction with \gls{vqe}~\cite{Farhi2014qaoa}.
Recent studies indicate that solving hard optimization problems requires many \gls{qaoa} layers~\cite{Zhou2020lukinQAOA, Farhi2020qaoaLargeP}.
At the same time, the optimally converged parameter values of \gls{qaoa} in the large-layer limit reproduce annealing schedules, meaning \gls{qaoa} implements digitized \gls{qa} with variationally found annealing schedules.
It is therefore important to better understand digitized \gls{qa}.
A relevant question is whether the algorithmic error stemming from a finite time step is detrimental or possibly even beneficial, similar to what happens in simulated \gls{qa}~\cite{Heim2015spinGlass}.
Here, we identify an optimal working point with respect to the time step and number of circuit layers to minimize the residual energy given the finite hardware resources and how it depends on the system's minimum energy gap.

In summary, we explore two directions -- benchmarking and optimization -- and show how the former can be used to guide the design of the latter.
\Cref{sec: quantum annealing} provides the necessary technical background to \gls{qa} and the \gls{qkzm}.
In \Cref{sec: benchmarking}, we introduce and implement our qualitative application-oriented benchmark of quantum simulation against universal behavior.
Here, we present the main results of this work -- a comparison of different levels of \gls{ems} on two quantum processors with up to 133 qubits using our \gls{qkzm}-based benchmarking.
 \Cref{sec: optimization} studies digitized \gls{qa}, using the same underlying circuits as in the benchmarking experiments, for solving optimization problems and showcases how results from \Cref{sec: benchmarking} are transferable to related applications.
We conclude in \Cref{sec: benchmarking conclusion}.

\section{Digitized quantum annealing and defect production}
\label{sec: quantum annealing}

Given a (mixing) Hamiltonian $\HM$ with an easy-to-prepare ground state and a problem Hamiltonian $\HP$ whose ground state we wish to compute, we construct a time-dependent combination of the two,
\begin{equation}
    H(s) = A(s) \HM + B(s) \HP \ .
\label{eq: hamiltonian annealing undefined}
\end{equation}
Here, $s = t / \tf$ is the time $t \in [0, \tf]$ rescaled by the total annealing time $\tf$.
$A(s)$ and $B(s)$ are the annealing schedules such that $H(0) = \HM$ and $H(1) = \HP$.
The system is initially prepared in the ground state of $\HM$ and evolved for time $\tf$.
If this evolution is adiabatically slow, i.e., for large enough $\tf$, the system remains in the ground state of the instantaneous Hamiltonian and ends in the desired ground state of $\HP$ at $t = \tf$~\cite{Albash2018annealingRMP}.

The prototypical Hamiltonian considered in \gls{qa} is an $N$-site \gls{tfim} with nearest-neighbor interactions,
\begin{equation}
    \HM = - \sum_{i} \sx_i
    \ , \quad
    \HP = - \sum_{\langle i,j \rangle} J_{ij} \sz_i \sz_{j} \ ,
\label{eq: hamiltonian ising}
\end{equation}
with Pauli matrices $\sx_i$ and $\sz_i$ acting on site $i$ and couplings $J_{ij}$ between neighboring sites indicated by $\langle i,j \rangle$.
The ground state of $H(0) = \HM$ is $\ket{+}^N$, with $\ket{+}=(\ket{0}+\ket{1})/\sqrt{2}$ (we will use $\ket{i}^N$ in the remainder of this thesis for short-hand notation of $\ket{i}^{\otimes N}$).
We apply linear schedules $A(s) = (1-s)$ and $B(s) = s$ for simplicity, resulting in
\begin{equation}
    H(s) = - A(s) \sum_{i} \sx_i - B(s) \sum_{\langle i,j \rangle} J_{ij} \sz_i \sz_j \ .
\label{eq: hamiltonian ising annealing}
\end{equation}
Unless otherwise stated, we consider an ordered Ising model with uniform couplings $J_{ij} \equiv J$.

The \gls{tfim} is the simplest model exhibiting a \gls{qpt}~\cite{Sachdev2011book} in the thermodynamic limit $N \rightarrow \infty$.
In the simple case of uniform couplings $J>0$ ($J<0$), the \gls{cp} at $J_c = A / B$ separates the paramagnetic phase ($A \gg J B$) from the doubly-degenerate (anti-)ferromagnetic phase ($A \ll JB$) with all spins (anti-)aligned.
In the disordered case, the latter is substituted by a glassy phase, i.e., an energy landscape with many deep local minima~\cite{Young2010firstOrderPT, Joerg2010firstOrderPT, Knysh2016spinGlass}.

\subsection{The Kibble-Zurek mechanism}

The \gls{qkzm} describes the non-equilibrium dynamics of the systems in the realistic setting of crossing this \gls{cp} with finite annealing time $\tf$, leading to defects in the final solution.
Ref.~\citenum{Zurek2005dynamicsQPT} provides an excellent account of the \gls{qkzm} and the resulting density of defects scaling in a \gls{tfim}, both by employing the original (classical) reasoning of the \gls{kzm} and in terms of a fully quantum description based on \gls{lz} theory.
Here, a defect refers to the wrong orientation of a spin and the density of defects to the number of defects averaged over all sites.
In the ferromagnetic phase specifically, defects are domain walls between anti-aligned spins.
We now summarize the classical description of the \gls{kzm} following Ref.~\citenum{Zurek2005dynamicsQPT}.
Classically, the \gls{cp} is characterized by a diverging correlation length $\xi$ and relaxation time $\tau$.
Far away from the \gls{cp}, $\tau$ is small enough for the system to instantaneously relax to equilibrium.
As the system approaches the \gls{cp}, $\tau$ becomes equal to and, eventually, grows beyond the time scale on which $H$ changes.
When this happens, the system can no longer relax to its instantaneous ground state, with reactions to changes slowing down until they halt completely.
This is usually referred to as the critical slowdown and subsequent freezing out.
Past the \gls{cp}, as $\tau$ decreases again, the system unfreezes and, crucially, continues evolving from approximately the frozen-out state, resulting in defects in the final state.
In summary, the growth of the system's relaxation time at the \gls{cp}, which is finite for finite system sizes, determines the necessary rate of change of the system $\diff s / \diff t = 1/\tf$ required to track the instantaneous ground state.
This is the essence of the \gls{kzm}.
It predicts the density of defects as a function of a finite annealing time $\tf$  with a power-law decay~\cite{Bando2020dwave,Keesling2019rydbergKZM}
\begin{equation}
    \ndef \propto \tf^{ - \frac{d \nu}{1 + z \nu}} \ ,
\label{eq:kzm defect density}
\end{equation}
fully determined by the system dimension $d$ and its critical exponents $\nu$ and $z$.
In 1D, with $d=1, z=1, \nu=1$, this becomes
\begin{equation}
     \ndef \propto \tf^{ - 1/2} \ .
\label{eq: kz scaling 1D}
\end{equation}

In the quantum description, the existence of a \gls{qpt} and all its related quantities originate from the closing of the system's energy gap at the \gls{cp}.
For finite system sizes $N < \infty$, the minimum gap between the ground state and the first excited state is small but finite and decreases as $\propto 1/N$.
It closes in the thermodynamic limit $N \rightarrow \infty$, causing both the correlation length $\xi$ and relaxation time $\tau$ to diverge.
Assuming annealing times fast enough to produce at least one defect on average, \gls{lz} theory yields the same \gls{kz} scaling in a quantum setting~\cite{Zurek2005dynamicsQPT}.

For uniform couplings $J$, defects in the ferromagnetic solution after annealing appear as domain walls between anti-aligned spins and can be measured by $\sigma_i^z\sigma_j^z$ correlators over all $\Ne$ edges of the spin-lattice, i.e., all nearest-neighbor correlators.
The density of defects is measured as the density of spin misalignments (domain walls) in 1D,
\begin{equation}
    \ndef = \frac{1}{2 \Ne} \sum_{\langle i, j \rangle}^{\Ne} \bigl( 1 - \sigma_i^z \sigma_j^z \bigr) \ .
\label{eq: defect density pauli op}
\end{equation}
For example, an open chain of $N$ spins has $\Ne=N-1$ edges and a periodic chain has $\Ne=N$ edges.
In arbitrary higher-dimensional systems, the scaling of the number of misaligned spins may not exactly match the scaling of the number of defects, i.e., domains of spin alignment.
This is particularly true when the system features few but large domains.
In that case, one could directly count the number of defects, taking into account the graph topology or, for instance, measure the average number of domains from the number of spin-misalignments along different directions~\cite{Du2023kzmIsingDomains}.

\subsection{Density of defect scaling across various regimes}
\label{subsec: beyond kzm}

As mentioned above, the \gls{qkzm} is not expected to hold for every possible annealing time $\tf$.
Deviations from the \gls{kz} scaling occur at both very short and very long $\tf$~\cite{Zeng2023kzmFastQuenches, Schmitt2022kzm2D,Chandran2012kibble}.
For finite system sizes and very slow anneals, i.e., large $\tf$, Ref.~\citenum{Zurek2005dynamicsQPT} describes an exponential drop in defect density following the regime of \gls{kz} scaling that is captured by \gls{lz} dynamics.
In that case, when the anneal is slow enough to never freeze out and to produce less than one defect in the chain on average, the number of defects is proportional to the \gls{lz} probability $p_\mathrm{LZ}$ of exciting the system,
\begin{equation}
    \ndef \propto p_\mathrm{LZ} \approx \exp \Bigl( - b\frac{\tf}{N^2} \Bigr) \ ,
\label{eq: lz scaling}
\end{equation}
with $b$ a constant.
Following Refs.~\citenum{Morita2008annealingMathFoundations,Caneva2008beyondKZM}, for adiabatically slow anneals, i.e., even larger $\tf$, the scaling reads
\begin{equation}
    \ndef
    \approx p_\mathrm{LZ}(\tf) + \frac{1 - 2 p_\mathrm{LZ}(\tf)}{ a \tf^2},
\label{eq: ftlz scaling}
\end{equation}
where $a$ is a constant that depends on the derivatives of the annealing schedules~\cite{Morita2008annealingMathFoundations}.
The expressions of constants $a$ and $b$ can be found in Refs.~\citenum{Zurek2005dynamicsQPT,Caneva2008beyondKZM}.

The theoretical predictions for the three scaling regimes in \Cref{eq: kz scaling 1D,eq: lz scaling,eq: ftlz scaling} are shown in \Cref{fig: schematic and statevector KZM}\textbf{b}.
Density of defects scalings obtained from ideal statevector simulations in \Cref{fig: schematic and statevector KZM}\textbf{c} with system sizes up to $N=20$ spins show that the beyond-\gls{kzm} scaling regimes, i.e., \Cref{eq: lz scaling,eq: ftlz scaling} are finite-size effects and that their onset is controlled by the system size.

\section{Application-oriented benchmarking through universal scaling}
\label{sec: benchmarking}

Here, we suggest a direct approach to benchmark close to applications.
We simulate the time evolution under a time-dependent Hamiltonian and benchmark the accuracy with which a known universal scaling is replicated.
A schematic of our method is shown in \Cref{fig: schematic and statevector KZM}\textbf{a}.
We propose to measure, for example, the density of defects after digitized \gls{qa} as a function of total annealing time $\tf$.
Given a fixed time step, the time $\tf$ directly corresponds to the number of circuit layers in the Trotterized time evolution circuit.
In the \gls{kz} regime, the density of defects will decrease according to the predicted scaling up to a certain threshold $\tf$, after which hardware noise becomes dominant and leads to a deviation from the \gls{kz} scaling.
The number of circuit layers for which the expected scaling of $\ndef$ is observed is the number of circuit layers for which reliable simulation was achieved.

In particular, decoherence of the system will lead to a deviation of the defect density from the predicted \gls{kz} scaling.
This has been previously confirmed in several numerical and analog experiments that studied the effect of dissipation on the \gls{kz} scaling~\cite{Dutta2016antiKZM,Arceci2018optimalWorkingPoint,Bando2020dwave,King2022dwave2000IsingChain}.
These studies show that different kinds of noise can cause decoherence of the system.
However, the behavior of the density of defects as noise accumulates, and whether it increases or decreases, depends on the noise.

In this regard, it is important to note that observing a $\tf^{-1/2}$ scaling may not be a sufficient condition to claim coherent quantum dynamics. 
For instance, a purely classical diffusion model may reproduce the same density of defects scaling~\cite{Mayo2021kinkDistr,King2022dwave2000IsingChain}.
This is particularly important for analog quantum annealers. 
In the presence of sufficiently strong thermal noise that destroys coherence, the density of defects could still decrease with  $\tf$, since the machine may behave like a classical thermal annealer (at low temperature) or as a machine featuring incoherent quantum tunneling events~\cite{Boixo2014evidence,Denchev2016finiteRangeTunneling,Isakov2016quantumTunnelingQMC}.
These processes could still lower the density of defects.
Indeed, this effect has been observed in Ref.~\citenum{King2022dwave2000IsingChain}, necessitating additional analysis to reasonably prove the existence of coherent annealing~\cite{Dziarmaga2022kinkcorrelations}.

In our digital quantum setting, the situation is different: the presence of hardware noise never yields a behavior that resembles a classical thermal limit.
To prove this, in \Cref{sec: benchmarking noisy simulations}, we study the impact of simulated hardware noise with varying strength.
We observe that introducing hardware noise qualitatively changes $\ndef$ as a function of $\tf$.
Even for large $\tf$ and accumulating noise, we do not recover a decreasing $\ndef(\tf)$, contrary to the noisy analog quantum annealing case~\cite{King2022dwave2000IsingChain}.
Therefore, the density of defects following the expected scaling constitutes evidence of a noise-free digital simulation up to a threshold $\tf$.
This argument is further corroborated by the fact that the noise model is in very good agreement with hardware results (see \Cref{sec: benchmarking noisy simulations}).

The proposed benchmark has several advantages.
First and foremost, it yields a concrete, intuitive, and scalable metric; the number of simulable circuit layers as they can be found in countless applications.
Even more so since Hamiltonian simulation is a prime application for quantum computers and a building block for many other applications, for example, quantum phase estimation~\cite{Motta2021emergingWIRES} and sampling algorithms~\cite{Layden2023quantumMCMC}.
Second, our method can benchmark hardware and \gls{ems} algorithms separately or in combination.
Third, our method is scalable since no classical verification is required.
Fourth, since the \gls{kz} scaling is defined in the thermodynamic limit $N \rightarrow \infty$, it is particularly well-suited to benchmark large digital quantum computers without scaling issues.
Finally, since finite-size scaling regimes are also well-understood (cf. \Cref{subsec: beyond kzm} and \Cref{fig: schematic and statevector KZM}\textbf{b},\textbf{c}), the method is applicable in the setting of scaling to large circuit depths at qubit counts far below the thermodynamic limit.

\subsection{Influence of hardware noise}
\label{sec: benchmarking noisy simulations}

\begin{figure}[t]
    \center
    \includegraphics[width=0.6\textwidth]{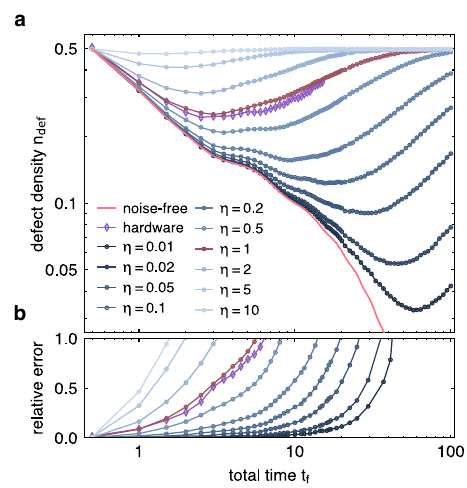}
    \mycaption{Density of defects scaling in the presence of simulated hardware noise}{
        \textbf{a} All curves show simulated Trotterized time evolution of a 12-qubit chain with $\Delta t = 0.5$.
        The noise-free reference curve (pink line) represents a statevector simulation.
        The hardware curve (purple diamonds) corresponds to results obtained from \sherbrooke{} using only \gls{rem}.
        All other curves show results of sampling $10^5$ measurements from a simulated circuit and include simulated hardware noise that corresponds to the full noise model of \sherbrooke{}.
        \textbf{b} Error of the density of defects relative to the noise-free results.
    }
    \label{fig: noisy simulations}
\end{figure}

The effects of noise, such as a finite temperature or bath-couplings, on \gls{qa} have been previously studied both numerically~\cite{Arceci2018optimalWorkingPoint, Bando2020dwave} and experimentally~\cite{King2022dwave2000IsingChain} in analog settings.
The result is a deviation from the \gls{kz} scaling and a subsequent increase in defect density.
As discussed in \Cref{sec: benchmarking}, the effect of the noise encountered in a digital simulation is expected to be fundamentally different from that in an analog simulation.
Specifically, digital hardware noise includes an accumulation of individual basis gate errors, mainly two-qubit gate errors, throughout the circuit and will not yield a classical thermal limit.

Here, we perform Trotter simulations of digitized \gls{qa} including a tunable noise model resembling the digital noise we encounter on our devices.
Since noise generally leads to decoherence of the system, this further validates our claim that observing a density of defects that follows the \gls{kz} scaling implies coherent evolution.
Scaling the average device errors across several orders of magnitude allows for a detailed understanding of their influence on the density of defects.
In other words, the noisier the system, the sooner we expect decoherence and, in turn, a deviation from the expected scaling.

The noise model is constructed from error rates of \sherbrooke{} obtained through standard calibrations~\cite{IBMQuantumPlatform}.
Specifically, we used \textsc{Qiskit Aer}'s method \linebreak \texttt{NoiseModel.from\_backend\_properties()}~\cite{qiskit2024} that generates a simplified noise model based on calibration data of a specific device.
The resulting noise model includes readout errors on measurements as well as individual basis gate errors consisting of depolarizing noise (tuned through gate errors obtained from calibrations), followed by thermal relaxation noise (tuned through decoherence times $T_1, T_2$ and gate execution times).
To simplify the scaling of the noise, we use error rates obtained from the average of individual gate and qubit properties.
Concretely, our model applies the following noise to all qubits,
\bea{C}
\begin{IEEEeqnarraybox}[][c]{C}
T_1 = T_1^{\mathrm{ave}} / \eta 
\ , \quad
T_2 = T_2^{\mathrm{ave}} / \eta \\
e_{\mathrm{1q}} =  \eta e_{\mathrm{1q}}^{\mathrm{ave}}
\ , \quad
e_{\mathrm{2q}} =  \eta e_{\mathrm{2q}}^{\mathrm{ave}}
\ , \quad
e_{\mathrm{ro}} =  \eta e_{\mathrm{ro}}^{\mathrm{ave}} \ .
\label{eq: benchmarking scaled noise}
\end{IEEEeqnarraybox}
\eea
Here, $T_1, T_2$ are relaxation time and dephasing time, $e_{\mathrm{1q}}, e_{\mathrm{2q}}, e_{\mathrm{ro}}$ are single-qubit gate, two-qubit gate, and readout errors, respectively, and $\eta$ is a scaling factor.
The average values are $T_1^{\mathrm{ave}} = \qty{266.37}{\us}$, $T_2^{\mathrm{ave}} = \qty{178.71}{\us}$, $e_{\mathrm{1q}}^{\mathrm{ave}} = 1.25 \times 10^{-3}$, $e_{\mathrm{2q}}^{\mathrm{ave}} = 1.10 \times 10^{-2}$, and $e_{\mathrm{ro}}^{\mathrm{ave}} = 2.41 \times 10^{-2}$.
Moreover, all gate and measurement times are left unchanged and correspond to those of \sherbrooke{}.

\Cref{fig: noisy simulations}\textbf{a} shows the density of defects obtained from \gls{qa} of a periodic 12-qubit system with simulated noise scaled by $\eta \in \{ 0.01, \ldots,10 \}$.
The noise-free line is a statevector simulation of the same system.
In addition, we plot real hardware results of simulating the same 12-qubit system on \sherbrooke{}, using only \gls{rem}.
The results of this experiment match the noisy simulations with unscaled noise $\eta=1$, thereby validating our choice of noise model.
Both the noisy and the noise-free simulations, as well as the hardware experiments, were done with the same time step of $\Delta t = 0.5$ as in \Cref{sec: benchmarking}.

Since the system size that we can simulate exactly with noise is small, the behavior of the density of defects over time appears less smooth compared to larger sizes.
These finite size effects manifest as a drop in the noise-free density of defects after roughly $\tf \approx 10$.
The noise-free simulations indicate a monotonically decreasing density of defects with $\tf$.
Instead, in the noisy cases, the density of defects starts increasing with time after a threshold time $\tf^*$. 
As the noise strength decreases, $\tf^*$ increases.
This is further underlined by \Cref{fig: noisy simulations}\textbf{b}, showing the relative error of the density of defects with respect to the noise-free curve, $|(\ndef(\eta) - \tilde{n}_\mathrm{def}) / \tilde{n}_\mathrm{def} |$, with $\tilde{n}_\mathrm{def}$ the noise-free defect density.
While this behavior is compatible with earlier numerical studies on dissipative quantum annealing and analog hardware experiments~\cite{Arceci2018optimalWorkingPoint,King2022dwave2000IsingChain,Bando2020dwave}, the main difference here is that the density of defects always keeps increasing with time after $\tf^*$. 
In contrast, dissipative quantum dynamics or annealing experiments~\cite{King2022dwave2000IsingChain,King2023spinGlass}, indicated that the density of defects may decrease again, possibly with the same $-1/2$ exponent, even beyond the coherence limit.

\subsection{Experimental setup}
\label{subsec: benchmarking experimental setup}

We now discuss benchmarking different levels of \gls{ems} on two quantum processors through digitized \gls{qa} and measuring defect densities.
An $N$-spin Ising model with uniform couplings $J=1$ and linear schedules (cf. \Cref{sec: quantum annealing}) is mapped to $N$ qubits.
The qubit register is initialized as $\psi(t=0) = \ket{+}^N$, the ground state of $H(t=0) = \HM$, and time evolved under the time-dependent Hamiltonian in \Cref{eq: hamiltonian ising annealing}.
The time evolution is implemented using a first-order \gls{pf} as in \Cref{eq: full PF-1 time-dep} with a time step of $\Delta t$.
Throughout this section, we use a time step of $\Delta t = 0.5$.
We justify the choice of this time step in \Cref{app: time step}, showing results of noiseless statevector simulations with different time steps and comparing them to continuum results.
Although smaller time steps would yield higher accuracy, more Trotter steps, i.e., deeper circuits, would be required to explore the relevant \gls{kz} regime.
Our results show that $\Delta t = 0.5$ is a reasonable compromise to avoid significant algorithmic errors and allow for sufficiently large annealing times.
Hardware improvements will eventually allow for smaller time steps and thereby increase the overall accuracy of the results, particularly at short time scales.
After annealing for a total time $\tf$, we measure the defect density via \Cref{eq: defect density pauli op} to compare to the expected scaling in \Cref{eq: kz scaling 1D}.
Expectation values of observables are estimated from measurements with \estimator{} primitive~\cite{qiskit2024}.

\begin{figure}[t]
    \centering
    \includegraphics{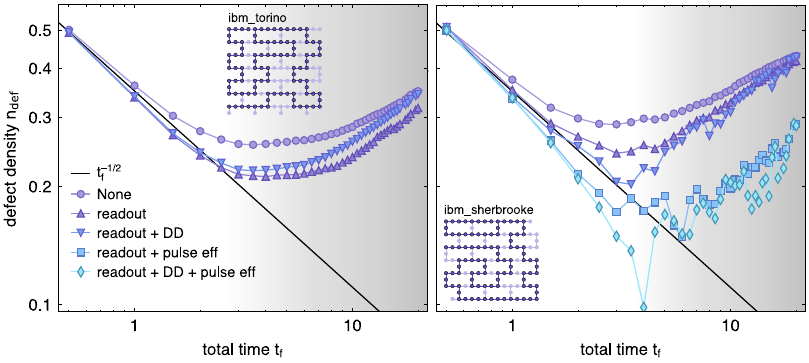}
    \mycaption{Density of defects scaling in 1D from 100-qubit circuits}
    {Hardware results comparing the density of defects scaling on subsets of 100 qubits on \torino{} (left) and \sherbrooke{} (right) employing different levels of EMS.
    Each point corresponds to one additional Trotter layer or time step of $\Delta t = 0.5$.
    Raw simulation results with no EMS (None) are compared to results using combinations of REM, DD error suppression, and pulse-efficient transpilation.
    Insets show the qubit layout of the respective processor with the chosen qubit subset highlighted.
    The grey shading indicates decoherence and deviation from the expected KZ scaling shown by the black solid line.}
    \label{fig: kzm device 1D}
\end{figure}

We employ the IBM Heron processor \torino{} and the IBM Eagle processor\linebreak \sherbrooke{} for our simulations, with 133 and 127 qubits, respectively, and heavy-hexagonal qubit connectivity~\cite{IBMQuantumPlatform}.
Hardware characteristics such as error rates are given in \Cref{app: hardware properties}.
Eagle and Heron processors differ in their native two-qubit gates and how they are physically realized.
\sherbrooke{}'s native two-qubit gate is an \gls{ecr} gate implemented via a dispersive coupling mediated by a fixed-frequency resonator~\cite{Rigetti2010ecrGate, Sheldon2016ecrGate}.
By contrast, \torino{}'s native two-qubit gate is a \cz{} gate realized through tunable-frequency couplers.
In a tunable coupler architecture, the coupling element is frequency tunable, and driving it can create different two-qubit gates~\cite{Mckay2016universal, Ganzhorn2020benchmarking}.
Most importantly, this results in \torino{} having a roughly $6 \times$ shorter two-qubit gate time compared to \sherbrooke{} and a higher two-qubit gate quality~\cite{Mckay2023layerFidelity}.

On both devices, we compare simulations with no \gls{ems} to \gls{rem}~\cite{Berg2022trex,Nation2021m3} as well as to \gls{rem} combined with \gls{dd}~\cite{Barron2024qaoa}.
In addition, \sherbrooke{} allows for pulse-efficient transpilation of our circuits, another method of error suppression~\cite{Earnest2023pulse}.
Pulse-efficient transpilation scales hardware native cross-resonance pulses and their echoes (combined, this makes up an \gls{ecr} gate on \sherbrooke{}) and is thus not compatible with \torino{}'s native \cz{} gates.
\Cref{sec: error mitigation} gives a brief technical summary of each method.

\subsection{Results: 1D chain}
\label{subsec: benchmarking 1D results}

The Trotter circuit for the time evolution of a spin chain requires at least two layers of $\rzz$ gates per Trotter layer, or time step.
Since each $\rzz$ gate is transpiled to two hardware native two-qubit gates (ECR or CZ), the final circuit for a 1D Ising model has a two-qubit gate depth of $4 \times N_t$ where $N_t$ is the number of time steps.
\Cref{fig: kzm device 1D} shows the density of defects scaling obtained from digitized \gls{qa} on subsets of 100 qubits connected through an open line (see insets), which was chosen to minimize cumulative two-qubit gate errors along the line (cf. \Cref{app: hardware properties}).
Owing to noise, the experimental results differ from the noiseless statevector results in \Cref{fig: schematic and statevector KZM}\textbf{c}, which show a monotonic decrease in defect density with increasing $\tf$.
Hardware errors only make it possible to follow the KZ scaling and subsequent scaling regimes up to a certain threshold time.
Beyond this point (indicated by grey shading), the defect density deviates from the $t^{-1/2}$ scaling and eventually increases, as the noise accumulates with increasing circuit depth, as predicted by the noisy emulation of the algorithm.

This threshold time depends on the hardware and the \gls{ems} employed.
The top panel of \Cref{fig: kzm device 1D} shows the results for \torino{} comparing no \gls{ems} with only \gls{rem}, and with \gls{rem} combined with \gls{dd}.
While we can simulate two Trotter layers with a two-qubit gate depth of 8 and a total of 396 CZ gates without any level of \gls{ems}, adding \gls{rem} already significantly improves the results.
We can reliably simulate up to five Trotter layers with a two-qubit gate depth of 20 and 990 CZ gates in total.
Assuming \torino{}'s median \cz{} gate time of $84 \, \mathrm{ns}$ for all two-qubit gates, this corresponds to a circuit execution time of $\qty{1.7}{\us}$.
This is remarkable given that \gls{rem} is the simplest and easiest-to-implement form of \gls{ems}.
We attribute the negligible impact of the \gls{dd} sequence on the \torino{} results to the reduced cross-talk of the device compared to \sherbrooke{}.

The picture is different on \sherbrooke{}, shown in the bottom panel of \Cref{fig: kzm device 1D}.
We compare the same \gls{ems} methods as on \torino{} in addition to pulse-efficiently transpiled circuits.
The results without any \gls{ems} and only \gls{rem} are slightly worse than on \torino{}.
However, adding \gls{dd} to counter static $ZZ$ cross-talk results in substantial gains on this device.
According to our metric, we achieve a reliable simulation of seven circuit layers with a two-qubit gate depth of 28 and 1386 ECR gates in total.
Assuming \sherbrooke{}'s median ECR gate time of $\qty{533}{\ns}$~\cite{IBMQuantumPlatform} for all two-qubit gates, this equates to a circuit execution time of roughly $\qty{15}{\us}$.
Moreover, we show in \Cref{app: kink-kink correlator} that the correlations between defects show the characteristic non-monotonic fingerprint of a genuine quantum \gls{kzm}~\cite{Nowak2021quantum,Dziarmaga2022kinkcorrelations}.
Pulse-efficient transpilation fares similarly, though we observe that the curve dips below the expected scaling.
We attribute this to rotation errors in the scaled two-qubit pulses at small angles~\cite{Earnest2023pulse} (cf. \Cref{sec: error mitigation}).
As the two-qubit gate angles $\theta = -2J B(m \Delta t / \tf) \Delta t = -2J m \Delta t^2 / \tf$, with $m\Delta t = t$, are proportional to the annealing schedule, increasing $\tf$ means smaller and smaller angles at the beginning of an anneal.
This causes a growing disparity between the target operation of the circuit before transpilation and the final sequence of gates with growing $\tf$.
Fixing these rotation errors requires custom calibrations~\cite{Vazquez2023wellconditioned}, which are difficult at scale through a cloud-based quantum computing service.
Since our goal is to compare out-of-the-box \gls{ems} methods, we did not conduct any custom calibrations.

\begin{figure}[t!]
    \centering
    \includegraphics[width=\textwidth]{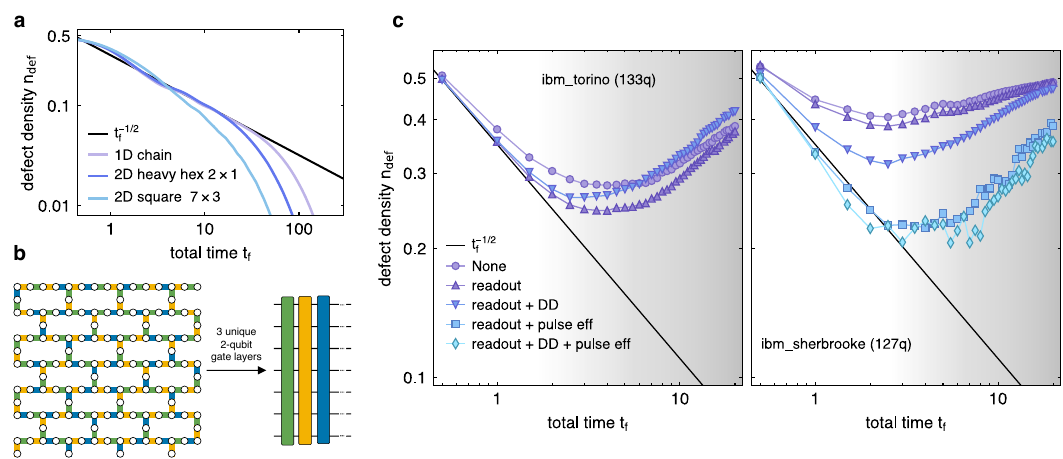}
    \mycaption{Density of defects scaling on a heavy-hexagonal lattice from 133- and 127-qubit circuits}
    {\textbf{a} Statevector results comparing the density of defects scaling in a periodic chain, a heavy-hexagonal lattice consisting of two heavy-hex cells ($2 \times 1$), and a $7 \times 3$ square lattice, all with $N=21$ spins.
    The heavy-hexagonal lattice exhibits the same $t^{-1/2}$ scaling as predicted for 1D systems, whereas defects in a true 2D lattice show a steeper scaling.
    \textbf{b} The couplings on the heavy-hexagonal lattice of \torino{} and \sherbrooke{} are grouped such that each Trotter layer consists of three layers of $R_{ZZ}$ gates.
    \textbf{c} Hardware results comparing the density of defects scaling on \torino{} (left) and \sherbrooke{} (right) employing different levels of EMS.
    Each point corresponds to one additional Trotter layer or time step of $\Delta t = 0.5$.
    Raw simulation results with no EMS (None) are compared to results using combinations of REM, DD error suppression, and pulse-efficient transpilation.
    The grey shading indicates decoherence and deviation from the expected KZ scaling shown by the black solid line.}
    \label{fig: kzm device heavy-hex}
\end{figure}

\subsection{Results: heavy-hexagonal lattice}
\label{subsec: benchmarking heavy hex results}

Our approach is not limited to one-dimensional spin systems.
In fact, such critical behavior and the resulting universal scaling manifest in lattice geometries beyond 1D~\cite{Schmitt2022kzm2D} and can hence benchmark any qubit connectivity, provided that critical exponents and scaling are known.

The processors we use have a 2D heavy-hexagonal qubit connectivity.
In this case, theoretical comparisons are less straightforward than in the 1D case.
\Cref{fig: kzm device heavy-hex}\textbf{a} shows noiseless statevector simulations comparing the density of defects scaling of a 1D periodic chain, a heavy-hex lattice, and a 2D square lattice with $N=21$ sites each.
Here, the density of defects in 1D and heavy-hex lattices coincide on the $t^{-1/2}$ scaling before the exponential drop-off at larger $\tf$, attributed to finite-size effects, while the 2D square lattice shows a steeper scaling.
However, this could be an artifact of the small system sizes since a heavy-hex lattice of 21 sites might not be large enough to show genuine 2D scaling.
We consider the $t^{-1/2}$ scaling line as a reference line also for experiments on the heavy-hex lattice, even though deviations from the 1D scaling are possible for this geometry.
Given also that this 2D system is only sparsely connected, with an edges-over-sites ratio of 1.2, meaning correlations will spread slowly across the lattice and may not acquire 2D character within a few time steps, we still adopt \Cref{eq: defect density pauli op} to measure the density of defects.

The Trotter circuit for a Hamiltonian \Cref{eq: hamiltonian ising annealing} on a heavy-hex lattice is made depth-optimal by grouping the couplings according to the edge-coloring shown in \Cref{fig: kzm device heavy-hex}\textbf{b}.
Each Trotter layer in the circuit therefore consists of three layers of $\rzz$ gates.
Again, each $\rzz$ gate is transpiled into two hardware-native two-qubit gates, resulting in a total two-qubit gate depth of $6 \times N_t$.

Utilizing all 133 and 127 qubits of \torino{} and \sherbrooke{}, respectively, we see an even starker difference between both chips in defect density (\Cref{fig: kzm device heavy-hex}\textbf{c}).
Taking the 1D scaling as a reference line as argued above, we can reliably simulate up to three Trotter layers on \torino{} employing only \gls{rem}.
With 150 edges, this equals a two-qubit gate depth of 18 and 900 CZ gates in total.
Again, assuming \torino{}'s median \cz{} gate time of $\qty{84}{ns}$ for all two-qubit gates, this implies a circuit execution time of $\qty{1.5}{\us}$ using merely \gls{rem}.
\sherbrooke{}, on the other hand, requires more involved \gls{ems}, which becomes even more apparent when using all qubits of the device.
Although \gls{dd} again provides sizable improvements, we do not reproduce the expected \gls{kz} scaling.
However, pulse-efficient transpilation combined with \gls{rem} extends the circuit depth up to five Trotter layers before noise becomes prevalent.
This is equivalent to a Trotter circuit with a CNOT gate depth of 30 and 1440 CNOT gates in total.
In this case, the circuit duration reduces to roughly $\qty{6}{\us}$ due to the pulse-efficient transpilation.

\section{Digitized quantum annealing for optimization}
\label{sec: optimization}

The proposed benchmark provides a reasonable prediction of hardware and error mitigation capabilities for a set of problems broader than the non-equilibrium quantum dynamics of a ferromagnetic Ising model.
Estimating how many circuit layers can be reliably executed before noise prevails is useful in many applications.
For instance, variational algorithms~\cite{Wu2024variationalBenchmarks, Astrakhantsev2023phenomenological} for many-body spin systems feature circuits very similar to the ones considered in the previous section.
Circuits with the same structure may also be used in quantum machine learning~\cite{Havlivcek2019quantumFeatureMaps, Melo2023pulseQML, Abbas2021power}.
Finally, the entangling blocks used previously can be adapted to \gls{qaoa} or digitized \gls{qa} of disordered spin models.

This section studies digitized \gls{qa} in the context of solving combinatorial optimization problems, using the same underlying circuits as in the benchmarking experiments, and showcases how results from \Cref{sec: benchmarking} are transferable to related applications.
Specifically, we now allow for non-uniform couplings $J_{ij}$ in \Cref{eq: hamiltonian ising} that can be arbitrary in magnitude and sign.
Optimization problems defined by this type of Hamiltonian may fall within the NP-complete complexity class, depending on the connectivity of the underlying graph, and are frequently employed as benchmarks for quantum optimization~\cite{Ronnow2014quantumSpeedup, Albash2018scalingAdvantageAnnealing}.
It is therefore of interest to determine whether the results obtained from benchmarking the \gls{qkzm} and identifying an optimal working point are transferable to an optimization context.
This is not straightforward since the \gls{qkzm} is connected with second-order phase transitions, while the bottlenecks in quantum annealing of hard optimization problems are due to first-order transitions in the glassy phase~\cite{Young2010firstOrderPT, Joerg2010firstOrderPT, Knysh2016spinGlass}.

\Gls{qa} has been little explored on digital quantum computers~\cite{Keever2023adiabaticQC}.
Instead, the focus has been mostly on \gls{qaoa}~\cite{Farhi2014qaoa}, an annealing-inspired variational ansatz with variational layers representing time steps in digitized \gls{qa}, used in conjunction with \gls{vqe}.
Due to its variational nature and the related hope of smaller circuit depths, \gls{qaoa} has become the most popular near-term algorithm for quantum optimization~\cite{Barron2024qaoa, Sack2024qaoa, Hadfield2019qaoa}.
However, recent studies suggest that solving hard optimization problems requires many \gls{qaoa} layers~\cite{Zhou2020lukinQAOA, Farhi2020qaoaLargeP}.
At the same time, optimizing the variational parameters in \gls{qaoa} is costly in itself~\cite{Scriva2024qaoa} and, once found, the optimal parameter values of \gls{qaoa} in the large-layer limit reduce to annealing schedules~\cite{Zhou2020lukinQAOA}, meaning \gls{qaoa} implements digitized \gls{qa} with variationally found annealing schedules.
This does not mean that the optimal \gls{qaoa} parameters can be directly derived from annealing schedules.
Rather, the variational optimization results in optimized annealing schedules.
Nonetheless, precisely because of this relationship, annealing schedules can be used to initialize \gls{qaoa} parameters, thereby reducing the training cost of the variational ansatz~\cite{Scriva2024qaoa, Sack2019qaoa}.
It is therefore sensible to study digitized \gls{qa} in more detail.
Previous experimental realizations of digitized \gls{qa} featured only up to nine qubits~\cite{Barends2016digitizedQA}.

Since digitized \gls{qa} requires discretizing the time evolution, a central question is what effect does the choice of the time step have on the success of the annealing.
In principle, a large time-step error could, counterintuitively, even improve the performance of digitized \gls{qa}.
This effect has been observed in classical path-integral simulations of \gls{qa}, where an unconverged, finite Trotter step is beneficial for tunneling between configurations~\cite{Heim2015spinGlass}, reconciling earlier expectations of quantum speed-up~\cite{Santoro2002spinGlass} with observations~\cite{Ronnow2014quantumSpeedup}.
While investigating this aspect, we identify an optimal time step for digitized \gls{qa} given a fixed number of circuit layers or time steps $N_t$.
Fixing $N_t$ is meaningful since, in reality, one can only access a limited amount of resources such as the expendable number of gates, determining both the computational cost and, more importantly, before achieving fault-tolerance, the amount of error introduced by hardware noise.
The success of \gls{qa} relies on as large as possible annealing times $\tf$, which is achieved by increasing the time step $\Delta t$ for a fixed number of time steps $N_t$.
On the other hand, the errors of time evolution algorithms typically scale with $\Delta t$~\cite{Miessen2023perspective}.
\Glspl{pf} are derived on the assumption of small time steps $\Delta t \ll 1$ and algorithmic errors scale with $\mathcal{O}(\Delta t^2)$~\cite{Childs2021trotterError}.
This naturally introduces a trade-off between \gls{qa} performance and the algorithmic error due to $\Delta t$~\cite{Sack2019qaoa}.
Note that, despite this trade-off, having a converged time step is no requirement here.
Instead, the sole objective is to obtain the best possible solution, i.e., the lowest energy.
This differs from \Cref{sec: benchmarking} where we chose the largest possible time step that is still reasonably close to the continuum while allowing us to reach as-large-as-possible $\tf$.

Since the solution to a classical optimization problem is a single bitstring, we are usually interested in individual measurement samples rather than expectation values.
Therefore, we use \qiskit{}'s Sampler primitive~\cite{qiskit2024} throughout this section to obtain individual measurement samples, i.e., bitstrings of measurement outcomes.
Many \gls{em} techniques apply only to expectation values.
By contrast, error suppression methods such as \gls{dd} and pulse-efficient transpilation apply to sampling.
However, as discussed in \Cref{sec: benchmarking}, \gls{dd} does not yield substantial improvements on \torino{}. 
Furthermore, IBM Heron processors, such as \torino{}, are based on tunable couplers and thus do not allow for pulse-efficient transpilation, which was designed for cross-resonance-based hardware~\cite{Earnest2023pulse}.
To make results comparable across different devices, we therefore only use \gls{rem} in this section.
Furthermore, we will compute the residual energy,
\begin{equation}
    \Eres = E(\tf) - E_0 \ ,
\end{equation}
where $E(\tf) = \braket{\psi(\tf) | H(\tf) | \psi(\tf)}$ and $E_0$ is the exact ground state obtained through exact diagonalization for small system sizes and using CPLEX~\cite{cplex} for large systems.

\subsection{Dependence of the residual energy on the time step and spectral gap}
\label{subsec: optimization auckland}

\begin{figure}[t]
    \centering
    \includegraphics[width=\textwidth]{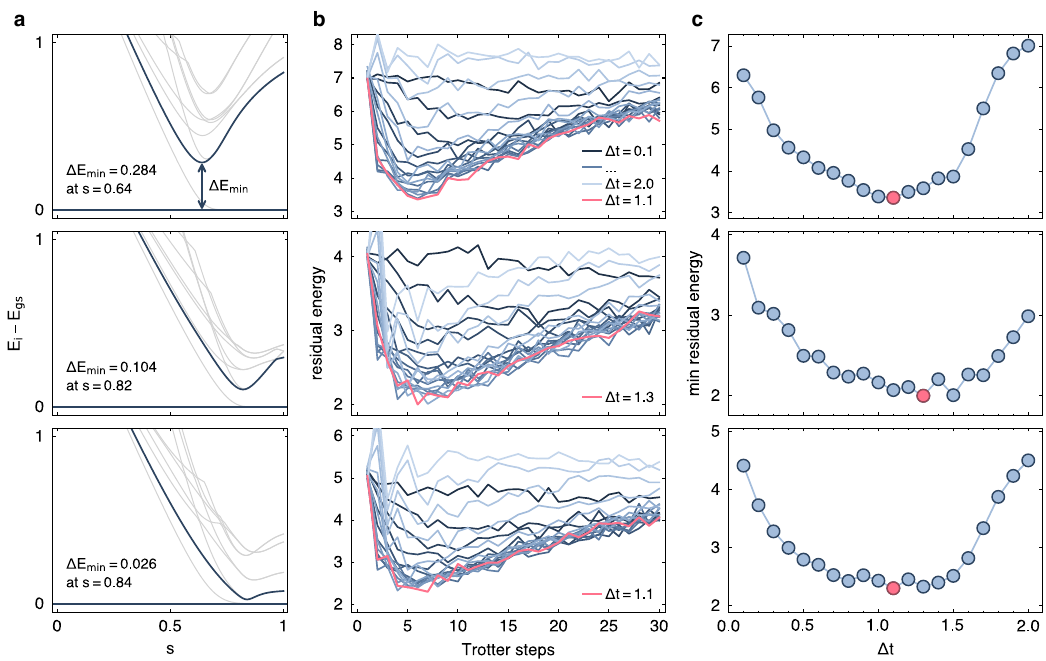}
    \mycaption{Hardware results of the residual energy dependence on time step and spectral gap}{
        \textbf{a} Spectra relative to the ground state of three different instances of disorder in a 12-qubit periodic chain with coupling coefficients uniformly sampled from $J_{ij} \in [-1, 1]$.
        The instances were specifically chosen with three different minimum energy gaps, differing by an order of magnitude between the largest (top) and the smallest gap (bottom).
        \textbf{b} Residual energy averaged over $400$ samples obtained from QA of the system in \textbf{a} on \auckland{} using only REM with fixed time steps $\Delta t \in \{ 0.1, \ldots, 2.0 \}$ as a function of circuit depth (number of time steps).
        \textbf{c} Minimum residual energy from \textbf{b} as a function of the time step, i.e., each point corresponds to the minimum over one curve in \textbf{b}.
    }
    \label{fig: residual energy auckland}
\end{figure}

We first study the dependence of the digitized \gls{qa} result on the time step on a small system of a periodic 12-qubit chain.
Three instances of disorder with couplings randomly sampled from $J_{ij} \in [-1, 1]$ are chosen with varying minimum energy gaps between the ground and the first parity-preserving excited state, since a smaller gap entails longer annealing times and makes finding a solution generally more difficult.
Note that the Hamiltonian \Cref{eq: hamiltonian ising annealing} and therefore its spectrum is a function of $s$.
The minimum energy gap takes on its smallest value at the critical point.
Furthermore, the first excited state of opposite parity becomes degenerate with the ground state as $s \rightarrow 1$.
The spectra of the three different Hamiltonians as a function of $s$ are shown in \Cref{fig: residual energy auckland}~\textbf{a} with minimum gaps ranging across one order of magnitude.
\Cref{fig: residual energy auckland}~\textbf{b} displays the residual energy after annealing the respective system from \textbf{a} on \auckland{}~\footnote{\auckland{} is one of IBM's by now retired 27-qubit Falcon chips} with fixed time step $\Delta t \in \{ 0.1, \ldots, 2.0 \}$ as a function of the number of time steps (depth).
Here, we observe that the smallest residual energies are reached after approximately $5-10$ Trotter layers, owing to decoherence.
The observation of a minimum is compatible with both numerical predictions~\cite{Arceci2018optimalWorkingPoint} and experimental results from analog simulation~\cite{King2022dwave2000IsingChain, Ebadi2022optimizationRydberg}.
Furthermore, it is analogous to what was observed in the benchmarking experiment \Cref{fig: kzm device heavy-hex}~\textbf{c}, where we also located a minimum defect density after roughly 5-10 Trotter layers.

For each $\Delta t$, the minimum residual energy achieved over all circuit depths is plotted in \Cref{fig: residual energy auckland}~\textbf{c}.
For all three systems, the lowest residual energy is obtained with a time step of $\Delta t > 1$, specifically $1.0 < \Delta t < 1.5$.
Noiseless statevector simulations confirm that this is not merely a consequence of hardware noise, as seen in \Cref{app: optimization statevector}, yielding an optimal time step of $1.2 < \Delta t < 1.4$.
However, in statevector simulation, the existence of a finite optimal time step is likely due to the finite range of depths chosen.
Therefore, choosing an infinitesimal time step seems to be both (i) inefficient in noise-free simulations and (ii) impractical in real experiments.
On the other hand, a too-large time step implies algorithmic errors.

To summarize, we observe a trade-off between realizing the largest possible annealing times while keeping the algorithmic error under control.
Choosing an unconverged, large time step is indeed advantageous when using digitized \gls{qa} for optimization.
This is of further relevance when initializing \gls{qaoa} variational parameters mentioned previously.
In the context of directly studying \gls{qaoa} performance, Ref.~\cite{Sack2019qaoa} finds an optimal time step of $0.75$ for initializing variational parameters.
Increasing $\Delta t$ even further, however, does not provide any benefits as shown by the statevector results.
This observation is consequential for practical noisy settings and allows us to optimize the number of Trotter steps needed to reach a target annealing time.

\subsection{Optimization of disordered heavy-hexagonal graph}
\label{subsec: optimization torino}

\begin{figure}[t]
    \centering
    \includegraphics[width=\textwidth]{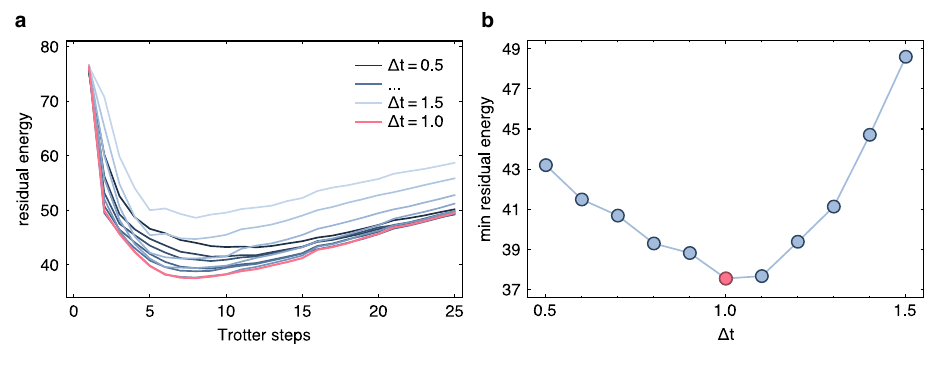}
    \mycaption{Hardware results of the residual energy dependence on the time step}{
        \textbf{a} Residual energy of a heavy-hexagonal 133-qubit graph with coupling coefficients uniformly sampled from $J_{ij} \in [-1, 1]$, averaged over $10^5$ samples obtained from QA on \torino{} with fixed time steps $\Delta t \in \{ 0.5, \ldots, 1.5 \}$ as a function of circuit depth (number of time steps).
        \textbf{b} Minimum mean residual energy from \textbf{a} as a function of the time step, i.e., each point corresponds to the minimum over one curve in \textbf{a}.
    }
    \label{fig: residual energy torino}
\end{figure}

We now study a heavy-hexagonal spin-lattice of 133 qubits, i.e., \torino{}'s full chip.
We consider just one realization of disorder with uniformly sampled couplings $J_{ij} \in [-1, 1]$, and compute $\Eres$ as a function of circuit depth and time step as before.
The ground state is again computed using CPLEX.
This time, the system is too large to compute its instantaneous higher excited states, which is why we have no information about the size of its minimum energy gap.
However, the comparison of different systems in the previous section indicates that the dependence of the optimal time step on the spectral gap is small.
\Cref{fig: residual energy torino}~\textbf{a} shows the results for time steps $\Delta t \in \{0.5, \ldots 1.5 \}$.
Again, we observe that, for each time step, the minimum residual energy is obtained at around $5-10$ Trotter layers, or time steps, after which decoherence begins to dominate.
\Cref{fig: residual energy torino}~\textbf{b} reports the minimum $\Eres$ for each value of the time step.
Also here, we observe a continuous decrease in $\Eres$ with increasing $\Delta t$ up to a time step of $1.0$ to $1.1$, after which $\Eres$ sharply increases.
In conclusion, this suggests an unconverged, large time step of $\Delta t \gtrsim 1.0$ to be optimal for digital \gls{qa} regardless of the system instance and graph connectivity.
Further studies will be required to explore the generality of this empirical finding.

\subsection{Consistency with the KZ benchmark}
\label{subsec: relationship optimization}

Here, we briefly summarize the relation of our large-scale optimization results in \Cref{subsec: optimization torino} to the results of \Cref{sec: benchmarking}.
In \Cref{sec: benchmarking}, the 1D ferromagnetic system (i.e., uniform couplings) simulation on \torino{}, see \Cref{fig: kzm device 1D}, indicates that about 5 Trotter steps can be executed reliably before the density of defects substantially deviates from the expected scaling (referring to the \gls{rem} curve for consistency with the optimization experiments).
Subsequently, at about 8 Trotter steps, the density of defects reaches a minimum before it increases due to the accumulated noise.
Similar values of 3 and 8 Trotter steps, respectively, are obtained on the heavy-hex lattice, see \Cref{fig: kzm device heavy-hex}.

The optimization experiment shares the same connectivity as these benchmarks but has random couplings and different time steps.
Depending on the time step used, the minimum energy we measure (which is the disordered analog to the defect density) lies again in the range of 5-10 Trotter steps, see \Cref{fig: residual energy torino}~\textbf{a}.
This finding is consistent with our results from \Cref{sec: benchmarking}.
When the goal is solely to minimize an expectation value, as in applications such as optimization or variational methods, the location of the minimum in $\ndef(\tf)$ can be relevant in addition to the point where it starts deviating from the expected scaling.

This demonstrates the usefulness of our benchmarking routine in an application context.
Suppose, for example, that one is interested in using \gls{qaoa} rather than digitized \gls{qa}.
Since \gls{qaoa} features the same circuit structure in a closed-loop classical parameter optimization, it is expensive to discover the optimal circuit depth empirically.
Running the proposed benchmark beforehand allows one to directly select the optimal \gls{qaoa} depth according to the hardware capabilities.

\section{Conclusion}
\label{sec: benchmarking conclusion}

In this chapter, we presented an application-oriented method to jointly benchmark hardware and \gls{ems} algorithms.
Developing such a method to predict simulation quality and, importantly, more closely resemble device and algorithm capabilities, is critical.
Particularly so in times of rapid algorithmic and hardware advancements.
The simple figure of merit that we introduce is the number of Trotter steps, i.e., the threshold depth, before which errors induce deviations from a universal scaling law.
Interestingly, the method has already seen adoption by researchers across a range of computational platforms and has been utilized to benchmark digital~\cite{Visuri2025kipuDCA} and analog~\cite{McGeoch2024dwaveIBMcomparison} quantum simulation, as well as classical methods~\cite{Lerch2024efficientClassicalSimulation}, emphasizing its versatility and practical relevance.
While we propose to use digitized \gls{qa} and the universal \gls{kz} scaling in our work, we emphasize the central idea: benchmark against the prediction of a known universal behavior.
Other scaling laws, model Hamiltonians, or lattice geometries could be employed as well.

We observed that different machines, equipped with different \gls{ems}, provide different threshold depths.
Importantly, the simple benchmark we propose can measure the continuous improvements of hardware and algorithms in the near future.
Moreover, it is tailored to Hamiltonian simulation, which is among the most anticipated applications with provable quantum advantage in physics and a building block in many other applications.

We tested our method on hardware native geometries and the resulting circuits contain only dense layers of $\rzz$ gates.
However, this does not limit its predictive power since all two-qubit gates are eventually transpiled to the same hardware-native two-qubit gate.
One can therefore directly transfer results in terms of two-qubit-gate depth to the simulation of more complicated models.
Nonetheless, future research could devise more varied models exhibiting universal behavior that could be used analogously to the \gls{kz} scaling.
For instance, circuits with interleaved two-qubit gates of different kinds that do not commute with each other, such that they cannot be arranged in a dense layer of gates, as encountered for example in fermionic models.
Furthermore, our method could be used to select best-qubit subsets.
By measuring all nearest-neighbor $\sz_i \sz_j$ correlators on a given quantum processor, the defect densities of different qubit-subsets can be reconstructed in post-processing.
For example, computing the density of defects for all qubit subsets of size $N$ in this way would allow selecting the subset that best matches the \gls{kz} scaling.
Moreover, the known finite-size scaling regimes for large annealing times could one day be used to benchmark very deep quantum circuits once they become accessible with advanced \gls{ems} techniques or, eventually, \gls{qec}.

Our benchmark extends beyond quantum many-body simulations to quantum optimization and variational algorithms, as we show in \Cref{sec: optimization}.
Importantly, we observe consistency between the threshold number of Trotter steps before the noise becomes prevalent, identified using the \gls{kz} scaling in \Cref{sec: benchmarking}, and the optimal depth of digitized \gls{qa} for combinatorial optimization in \Cref{sec: optimization}.
We demonstrate that, counter-intuitively, the time step should indeed be chosen at a value $\gtrsim 1.0$, resulting in significantly improved residual energy values after annealing.
This means in practice that digitized \gls{qa} could be a competitive quantum optimization framework as it (i) avoids a classical optimization loop and (ii) greatly reduces the runtime of \gls{qaoa}~\cite{Weidenfeller2022scalingofquantum}, avoiding the costly iterative optimization, affected by hardware and shot noise~\cite{Scriva2024qaoa, Mazzola2024quantum}.
More generally, our results suggest that current hardware with simple \gls{ems} may support variational ansatze (for a similar class of Hamiltonians) comprising about ten Trotter-circuit layers.
Our results seem to be robust against several instances of disorder with varying minimum spectral gaps as well as on systems of different sizes and connectivities.
Nonetheless, future work should aim at exploring other settings, such as solving dense graphs~\cite{Sack2024qaoa}.

In introducing this benchmark, we report some of the largest-scale digital quantum simulations as of today~\cite{Kim2023utility, Farrell2023scalable, Pelofske2024scalingQAOA}. 
Compared to Ref.~\citenum{Kim2023utility}, we move closer to a physical system by incorporating Trotter error and time-dependent, general two-qubit gate angles, requiring a full gate decomposition of the $\rzz$ gate into hardware native two-qubit gates.
However, we do not introduce sophisticated \gls{ems} such as probabilistic error amplification, which could be a goal of future work.

Finally, we emphasize that the purpose of this paper is not to claim direct quantum advantage with the featured experiments, including the optimization experiments in \Cref{sec: optimization}.
For one, the sparse-graph optimization problem studied here is not hard. 
Other quantum or classical platforms may demonstrate better performance than we do~\cite{Rehfeldt2023maxCut}.
Moreover, we do not claim that the digitized quantum annealing method is the best-performing digital algorithm.
Indeed, warm-start methods~\cite{Egger2021warmstart, Tate2023warmstart} and digitized counterdiabatic approaches~\cite{Chandarana2022counterdiabatic} all improve the performance of quantum approximate optimization.
Instead, the value of our optimization experiments lies in understanding its correlation with the benchmark proposed in \Cref{chap: benchmarking} and to guide better informed choices when it comes to parameter initialization in \gls{qaoa} or choosing the right time step in digitized \gls{qa}.
Nonetheless, the benchmarks introduced here will be instrumental to verify future quantum simulations on sufficiently large hardware, beyond the reach of approximate classical methods.

%% file: ch5_pea_oqs.tex
\chapter{Open quantum dynamics through partial probabilistic error amplification}
\label{chap: pea oqs}

\begin{zitat}{}
    This chapter is based on unpublished work that may, however, be published at a later time.
    \Cref{sec: pea oqs pea} provides a brief but technically detailed recap of probabilistic error amplification (PEA).
    In \Cref{sec: pea different noise model}, we present the main contribution of this chapter and propose a novel method to extrapolate from a learned hardware noise model to a different noise model, which one seeks to simulate.
    We evaluate random errors inferred by the extrapolation and derive an analytical error bound in \Cref{sec: pea oqs random error}.
    Subsequently, in \Cref{sec: pea oqs numerical scaling}, we validate this error bound numerically.
    Lastly, in \Cref{sec: peq oqs diss ising}, we present a first proof-of-concept, extrapolating between two different noise models to simulate a dissipative Ising model.
    Alexander Miessen carried out all calculations and simulations, implemented the code, and conducted the data analysis and visualization.
\end{zitat}

Currently available quantum devices are noisy, as described in previous parts of this thesis (see \Cref{sec: intro hardware,chap: benchmarking}).
Although manifold in its origin, most noise stems from an imperfect isolation of the system and a resulting interaction with an environment.
In essence, this means that the quantum system we want to employ for simulations is not an isolated, closed quantum system, but an open quantum system (see \Cref{chap: methods} for details).
Here, the open character of the system is typically unwanted and everything possible is done to decouple the system from the environment and mitigate errors caused through these interactions.

In certain cases, however, it is desirable to include specific environment interactions (also called a bath) and simulate open quantum systems, for example, in the study of non-equilibrium steady states or dissipative phase transitions~\cite{Hendrik2021openSystemsRMP, ORourke2020simplified, Kshetrimayum2017steadyStates, Helmrich2018openSystems, Fink2017photonBlockade}.
Applications in which the explicit modeling of an environment is necessary were described in detail in \Cref{chap: perspective}.
However, while full control over the system and all interactions is necessary to simulate such models, system-environment interactions present in a quantum processor can typically not be controlled by the user through standard instructions and operations.
This begs the question of whether the latter can be characterized and controlled to be incorporated in the simulation of open quantum systems.

Simulating open quantum dynamics with noise-free, unitary operations necessitates additional resources to implement the system-bath interaction and, often, the implementation of non-unitary dynamics (see also \Cref{chap: methods,subsec: algorithms open systems,sec: perpective - open systems}).
In addition, unwanted hardware noise would still be present in the simulation and have to be mitigated.
Gaining control over such hardware noise to implement parts or all of the bath interactions one seeks to simulate could alleviate some of these complications.
Most importantly, it would mean combining the steps of implementing additional operations and mitigation of noise, and lead to a reduction of computational resources, possibly including intricate non-unitary operations.

In a possible realization, this would imply a hybrid digital-analog scheme, where one digitally implements the time evolution of a system of interest, and utilizes the inherent -- analog -- hardware noise~\cite{Bluvstein2022hybridDigitalAnalog, Andersen2025analogDigitalKZ}.
Ideas to implement open quantum dynamics by engineering interactions in a quantum device to model specific system-environment interactions have emerged and been tested in various shapes in recent years~\cite{Lloyd2001engineeringQuantumDynamics, Verstraete2009dissipativeEngineering, Barreiro2011openSystems, Reiter2016dissipativePreparation, GarciaPerez2020openSystems, DelRe2020drivenDissipative, Harrington2022engineeredDissipation, Cattaneo2023dissipative, Guimaraes2023partialPEC, Guimaraes2024partialPECtimeDep, Google2024dissipativeEngineering}.
Within this field of research, most approaches fall into the category of dissipative engineering or noise engineering, where certain controls of the system, such as driving pulses or controllable interactions, are tuned to implement the interaction.

An alternative approach, as hinted at above, is to utilize the device noise algorithmically.
Ref.~\citenum{Guimaraes2023partialPEC} presented such a method based on \gls{pec}~\cite{Berg2023pec} (see \Cref{sec: methods pea}).
The central part of that algorithm is a variant of \gls{pec}, which first learns a noise model of a given hardware, and then partially mitigates it to implement another noise model.
This approach has two important limitations.
First, it is limited to Pauli errors as described in \Cref{sec: methods pea} of the form $P \rho P'$ with Pauli strings $P = P'$, such as for example dephasing or depolarizing channels.
However, the method cannot be used to implement more complicated noise channels, particularly for non-unital noise of the form $P \rho P'$ with $P \neq P'$, such as amplitude damping noise.
To implement non-unital noise, conventional methods such as dilation or mid-circuit measurements are necessary.
In addition, it is known that mitigating any noise channel using \gls{pec} introduces a sampling overhead that scales exponentially with the noise strength that should be mitigated.
This limitation remains even when only partially mitigating errors (see the corresponding discussion in Ref.~\citenum{Guimaraes2023partialPEC}).

Here, we propose to follow a similar direction, but using \gls{pea}~\cite{Kim2023utility} to partially mitigate noise and effectively modify the device noise to a different noise model.
\Gls{pea} was developed to overcome the exponential sampling overheads inferred by \gls{pec}.
It utilizes the same noise characterization pipeline as \gls{pec} but, instead of cancelling the noise directly in the circuit, amplifies it in a first step to then extrapolate back to zero noise in a second step (see \Cref{sec: methods pea} and below).
We detail a variant of \gls{pea} that only partially mitigates a learned noise model, effectively extrapolating from an amplified noise model to a different noise model instead of to zero noise.
This improves upon the proposal of Ref.~\citenum{Guimaraes2023partialPEC}, mitigating its exponential sampling overhead.
However, this approach is still limited to Pauli noise.

There exist numerous possible applications in the domain of open quantum dynamics that could benefit from quantum simulation, which we summarized in detail in \Cref{sec: perpective - open systems}.
In the context of unital Pauli noise, investigating models of exciton transfer represents a particularly interesting direction~\cite{potocnik2018lightHarveting}.
On a high level, these systems aim to model the complex processes underlying photosynthesis by transferring a localized excitation through a system.
Crucially, it is believed that dissipation arising from an interaction of the system with an environment plays a central role in increasing transport efficiency~\cite{Landi2022boundaryDrivenQuantumSystemsRMP, MendozaArenas2013dephasingEnhancement, MendozaArenas2013noiseAssistedTransport, Maier2019noiseAssistedTransport}.
A broad class of more simplistic toy models for these kinds of systems are given by so-called boundary-driven quantum systems, or quantum batteries~\cite{Landi2022boundaryDrivenQuantumSystemsRMP}.
Such models represent ideal testbeds for our method and will be studied in future work.

This chapter is organized as follows.
We first summarize the technical details of regular \gls{pea} in \Cref{sec: pea oqs pea}, including a detailed derivation of the functional form of an expectation value as a function of the amplified noise.
Subsequently, in \Cref{sec: pea different noise model}, we introduce our method of partial \gls{pea}, discussing all possible cases one can encounter in extrapolating between different noise models.
Here, we also derive a functional form of expectation values measured within this protocol as a function of the noise amplification.
This is followed by a rigorous analysis of the random errors occurring in extrapolating these expectation values and the derivation of an error bound.
We validate this bound through numerical simulations.
Lastly, we simulate the open quantum dynamics of a dissipative Ising model by extrapolating from one to another noise model in \Cref{sec: peq oqs diss ising}.

\section{A brief recap of probabilistic error amplification}
\label{sec: pea oqs pea}

We gave a high-level overview of the working principles of \gls{pea} and \gls{zne} in \Cref{sec: methods pea}, following mostly Refs.~\citenum{Kim2023utility} and \citenum{Berg2023pec}.
In this chapter, however, we will describe \gls{pea} in more technical detail.
We will expand on the previously used notation and largely follow Ref.~\citenum{Filippov2024peaTheory}.

Suppose we want to estimate the expectation value of an observable $\mathcal{O}$ at the end of an $N$-qubit quantum circuit that is given by a unitary channel $\mathcal{U}$, consisting of layers $\mathcal{U}_l$.
We assume that noise enters the circuit as a noisy channel $\Lambda_l$ preceding every ideal unitary layer $\mathcal{U}_l$, resulting in a noisy gate layer $\Lambda_l \circ \mathcal{U}_l$.
Noise predominantly originates from two-qubit gates, and so $\mathcal{U}_l$ are in practice two-qubit gate layers.
\Gls{pea} relies on the assumption of a learnable sparse Pauli noise model.
Once learned, the noise model can be sampled to amplify the noise by a power (the noise gain) $G > 1$.
Specifically, an additional noisy layer $\Lambda_l^{G-1}$ is sampled into the circuit, resulting in an effective noise layer $\Lambda_l^G$.
\Gls{pea} prescribes to evaluate the expectation value $\braket{\mathcal{O}}^{(G)}$ at different noise gains $G$, fit the functional form $F(G)$ of the expectation value, and extrapolate to zero noise, $F(0)$.

\subsection{Learning the noise model}

When constructing the sparse Pauli noise model, only single- and two-qubit Pauli channels are considered, and higher-order terms in the full Pauli noise model of $4^n$ terms are neglected.
Concretely, there are three single-qubit error channels, $\{X, Y, Z \}$ for each qubit and nine two-qubit error channels $\{ XX, YY, ZZ, XY, XZ, YX, YZ, ZX, ZY \}$ for each qubit pair with corresponding error rates $\lambda_\alpha$.
The noise learning aims to learn the error rates $\lambda_\alpha^{(l)}$ for every Pauli error channel $P_\alpha$ in the noise channel $\Lambda_l$ of each unique layer $l$.
Here, the Pauli channel $P_\alpha = \bigotimes_{q=0}^{N-1} P_{\alpha_q}$ is a string of Paulis $P_{\alpha_q} \in \{ I, X, Y, Z \}$ with $\alpha = (\alpha_0, \ldots, \alpha_{N-1})$.
The resulting learning model is a set of Pauli rates that we denote $\mathrm{SPL}(\Lambda_l)$.

The noise model is constructed from the learned rates as
\begin{equation}
    \Lambda_l [\cdot] = \exp \biggl[ \sum_\alpha \lambda_\alpha^{(l)} (P_\alpha \cdot P_\alpha - \cdot) \biggr] \ ,
\label{eq: pea unamplified noise map}
\end{equation}
Importantly, any Pauli observable $\mathcal{O} = P_\beta$ is an eigenvector of the noise model $\Lambda_l$,
\begin{equation}
    \Lambda_l[P_\beta] = f_{l \beta} P_\beta \ .
\label{eq: Pauli noise map eigenvalue}
\end{equation}
The corresponding eigenvalue is the Pauli fidelity,
\begin{equation}
    f_{l \beta} = \prod_{\lambda_\alpha \in \mathrm{SPL(\Lambda_l)}: \{ P_\beta, P_\alpha \} = 0} e^{-2 \lambda_\alpha} \ .
\label{eq: pauli fidelities}
\end{equation}
Concretely, this means, Pauli observables are sensitive to all Pauli noise channels that anti-commute with the observable.
Here, the product is over all rates $\lambda_\alpha$ that make up the respective noise model for layer $l$, $\mathrm{SPL(\Lambda_l)}$.

\subsection{Evaluating expectation values at amplified noise}
\label{sec: pea expectation values}

Amplifying the noise to a noise gain $G > 1$ amounts to sampling an additional noise channel $\Lambda_l^{\eta}$ into each noisy layer $\Ut_l = \Lambda_l \circ \U_l$ with $\eta = G - 1 > 0$.
In practice, this is done via the routine described in \Cref{sec: methods pea}, which we repeat here for the reader's convenience.
The noise model $\Lambda_l$ applied to a state $\rho$ can then be written as a composition in the form of
\begin{equation}
	\Lambda_l [\rho] = \prod_\alpha \Bigl( w_\alpha^{(l)} \cdot + (1 - w_\alpha^{(l)}) P_\alpha \cdot P_\alpha^\dagger \Bigr) \rho \ .
\end{equation}
Specifically, the respective error channel $P_k$ occurs with probability $1-w_k$, where
\begin{equation}
	w_\alpha^{(l)} = \frac{1 + e^{-2 \lambda_\alpha^{(l)}}}{2} \ .
\label{eq: methods pea probability}
\end{equation}
As a result, since we can write
\begin{equation}
    \mathcal{U}_l \circ {\Lambda}_l \circ {\Lambda}_l^\eta = \mathcal{U}_l \circ {\Lambda}_l^{\eta + 1} \ .
\end{equation}
Amplifying the noisy channel from $1 \rightarrow \eta + 1$ amounts to sampling additional error channels with probabilities determined via scaled rates $\eta \lambda_\alpha^{(l)}$ into each circuit layer $l$.
The scaled noise model then reads
\begin{equation}
    \Lambda_l^G [\cdot] = \exp \biggl[ \sum_\alpha \lambda_\alpha^{(l)} ( \eta + 1) (P_\alpha \cdot P_\alpha - \cdot) \biggr] \ ,
\end{equation}
Therefore, scaling the noise amounts to scaling the Pauli rates $\lambda_\alpha^{(l)}$ that determine the distribution from which Pauli error terms are sampled.

We are interested in measuring an expectation value $\braket{\O} = \tr \{ \O \rho_\mathrm{final} \}$ under noise gain $G$.
Here, $\rho_\mathrm{final}$ is the outcome of the simulation after evolving an initial state $\rho_0$ through a series of noisy gate layers.
We adopt here the notation of Ref.~\citenum{Filippov2024peaTheory}, defining left-to-right composition $\bigcirc_l^{\rightharpoonup} A_l = A_0 \circ \ldots \circ A_L$ and right-to-left composition $\bigcirc_l^{\leftharpoonup} A_l = A_L \circ \ldots \circ A_0$.
Starting from an initial state $\rho_0 = \ket{0}\bra{0}^N$ and evolving through a total of $L$ circuit layers, the expectation value under amplified noise can be rewritten as (see \Cref{app: pea expectation values} and Ref.~\citenum{Filippov2024peaTheory} for a detailed derivation)
\bea{rCl}
\braket{\O}^{(G)}_{\rm noisy} & = &
    \tr \Bigl\{
    \mathcal{O} \cdot \Bigl( \overset{\leftharpoonup}{\bigcirc}_l \U_l \circ \Lambda_l^G \Bigr) [\rho_0]
    \Bigr\} \nonumber \\[4pt]
& = & 
    \tr \Bigl\{
    \overset{\rightharpoonup}{\bigcirc}_l \Tilde{\Lambda}_l^{G\dagger} [\mathcal{O}]
    \cdot \Tilde{\rho}_\mathrm{ideal}
    \Bigr\} \ .
\label{eq: noisy expectation value}
\eea
Here, the second factor represents the noise-free Schrödinger evolution of the initial state through all ideal circuit layers $\U_l$,
\begin{equation}
    \Tilde{\rho}_\mathrm{ideal} := \overset{\leftharpoonup}{\bigcirc}_l \U_l [\rho_0] \ .
\end{equation}
The first factor is the Heisenberg propagation of all noisy layers $\Lambda_l^{G\dagger}$, as derived in \Cref{app: pea expectation values} and defined as
\begin{equation}
	\Tilde{\Lambda}_l^{G\dagger}
    :=
    \U_{\geq l} \circ {\Lambda_l^G}^\dagger \circ \U_{\geq l}^\dagger
    \ ,
\end{equation}
with $\U_{\geq l} = \overset{\leftharpoonup}{\bigcirc}_{m \geq l} \U_m$.

We now need to distinguish between Clifford and non-Clifford circuits~\cite{NielsenChuang2010}.
A Clifford circuit consists of only Clifford gates that are defined as leaving a Pauli observable unchanged upon evolution, $C P C^\dagger = P'$ for any Pauli strings $P, P'$.
However, Clifford circuits are efficiently classically simulable~\cite{Aaronson2004cliffordSimulation} and circuits in practice are rarely fully Clifford.
As we will see in the following, whether or not the unitary layers $\U_l$ are Clifford determines the functional form of the expectation value as a function of $G$.

Specifically, a Pauli-observable $\O = P_\beta$ will be propagated (in the Heisenberg picture) through each unitary $\U_{\geq l}^\dagger$ differently, depending of whether or not the circuit is Clifford.
Evolving $\O$ through circuit layers $\U$ consisting only of Clifford operations results in a single Pauli string,
\begin{equation}
    \U_{\geq l}^\dagger [P_\beta] = P_{\beta(l)} \ ,
\end{equation}
where $P_{\beta(l)}$ is the Pauli string resulting from propagation through Clifford layers  $l \ldots L$.
Evolution through all noisy layers and using \Cref{eq: Pauli noise map eigenvalue} gives $\Lambda_l^{G \dagger} [P_\beta] = f_{l \beta}^G P_\beta$, resulting in (see \Cref{app:  pea expectation values})
\begin{equation}
    \overset{\rightharpoonup}{\bigcirc}_l \Tilde{\Lambda}_l^{G\dagger} [P_\beta] = \prod_l f_{l \beta(l)}^G P_{\beta} \ .
\end{equation}
The expectation value \Cref{eq: noisy expectation value} then evaluates to
\begin{equation}
	\braket{\O}^{(G)}_{\rm noisy} = \Bigl( \prod_l f_{l \beta(l)} \Bigr)^G \braket{\O}_{\rm ideal} \ ,
\label{eq: noisy expectation value clifford}
\end{equation}
where $\braket{\O}_{\rm ideal} = \tr \bigl\{ P_{\beta} \cdot \Tilde{\rho}_\mathrm{ideal} \bigl\}$ is the noise-free expectation value.
Hence, in the case of Clifford circuits, the functional form of a noisy expectation value as a function of the noise gain $G$ is a single exponential.

Instead, in the more general setting of non-Clifford circuits, evolving a Pauli observable through the circuit results in a linear combination of Paulis,
\begin{equation}
    \U_{\geq l}^\dagger [P_\beta] = \sum_\alpha c_{\beta \alpha}^{(l)} P_\alpha \ .
\end{equation}
Evolving through the noise layers again weights each Pauli with its respective fidelity.
Therefore, the result of evolution through general, noisy, non-Clifford circuit layers is a complex sum of rescaled Pauli strings.
This means, the functional form of the expectation value as a function of the noise gain, $F(G)$, will be a sum of exponentials of $G$~\cite{Filippov2024peaTheory}.

\section{Extrapolating between different Pauli noise models}
\label{sec: pea different noise model}

\Gls{pea} extrapolates to zero noise by fitting the noise-amplified expectation values and extrapolating $G = \eta + 1 \rightarrow 0$, or, equivalently, $\eta \rightarrow -1$.
In the extrapolation limit, this amounts to scaling all Pauli rates $\lambda_k$ of the approximate noise model $\mathrm{SPL}(\Lambda_l)$ to $0$.
Instead of extrapolating to zero noise, here we want to extrapolate to a different Pauli noise model $\mathrm{SPL}(\tilde{\Lambda}_l)$ with modified rates $\Tilde{\lambda}_\alpha^{(l)}$.
This means we are not globally scaling the noise model, i.e., scaling all rates at once.
Instead, we need to apply local scaling factors $\delta_\alpha$ to each Pauli rate individually,
\begin{equation}
    \lambda_\alpha^{(l)} \rightarrow \lambda_\alpha^{(l)} (\delta_\alpha \eta + 1) \ .
\label{eq: pea local scaling}
\end{equation}
Requiring that extrapolation to $\eta \rightarrow - 1$ yields the new noise model $\Tilde{\lambda}_\alpha^{(l)}$, we can distinguish several cases that determine our choice of $\delta_\alpha$.
More specifically, given a set of Pauli error channels up to order two that can be present in both the original and the target noise model, we can think of five scenarios for changing the noise model accordingly:

\begin{enumerate}
    \item[1.] $P_\alpha \in \mathrm{SPL}(\Lambda_l), P_\alpha \in \mathrm{SPL}(\tilde{\Lambda}_l)$
    \begin{enumerate}
        \item[a)] $\tilde{\lambda}_\alpha < \lambda_\alpha \ \rightarrow \ \delta_\alpha = 1 - \frac{\tilde{\lambda}_\alpha}{\lambda_\alpha} \in (0, 1)$\\
		These channels are amplified according to local scaling factors $\delta_\alpha$.
		\item[b)] $\tilde{\lambda}_\alpha = \lambda_\alpha \ \rightarrow \ \delta_\alpha = 0$\\
		The original hardware noise matches the target noise.
		The local scaling factors are automatically $0$ and no additional noise is sampled.
		\item[c)] $\tilde{\lambda}_\alpha > \lambda_\alpha \ \rightarrow \ \delta_\alpha = \frac{1}{\eta} \bigl( \frac{\tilde{\lambda}_\alpha}{\lambda_\alpha} - 1 \bigr)$\\
		The respective error channel only needs to be amplified and no scaling through $\eta$ is necessary.
    \end{enumerate}
    \item[2.] $P_\alpha \notin \mathrm{SPL}(\Lambda_l), P_\alpha \in \mathrm{SPL}(\tilde{\Lambda}_l)$ \\
    This is a special case of (1.c), with $\lambda_\alpha = 0$.
    The error channel is not present in the original noise model and so needs to be sampled into the circuit.
    No scaling is needed.
    \item[3.] $P_\alpha \in \mathrm{SPL}(\Lambda_l), P_\alpha \notin \mathrm{SPL}(\tilde{\Lambda}_l) \rightarrow \delta = 1$ \\
    The error channel is not present in the target noise model and needs to be fully mitigated.
    This case reduces to regular \gls{pea}.
\end{enumerate}

These cases cover all possible combinations of changing between two different sparse Pauli noise models.
It is important to note that, while the respective Pauli rates of each channel are scaled locally, the extrapolation will be done globally, averaging over all Pauli channels as in conventional \gls{pea}.

\subsection{Evaluating expectation values at locally amplified noise}
\label{sec: pea oqs expectation values}

Analogously to \Cref{sec: pea expectation values} and still following the machinery of Ref.~\citenum{Filippov2024peaTheory}, we now derive the functional form of an expectation value measured under locally amplified noise as a function of the noise gain.
A locally amplified noise channel reads (labeled now by $\delta \eta$ to indicate the difference to a global noise gain $G$)
\begin{equation}
    \Lambda_l^{(\delta\eta)} [\cdot] = \exp \biggl[ \sum_\alpha \lambda_\alpha^{(l)} (\delta_\alpha \eta + 1) (P_\alpha \cdot P_\alpha - \cdot) \biggr] \ .
\end{equation}
The new eigenvalues of this map are locally scaled Pauli fidelities,
\begin{equation}
    \Lambda_l^{(\delta\eta)} [P_\beta]
    = \prod_{\{ l ; \alpha ; \beta \}}
    e^{-2 \lambda_\alpha (\delta_\alpha \eta + 1)} P_\beta
    \ ,
\end{equation}
where we abbreviated the set over which the product is defined as $\{ l ; \alpha ; \beta \} \equiv \{ \lambda_\alpha \in \mathrm{SPL(\Lambda_l)}: \{ P_\beta, P_\alpha \} = 0 \}$.
Again, we assume a Clifford circuit in propagating a Pauli observable $\O = P_\beta$ through all noisy circuit layers and, in complete analogy to \Cref{sec: pea expectation values}, arrive at
\bea{rCl}
    \overset{\rightharpoonup}{\bigcirc}_l \Tilde{\Lambda}_l^{(\delta\eta)\dagger} [P_\beta]
    =
    \Bigl(\prod_l \prod_{\{ l ; \alpha ; \beta(l) \}}
    e^{-2 \lambda_\alpha (\delta_\alpha \eta + 1)} \Bigr)
    P_{\beta} \ .
\eea
The preliminary expression for the noisy expectation value is then
\begin{equation}
    \braket{\O}^{(G)}_{\rm noisy}
    =
    \prod_l \prod_{ \{ l ; \alpha ; \beta(l) \} } e^{-2 \lambda_\alpha (\delta_\alpha \eta + 1)}
    \braket{\O}_{\rm ideal} \ .
\label{eq: pea oqs expectation value general}
\end{equation}

For each of the above cases, we can evaluate this product as is shown in \Cref{app: pea oqs cases functional form}.
In the case of (1.a) above, i.e., the respective Pauli channel $P_\alpha$ being present both in the original and the target noise model with $\tilde{\lambda}_\alpha < \lambda_\alpha$, we have $\delta_\alpha = 1 - \frac{\tilde{\lambda}_\alpha}{\lambda_\alpha}$.
Similar to the Pauli fidelities \Cref{eq: pauli fidelities}, we can define
\begin{equation}
    \Tilde{f}_{l\beta}
    = \prod_{ \{ l ; \alpha ; \beta \} }
    e^{-2 \tilde{\lambda}_\alpha} \ .
\end{equation}
Crucially, $\Tilde{f}_{l\beta}$ here denotes the Pauli fidelities with respect to target rates $\tilde{\lambda}$ but defined over the original noise model $\mathrm{SPL(\Lambda_l)}$, meaning the set over which the product is defined \textit{never changes},
Together with $\eta = G - 1$, we obtain
\begin{equation}
    \braket{\O}^{(G)}_{\rm noisy}
    = \Bigl(\prod_l \prod_{ \{ l ; \alpha ; \beta(l) \} } e^{-2 (\lambda_\alpha - \Tilde{\lambda}_\alpha )} \Bigr)^G 
    \Bigl( \prod_l \Tilde{f}_{l\beta(l)} \Bigr)
    \braket{\O}_{\rm ideal} \ .
\end{equation}
Further defining $\prod_l f_{l\beta(l)} =: K$ and $\prod_l \Tilde{f}_{l\beta(l)} =: \Kt$, the expression can be written more concisely
\begin{equation}
    \braket{\O}^{(G)}_{\rm noisy} = \biggl( \frac{K}{\Kt} \biggr)^G \Kt \braket{\O}_{\rm ideal} \ .
\label{eq: pea oqs expectation value final}
\end{equation}

From this and \Cref{app: pea oqs cases functional form}, we can conclude that the functional form of the expectation value measured on the output of a Clifford circuit is again either a single exponential function in $G = \eta - 1$ or a constant.
This is analogous to regular \gls{pea}, the difference being the extra factors of $\Tilde{K}$.
For non-Clifford circuits, this generalizes to a sum of exponentials analogously to regular \gls{pea}.

Note that, most generally, the noise channel will be composed of all of the above cases.
Since the noise map is a composition of all individual error channels, the expectation value will be a product of the different cases evaluated in \Cref{app: pea oqs cases functional form}.
This does not present a problem for the fitting and extrapolation in practice.
In fact, any Pauli error channel will contribute either an exponential in $G$ or a constant pre-factor, meaning the overall functional form will remain exponential in $G$ (single or multi-exponential, depending on whether or not the circuit is Clifford).

\subsection{Random error propagation and sampling overheads}
\label{sec: pea oqs random error}

Next, we want to estimate the random error in the final extrapolated value $F(G=0)$ stemming from imprecisely estimating the expectation values.
We consider a Clifford circuit and a stabilizing observable $O\ket{\psi} = \ket{\psi}$ such that $\braket{\O}_{\rm ideal} = 1$.
Then, the expectation value most generally takes the form \Cref{eq: pea oqs expectation value final}.
The following derivation is analogous to that presented in Ref.~\citenum{Filippov2024peaTheory} but generalized to our case of partial \gls{pea} and presented in more detail.

In the amplification step, $\braket{\O}^{(G)}_{\rm noisy}$ is measured at several noise gains $\{ G_i \}$.
We cannot measure the expectation value $\braket{\O}^{(G)}_{\rm noisy}$ to infinite accuracy.
Instead, for each noise gain $G_i$, we estimate the expectation value with $S_i$ measurement shots.
The resulting estimator $\Bar{\mathcal{O}}(G_i)$ will fluctuate around $(K / \Kt)^G \Kt$ with error $\Delta \Bar{\mathcal{O}}(G_i) \propto S_i^{-1/2}$.
The functional form of the expectation value is a single exponential.
To exponentially extrapolate these estimated expectation values to $G_i \rightarrow 0$, we perform a linear fit on the logarithm of the exponential,
\bea{rCl}
    Y(G_i) = \ln \bigl[ \Bar{\mathcal{O}}(G_i) + \xi_i \Delta\Bar{\mathcal{O}}(G_i) \bigr]
    & \overset{\Delta\Bar{\mathcal{O}} \ll \Bar{\mathcal{O}}}{\approx} &
    \ln \bigl[ \Bar{\mathcal{O}}(G_i) \bigr] + \xi_i \Delta\Bar{\mathcal{O}}(G_i) / \Bar{\mathcal{O}}(G_i) \\
    & = & y_i + \xi_i \Delta y_i \ .
\eea
Here, for each $i$, $\xi_i$ is a normally distributed random variable with mean $0$ and variance $1$.
Note that $\xi_i$ is not averaged over all indices $i$ but instead each $\xi_i$ for fixed $i$ has mean $0$.
In other words, $i$ does not index different realizations of $\xi_i$ but refers to the noise gain $G_i$.
For the extrapolation function $y_i(x_i)$ with error $\Delta y_i$, we have
\begin{equation}
x_i = G_i \quad , \quad \quad y_i = G_i \ln \frac{K}{\Kt} + \ln \Kt \quad , \quad \quad \Delta y_i = \biggl( \frac{\Kt}{K} \biggr)^{G_i} \frac{1}{ S_i^{1/2} \Kt} \ .
\label{eq: regression variables}
\end{equation}

Linear extrapolation of $Y(x) = ax + b$ to $x=0$ means evaluating $Y(0) = b$.
The spread in $\xi_i$ introduces an error $\Delta b$ also in the fit parameter $b$.
From the theory of linear regression, we know that $b = \frac{\sum_i x_i^2 \sum_j Y_j - \sum_i x_i \sum_j x_j Y_j}{R \sum_i x_i^2 - (\sum_i x_i)^2}$ with $R$ values of $x_i$.
Inserting $Y_i=y_i + \xi_i \Delta y_i$, this becomes
\bea{rCl}
    b & = & \frac{\sum_j y_j \sum_i x_i (x_i - x_j) + \sum_j \xi_j \Delta y_j \sum_i x_i (x_i - x_j)}{R \sum_i x_i^2 - (\sum_i x_i)^2} \ .
\eea
Since $\mathbb{E}\xi_i = 0$ and $\mathrm{Var}(\xi_i) = 1$, the expectation value of $b$ with respect to the independent and identically distributed random variable $\xi_i$ is
$\mathbb{E}b = \frac{\sum_j y_j \sum_i x_i (x_i - x_j)}{R \sum_i x_i^2 - (\sum_i x_i)^2}$.
An expression for the  standard deviation of $b$ with respect to the spread in $\xi_i$ is obtained using the properties of the variance, $\mathrm{Var}(\sum X_i) = \sum \mathrm{Var}(X_i)$ and $\mathrm{Var}(aX) = a \mathrm{Var}(X)$ for constant $a$ and random variable $X$ (note that $\Delta y_i, x_i$ are constants with respect to the random variables $\xi_i$),
\begin{equation}
    \Delta b = \sqrt{\mathrm{Var}(b)}= \frac{ \Bigl( \sum_j \Delta y_j^2 \bigl[ \sum_i x_i (x_i - x_j) \bigr]^2 \Bigr)^{1/2} } {R \sum_i x_i^2 - (\sum_i x_i)^2} \ .
\end{equation}
With \Cref{eq: regression variables}, we obtain
\bea{rCl}
    \mathbb{E}b = \ln \Kt
    \quad \quad \mathrm{and} \quad \quad
    \Delta b = \frac{
    		\Bigl( \sum_j \bigl[ \sum_i G_i (G_i - G_j) \bigr]^2 \bigl( \frac{\Kt}{K} \bigr)^{2G_j} \frac{1}{ S_j \Kt^2} \Bigr)^{1/2}
		}{
			R \sum_i G_i^2 - (\sum_i G_i)^2
		}
    \ .
\label{eq: pea oqs error b}
\eea
Evaluating the function $F(G) = \exp(aG + b)$ at $G=0$ then yields
\bea{rCl}
F(0) = \exp(\mathbb{E}b + \xi \Delta b) = \exp(\ln \Kt + \xi \Delta b) = \Kt e^{\xi \Delta b} \approx \Kt (1 + \xi \Delta b) \ .
\eea
This means, the extrapolated value fluctuates around the noise-free value $\mathbb{E}F(0)$ with random error $\Delta F(0)$ given by
\begin{equation}
    \mathbb{E}F(0) = \Kt
    \quad \quad \mathrm{and} \quad \quad
    \Delta F(0) = \frac{
    		\Bigl( \sum_j \bigl[ \sum_i G_i (G_i - G_j) \bigr]^2 \bigl( \frac{\Kt}{K} \bigr)^{2G_j} \frac{1}{ S_j} \Bigr)^{1/2}
		}{
			R \sum_i G_i^2 - (\sum_i G_i)^2
		}
    \ .
\label{eq: pea oqs error F}
\end{equation}

Our calculation remains completely parallel to Ref.~\citenum{Filippov2024peaTheory}.
To find the minimum random error, we need to minimize $\Delta F(0)$ over the two sets of free parameters\footnote{Here, we minimize sequentially, i.e., fixing one to minimize over the other.
Note that this is an approximation that might result in local minima.}, $\{ S_i \}_{i=1}^R$ and $\{ G_i \}_{i=1}^R$.
First, we fix $\{ G_i \}$ to find the optimal shot allocation given a fixed total shot budget, $\sum_i S_i = M$.\\
Using the method of Lagrange multipliers (see \Cref{app: pea oqs random error} for detailed derivations), we first find the optimal shot allocation per noise gain,
\begin{equation}
    S_j^* = \frac{M |A_j|}{\sum_l |A_l|} \ .
\label{eq: pea oqs optimial shots}
\end{equation}
Here, $A_j = \sum_i G_i (G_i - G_j) \bigl( \frac{\Kt}{K} \bigr)^{G_j} \frac{1}{\Kt}$.
With $x^2/|x| = |x|$, the random error becomes
\begin{equation}
    \Delta F(0, S_j^*) = \frac{1}{\sqrt{M}} \frac{ \sum_j \bigl| \sum_i G_i (G_i - G_j) \bigr| \bigl( \frac{\Kt}{K} \bigr)^{G_j} }{ R \sum_i G_i^2 - (\sum_i G_i)^2 } \ .
\label{eq: pea oqs error F optimial S}
\end{equation}

Next, we minimize over $\{ G_i \}$ fixing $\{ S_i \} = \{ S_i^* \}$.
Assuming two noise gains $G_2 > G_1 \geq 1$, in which case analytical expressions can be found minimizing $\Delta F(0)$, we can fix the first noise gain at $G_1^*=1$.
As shown in \Cref{app: pea oqs random error}, the second gain is found to be
\begin{equation}
    G_2^* = 1 + \frac{W \bigl(\frac{1}{e}\bigr) + 1}{\ln \frac{\Kt}{K}} \ .
\label{eq: pea oqs optimal gain}
\end{equation}
Inserting $G_1^*, G_2^*$ back into \Cref{eq: pea oqs error F optimial S} with $R=2$ yields for the minimum random extrapolation error (see \Cref{app: pea oqs random error})
\begin{equation}
    \Delta F(0, S_j^*, G_j^*) = \frac{\Kt}{K \sqrt{M}} \biggl( 1 + \frac{\ln(\Kt / K)}{W(\frac{1}{e})} \biggr) \ .
\label{eq: pea oqs error F min}
\end{equation}

\subsection{Numerically verifying the error bound}
\label{sec: pea oqs numerical scaling}

\renewcommand{\arraystretch}{1.2}
\begin{table}[t]
\begin{center}
        \begin{tabular}{C{0.7cm}C{0.7cm}C{0.7cm}C{0.7cm}C{0.7cm}C{0.7cm}C{0.7cm}C{0.7cm}C{0.7cm}}
            \hline\hline
            $P_\alpha$ & XI & YI & ZI & XX & YX & ZX & YY & ZY \\
            \\[-12pt] \hline \\[-12pt]
			$\lambda_\alpha$ & 0.05 & 0.07 & 0.01 & 0.06 & 0.07 & 0.03 & 0 & 0 \\
			$\tilde{\lambda}_\alpha$ & 0.03 & 0 & 0.07 & 0.01 & 0.03 & 0 & 0.03 & 0.02 \\
            \hline\hline
        \end{tabular}
    \caption[Example sparse Pauli noise models used for random error analysis]{Arbitrarily chosen examples of two sparse Pauli noise models used for numerical verification of the random error scaling when extrapolating from one to another noise model.}
    \label{tab: noise models scaling}
\end{center}
\end{table}

Next, we numerically validate this bound, ensuring that the assumptions of a Clifford circuit and stabilizing observable under which the bound was derived are satisfied.
Specifically, we compute the noisy time evolution of a simple two-qubit Ising model with Hamiltonian $H = - J \sz_0 \sz_1$.
The time evolution operator for one time step is then
\begin{equation}
    e^{ - i H \Delta t } = e^{ i J \Delta t \sz_0 \sz_1 } = R_{ZZ}( - 2 J \Delta t) \ .
\label{eq: pea oqs clifford time evo}
\end{equation}
To make the time evolution circuit fully Clifford, we choose $J = \pi / (2 \Delta t)$, yielding $R_{ZZ}( - \pi )$
We set the initial state to $\ket{\psi_0} = \ket{00}$, evolve it over five time steps of $\Delta t = 0.2$, and measure the expectation value $\braket{\sz\sz}$ throughout the time evolution.
We impose a fictitious hardware noise model defined by the rates $\{ \lambda_\alpha \} = \mathrm{SPL}(\Lambda)$ given in \Cref{tab: noise models scaling} onto the time evolution circuit by sampling the respective Pauli channels into the circuit (see also \Cref{sec: methods pea}).
The target noise model, to which we extrapolate, is given in the same table with the corresponding rates $\{ \tilde{\lambda}_\alpha \} = \mathrm{SPL}(\tilde{\Lambda})$.
Note that all of these rates are chosen arbitrarily in the order of magnitude of what can be expected on real hardware.
Furthermore, the target rates are chosen such that all of the previously discussed cases are present.
For each time step, we amplify $\Lambda$ to the optimal noise gains $G^*$ given in \Cref{eq: pea oqs optimal gain}, fit a single exponential function to the two data points (expectation values at two noise gains), and extrapolate to $G=0$.
Specifically, the amplification is done by scaling all rates $\lambda_\alpha$ according to the cases at the beginning of \Cref{sec: pea different noise model}.
The respective Pauli error is then sampled into the circuit at random according to the probabilities \Cref{eq: methods pea probability}.
Each circuit sample represents a potentially different realization of the (amplified) noise model.
For each sample, after the circuit is constructed, including the error terms, the expectation value is evaluated exactly from the statevector resulting from the circuit.
This is repeated for several total shot budgets ranging from $10^2$ to $5 \times 10^5$ samples.
These shot budgets are distributed over the noise gain values according to the optimal distribution of samples in \Cref{eq: pea oqs optimial shots}.

\begin{figure}[t!]
    \centering
    \includegraphics[]{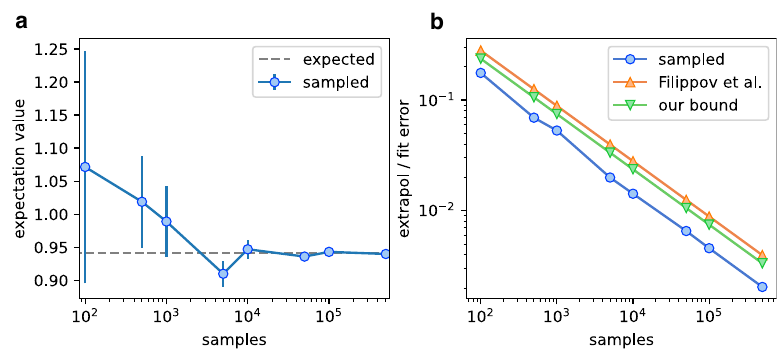}
    \mycaption{Scaling of the PEA extrapolation error}
    {
    \textbf{a} Extrapolated expectation values $\braket{ZZ}$ as a function of the total number of circuit samples $M$ distributed over several noise gains.
    The expectation value is measured after five time steps as described in the main text.
    Every point is the result of extrapolation to $G=0$ after fitting expectation values for the two noise gain values $G_1^*, G_2^*$ and respective allocation of circuit samples $S_1^*, S_2^*$ computed in the main text.
    Here, each sample is also a different random realization of the noise model.
    The grey dashed line indicates the exact expectation value after five time steps.
    \textbf{b} Fit error $\Delta b$ as a function of the number of samples.
    Here, ``sampled'' refers to the random error obtained in fitting $\{G_i^*\}$ in \textbf{a} (every point is one extrapolation), ``our bound'' refers to \Cref{eq: pea oqs error F min}, and ``Filippov et al.'' to the result derived for regular PEA in Ref.~\citenum{Filippov2024peaTheory} (which applies to a different case as our result and is not to be understood as a looser or worse bound than our result, see main text).
    Note that this is not to be mistaken with the relative error between the expectation value and the exact reference line in \textbf{a}.
    }
    \label{fig: pea scaling error}
\end{figure}

The results are shown in \Cref{fig: pea scaling error}~\textbf{a}.
Here, we plot the extrapolated expectation value at $G=0$ after the last time step as a function of the total number of samples $M$.
The standard error is taken to be $1/\sqrt{M}$.
We see that the expectation value converges to the exact reference value, indicated by the dashed line, for increasing sample sizes.
Importantly, the reference value was computed using QuTiP's~\cite{Lambert2024qutip} exact master equation solver with rates and Pauli errors as jump operators.

\Cref{fig: pea scaling error}~\textbf{b} shows the fit error $\Delta b$ obtained in \textbf{a} (``sampled''), i.e., for fitting the noise amplified values at the last time step for different total sample counts $M$.
This is compared to the analytical bound for the random error \Cref{eq: pea oqs error F min} (``our bound'') and the one derived for regular \gls{pea} in Ref.~\citenum{Filippov2024peaTheory} (``Filippov et al.'').
Note that both bounds apply to different cases (regular versus partial \gls{pea}) and are not to be understood as one improving upon or representing a tighter bound than the other.
Furthermore, we plot only bare scalings here, irrespective of their pre-factors, which could shift both curves.
Remarkably, the scaling of the random errors with the total number of samples $M$ calculated from circuit sampling matches the analytical bounds perfectly.
It is also worth noting that our result \Cref{eq: pea oqs error F min} and that of Ref.~\citenum{Filippov2024peaTheory} are merely shifted against each other.
This was to be expected as extrapolation to a different noise model instead of zero noise merely introduces an additional factor $\Kt$ in the exponential, i.e., changes the basis of the exponential.

\subsection{Open quantum dynamics in the dissipative Ising model}
\label{sec: peq oqs diss ising}

As a first example and proof-of-principle simulation, we study the dissipative Ising model.
In the previous section, we simulated pure Clifford circuits based on a classical Ising model, i.e., having only a $\sz \sz$ interaction but no transverse field.
Here, we extend this study from Clifford to non-Clifford circuits.
For this purpose, we consider a two-qubit \gls{tfim} 
\begin{equation}
	H = - J \sz_0 \sz_1 - h (\sx_0 + \sx_1) \ ,
\end{equation}
with coupling $J = 1$ and transverse field $h = 0.4$.
The resulting Trotter circuit implementing the time evolution consists of a product of $R_{ZZ}(\theta_J)$ and $R_X(\theta_h)$ gates with angles $\theta_J = - 2 J \Delta t$ and $\theta_h = - 2 h \Delta t$, respectively, which are no longer Clifford gates.
We compute the Trotterized time evolution for five time steps, again with $\Delta t = 0.2$, and measure $\braket{\sz\sz}$ throughout.
The optimal noise gain values $G^*$ above were derived under the assumption of a Clifford circuit.
Here, for a non-Clifford circuit, we therefore choose different values $G_i \in \{ 1.0, 1.2, 1.4, 1.6 \}$.
At every time step, we take $10^4$ circuit samples in total, distributed over the different noise gain values.

\begin{figure}[t!]
    \centering
    \includegraphics[]{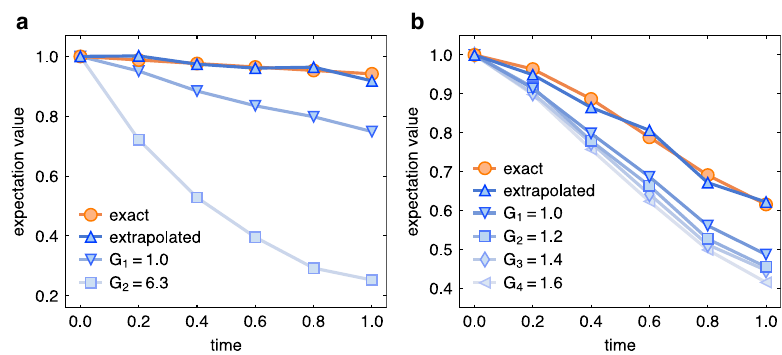}
    \mycaption{Noise model extrapolation Clifford vs. non-Clifford}
    {
    \textbf{a} Time evolution of a dissipative Ising model without transverse field.
    These results correspond to the ones shown in \Cref{fig: pea scaling error}.
    Since the circuit is fully Clifford, the optimal noise gain and distribution of samples derived in \Cref{sec: pea oqs random error} are used to amplify the fictitious noise model ${\Lambda}$ given in \Cref{tab: noise models scaling}.
    The extrapolated values correspond to the ones shown in \Cref{fig: pea scaling error}~\textbf{a} and represent the second Pauli noise model $\tilde{\Lambda}$ defined in \Cref{tab: noise models scaling} and fall onto the reference points computed with an exact solver.
    \textbf{b} Time evolution of a dissipative transverse field Ising model as defined in the main text.
    Here, the circuit is no longer Clifford and, therefore, different noise gains are used than in \textbf{a}.
    Nonetheless, extrapolation from the amplified noise model (see \Cref{tab: noise models scaling}) to the other noise yields expectation values that closely match those of the exact solver.
    }
    \label{fig: pea dissipative ising}
\end{figure}

\Cref{fig: pea dissipative ising} shows the results of both the Clifford case (\textbf{a}) as well as the non-Clifford case (\textbf{b}).
In both cases, we used the noise models defined in \Cref{tab: noise models scaling}.
The Clifford case in \textbf{a} corresponds to the same simulation already reported in \Cref{fig: pea scaling error}.
Using the optimal noise gain and sample distribution derived in \Cref{sec: pea oqs random error}, we see that the extrapolated value coincides with the reference values computed through QuTiP's exact master equation solver~\cite{Lambert2024qutip}.
The other case in \textbf{b} corresponds to the non-Clifford case of the \gls{tfim}.
Here, the optimal noise gain and sample distribution are no longer valid, and different values are chosen.
Moreover, the total of $10^4$ circuit samples is distributed evenly over the different noise gains we measure.
We fit against a single exponential function, which is also why we chose to measure the expectation value at four instead of two noise gain values to make the fitting more robust against the multi-exponential functional form that is expected for non-Clifford circuits.
Again, the extrapolated values match those of the exact solver.

\section{Conclusion}

We extended the formalism of \gls{pea} from pure error mitigation to the simulation of open system dynamics.
Concretely, we proposed to use a variant of the method to amplify different noise channels with different noise gains to extrapolate to another noise model, instead of to zero noise.
This serves the purpose of implicitly implementing the time evolution of a system subject to environment-couplings that are modeled through the extrapolated noise model.
In doing so, we mitigate exponential sampling overheads inferred by related methods based on \gls{pec}~\cite{Guimaraes2023partialPEC, Guimaraes2024partialPECtimeDep}.
Moreover, we presented a rigorous error analysis.
This resulted in a bound for the random error, stemming from a finite sample size that propagates through to the extrapolated value.
The fact that we showed agreement between the bound and numerical tests implies tight control over the final error of the simulation outcome and paves the way for a proof-of-principle demonstration of the method.
Lastly, we demonstrated the functioning of the method on the example of a dissipative Ising model.
Here, studying both the Clifford and the non-Clifford case, we successfully interpolated between two different models.
The extrapolated values matched those of exact reference calculations in both cases.
As next steps, we will therefore extend this investigation to larger, more realistic models.

As mentioned at the beginning of this chapter, exciton transfer models are of high practical relevance and present interesting candidate applications for an implementation of this method~\cite{potocnik2018lightHarveting}.
Several studies on more simplistic toy models for exciton transport have shown the existence of a regime in which transport properties of the model are enhanced by including noise of intermediate strength acting on the bulk of the system~\cite{Landi2022boundaryDrivenQuantumSystemsRMP, MendozaArenas2013noiseAssistedTransport, MendozaArenas2013dephasingEnhancement, Maier2019noiseAssistedTransport}.

%% file: ch6_ground_states.tex
\chapter{State preparation and phase characterization beyond dynamics}
\label{chap: ground states}


Besides time-dependent problems, the study of time-\textit{in}dependent states and phenomena is of key interest and presents an active and challenging field of research in itself, as outlined in \Cref{sec: intro applications}.
In fact, understanding and being able to prepare eigenstates of a system is important even in the context of quantum dynamics.
The initial state of a quantum simulation aimed at modeling a realistic experiment might not be an easy-to-prepare product state.
Instead, for a realistic description of a quantum process, a suitable initial state must be prepared, for example, a thermal state or the ground state of another Hamiltonian~\cite{Bauer2020qcMaterialScienceReview, Google2024dissipativeEngineering}.
Beyond state preparation as input states for quantum dynamics, understanding stationary properties of quantum systems is crucial to many areas of research~\cite{Cao2019quantumChemistryQCreview, McArdle2020qcChemistryReview, Bauer2020qcMaterialScienceReview}.
Even though a possible quantum advantage may be more difficult to realize and less obvious to identify than with quantum dynamics~\cite{Lee2023evidenceQuantumAdvantage}, the study of ground and excited states remains one of the most actively researched applications of quantum hardware~\cite{Ollitrault2021molecular, RobledoMoreno2024sqd, Alexeev2024materialsWhitePaper}.

In this chapter, we present two topically related works focusing on preparing ground states of many-body quantum systems and understanding their properties.
First, in \Cref{sec: elph}, we present a variational, hybrid quantum-classical algorithm to prepare ground states of electron-phonon systems, specifically, the Hubbard-(extended-)Holstein model~\cite{Denner2023phonon}.
Electron-phonon couplings, i.e., the interaction between electrons in a lattice and the collective motion of the lattice sites, are responsible for many interesting properties of real-world materials~\cite{He2018rapid, Kang2019bilayerGraphene, Hejazi2019bilayerGraphene}.
Like many other models of strongly correlated many-body systems, the Hubbard-(extended-)Holstein model exhibits a rich phase diagram, featuring phases of both weakly and strongly correlated ground states.

Understanding these phases and their properties is generally important in many areas of research, such as condensed matter physics, materials science, and high-energy physics~\cite{Sachdev2011book}.
However, it is not necessarily straightforward to distinguish phases of quantum states~\cite{Carrasquilla2017machineLearningPhases}.
Therefore, in \Cref{sec: qcnn}, we present a tool to classify phases of quantum states prepared with a quantum computer, e.g., in quantum simulation.
Specifically, we employ \glspl{qcnn} within the framework of \gls{qdl}, i.e., \gls{qml} with genuine quantum states as input data.
We demonstrate this framework on input states of two different model systems, by training and successfully testing the \gls{qcnn} in each of the two settings separately.
First, on ground states of the Schwinger model, a paradigmatic toy model in high-energy physics, that undergo a phase transition as a function of a model parameter.
The second test case is that of classifying phases of time-evolved states in a $\mathbb{Z}_2$ model, another popular toy model in high-energy physics.
Here, phase classification is done according to their (de-)confinement, which depends on the presence of an external field.

\section{State preparation in strongly correlated electron-phonon systems}
\label{sec: elph}

\begin{zitat}{}
    This section is reproduced with permission in parts from M. Michael Denner, Alexander Miessen, Haoran Yan, Ivano Tavernelli, Titus Neupert, Eugene Demler, and Yao Wang, ``A hybrid quantum-classical method for electron-phonon systems'', Communications Physics \textbf{6} (1), 233 (2023)~\cite{Denner2023phonon}.
    \Cref{sec: elph methods} introduces the hybrid quantum algorithm for solving ground states of electron-phonon systems.
    We benchmark the performance of the algorithm across the phase diagram of a Hubbard-Holstein model in \Cref{sec: elph phase diagram} and the dependence of its error on the variational circuit depth in \Cref{sec: elph circuit depth}.
    Finally, we assess the feasibility of a hardware implementation of our algorithm and study the influence of noise in \Cref{sec: elph circuit transpilation,sec: elph noisy simulations}.
    Alexander Miessen contributed to the implementation, testing, and optimization of the algorithm.
    In particular, he devised the necessary tools and carried out the noisy simulations and resource estimates for a hardware implementation, presented in \Cref{sec: elph circuit transpilation,sec: elph noisy simulations}.
\end{zitat}

Understanding strongly correlated many-body systems is vital to many areas of science and technology, such as the development and analysis of functional quantum materials~\cite{Keimer2015highTsuperconductivity}.
Due to the entanglement induced by correlations, macroscopic properties of quantum materials are often unpredictable from reductive single-particle models.
Theoretical analysis and understanding of macroscopic properties of materials requires the analysis of sufficiently large model systems, which cannot be done accurately with classical computers.
Quantum computing technologies constitute an intriguing new direction for studying strongly correlated or large-scale quantum many-body systems and especially quantum materials.

\begin{figure}[t]
	\centering
    \includegraphics[width=\textwidth]{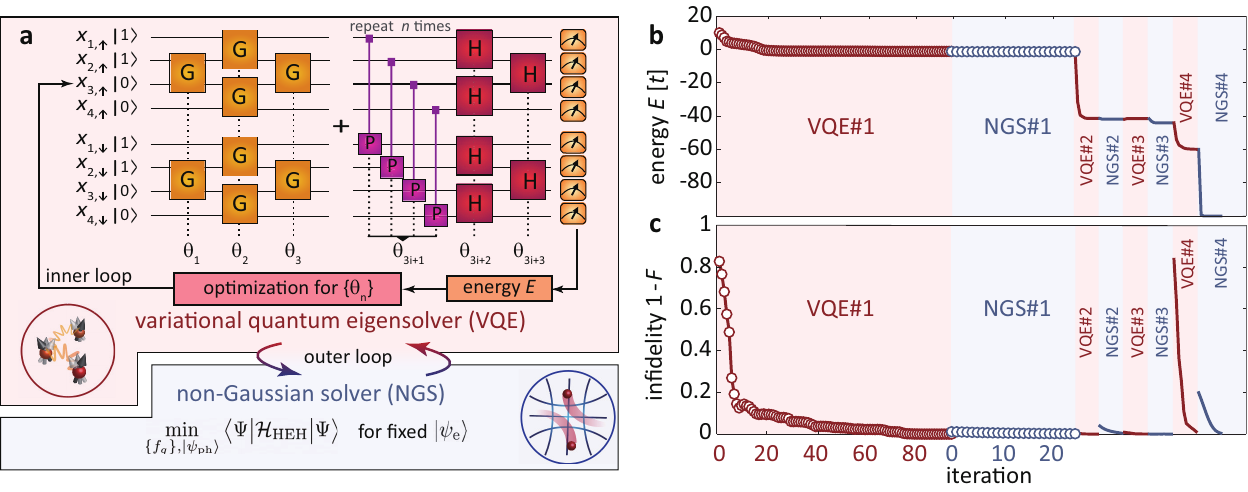}
    \mycaption{Hybrid quantum-classical electron-phonon solver}{
    \textbf{a} The hybrid quantum algorithm iterates between a variational quantum eigensolver (VQE) for the electronic part and a non-Gaussian solver (NGS) for the phonon part of the many-body ground state.
    The quantum circuit structure for a 4-site example at half filling contains Givens rotations \textsf{G}, on-site gates \textsf{P}, and hopping gates \textsf{H}.
    The \textsf{P} and \textsf{H} layers are repeated $n$ times to express the ground state wavefunction.
    Within each layer, gates share the same variational parameters $\theta_i$, which are optimized on a classical computer inside each VQE iteration.
    \textbf{b} Convergence of the NGS-VQE algorithm, reflected by the total energy as a function of inner-loop (NGS or VQE) iteration steps for a 4-site Hubbard-Holstein model with $u = 10$, $\lambda = 10$, and $\omega = 1$.
    VQE steps were performed with quantum circuit statevector simulations and a circuit depth of $n=5$.
    Alternative outer-loop iterations are colored red (for VQE) and blue (for NGS), and the data points are compressed after NGS \#1, for illustration purposes.
    \textbf{c} Convergence of the ground state infidelity $1-F$ during each iteration.
    The reference state chosen for each outer-loop iteration was obtained by exact diagonalization on classical computers.}
    \label{fig: elph circuit vqe}
\end{figure}

However, as already established in \Cref{chap: perspective}, the resources necessary to implement realistic simulations of materials or chemical compounds are far too many for current-day noisy hardware.
Variational algorithms~\cite{Cerezo2020variational, Bharti2022nisqAlgorithms, Endo2021hybridAlgorithmsErrorMitigation}, including the \gls{vqe}~\cite{Peruzzo2014vqe}, could in part help mitigate these resource requirements by splitting the workload between the quantum processor and a classical co-processor. 
An example of such a variational protocol is shown in the upper panel of \Cref{fig: elph circuit vqe}~\textbf{a}.
A variational wavefunction is prepared with a parameterized quantum circuit, after which the expectation value of the Hamiltonian is measured and the parameters are iteratively optimized classically.
\Gls{vqe} has been used to study small instances of molecules~\cite{Kandala2017hardwareEfficientVQE, Colless2018molecularSpectraVQE, Takeshita2020quantumChemistryVQE, Ollitrault2020qeom} and solid-state systems, including quantum magnets and Mott insulators~\cite{Uranov2020vqeFrustratedSystems, Cade2020vqeFermiHubbard, Stanisic2022vqeFermiHubbard}. 

However, realistic materials usually contain more complex interactions than simplified electronic models, such as the Hubbard model, which only features local Coulomb interaction.
The interactions between mobile electrons and the ionic lattice in solids, so-called \glspl{epc}, underlie many electrical and mechanical properties of materials.
Notably, it has been suggested that the interplay between the electron-phonon interaction and the electronic Coulomb repulsion is crucial for many novel quantum phases, such as unconventional superconductivity in cuprates~\cite{Scalapino1986dwaveSuperconductivity, Gros1987superconducting, Shen2004cuprateSuperconductors, Lanzara2001evidence, Reznik2006electron, He2018rapid} and twisted bilayer graphene~\cite{Kang2019bilayerGraphene, Hejazi2019bilayerGraphene, Guinea2018bilayerGraphene, Lian2019bilayerGraphene, Wu2018bilayerGraphene, Peltonen2018bilayerGraphene}.
Achieving predictive control of these quantum phases calls for developing reliable theoretical models describing materials with \gls{epc}~\cite{Fausti2011light, Hu2014optically, Boschini2018collapse}, which has motivated studies based on small clusters~\cite{Rosch2004electronPhonon, Khatami2008effect, Wang2018dWaveSuperconductivity} or perturbative couplings~\cite{Murakami2013hubbardHolstein, Kemper2015direct, Sentef2016phononSuperconductivity, Babadi2017electronPhonon}.
However, quantum simulation of materials with strong \glspl{epc} remains challenging due to the unbounded bosonic Hilbert space of the phonons~\cite{Macridin2018electronPhonon, Li2013electronPhononQuantumSimulation, Macridin2018fermionBosonQuantumSimulation, Pavosevic2021polaritonic}.
Therefore, even with a single electronic band and only a single phonon mode, the system has much higher computational complexity compared to electrons alone.
Inclusion of phonons in an $L$-site spinful system increases the Hilbert-space size from $4^L$ (for electrons only) to $4^L(m+1)^L$ where $m$ is the (truncated) maximal local phonon occupation.
Hence, for materials with non-negligible \glspl{epc}, the required $m \gg 1$ leads to an unreasonably large, or even unbounded, Hilbert space.
This issue prohibits not only classical simulations but also an efficient encoding on a quantum machine.

To this end, we design a hybrid quantum algorithm that leverages the capability of \gls{vqe} to address the electronic part with quantum computers and the variational non-Gaussian description of non-perturbative polaronic dressing for the phonon and coupling part~\cite{Shi2018nonGaussianStates, Wang2020hubbardHolstein, Wang2021timeDepNonGaussian}.
We prove the validity of our approach using the one-dimensional Hubbard-Holstein model and its variants, which is summarized together with the specifics of the algorithm in \Cref{sec: elph methods}.
We then show in \Cref{sec: elph phase diagram} that our hybrid quantum algorithm can reliably capture the ground-state properties of the paradigmatic Hubbard-Holstein model in all regions of the phase diagram, when compared to \gls{ngsed} results.
Our algorithm does not require any additional qubit overhead stemming from unbounded phononic degrees of freedom or the truncation to a low phonon occupation~\cite{Macridin2018electronPhonon, Macridin2018fermionBosonQuantumSimulation, Pavosevic2021polaritonic}, as we implicitly sample the phonon Hilbert space.
This makes it possible to analyze electron-phonon systems over a broad range of parameters.
Next, we investigate the scaling of the algorithm's performance with respect to the system size in \Cref{sec: elph circuit depth}.
To assess the feasibility of hardware simulations of our algorithms, we analyze the resource overheads inferred by transpilation to an IBM quantum device in \Cref{sec: elph circuit transpilation}.
Subsequently, we perform simulations under the influence of realistic hardware noise in \Cref{sec: elph noisy simulations}

\subsection{The variational non-Gaussian VQE method}
\label{sec: elph methods}

We consider a prototypical correlated system with electrons interacting via local Coulomb interactions (the Hubbard model), coupled to bosonic phonon modes through a Fröhlich-type interaction.
The resulting model is the \gls{heh} model with Hamiltonian 
\begin{equation}
    \begin{aligned}
        \mathcal{H}_{\rm HEH} = &-t \sum_{\langle i,j \rangle, \sigma} \left(c_{i,\sigma}^\dagger c_{j,\sigma}^{} + \mathrm{h.c.}\right) + U \sum_i n_{i,\uparrow}n_{i,\downarrow} \\
        &+ \sum_{i,j,\sigma}g_{ij} \left(a_i^{} + a_i^\dagger\right)n_{j,\sigma} + \omega_0 \sum_i a_i^\dagger a_i^{}. 
    \end{aligned}
\label{eq: hubbard holstein hamiltonian}
\end{equation}
Here, $c_{i,\sigma}^{}$ ($c_{i,\sigma}^{\dagger}$) annihilates (creates) an electron with spin $\sigma$ at site $i$, with associated density operators $n_{i,\sigma} = c_{i,\sigma}^\dagger c_{i,\sigma}^{}$, and $a_i^{}$ ($a_i^{\dagger}$) annihilates (creates) a phonon at site $i$.
Among the model parameters, $t$ sets the (nearest-neighbor $\langle i,j \rangle$) electronic hopping strength, $U$ sets the electronic on-site repulsive interaction, $\omega_0$ denotes the phonon energy, and $g_{ij}$ is the coupling strength between the phonon displacement at site $i$ and electron density at site $j$.
While our method can tackle any distribution of \glspl{epc}, we restrict ourselves to local, $g_{ii} = g$, and nearest-neighbor couplings, $g_{i,i \pm 1}=g'$, in one dimension.

For purely local \gls{epc} $g_{ij} = g \delta_{ij}$ (i.e., $g'=0$), the \gls{heh} model is reduced to the Hubbard-Holstein model.
The physical properties of the Hubbard-Holstein model have been studied with various numerical methods, in one-dimensional (1D) systems\,\cite{Tezuka2007holsteinPhaseDiagram, Fehske2008metallicityHolstein, Clay2005holstein, Greitemann2015holstein}, two-dimensional (2D) systems~\cite{Nowadnick2012holstein2D, Karakuzu2017holstein2D, Ohgoe2017holstein2D, Hohenadler2019holstein2D}, and infinite dimensions~\cite{Georges1996holsteinInfiniteDim, Werner2007holsteinInfiniteDim}. 
The phase diagram of the Hubbard-Holstein model is controlled by the three dimensionless parameters $u = U/t$ for electronic correlations, $\lambda = g^2/\omega_0 t$ for the effective \gls{epc}, and $\omega = \omega_0 / t$ for phonon retardation effects.
The presence of nonlocal \glspl{epc} has been studied recently, motivated by the observed attractive nearest-neighbor interactions in cuprate chains~\cite{Chen2021anomalously}.
For this reason, numerical studies of the \gls{heh} model were primarily focused on 1D systems~\cite{Wang2021phononLongRange, Tang2023traces}.
Here, we also restrict to periodic 1D systems, while the presented algorithm can be naturally extended to 2D.

To handle the strongly entangled electronic wavefunction and unbounded phonon Hilbert space simultaneously, we employ a variational, non-Gaussian construction of the many-body wavefunction~\cite{Shi2018nonGaussianStates, Shi2020finiteTemperatureNGS}.
A universal electron-phonon wavefunction can always be written as
\begin{equation}
    | \Psi \rangle  = U_{\rm NGS}(\{f_q\}) | \psi_{\rm ph} \rangle \otimes| \psi_{\rm e} \rangle \ .
\label{eq: elph full ansatz}
\end{equation}
Here, the right-hand side is a direct product of electron and phonon states, $|\psi_{\rm e}\rangle$ and $|\psi_{\rm ph}\rangle$, respectively, and a variational non-Gaussian entangling transformation $U_{\rm NGS}=e^{i\mathcal{S}}$.
The Hermitian operator $\mathcal{S}$ is a polynomial function of $c$, $c^\dagger$, $a$, and $a^\dagger$ operators.
For Holstein-type couplings, it has been shown that $\mathcal{S}$ truncated to lowest order yields sufficiently fast convergence~\cite{Wang2020hubbardHolstein, Wang2021timeDepNonGaussian} (see \Cref{app: elph effective hamiltonian} and Ref.~\citenum{Denner2023phonon}).
In this case, the lowest-order coefficients $\{f_q\}$ ($q$ denoting the momentum for periodic systems), fully determine the non-Gaussian transformation $U_{\rm NGS}(\{f_q\})$.

Using this ansatz, we solve the \gls{heh} problem by minimizing the energy
\begin{equation}
    E(\{f_q\}, |\psi_{\rm ph}\rangle, |\psi_{\rm e}\rangle) =\big\langle\Psi\big| \mathcal{H}_{\rm HEH} \big|\Psi\big\rangle
\label{eq: elph energy expectation value}
\end{equation}
self-consistently with respect to the unrestricted electronic state $|\psi_{\rm e}\rangle$ and the variational parameters in $U_{\rm NGS}(\{f_q\})$ and $|\psi_{\rm ph}\rangle$.
A schematic representation of the algorithm is shown in \Cref{fig: elph circuit vqe}~\textbf{a}.
Within each iteration, the variational parameters $\{f_q\}$ and the phonon wavefunction $|\psi_{\rm ph}\rangle$ (here restricted to be a Gaussian state) are optimized using imaginary-time evolution~\cite{Wang2020hubbardHolstein}.
This is referred to as the \gls{ngs}, whose computational complexity scales polynomially with the system size $L$.
On the other hand, the electronic part of the wavefunction $|\psi_{\rm e}\rangle$ is represented by a variational quantum circuit that is optimized by regarding the $|\psi_{\rm ph}\rangle$ and $\{f_q\}$ as fixed when minimizing the total energy.
The latter step is equivalent to solving the electronic ground state of an effective Hubbard Hamiltonian,
\bea{rCl}
    \mathcal{H}_{\rm eff} & = & \langle \psi_{\rm ph} | U_{\rm NGS}^\dagger \mathcal{H}_{\rm HEH} U_{\rm NGS} | \psi_{\rm ph} \rangle \nonumber \\[5pt]
    & = &-\tilde{t} \sum_{\langle i,j \rangle, \sigma} \left(c_{i,\sigma}^\dagger c_{j,\sigma}^{} + \mathrm{h.c.}\right) + U \sum_i n_{i,\uparrow}n_{i,\downarrow} \\
    & & + \sum_{i,j} \sum_{\sigma,\sigma'} \tilde{V}_{ij} n_{i,\sigma} n_{j,\sigma'}+ \tilde{E}_{\rm ph} \ . \nonumber
\label{eq: elph effective hamiltonian}
\eea
Physically, $\mathcal{H}_{\rm eff}$ describes the behavior of polarons, i.e., phonon-dressed electrons.
The phonon dressing gives rise to a heavier effective mass through a modified hopping strength $\tilde{t}$ and mediates a long-range attraction $\tilde{V}_{ij} < 0$ between polarons (see \Cref{app: elph effective hamiltonian} for detailed expressions).

Within each self-consistent iteration, the difficulty of solving the electron-phonon problem has thus been converted into solving a purely electronic Hamiltonian $\mathcal{H}_{\mathrm{eff}}$ with extended Hubbard interactions $\tilde{V}_{ij}$.
This electronic problem can be efficiently embedded on a quantum chip by using a suitable fermion-to-qubit mapping.
Here, we employ the Jordan-Wigner transformation, which maps each electron with a given spin orientation to one qubit.
In particular, $N_s$ lattice sites that can be occupied by an electronic mode are mapped to $2 N_s$ qubits, to encode occupation of spin-up and spin-down states in one qubit each.
They are ordered according to their spin-state, as seen in \Cref{fig: elph circuit vqe}~\textbf{a}.
As we study all models at half-filling in this work, the qubit register is initialized as $\ket{1}_\uparrow^{N_s/2} \ket{0}_\uparrow^{N_s/2} \ket{1}_\downarrow^{N_s/2}\ket{0}_\downarrow^{N_s/2}$.
The state $|\psi_{\rm e} (\{\theta_i\})\rangle$ is construcuted via a variational quantum circuit and variational parameters $\{\theta_i\}$ are optimized iterating between quantum and classical processor using \gls{vqe} to minimize the energy $\langle \psi_{\rm e} (\{\theta_i\}) | \mathcal{H}_{\mathrm{eff}}|\psi_{\rm e} (\{\theta_i\})\rangle$.
The solution of \gls{vqe} then approximates $|\psi_{\rm e} \rangle$ in \Cref{eq: elph full ansatz}.
Unless explicitly specified otherwise, we conduct the \gls{vqe} step of the \gls{ngs}-\gls{vqe} iterations with exact statevector simulations using \qiskit~\cite{qiskit2024}.

Our variational quantum circuit for $|\psi_{\rm e} (\{\theta_i\})\rangle$ consists of two parts that are shown in \Cref{fig: elph circuit vqe}~\textbf{a}.
The first part is a set of Givens rotations that efficiently prepare the ground state of a non-interacting Hubbard model~\cite{Cade2020vqeFermiHubbard, Stanisic2022vqeFermiHubbard, Jiang2018quantumAlgorithmsFermions}.
For the second part, we employ the \gls{hva}~\cite{Wecker2015hva} of the Hubbard model that parameterizes the Trotterized time evolution operators corresponding to the different Hamiltonian terms, i.e., blocks of $\textsf{P}$ and $\textsf{H}$ gates in \Cref{fig: elph circuit vqe}~\textbf{a} representing onsite and hopping terms, respectively (see \Cref{app: elph circuit gates}). 
Since the ground state of a finite-size periodic system preserves translational symmetry, we assume the gates in the same layer to share the same parameter (denoted as $\theta_n$ in \Cref{fig: elph circuit vqe}~\textbf{a}).
The expressibility of the ansatz is controlled by the number of times $n$ that the \gls{hva} is repeated.
We investigate the scaling of the ansatz depth $n$ with the system size in \Cref{sec: elph circuit depth}.

Figure \ref{fig: elph circuit vqe}~\textbf{b} shows an example of the \gls{ngs}-\gls{vqe} simulation for a 4-site (8 qubits) Hubbard-Holstein model with $u=10$ and $\lambda=10$ and the phonon wavefunction initially set to the vacuum.
Consequently, the first outer-loop iteration with \gls{vqe} (VQE\#1 in the panel) starts with a pure Hubbard model, followed by the adjustment of phonon states and \gls{ngs} parameters (NGS\#1 in the panel).
After the first outer-loop iteration (VQE\#1 and NGS\#1), the electronic state $|\psi_{\rm e}\rangle$ lies in an \gls{afm} state as a solution for the pure Hubbard model, while the phonon state $|\psi_{\rm ph}\rangle$ induces a large attractive potential (see \Cref{eq:effInteraction}).
This phonon-mediated interaction tends to stabilize a \gls{cdw}, which contradicts the \gls{afm} state (see \Cref{sec: elph phase diagram}).
As a result, the electronic state rapidly evolves once the second self-consistent iteration (VQE\#2) starts. 
In addition to the energy evolution, \Cref{fig: elph circuit vqe}~\textbf{c} shows the infidelity $1 - F = 1 - \left|\langle \Psi_{\rm VQE} | \Psi_{\rm ED}\rangle\right|^2$ of the state through the \gls{ngs}-\gls{vqe} iterations.
Importantly, the reference state $| \Psi_{\rm ED}\rangle$ is chosen as the \gls{ed} solution within each \gls{vqe} and \gls{ngs} step of the outer-loop iteration, i.e., for the respective $\mathcal{H}_\mathrm{eff}$ in \Cref{eq: elph effective hamiltonian}.
Even though the infidelity is usually inaccessible and cannot be used as the target function of the iteration, comparing energy and infidelity side by side provides insights into convergence mechanisms.
Specifically, a slow energy convergence during a comparatively rapid change of infidelity might be indicative of a barren plateau~\cite{Larocca2025barrenPlateaus}.

In contrast to solving a Hubbard model, the \gls{ngs}-\gls{vqe} method involves a self-consistent outer loop between electrons and phonons.
To mitigate optimization issues of the variational quantum circuit, for instance, the barren plateau or a multitude of local minima, we employ a multi-step optimization.
First of all, since all gates within a single ansatz layer share the same variational parameter $\theta_i$, we can reuse parameters across circuits for different system sizes $L$.
We therefore pre-run the \gls{vqe} with smaller system sizes to initialize the circuit of the target system with these converged variational parameters.
Moreover, the ground state evolves adiabatically for small changes in the model parameters ($u$, $\lambda$, and $\omega$) within the same phase.
This is why (similar to \Cref{sec: qcnn schwinger}) we further recycle converged parameters if results for similar model parameters exist.
More details on this and on adaptively adjusting the convergence criteria of the \gls{ngs}-\gls{vqe} can be found in Ref.~\citenum{Denner2023phonon} and its Supplementary Material.

\subsection{Charge and spin phases in the Hubbard-Holstein model}
\label{sec: elph phase diagram}

\begin{figure}[t!]
	\centering
    \includegraphics[width=\textwidth]{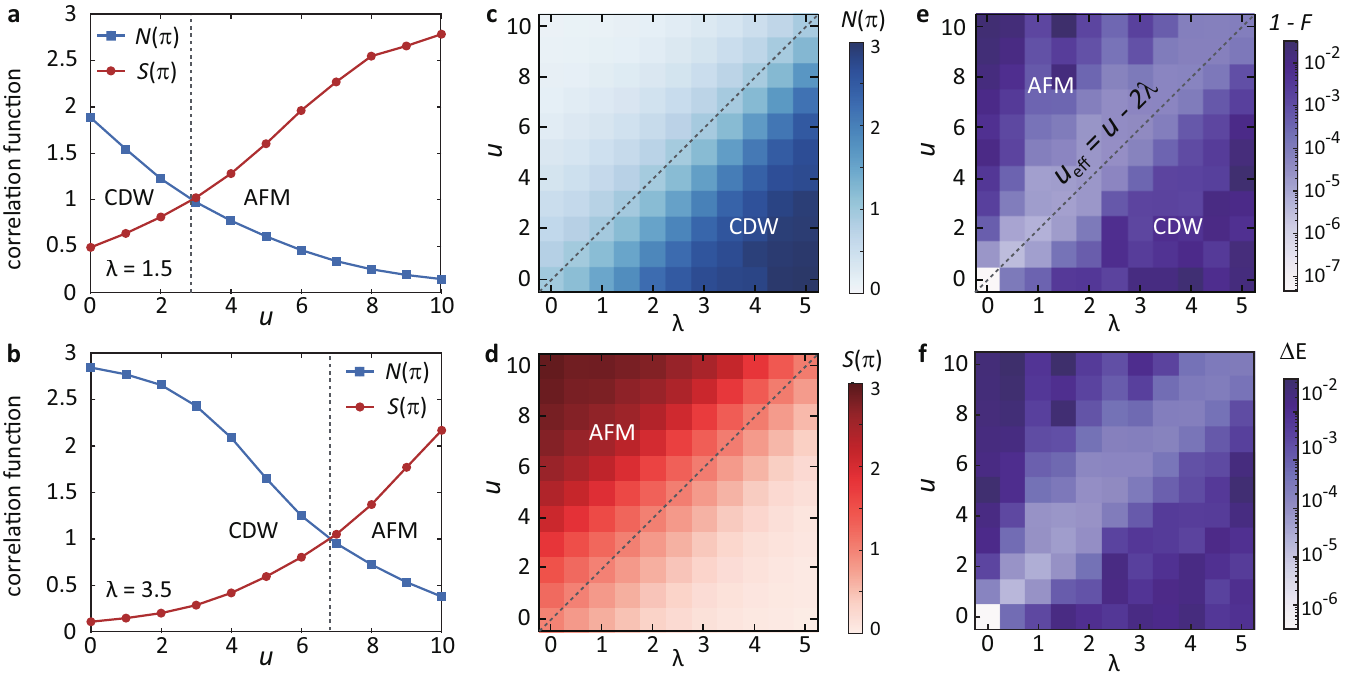}
    \mycaption{Phase diagram of the one-dimensional Hubbard-Holstein model}{
        \textbf{a}, \textbf{b} Charge $N(\pi)$ and spin $S(\pi)$ structure factors for the ground state of a 6-site Hubbard-Holstein model, simulated with the \gls{ngs}-\gls{vqe} algorithm as a function of $u$ for fixed \textbf{a} $\lambda = 1.5$ and \textbf{b} $\lambda = 3.5$.
        The charge-density-wave (CDW) and antiferromagnetic (AFM) regimes are marked.
        The phonon frequency is set as $\omega = 10$ and the quantum circuit depth is $n = 9$.
        \textbf{c}, \textbf{d} Distribution of the static \textbf{c} charge and \textbf{d} spin structure factors in the $u-\lambda$ parameter plane, for the same conditions as \textbf{a} and \textbf{b}.
        The dashed line indicates the anti-adiabatic phase boundary $u = 2 \lambda$ between AFM and CDW. \textbf{e} Ground state infidelity and \textbf{f} (absolute-value) energy error with respect to ED results for the phase diagram in \textbf{c} and \textbf{d}.
    }
    \label{fig: elph phase diagram}
\end{figure}

At half-filling and in 1D, the Hubbard-Holstein model has a rich phase diagram, hosting an AFM, CDW, and a narrow intermediate phase~\cite{Tezuka2007holsteinPhaseDiagram, Fehske2008metallicityHolstein, Clay2005holstein, Greitemann2015holstein}, providing a rich playground to study the interplay of electronic correlations and \glspl{epc}. 
To demonstrate the accuracy and efficiency of the \gls{ngs}-\gls{vqe} algorithm, we first restrict to the Hubbard-Holstein model with $g'=0$ and measure the spin and charge structure factors of the ground state for different model parameters.
The (static) spin structure factor is defined as
\begin{equation}
    S(q) = \frac1L \sum_{ij} \langle  (n_{i\uparrow}- n_{i\downarrow}) (n_{j\uparrow}- n_{j\downarrow})\rangle e^{-i q\cdot (r_i-r_j)} \ ,
\label{eq: elph spin structure factor}
\end{equation}
while the (static) charge structure factor is defined as
\begin{equation}
    N(q) = \frac1L \sum_{ij} \langle  (n_{i\uparrow}+ n_{i\downarrow}) (n_{j\uparrow}+ n_{j\downarrow})\rangle e^{-i q\cdot (r_i-r_j)} \ .
\label{eq: elph charge structure factor}
\end{equation}
Using the half-filled system with phonon frequency $\omega = 10$ as the benchmark platform, we compute both structure factors at $q=\pi$, shown in \Cref{fig: elph phase diagram}. 
In the regime $u \gg \lambda$, where electronic interactions dominate, the spin structure factor $S(\pi)$ prevails over the charge structure factor, reflecting an \gls{afm} state in a finite cluster (see \Cref{fig: elph phase diagram}~\textbf{a}, \textbf{b}, and \textbf{d}).
With the increase of $u-2\lambda$, $N(\pi)$ gradually vanishes as charge degrees of freedom are frozen with a substantial energy penalty for double occupations.
In the other limit where \glspl{epc} dominate ($\lambda \gg u$), the charge structure factor $N(\pi)$ dominates over the spin structure factor $S(\pi)$ (see \Cref{fig: elph phase diagram}~\textbf{a}, \textbf{b}, and \textbf{c}).
This reflects the onset of a \gls{cdw} state, although a more rigorous identification requires either scaling to larger system sizes or excited-state analysis.
We summarize the dependence of both spin and charge structure factors on the two interaction parameters $u$ and $\lambda$ in \Cref{fig: elph phase diagram}~\textbf{c} and \textbf{d}.
The trends of these two observables reflect the two dominant phases, qualitatively consistent with physical intuition.
Due to the underlying finite-size system, the two phases are separated by a continuous crossover instead of a sharp phase boundary.
Recent studies have shown the presence of an intermediate Luther-Emery liquid phase for $u \approx 2 \lambda$, whose width is controlled by the phonon frequency~\cite{Greitemann2015holstein}.
The discussion of this phase requires finite-size scaling and is beyond the scope of this work and Ref.~\citenum{Denner2023phonon}.

To assess the accuracy of our algorithm quantitatively, we plot the final energy error (\Cref{fig: elph phase diagram}~\textbf{e}) and the infidelity (\Cref{fig: elph phase diagram}~\textbf{f}) of the converged ground state for each set of model parameters against reference states obtained from \gls{ngsed}~\cite{Wang2020hubbardHolstein, Costa2020phase, Wang2021timeDepNonGaussian}.
The largest infidelity across all parameter ranges is of order $\mathcal{O}(10^{-2})$.
These errors do not change significantly with increasing system size, as outlined in \Cref{sec: elph circuit depth}.
Interestingly, the most accurate solutions (with infidelity of order $10^{-5}$) are obtained near the boundary of the \gls{cdw} and \gls{afm} phases, i.e., along the diagonal $u \approx 2 \lambda$.
In this regime, the finite-system solution is more metallic due to the delicate balance between the electronic repulsion and phonon-mediated attraction.
Therefore, the true ground state of systems near the phase boundary can be efficiently captured with a Slater determinant prepared only by Givens rotations~\cite{Cade2020vqeFermiHubbard, Stanisic2022vqeFermiHubbard, Jiang2018quantumAlgorithmsFermions, Wecker2015hva}. 
In contrast, even though the \gls{ngs}-\gls{vqe} algorithm yields quantitatively accurate results throughout the phase diagram, the infidelity increases when the system evolves into \gls{cdw} or \gls{afm} states.
This observation contrasts with the intuition that the \gls{afm} or \gls{cdw} states are more classical.
Instead, these states are cat states in these small and low-dimensional systems.
Therefore, an accurate representation of these highly entangled states with long-range correlations requires deeper quantum circuits, as discussed in \Cref{sec: elph circuit depth}.

Note that, so far, we have studied only relatively large phonon frequencies $\omega=10$, where the competition between \gls{cdw} and \gls{afm} states is primarily controlled by the effective local interaction $u_{\rm eff} = u - 2\lambda$.
The dependence of charge and structure factors on different phonon frequencies and system sizes is discussed in the Supplementary Material of Ref.~\citenum{Denner2023phonon}.
However, phonon frequencies in typical correlated materials are usually comparable to the electronic bandwidth, if not even reaching the adiabatic limit ($\omega \rightarrow 0$).
In the thermodynamic limit, smaller phonon frequencies usually lead to a steeper crossover between the two phases~\cite{Tezuka2007holsteinPhaseDiagram, Fehske2008metallicityHolstein}, with both $S(\pi)$ and $N(\pi)$ dropping more rapidly when approaching the phase boundary.
Note, however, that this intermediate phase cannot be resolved in small system sizes. 

\subsection{Scaling in circuit depth and system size}
\label{sec: elph circuit depth}

\begin{figure}[t!]
	\centering
    \includegraphics[width=\textwidth]{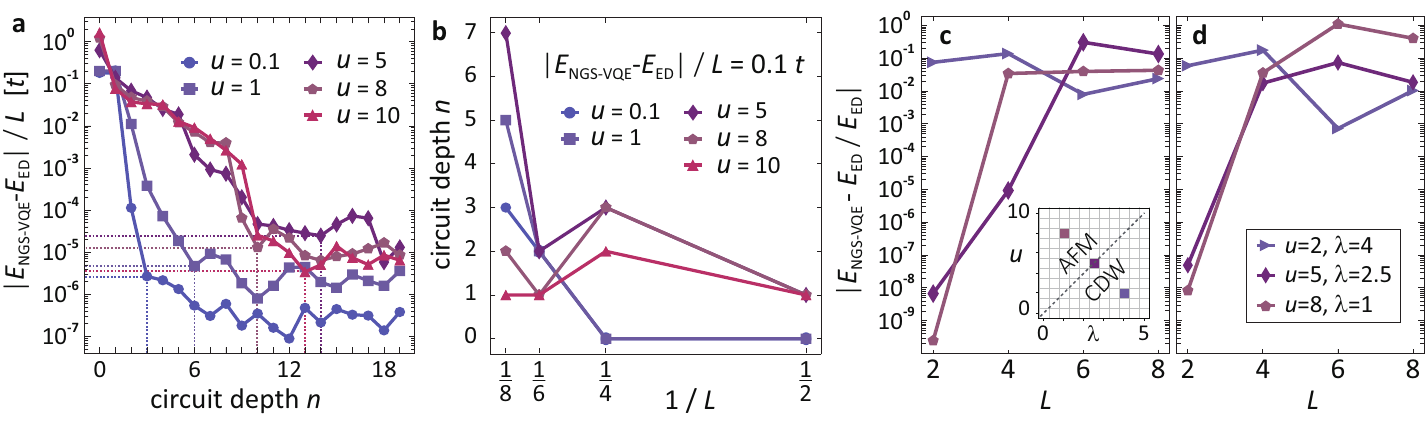}
    \mycaption{Scaling behavior of the hybrid electron-phonon solver}{
    \textbf{a} Simulation error for the ground-state energy $E_{\mathrm{\gls{ngs}-\gls{vqe}}}$ of the circuit ansatz in \Cref{fig: elph circuit vqe} as a function of circuit depth $n$ and various on-site interaction strengths $u$, compared to ED ($\lambda = 0$) for a 6-site Hubbard model (i.e., $\lambda = 0$).
    \textbf{b} Scaling of the circuit depth $n$ with system size $L$ necessary to achieve a fixed accuracy of $|E_{\mathrm{\gls{ngs}-\gls{vqe}}}-E_{\mathrm{ED}}|/L = 0.1t$.
    The required circuit depth changes only slightly with the system size for various on-site interaction strengths $u$ and $\lambda = 0$.
    \textbf{c}, \textbf{d} Relative error of the converged ground-state energy for three distinct parameter sets for the Hubbard-Holstein model ($\lambda \neq 0$) as a function of system size $L$ (the inset highlights the CDW and AFM phase) for phonon frequencies \textbf{c} $\omega = 10$ and \textbf{d} $\omega = 1$.
    The VQE was performed with the ansatz depth obtained in panel \textbf{b}, and we see that the error remains approximately constant when increasing $L$ from 4 to 8.
    }
    \label{fig: elph ansatz errors}
\end{figure}

The results presented in \Cref{fig: elph phase diagram} demonstrate that our algorithm can produce quantitatively accurate results across various parameter regimes in $u$ and $\lambda$ (and $\omega$, see Ref.~\citenum{Denner2023phonon}).
To further analyze the accuracy of our algorithm, we investigate the influence of different system sizes  $L$ and parameter regimes on the required circuit depth.
The depth $n$ of the quantum circuit controls the expressibility of the variational state, potentially allowing for a more accurate approximation of the electronic ground state by increasing $n$.
However, for an efficient encoding, the circuit depth of the variational ansatz should not scale exponentially with the system size $L$.
To this end, we first study the interplay between ansatz depth and accuracy only for the electronic problem, i.e., $\lambda = 0$, which reduces \Cref{eq: hubbard holstein hamiltonian} to the Hubbard model.
In this case, as shown in \Cref{fig: elph ansatz errors}~\textbf{a}, the ground states for small $u$ can be efficiently expressed by an ansatz close to a Slater determinant, as prepared by the Givens rotations \textsf{G} in the first part of the variational ansatz.
Thus, only a few layers are needed to reach ground state energy errors below $10^{-6}$ when compared to \gls{ed}.
However, a larger $u$ requires deeper circuits and the energy error eventually plateaus around $10^{-4}$ to $10^{-5}$.
To investigate the scaling of ansatz depth $n$ necessary to reach a certain target accuracy as a function of the system size $L$, we consider a fixed error in the electronic ground state energy of $|E_{\mathrm{VQE}}-E_{\mathrm{ED}}| / L = 0.1t$.
The circuit depth required to achieve this performance as a function of $L$ is shown in \Cref{fig: elph ansatz errors}~\textbf{b}, highlighting a moderate increase in required depth for small and large $u$.
Intermediate values for $u$, however, require significantly deeper circuits, as quantum fluctuations are larger around the crossover between metallic and \gls{afm} phase.

Next, we take into account also the phonon problem and \gls{epc} and study the full \gls{heh} model, including not only local couplings $g$ but also nearest-neighbor couplings with $g' = 1/\sqrt{5} g$.
Specifically, we consider the relative error in the ground-state energy of the converged extended-Hubbard Hamiltonian \Cref{eq: elph effective hamiltonian}, containing the phonon dressing of kinetic hopping and long-range interactions, in different regions of the phase diagram.
For the electronic part, we use the ansatz depth $n$ found in \Cref{fig: elph ansatz errors}~\textbf{b} for the purely electronic problem and a fixed error in the electron ground-state energy of $|E_{\mathrm{NGS-VQE}}-E_{\mathrm{ED}}|/L = 0.1t$.
\Cref{fig: elph ansatz errors}~\textbf{c} and \textbf{d} show the relative energy error for phonon frequencies $\omega=10$ and $\omega=1$, respectively. 
Our results indicate that the combined \gls{ngs}-\gls{vqe} simulation errors are typically smaller than those of the electronic solver.
This means errors in the \gls{vqe} solutions do not necessarily accumulate, but can actually be mitigated by the phonon solver.
The largest errors are obtained for small phonon frequencies (\Cref{fig: elph ansatz errors}~\textbf{d}, $\omega = 1$), where the absence of quantum fluctuations hinders the phonon solver from escaping local minima during the self-consistent iteration~\cite{Wang2020hubbardHolstein}.
Warm-up iterations with larger phonon frequencies can help to alleviate this issue~\cite{Wang2020hubbardHolstein}.
Moreover, the relative errors do not increase for systems larger than $L = 4$, indicating quantitatively accurate results across different system sizes and phases.
The only exception appears in the \gls{cdw} phase at small phonon frequencies $\omega = 1$, where the relative error oscillates with $L$, likely due to the degeneracy of ground states.
The ability to mitigate errors of the quantum solver also provides a promising path to experimental realizations.
Hardware implementations, irrespective of the specific platform, suffer from decoherence, making noise-resilient algorithms of key importance.
Our hybrid algorithm is able to improve \gls{vqe} results over a range of phonon frequencies, phase regions, and noise levels, suggesting efficient hardware realizations.

\subsection{Circuit transpilation for hardware experiments}
\label{sec: elph circuit transpilation}

\begin{figure}[t!]
	\centering
    \includegraphics[width=\textwidth]{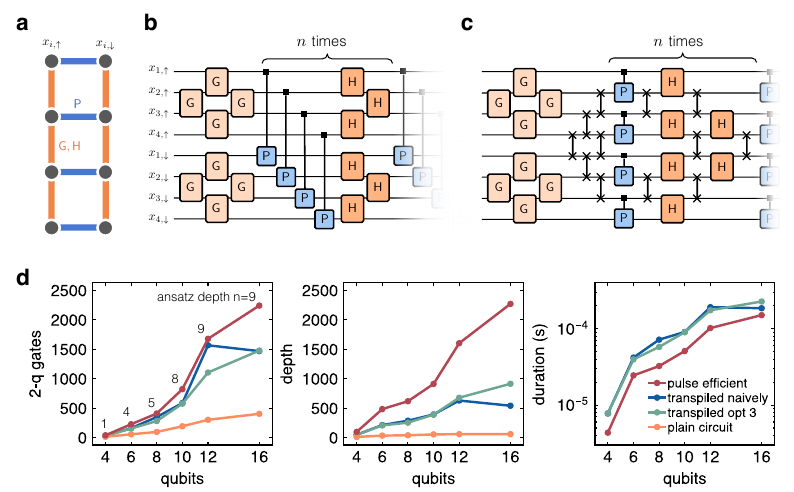}
    \mycaption{Transpilation of variational circuit to linear connectivity}{
    \textbf{a} Required connectivity for the \gls{hva} circuit of a 1D Hubbard lattice in \textbf{b} (same as in \Cref{fig: elph circuit vqe}).
    Shown here is the example of a 4-site system encoded in 8 qubits.
    This analysis generalizes to arbitrary system sizes.
    Circles represent qubits encoding spin-up $x_{i,\uparrow}$ and spin-down $x_{i,\uparrow}$ degrees of freedom on each lattice site, and color-coded edges indicate between which qubits the two-qubit gates present in the variational ansatz act.
    \textbf{b} The variational circuit from \Cref{fig: elph circuit vqe} could be mapped to qubits as is if the topology of the underlying quantum chip was a square lattice.
    \textbf{c} The same variational circuit as in \textbf{b} but including SWAP gates that allow to map the circuit to a linear chain of qubits, necessary, for example, to map the circuit to IBM's heavy-hexagonal qubit connectivity.
    \textbf{d} Absolute number of two-qubit gates, gate depth, and circuit duration (left to right) for various levels of transpiling the circuit in \textbf{b} to \kolkata{} with heavy-hex connectivity~\cite{IBMQuantumPlatform}.
    Data is shown for various system sizes $N=2, \ldots 8$ with $4, \ldots 16$ qubits and the ansatz depth $n$ used for each system size indicated in the left-most plot.
    The three curves correspond to different levels of transpilation.
    ``plain circuit'' indicates no insertion of swap gates, basis gate transpilation, etc.
    The two out-of-the-box transpilation routines ``naively'' and ``opt3''  refer to different levels of circuit optimization found in \qiskit{}~\cite{qiskit2024}.
    Pulse-efficient transpilation refers to the method presented in \Cref{sec: pulse efficient}.
    The circuit duration can be computed only for circuits transpiled to hardware-native basis gates and scheduled pulse-sequences, which is why ``plain circuit'' is missing in the last panel of \textbf{c}.
    }
    \label{fig: elph circuit transpilation}
\end{figure}

In this section, we study the feasibility and related resource requirements of executing the electronic solver on a quantum device, i.e., to execute the variational circuit from \Cref{fig: elph circuit vqe}~\textbf{a} on an IBM quantum computer.
Specifically, the circuit needs to be transpiled to fit the quantum chip's qubit connectivity, and all gates in the circuit need to be transpiled to the respective basis gate set.
We do not include error mitigation and suppression techniques (see \Cref{sec: error mitigation}) in the following analysis, which can further increase the computational overhead, depending on the methods employed.
It should be noted that similar circuits have been previously executed on hardware~\cite{Stanisic2022vqeFermiHubbard} to approximate ground states of a pure Hubbard model.
However, Ref.~\citenum{Stanisic2022vqeFermiHubbard} had access to a qubit lattice with square connectivity, which is optimal for the given variational circuit, as detailed below.

As previously outlined, the qubit register is split into a spin-up and a spin-down sub-register, with one qubit per spin-up and spin-down on each spatial lattice site.
Effectively, due to this mapping of spin modes to separate qubits,
the \gls{hva} circuit requires a qubit connectivity of one dimension higher than the lattice dimension.
For example, encoding a 1D chain, like we study in this work, requires a square (2D) qubit connectivity, as can be seen in \Cref{fig: elph circuit transpilation}~\textbf{a}, where this is shown on the example of an 8-qubit system ($L=4$).
Here, each point represents a qubit, while each edge represents one of the two-qubit gates present in the variational circuit.
In this case, the circuit can be transpiled to the chip's respective basis gate set without insertion of SWAP layers (see \Cref{fig: elph circuit transpilation}~\textbf{b}).
Analogously, encoding a 2D lattice effectively requires a three-dimensional (3D) qubit connectivity to make all two-qubit gates in the variational circuit nearest-neighbor, and so forth.
However, if a given device does not support the required qubit connectivity of the circuit, the quantum circuit necessarily contains non-local two-qubit gates, i.e., two-qubit gates that are applied on non-neighboring qubits.
This inevitably increases the circuit depth as SWAP gates must be included between the different gate layers to compensate.
The heavy-hexagonal qubit lattice of an IBM device is much more sparsely connected than a square lattice.
This means we would need to include SWAP gates in our circuit to execute all gates in the circuit.
One such SWAP strategy, i.e., one way to execute our circuit on a linearly connected chain of qubits using SWAP gates, is shown in \Cref{fig: elph circuit transpilation}~\textbf{c}.
For the 8-qubit example, we need six SWAP gates between the \textsf{G} and the \textsf{P} layer.
Then, in each of the $n$ circuit layers of the \gls{hva}, two SWAPs are needed between \textsf{P} and the odd-site \textsf{H} layers, three SWAPs between the odd and the even-site \textsf{H} layers, and one SWAP between the even-site \textsf{H} layers and the \textsf{P} layers.
In total, this means we need $6 + 6 n - 1$ SWAP gates for an $n$-layer 8-qubit ansatz.
Given that each SWAP gate is implemented using CNOT gates and that two-qubit gates are the primary source of error, this generates significant gate-overhead and, consequently, additional errors.

\Cref{fig: elph circuit transpilation}~\textbf{d} summarizes (from left to right) the two-qubit gate counts, gate-depth, and circuit duration of different variational circuits (different ansatz depths $n$) for several system sizes after various levels of transpilation.
Note that the gate-depth here means total gate-depth, including both one- and two-qubit gates.
The circuit duration is the physical time that would be needed to execute the circuit on the device and is obtained through a last transpilation step that translates all circuit operations to pulse instructions, called scheduling.
As such, circuit duration is the most significant of the three numbers in \Cref{fig: elph circuit transpilation}~\textbf{d}.
Here, ``plain circuit'' refers to the bare, untranspiled circuit, and ``transpiled naively'' and ``transpiled opt 3'' to optimization levels 1 and 3, respectively, using \qiskit{}'s transpilation routines.
The transpiler can optimize the circuit depth since it is sometimes possible to permute and combine certain gates.
However, these techniques often work stochastically, and optimization therefore increases the time of transpilation per circuit.
In addition, we also compare to pulse-efficient transpilation (see \Cref{sec: pulse efficient}).
The depth-5 variational circuit to solve the electronic part of an 8-qubit system includes 281 two-qubit gates when transpiled to \kolkata{}~\footnote{\kolkata{} is one of IBM's by now retired 27-qubit Falcon chips} (optimization level 3) and a circuit duration of \qty{49}{\us}.
As in \Cref{chap: benchmarking}, pulse-efficient transpilation reduces circuit duration by roughly 40\%, resulting in \qty{30}{\us} for the same circuit.
Note that this is despite an increased circuit depth, which is larger only because pulse-efficient transpilation includes additional basis rotations.
In \Cref{chap: benchmarking}, specifically \Cref{subsec: benchmarking 1D results} and \Cref{subsec: benchmarking heavy hex results}, we could observe coherent simulation of a 100-qubit chain for up to \qty{15}{\us}~\cite{Miessen2024benchmarking}.
Comparing these numbers to those in \Cref{fig: elph circuit transpilation}~\textbf{d}, we notice that we are at the limit of current hardware capabilities with only an 8-qubit circuit.
Here, it is important to keep two things in mind.
First, one is usually limited by the volume of the circuit, i.e., width (number of qubits) and depth, not only by its depth.
It is therefore conceivable that an 8-qubit circuit achieves coherence at larger circuit depths than a 100-qubit circuit.
Second, and more importantly, the hardware available at the time of preparing Ref.~\citenum{Denner2023phonon} (e.g., \kolkata{}) was far more error-prone than that available during the time of preparing \Cref{chap: benchmarking}(Ref.~\citenum{Miessen2024benchmarking}).
For this latter reason, as we were aware, it would require high levels of error mitigation, fine-tuning thereof, and experimental care, and due to time constraints, we were not able to conduct actual hardware experiments using our hybrid electron-phonon solver.
Instead, to still put our routine to a more realistic test and study the influence of noise, we chose to conduct noisy simulations.

\subsection{Noisy simulations}
\label{sec: elph noisy simulations}

\begin{figure}[t!]
	\centering
    \includegraphics[width=0.8\textwidth]{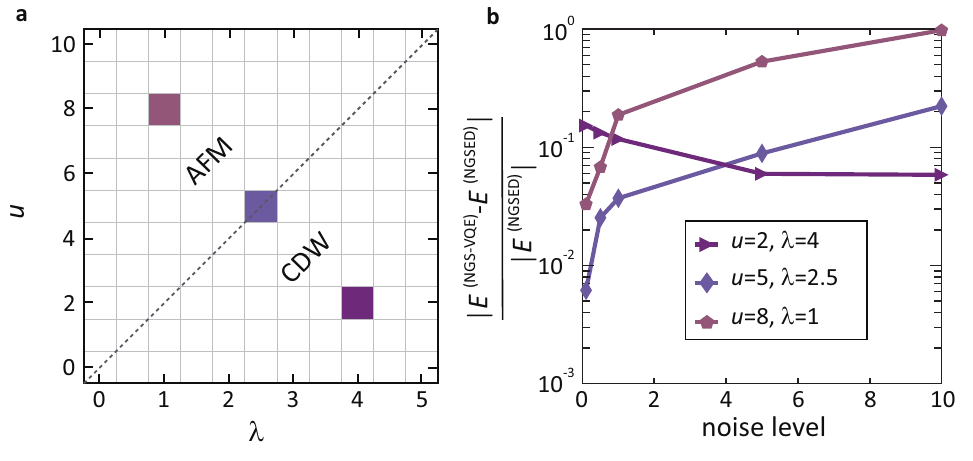}
    \mycaption{Influence of noise on the performance of the hybrid electron-phonon solver}{
    \textbf{a} Phase diagram of the 1D Hubbard-Holstein model.
    The highlighted points were used for the noise simulations of panel \textbf{b}. \textbf{b} Relative error of the converged ground state energy for three distinct points in the $u,\lambda$ phase diagram as a function of device noise strength in the VQE solver ($L = 4$, $\omega = 10$).
    A noise strength of 1 corresponds to a typical configuration of IBM's device \kolkata.
    VQE was performed with the circuit depth obtained in \Cref{fig: elph ansatz errors}~\textbf{b}.
    While the relative error in the ground state energy increases with increasing noise for the phase transition point and AFM phase, it decreases in the CDW phase.}
    \label{fig: elph noisy simulations}
\end{figure}

To assess the practicality of our approach with near-term quantum devices, we study how resilient it is to characteristic hardware noise.
To this end, we transpile an $8$-qubit circuit of the $L=4$ site \gls{heh} model to a linear chain of qubits on \kolkata{} and simulate the hardware-transpiled circuit, including both statistical noise ($10^5$ circuit samples per operator expectation value) and realistic device noise in our simulations.
Importantly, the circuit transpiled to only linearly connected qubits includes a series of swap gates (see next section \Cref{sec: elph circuit transpilation} and \Cref{fig: elph circuit transpilation}), increasing the gate-depth and therefore the impact of device noise significantly.
Analogous to \Cref{sec: benchmarking noisy simulations}, we use a simplified noise model based on \kolkata{}'s calibration data and generated through \textsc{Qiskit Aer}.
We use the same technique of scaling the average device errors across several orders of magnitude, with scaling factors $\eta \in \{ 0.1,0.5,1,5,10 \}$ as in \Cref{eq: benchmarking scaled noise}.
Here, we use the average noise values of \kolkata{}, $T_1^{\mathrm{ave}} = 106.1 \, \mathrm{\mu s}$, $T_2^{\mathrm{ave}} = 82.93 \, \mathrm{\mu s}$, $e_{\mathrm{1q}}^{\mathrm{ave}} = 3.78 \times 10^{-4}$, $e_{\mathrm{2q}}^{\mathrm{ave}} = 1.07 \times 10^{-2}$, and $e_{\mathrm{ro}}^{\mathrm{ave}} = 2.27 \times 10^{-2}$, which are the device's average relaxation time, dephasing time, one-qubit gate error, two-qubit gate error, and readout error, respectively.

We observe the expected behavior of an increasing simulation error with increasing levels of device noise in the \gls{afm} phase and at the phase transition (see \Cref{fig: elph noisy simulations}~\textbf{b}).
We obtain relative error rates $\Delta_\mathrm{rel}(E) = |E_{\mathrm{NGS-VQE}}-E_{\mathrm{NGSED}}|/|E_{\mathrm{NGSED}}|$ at $\eta = 1$ about one order of magnitude larger than in the statevector case (see \Cref{fig: elph ansatz errors}).
The error drops off for lower noise levels that are comparable to the ones on updated quantum devices, such as \torino{}, and anticipated with next-generation devices.
Interestingly, in the \gls{cdw} phase, we see a slight decrease in simulation error with increasing hardware noise.
This can occur when the optimal solution is such that the noise will naturally relax the system toward this solution, e.g., the qubit ground state $\ket{0\ldots0}$.
Furthermore, fluctuations have proven to be crucial to escape local minima~\cite{Wang2020hubbardHolstein}, possibly also explaining the increasing accuracy of the simulation results with noise.

\subsection{Conclusion}
\label{sec: elph conclusion}

Our \gls{ngs}-\gls{vqe} method provides a general framework for an accurate and efficient study of ground state properties of electron-phonon systems with arbitrary interaction strengths.
Using this method, we have studied the Hubbard-Holstein and \gls{heh} models, reproduced the \gls{cdw}-\gls{afm} crossover, and, in Ref.~\citenum{Denner2023phonon}, extracted the phonon-mediated long-range interactions across a wide range of phonon frequencies.
While we focused on paradigmatic (and experimentally relevant) cases, this method can be generally applied to any model with electronic Coulomb correlations and Fr\"ohlich-type electron-phonon couplings.
The algorithm is extendable -- at the cost of an increased computational complexity -- to other types of electron-boson interactions (like Su-Schrieffer-Heeger phonons and cavity QED) through a generalization of the non-Gaussian transformation $U_{\rm NGS}$ and its optimization strategy.
Anharmonic potentials can be tackled at the price of replacing $|\psi_{\rm ph}\rangle$ with more complicated many-body wavefunctions similar to the electronic ones.
Moreover, this framework could be extended to non-equilibrium dynamics~\cite{Wang2021timeDepNonGaussian}, employing a quantum algorithm for the long-time propagation of $|\psi_{\rm e}\rangle$~\cite{Miessen2023perspective, Miessen2021quantum}.
The study of non-equilibrium dynamics further allows for the study of excited-state spectra through Fourier analysis~\cite{Ollitrault2021molecular, Miessen2023perspective}. 

Moreover, our work shows that the phonon solver can mitigate potential errors of the quantum hardware, simplifying a future experimental implementation.
As highlighted in \Cref{sec: elph circuit transpilation,sec: elph noisy simulations}, an experimental realization of our protocol was not feasible at the time of publication of Ref.~\cite{Denner2023phonon}.
This was mostly due to the gate-overheads inferred by the non-local two-qubit gates in the circuit and the limited connectivities of available devices.
At the same time, however, simulated noise indicated that the algorithm is stable against low levels of noise.
Given the low error rates of modern quantum devices~\cite{IBMQuantumPlatform} and potentially higher connectivities in the coming years, an implementation on quantum hardware, paired with error mitigation, could be successful already.

\section{Phase classification in many-body quantum systems}
\label{sec: qcnn}

\begin{zitat}{}
    This section is reproduced with permission in parts from Lento Nagano, Alexander Miessen, Tamiya Onodera, Ivano Tavernelli, Francesco Tacchino, and Koji Terashi, ``Quantum data learning for quantum simulations in high-energy physics'', Physical Review Research \textbf{5}, (4), 043250 (2023)~\cite{Nagano2023datalearning}.
    \Cref{sec: qcnn method} introduces the general framework of quantum data learning and quantum convolutional neural networks (QCNNs).
    This methodology is then applied to two example models in \Cref{sec: qcnn schwinger,sec: qcnn z2}, where we employ the QCNN to classify phases of ground states in the Schwinger model and of time-evolved states in a $\mathbb{Z}_2$ model, respectively.
    Alexander Miessen contributed to the implementation, testing, and optimization of the algorithm.
    In particular, he was responsible for the selection of relevant physical system parameters and the variational state preparation in \Cref{sec: qcnn schwinger}.
\end{zitat}

\Gls{qml} has received considerable attention in recent years~\cite{Biamonte2017QuantumLearning, Cerezo2022challengesQML}, not least due to the general hype around classical machine learning and quantum computing.
On one hand, provable performance guarantees have been found for artificial problems~\cite{Liu2021speedupQML} and significant progress has been made in understanding the trainability and generalization power of \gls{qml} models based on \glspl{vqa}~\cite{Cerezo2020variational, Abbas2021power, Caro2022generalizationQML, Larocca2025barrenVQA}.
On the other hand, the latter are inherently heuristic, and it remains largely unclear whether, and, if so, to what extent, \gls{qml} could complement or even surpass classical models.
Even more so with increasing evidence pointing to an efficient classical simulability of \gls{qml} models that are based on variational training~\cite{Thanasilp2023subtletiesTrainability, Thanasilp2024exponentialConcentrationKernal, Cerezo2024barrenAbsenceClassical}.
While there is growing evidence that narrows hopes for a quantum advantage in \gls{qml} with classical input data, it is not yet clear how much this affects \gls{qml} with input data that is genuinely quantum mechanical.
There, quantum models might still hold an advantage in maintaining correlations in the input states that would be lost in classical models.

In this context, it appears more appealing to further investigate \gls{qml} for inherently quantum mechanical data, i.e., use genuine quantum states as input for the respective \gls{qml} model.
This approach, referred to as \gls{qdl}, eliminates the potentially costly loading of classical data and, instead, directly leverages the capabilities of \gls{qml} architectures to manipulate quantum states.
\Gls{qdl} encompasses paradigmatic problems such as phase recognition of strongly correlated systems~\cite{Cong2019qcnn, SanchoLorente2022kernalsPhases, Herrmann2022qcnnPhases, Monaco2023qnnPhases, Caro2022generalizationQML} and learning of quantum processes~\cite{Huang2022quantumLearning, Beer2020trainingQNN, Caro2023quantumLearning} and may therefore impact the research areas such as quantum many-body physics, high-energy physics, and quantum chemistry. 

We adopt the approach of supervised \gls{qml} with a parameterized quantum circuit~\cite{Benedetti2019parameterizedQML} to classify phases of quantum states.
Specifically, we employ a \gls{qcnn} (see \Cref{fig: qcnn scheme}), a special variant of a \gls{qnn} ansatz~\cite{Cong2019qcnn}, motivated by the successful application of classical \glspl{cnn} to classify phases of matter~\cite{Carrasquilla2017machineLearningPhases}.
In this regard, \glspl{qcnn} are closely related to the \gls{mera} in \glspl{tn} and are known to be particularly well suited for paradigmatic \gls{qdl} applications such as quantum phase recognition~\cite{Cong2019qcnn, Herrmann2022qcnnPhases, Monaco2023qnnPhases}.
Moreover, \glspl{qcnn} are expected to have reduced sampling costs compared to conventional methods for non-local order parameters~\cite{Cong2019qcnn}.
\Glspl{qcnn} consist of a naturally shallow circuit structure, necessitating only $O(\log(N))$ variational parameters, and are therefore believed to be beneficial for near-term quantum devices.
Furthermore, they are provably resilient to the phenomenon of barren plateaus, which hinders the trainability of more generic \gls{qnn} models~\cite{Pesah2021barrenAbsenceQCNN}.

Here and in Ref.~\citenum{Nagano2023datalearning}, we apply the \gls{qdl} framework with \glspl{qcnn}, which is detailed in \Cref{sec: qcnn method}, to two cases of phase recognition in quantum states in prototypical toy models in high-energy physics.
A schematic of the framework is shown in \Cref{fig: qcnn scheme}.
First, in \Cref{sec: qcnn schwinger}, we classify ground state phases of the Schwinger model, which we prepare variationally using \gls{vqe}.
Note that, while \gls{vqe} is likely limited to small system sizes, ground states resulting from any other post-\gls{vqe} or fault-tolerant algorithm could be used as input states to the \gls{qcnn} as well.
Second, in \Cref{sec: qcnn z2}, we classify phases of time-evolved states in a $\mathbb{Z}_2$ gauge theory obtained from Trotter evolution.
Ref.~\citenum{Nagano2023datalearning} further applies \gls{qdl} to learn the coupling parameters of an effective field theory Hamiltonian from many-particle input states.
All results presented in this section are obtained from noiseless statevector simulations.

\subsection{Quantum data learning via a quantum convolutional neural network}
\label{sec: qcnn method}

\begin{figure}[t]
    \centering
    \includegraphics[width=0.8\textwidth]{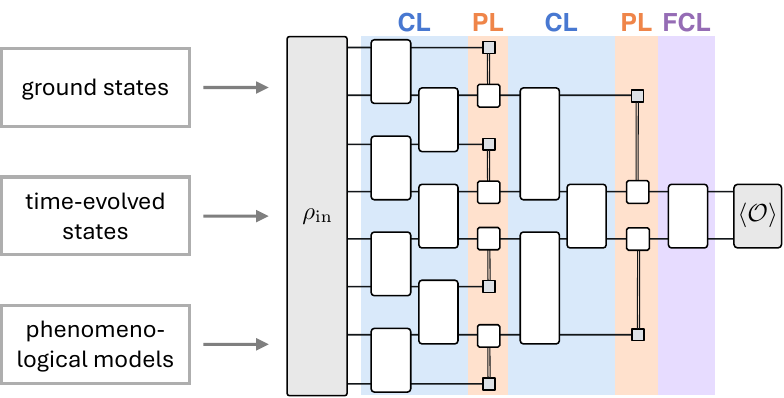}
    \mycaption{Architecture of the quantum convolutional neural network}{
	Structure of a prototypical \gls{qcnn} circuit, consisting of convolutional layers (CL), pooling layers (PL), and a fully connected layer (FCL).
	Within the \gls{qdl} framework, the input state $\rho_\mathrm{in}$ is a quantum state, prepared with a preceding preparation circuit.
	Here, this is either a ground state (prepared through \gls{vqe}) or a time-evolved state (prepared through a Trotter circuit). In Ref.~\citenum{Nagano2023datalearning}, a third type of input state representing particle number states of effective field theories is considered.
	More details on the concrete circuits can be found in Ref.~\citenum{Nagano2023datalearning}.
    }
    \label{fig: qcnn scheme}
\end{figure}

We will first briefly outline the learning process within the general framework of supervised \gls{qml} with a \gls{qnn}.
In the applications presented in this work, input quantum states may implicitly encode some classical parameters~${\bm{x}}$, such as certain values of coupling constants appearing in an underlying Hamiltonian.
Corresponding labels $y_{\bm{x}}$ denote, for example, the physical phase, properties of the quantum state, or Hamiltonian parameters that the model should learn to recognize or predict.
For input states to the \gls{qnn}, we generically write~$\rho_{\rm in} \equiv \rho(\bm{x})$, which are chosen from a training dataset~$\{ ( \rho(\bm{x} ), y_{\bm{x}}) \}_{\bm{x} \in \mathcal{T}_\mathrm{train}}$.
In practice, for any given ${\bm{x}}$, the corresponding input state $\rho(\bm{x})$ will be prepared by a unitary $U_{\rm prep} (\bm{x})$, i.e., a circuit preceeding the \gls{qcnn}.
Depending on the problem, $U_{\rm prep}(\bm{x})$ can take various forms as depicted in \Cref{fig: qcnn scheme}, including a ground state preparation for a given Hamiltonian $H(\bm{x})$, a time evolution circuit $U=e^{-iH(\bm{x})t}$, or another phenomenological quantum simulation circuit~\cite{Nagano2023datalearning}.
In a general \gls{qnn} setting, each input state is processed by a circuit $U_{\rm QNN}({\bm{\theta}})$ parameterized with a set of trainable variables ${\bm{\theta}}$.
This unitary evolves $\rho_\mathrm{in}$ to an output state
\begin{equation}
    \rho_{\rm out} ( \bm{x}, {\bm{\theta}} ) = U_{\rm QNN}({\bm{\theta}})\rho(\bm{x}) U_{\rm QNN}^\dagger({\bm{\theta}}) \ .
\end{equation}
Upon measurement of the expectation value of a given observable, $\langle \mathcal{O} \rangle = {\rm Tr}[\rho_{\rm out} \mathcal{O} ]$, the output state produces a model output $y_{\rm out}(\bm{x}, \bm{\theta})$.
The learning is then done by optimizing the set of \gls{qnn} parameters ${\bm{\theta}}$, aiming to minimize a specified cost function ${\cal L}(\bm{\theta})$.
Specifically, the cost function quantifies the distance of the model predictions $\{ y_{\rm out} ( \bm{x}, \bm{\theta} ) \}_{\bm{x} \in \mathcal{T}_\mathrm{train}}$ from the known, ideal training labels $\{y_{\bm{x}}\}_{\bm{x}\in\mathcal{T}_\mathrm{train}}$,
\begin{equation}
    \mathcal{L} (\bm{\theta}) = \frac{1}{\left | \mathcal{T}_{\mathrm{train}} \right| } \sum_{\bm{x} \in \mathcal{T}_{\text{train}}} (y_{\bm{x}} - y_{\rm out} (\bm{x}, \bm{\theta}))^{2} \ .
\label{eq: qnn cost function}
\end{equation}
Here, $\left|\mathcal{T}_{\text{train}}\right|$ is the number of training data.
The cost function is then minimized classically to find the optimal set of parameters,
\begin{equation}
    \bm{\theta}_{\text{opt}} = \arg \min_{\bm{\theta}} \mathcal{L} (\bm{\theta}) \ .
\end{equation}
Given an optimized set of parameters $\bm{\theta}_\mathrm{opt}$, the trained model should ideally be able to provide the correct prediction label $y_{\rm out}(\bm{x}, \bm{\theta}_\mathrm{opt})$ for the test data $\bm{x}\in\mathcal{T}_{\text{test}}$. 
Note, however, that perfect generalization of the model from training to test data is not guaranteed, and, hence, the parameters minimizing the cost function do not guarantee zero error on the test data.

As mentioned above, we employ a \gls{qcnn} as ansatz, a specific class of \gls{qnn} models.
A prototypical \gls{qcnn} circuit $U_{\text{QCNN}}$ is schematically shown in \Cref{fig: qcnn scheme}.
It consists of alternating convolutional layers (CL) and pooling layers (PL).
The former apply translationally-invariant unitary gates to local subsets of qubits, while the latter reduce the dimensionality of the state by local measurements and classical feed-forward operations.
Finally, a fully connected layer (FCL) is applied globally to the remaining qubits, followed by the measurement of an observable.
With a QCNN model $U_{\text{QCNN}}$, the supervised \gls{qdl} architecture outlined above can be further detailed as follows.
For an $N$-qubit quantum data input of the form $\rho_\mathrm{in}(\bm{x}) = \ket{\psi_{\bm{x}}} \bra{\psi_{\bm{x}}}$, the model output $y_{\rm out}(\bm{x}, \bm{\theta})$ is obtained by measuring the last qubit in the $Z$ basis, namely
\begin{equation}
    y_{\rm out}(\bm{x}, \bm{\theta})=\braket{\psi_{\bm{x}} | U_{\mathrm{QCNN}}^{\dag} c Z_{N-1}U_{\mathrm{QCNN}} | \psi_{\bm{x}}} \ .
\end{equation}
Here, $c$ is a scaling factor to scale the output into the interval $[-c,c]$.
In the following two sections, for phase recognition in the Schwinger and $\mathbb{Z}_2$ model, we set $c=1$.
The output is then fed into the cost function \Cref{eq: qnn cost function}.
We use the same cost function to train the \gls{qcnn} for both models discussed here\footnote{Note that, for the additional phenomenological model studied in Ref.~\citenum{Nagano2023datalearning}, a modified cost function was used.}
The number of trainable parameters ${\bm{\theta}}$ used here is $21 N_L$, where $N_L$ is the number of convolutional and pooling layers.
The concrete circuit structure of $U_\mathrm{QCNN}$ is detailed in \Cref{app: qcnn circuit structure}.

The number of trainable parameters ${\bm{\theta}}$ used here is $21 N_L$, where $N_L$ is the number of convolutional and pooling layers.
The concrete circuit structure of $U_\mathrm{QCNN}$ is detailed in \Cref{app: qcnn circuit structure}.

\subsection{Classifying ground state phases of the Schwinger model}
\label{sec: qcnn schwinger}

In this section, we employ the \gls{qcnn} for phase recognition in ground states of the Schwinger model.
The aim is to train the \gls{qcnn} with variationally prepared ground states provided as input data and their associated phases as labels, and to predict the phase of states for which it is unknown.
We first introduce the Schwinger model and its phases, then describe how we prepare the input dataset using \gls{vqe} and construct the QCNN model.
Finally, we present some numerical results.

\subsubsection{The Schwinger model}

The Schwinger model is a $(1+1)$-dimensional gauge theory, i.e., one spatial and one temporal dimension, with a U(1) gauge symmetry~\cite{Schwinger1962schwingerModel, Martinez2016realTimeLGT}. 
Its continuum Lagrangian reads
\begin{equation}
    \mathcal{L}_{\rm Schwinger} =  -\frac{1}{4} F_{\mu\nu} F^{\mu\nu} +i \bar{\psi} \gamma^\mu (\partial \mu + i gA_\mu -m) \psi + \frac{g\vartheta}{4\pi} \epsilon_{\mu\nu} F^{\mu\nu} \ .
\end{equation}
Here, the first term is the kinetic term for a gauge field $A_{\mu}$, the second term comprises the kinetic and mass term of a Dirac fermion~$\psi$ as well as its coupling to the gauge field, and, lastly, a topological term, also known as $\vartheta$-term.
We can turn this into a \gls{lgt} by discretizing the corresponding Hamiltonian~\cite{Mathis2020scalableLGT}.
The lattice Hamiltonian can be obtained via the staggered fermion formalism~\cite{Kogut1975wilsonFermions}, which assigns fermions to the sites of the lattice and gauge fields to its edges.
In $1$D, we can further integrate out all gauge fields, making use of the open boundary conditions and Gauss's law~\cite{Kokail2019schwinger}.
Applying the Jordan-Wigner mapping~\cite{Jordan1928jordanWigner} to map fermion to qubit operators, the Schwinger Hamiltonian reads
\begin{equation}
\begin{aligned}
    H = & J \sum_{n} \left[ \sum_{i=0}^{n} \frac{Z_i + (-1)^i}{2}
    + \frac{\vartheta}{2\pi} \right]^2 \\[5pt] 
    & + \frac{w}{2} \sum_{n} \big[X_n X_{n+1}+Y_{n}Y_{n+1} \big] 
    + \frac{m}{2} \sum_{n} (-1)^n Z_n 
     \ .
\end{aligned}
\label{eq: schwinger hamiltonian}
\end{equation}
Here, $N_s$ is the number of spatial lattice sites with spacing $a$, $J=ag^{2}/2$, and $w=1/(2a)$.

\begin{figure}[t]
    \centering
    \includegraphics[width=\textwidth]{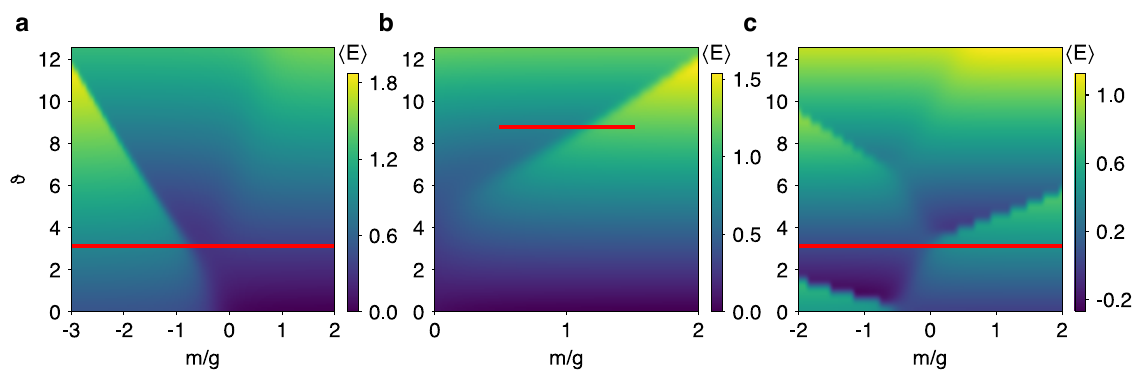}
    \mycaption{Schwinger model phase diagrams}{
	Average electric field in the $\vartheta$-$m/g$ plane, where $\vartheta$ acts as an external gauge field.
	From left to right, the plots show system sizes and lattice spacings of \textbf{a} $N_s=4, a=1$, \textbf{b} $N_s=6, a=1$, \textbf{c} $N_s=8, a=2$.
	The red line highlights the parameter range that is chosen to study the ground state variationally with \gls{vqe} and, in the case of $N=8$, to perform phase classification using the \gls{qcnn}.
	The parameter ranges are \textbf{a} $\vartheta=\pi, m/g \in [-3, 2]$, \textbf{b} $\vartheta=2.8\pi, m/g \in [-1.5, 1.5]$, \textbf{c} $\vartheta=\pi, m/g \in [-2, 2]$.
    }
    \label{fig: schwinger phase diagram}
\end{figure}

The continuum model is known to exhibit a phase transition at $\vartheta=\pi$ and a critical mass $(m/g)=(m/g)_{c} \approx 0.33$~\cite{Buyens2017schwingerPT}. 
A simple order parameter characterizing this phase transition is the average electric field,
\begin{equation}
    E = \frac{1}{N} \sum_{n} \sum_{i=0}^{n} \frac{Z_i + (-1)^i}{2} \ . 
\label{eq: schwinger electric field}
\end{equation}
At $\vartheta=\pi$, the electric field vanishes below the critical mass (symmetric phase), $\braket{E}=0$, and is non-zero above the critical mass (symmetry-breaking phase), $\braket{E}\neq0$.
When studying the discrete model, finite size effects and discretization errors are inevitable, in particular for smaller system sizes like the ones we study below.
As a result, the phase diagram will have a different structure for different system sizes.
This can be seen in \Cref{fig: schwinger phase diagram} for system sizes of $N=4, 6, 8$.
Here, we show the phase diagrams of the Schwinger model in the $\vartheta$-$m/g$-plane, with the expectation value of the electric field \Cref{eq: schwinger electric field} as order parameter.
Although the transition $\braket{E} = 0 \rightarrow \braket{E} \neq 0$ at $(m/g)_{c} \approx 0.33$ and $\vartheta = \pi$ expected in the continuum is not present, many other transitions are present that suffice to study our phase characterization scheme.
The red lines in \Cref{fig: schwinger phase diagram} depict parameter ranges we choose to study variationally to prepare input states for the \gls{qcnn}.

\subsubsection{Grounds state preparation as input data}

We generate an input dataset $\{ \ket{\psi_m}, y_m \}_{m \in \mathcal{M}}$ consisting of ground states of \Cref{eq: schwinger hamiltonian} for different masses $m \in \mathcal{M}$, which we prepare using \gls{vqe}.
For every value of the mass $m$, the energy of the system is minimized through classical optimization of the expectation value,
\begin{equation}
    \bm{\lambda}_\mathrm{opt} (m) = \arg\min_{\bm{\lambda}} \braket{\psi(\bm{\lambda}) | H(m) | \psi(\bm{\lambda})} \ .
\label{eq: schwinger energy minimization}
\end{equation}
Here, we explicitly write $H(m)$ to indicate that the mass is a parameter of the Hamiltonian.
The input data is then produced by $\ket{\psi_{m}} := \ket{\psi(\bm{\lambda}_\mathrm{opt}(m))}$.
As mentioned above, the order parameter \Cref{eq: schwinger electric field} is measured on the ground state, labeling the respective phase as
\begin{equation}
    y_{m} =
    \begin{cases}
        +1 \ , & \braket{E} \neq 0 \\ 
        -1 \ , & \braket{E} = 0 
    \end{cases}
    \ .
\label{eq: qcnn schwinger labels}
\end{equation}
We split the full data into training and test data as $\{ \ket{\psi_m}, y_m \}_{m \in \mathcal{M}_\text{train/test}}$ with $\mathcal{M} = \mathcal{M}_\text{train} \sqcup \mathcal{M}_\text{test}$ being the mass range that determines the disjoint union of the training and test data.
Importantly, the order parameter is measured only to produce labels for the training data.
If successfully trained on the training data, the \gls{qcnn} should be able to correctly predict the order parameter for unseen states (the test data), without explicitly providing an order parameter.
Before discussing training and testing the model, however, we will briefly detail the variational ground state preparation across different phases since finding optimal variational parameters to represent the ground state is especially difficult at the phase transition.

We use the \gls{hva} to prepare the variational state $\ket{\psi(\bm{\lambda})} = U_\mathrm{HVA} (\bm{\lambda}) \ket{\phi}$, with variational parameters $\bm{\lambda}$~\cite{Wecker2015hva}. 
Concretely, the \gls{hva} parameterizes the Trotterized time evolution operators corresponding to the Hamiltonian terms in \Cref{eq: schwinger hamiltonian},
\begin{equation}
    U_\mathrm{HVA}(\bm{\lambda}) = \prod_{l=0}^{d-1} \biggl[ \exp \Bigl( - i \lambda_l^{(0)}H_{Z} \Bigr)
    \exp \Bigl( - i \lambda_l^{(1)} H_{XY}^\mathrm{(odd)} \Bigr)
    \exp \Bigl( - i \lambda_l^{(2)} H_{XY}^\mathrm{(even)} \Bigr)
    \biggl] \ .
\end{equation}
Similarly to the ansatz used in \Cref{sec: elph}, the parameterization is such that parameters are unique for every sum of commuting Hamiltonian terms.
Analogous to splitting a Trotter formula into many time steps, the \gls{hva} can be repeated $d$ times for a finer parameterization, resulting in a depth-$d$ ansatz with $N_\lambda = 3d$ variational parameters in total.
Explicitly, the terms read
\bea{C}
\begin{IEEEeqnarraybox}[][c]{C}
     H_{Z} = \sum_{n} \biggl( \sum_{i=0}^{n} \frac{Z_i + (-1)^i}{2} + \frac{\theta}{2 \pi} \biggr)^2 + \frac{m}{2}\sum_{n} (-1)^n Z_n \ , \\[5pt] 
    H_{XY}^{(\mathrm{odd})} =
     \sum_{n \ \mathrm{odd}} \bigl[ X_{n} X_{n+1} + Y_{n} Y_{n+1} \bigr] \ ,
\end{IEEEeqnarraybox}
\eea
and analogously for $H_{XY}^{(\mathrm{even})}$.

We study three system sizes across the parameter regimes highlighted in \Cref{fig: schwinger phase diagram}.
Specifically, we study $\{N_s, ag, \vartheta \} = \{4, 1, \pi \}$, $\{6, 1, 2.8 \pi \}$, $\{8, 2, \pi \}$ across mass ranges $\mathcal{M} = [-3, 2]$, $[0.5, 1.5]$, $[-2, 2]$ with step sizes $\Delta m = 0.03, 0.01, 0.05$, respectively.
The system is always initialized within the zero-magnetization sector $\sum_i \braket{Z_i}_0 = 0$ with initial state $\ket{\phi_0} = \ket{01}^{N_s/2}$.
For the classical optimization, we use \textsc{SciPy}'s SLSQP optimizer~\cite{SciPy2020}.

\begin{figure}[t]
    \centering
    \includegraphics[width=\textwidth]{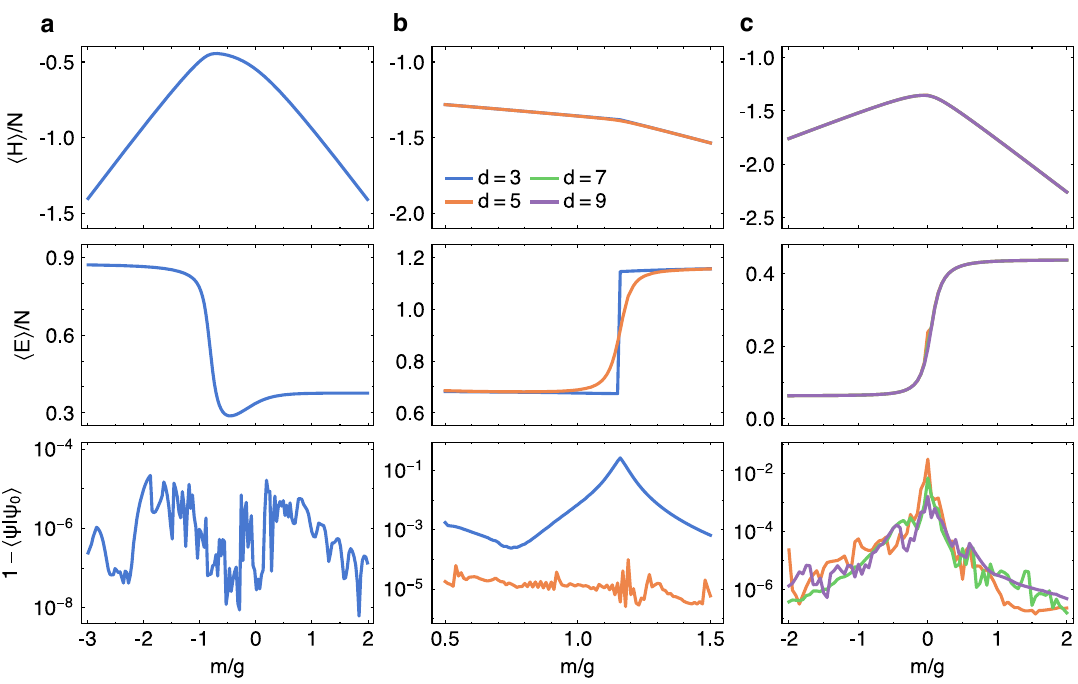}
    \mycaption{Variationally prepared ground states of the Schwinger model}{
    Results of variationally preparing the ground states of the Schwinger model within the parameter ranges $m/g$ corresponding to the highlighted regions in \Cref{fig: schwinger phase diagram}.
        The system sizes and Hamiltonian parameters are \textbf{a} $N_s=4, a=1, \vartheta=\pi$, \textbf{b} $N_s=6, a=1, \vartheta=2.8\pi$, \textbf{c} $N_s=8, a=2, \vartheta=\pi$, respectively.
        For each system size, different depths of the \gls{vha} circuit are compared.
	The top row shows the energy per site of the final variational state, the middle row the electric field per site (the order parameter), and the bottom row the infidelity of the final state with respect to the reference state obtained from exact diagonalization.
    }
    \label{fig: schwinger vqe}
\end{figure}

\Cref{fig: schwinger vqe} shows the results for all three system sizes and parameter sets for ansatz depths ranging from $d=3$ to $d=9$.
The top row displays the energy per site of the solution, the middle row the electric field per site, and the bottom row the infidelity, i.e., the error with respect to the exact ground state, $1 - \braket{\psi|\psi_0}$.
Finding a good solution to \Cref{eq: schwinger energy minimization} is generally straightforward when far away from the phase transition and the optimizer converges to the approximate ground state within a few iterations.
However, close to the phase transition, this becomes more and more challenging.
The initial choice of variational parameters plays a crucial role in this regard.
For the first point, i.e., the $m$-value furthest away from the phase transition, we initialize parameters to $0$.
Then, for the next $m$, the initial parameters are chosen as the optimal parameters found with \gls{vqe} at the previous value $m$.
Intuitively, if consecutive values in $m$ are close together, this initializes the optimizer closer to the ground state than with random or all-$0$ initial parameters.

This strategy helps to achieve reasonable good infidelities of $\mathcal{O}(10^{-5}) - \mathcal{O}(10^{-3})$, depending on the ansatz depth, up to the phase transition.
However, even this parameter initialization cannot reliably converge to the ground state at the phase transition and after, with the solutions either deviating entirely from the ground state after the transition or showing inaccuracies at the transition.
Starting the scan over all values $m$ in reverse, i.e., from $\max(\mathcal{M})$ down to $\min(\mathcal{M})$ yields a similar behavior in the other direction.
The solution to faithfully capture the ground state variationally across the entire mass range is to scan over all values $m$ in both directions (in increasing and decreasing order) and, for each $m$, take the minimum energy solution over both scans.
The results of this technique are shown in \Cref{fig: schwinger vqe}.
We see that, for $N=4$, a depth $d=3$ ansatz yields good accuracy across all $m$, while for $N=6$, the depth needs to be increased to $d=5$.
For $N=8$, the infidelity is not significantly improved when increasing the ansatz depth from $d=5$ to $d=9$, and some inaccuracy remains at the phase transition.
Nonetheless, the phase transition is still faithfully reproduced by taking the best solution obtained from two scans over the parameter range $\mathcal{M}$ with our initialization scheme.

\subsubsection{QCNN results}

After successfully preparing ground states across the phase transition, we train and test the \gls{qcnn} on the ground states obtained from \gls{vqe} with a depth $d=5$ ansatz and for the same parameters, $N_s=8, ag=2, \vartheta=\pi$ across a mass range $m/g \in \mathcal{M} = [-2, 2], \Delta m = 0.05$.
For these system parameters, the critical mass takes the value $(m/g)_\mathrm{c} \approx 0.143$ for $ag=2$ as obtained from exact diagonalization (see Appendix in Ref.~\citenum{Nagano2023datalearning}).
As for the \gls{qcnn} training, the pairing of the convolution and pooling layers is repeated three times ($N_L=3$), and the COBYLA optimizer is used for the classical optimization.
This training process is repeated 20 times, starting from different random initial \gls{qcnn} parameters.

\begin{figure}[t!]
  \centering
   \includegraphics[width=0.6\textwidth]{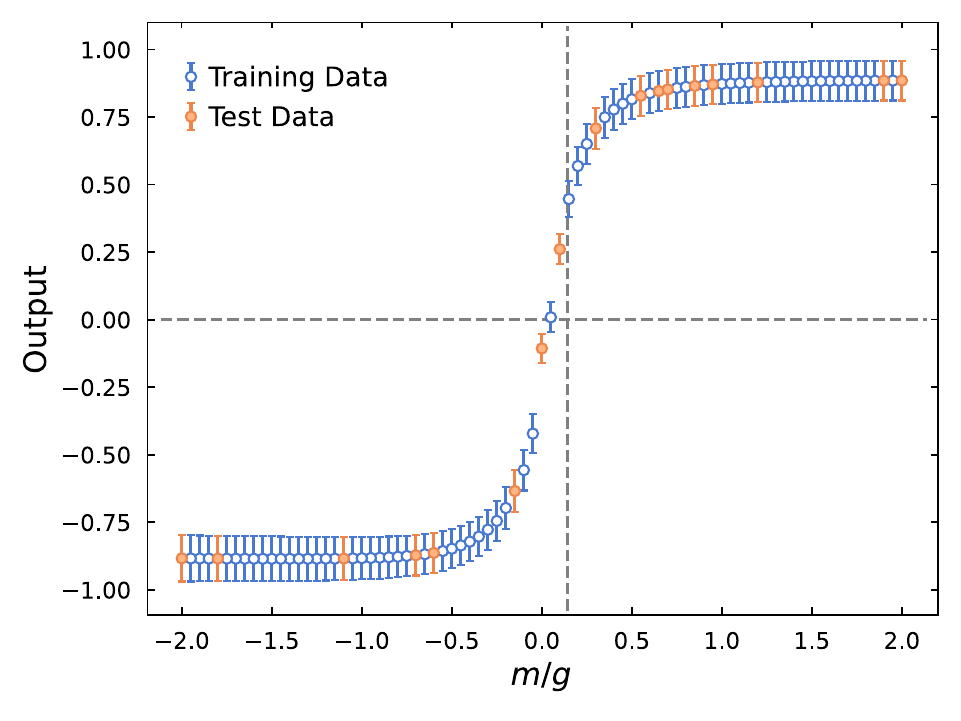}
  \mycaption{QCNN outputs for ground state phase recognition in the Schwinger model}{Labels output by the \gls{qcnn} after training on ground states of the Schwinger model with $N_s = 8$, $ag = 2$, $\vartheta = \pi$, averaged over $20$ trainings.
  The open blue and filled orange markers represent training and test data, respectively, with bars indicating standard deviations.
  Dashed lines are placed to highlight the phase transition at the critical mass value $(m/g)_\mathrm{c}$ obtained from exact diagonalization (vertical) and $y_m = 0$ (horizontal).
}
  \label{fig: qcnn schwinger classification}
\end{figure}

The outputs ${y}_\mathrm{out} (m, \bm{\theta})$ from the \gls{qcnn} circuit after training are shown in Fig.~\ref{fig: qcnn schwinger classification}.
The model correctly labels inputs with respect to the critical mass as expected from \Cref{eq: qcnn schwinger labels}.
This is made more explicit by the dashed lines, highlighting $(m/g)_\mathrm{c}$ (vertical) and the label separation line (horizontal).
Ideally, output labels would fall only into the lower left and upper right quadrants of the axes spanned by those dashed lines.
Except for two points close to the phase transition, this is the case.
Possible sources of these deviations near the critical point are errors in the \gls{qcnn} training, errors in the ground state preparation, and systematic numerical errors, such as discretization errors and finite volume effects.

Lastly, it is important to note the following.
On one hand, these proof-of-principle experiments demonstrate the applicability of \gls{qdl} methods to a paradigmatic learning task in high-energy physics -- particularly, the ability to recognize structure in the model phase space without explicitly introducing an order parameter.
On the other hand, however, a curve similar to the one in \Cref{fig: qcnn schwinger classification} could be easily reproduced by directly measuring the electric field as given in \Cref{eq: schwinger electric field}, which would require only single-qubit measurements.
On top of that, a similar profile is already observed in single-site magnetization $\braket{Z_n}$ in the bulk, as shown in Ref.~\citenum{Nagano2023datalearning}.
This implies that -- at least for this model -- there is no clear advantage in terms of sampling complexity since both phases can be distinguished with single qubit measurements.

\subsection{Phase classification of time-evolved states in a $\mathbb{Z}_2$ gauge theory} 
\label{sec: qcnn z2}

In this second part, we employ the \gls{qcnn} not to classify phases of ground states, but of time-evolved states.
Again, we consider a toy model that is known to exhibit two phases when time-evolved -- a one-dimensional $\mathbb{Z}_{2}$ gauge theory with staggered fermionic matter~\cite{Kogut1975wilsonFermions, Schweizer2019z2coldAtoms, Mildenberger2025z2quantumComputer}.

\subsubsection{The $(1+1)$D $\mathbb{Z}_2$ gauge theory}

As with the Schwinger model, also the $\mathbb{Z}_2$ model consists of fermions defined on lattice sites, while gauge fields are defined on the edges between sites.
Importantly, the degrees of freedom in this model can be directly mapped to qubits, with one qubit per lattice site and one qubit per edge.
With periodic boundary conditions, this corresponds to $2 N_s$ qubits.
The Hamiltonian of the $\mathbb{Z}_{2}$ gauge theory on a $1$D lattice with $N_s$ sites reads
\bea{rCl}
\begin{IEEEeqnarraybox}[][c]{rCl}
    H & = & -\frac{J}{2} \sum_{n} ( X_{n} Z_{n,n+1} X_{n+1} + Y_{n} Z_{n,n+1} Y_{n+1}) \\[5pt] 
    & & - f \sum_{n} X_{n,n+1} + \frac{m}{2} \sum_{n} (-1)^{n} Z_{n} \ . 
\end{IEEEeqnarraybox}
\label{eq: qcnn z2 hamiltonian}
\eea
Here, Pauli operators ${P}_n, {P}_{n, n+1} \in \{ X, Y, Z \}$, act on site $n$ and on edges between sites $n$ and $(n+1)$, respectively.
The first term couples fermionic and gauge fields with coupling strength $J$, the second represents a background $\mathbb{Z}_2$ gauge field, and the third term is the fermionic mass term.
We use periodic boundary conditions, $P_{N_s} := P_{0}$.
Physical states must satisfy the gauge condition,
\begin{equation}
    G_{n} \ket{\mathrm{phys}} = g_{n} \ket{\mathrm{phys}} \ , \quad \quad
    G_{n} := -{X}_{n-1,n}{Z}_{n}{X}_{n,n+1} \ ,
\label{eq: qcnn z2 gauge condition}
\end{equation}
with $g_{n}$ a constant.
To study the effects of a probe charge at site $n_\mathrm{probe}$, we set $g_{n_\mathrm{probe}}=+1$ and $g_{n}=-1$ for $n\neq n_\mathrm{probe}$ otherwise.

This model is known to exhibit two phases when time-evolved~\cite{Schweizer2019z2coldAtoms, Mildenberger2025z2quantumComputer}.
The matter fields are confined when a background field is present, i.e., $f \neq 0$, meaning that the effects of matter fields do not spread out.
In the absence of the background field, with $f=0$, they become deconfined and correlations spread out across the lattice.

\subsubsection{Time-evolved states as input data}

We train the \gls{qcnn} to classify time-evolved states according to the value of $f$.
Input states are of the form $\ket{\psi_{m,f}} = e^{- i H(m,f) T} \ket{\psi_\mathrm{init}}$ at a fixed time slice $T$, with the dependence on the Hamiltonian parameters of mass $m$ and background field $f$ made explicit.
To obtain the dataset, the time evolution is approximated through a first-order \gls{pf}.
The initial state satisfies Gauss' law \Cref{eq: qcnn z2 gauge condition} with a probe charge at $n_\mathrm{probe} = 1$
\begin{equation}
    \ket{\psi_\mathrm{init}} = \prod_{n} H_{n,n+1} \prod_{n \neq 1} X_n \ket{0}
    = \ket{1}\ket{+}\ket{0}\ket{+}\ket{1}\ket{+}\ket{1}\ldots \ .
\end{equation}
Here, $H_{n, n+1}$ is the Hadamard gate acting on the qubit representing the edge between sites $n$ and $n+1$.

The label $y_{m,f}$ associated with a state $\ket{\psi_{m,f}}$ is given by 
\begin{equation}
y_{m,f} =
    \begin{cases}
        +1 & f\neq 0 \quad (\mathrm{confined})
        \\
        -1 & f=0 \quad (\mathrm{deconfined})
    \end{cases} \ .
\end{equation}
As in \Cref{sec: qcnn schwinger}, the resulting dataset $\{ \ket{ \psi_{m,f} }, y_{m,f} \}_{ (m,f) \in \mathcal{M} \times \mathcal{F}}$ is split into training and test data.
We employ a \gls{qcnn} circuit with a similar structure to that in \Cref{sec: qcnn schwinger}, but with two important differences.
First, periodic boundary conditions mean convolutional gates connecting the first and the last qubit.
Second, convolutional and pooling layers are not parameterized with one variational parameter per layer, but with one parameter per individual gate, resulting in a circuit that is not translationally invariant (see \Cref{app: qcnn circuit structure} for details).
The reason for not keeping the circuit translationally invariant is that including a probe charge in our model explicitly breaks translational invariance of the model.

\begin{figure}[t!]
\centering
\includegraphics[width=0.6\textwidth]{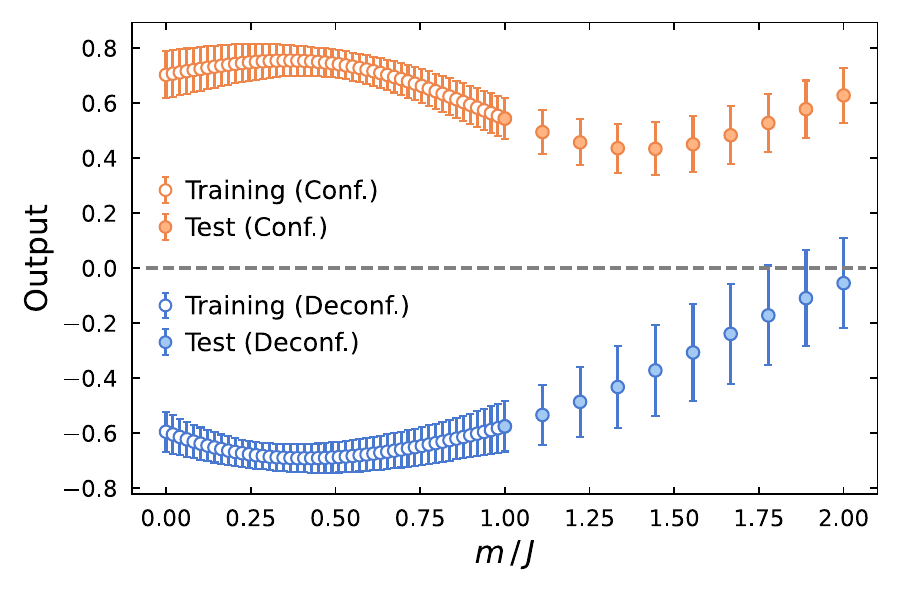}
\mycaption{QCNN outputs for classifying confinement in a $\mathbb{Z}_2$ gauge theory}{
Labels output by the \gls{qcnn} after training on time-evolved states of the $\mathbb{Z}_2$ gauge theory with $N_s=2$ at $T=2$, averaged over 20 trainings.
The open markers represent training data, while filled markers represent test data, with standard deviations indicated by bars.
The dashed line separates the two phases with positive and negative label outputs for the confined ($f/J=3$) and deconfinement ($f/J=0$) phase, respectively.
}
\label{fig: qcnn z2 output}
\end{figure}

\subsubsection{QCNN results}

We study a system of size $N_{s}=2$ at final evolution time $T=2/J$, and for two background field strengths, $f \in \mathcal{F}=\{0,3J\}$.
The model should show different phases depending on $f$.
Our goal is to classify these phases for different values of $m$.
Using periodic boundary conditions, the model is mapped to $N = 2 N_s = 4$ qubits (one per site and one per link).
Input data is generated through a first-order \gls{pf} with $20$ time steps.
To see the generalization ability to data with larger mass values than those of the training data, we use as the training and test datasets $\mathcal{M}_\mathrm{train} = \{ \mathrm{n}/49 \, | \, \mathrm{n} = 0, 1, \ldots, 48 \}$ and $\mathcal{M}_{\mathrm{test}} = \{ 1 + \text{n} / 9 \, | \, \mathrm{n} = 0, 1, \ldots, 9 \}$, respectively.
The \gls{qcnn} circuit has $N_L=3$ convolutional and pooling layers, and is trained starting from $20$ different initializations with $200$ iterations each.

The \gls{qcnn} outputs after the training are shown in \Cref{fig: qcnn z2 output}.
A clear separation in output labels is visible for the two phases of the training dataset.
The model also generalizes well for the test data up to $m \sim 1.75$.
For larger mass values, the separation becomes less evident.
This could be anticipated given increasingly different test and training data.
It is also worth remarking that, contrary to the Schwinger model example, there is no known simple local order parameter associated with symmetry breaking by which the two phases can be distinguished.

\subsection{Conclusion}
\label{sec: qcnn conclusion}

In this section, we applied the \gls{qdl} framework, i.e., \gls{qml} on quantum input data, in conjunction with a \gls{qcnn} to phase classification problems in two paradigmatic toy models for high-energy physics.
We classified ground state phases of the Schwinger model, successfully distinguishing phases around a critical point.
For this purpose, we variationally prepared the ground states of the Schwinger model across various phases using \gls{vqe}.
Those states were then used as input states to the \gls{qcnn}, both for training and testing.
However, variationally preparing these input states in itself proved to be challenging, particularly when close to the critical point.
A first step in overcoming these difficulties was to re-use the optimized variational parameters found in previous \glspl{vqe} for similar Hamiltonian parameters.
Second, scanning over a range of values for the Hamiltonian parameter that controls the phase transition (here, the mass) from different directions, eventually minimizing over different scans.
Both of these strategies combined proved successful in approximating the ground state near the critical point.
This is transferrable to similar problems (see \Cref{sec: elph}).
Furthermore, we classified phases of time-evolved states in a $\mathbb{Z}_2$ model, successfully distinguishing between a confined and deconfined phase.
Both applications demonstrate the non-trivial learning and generalization capabilities of the \gls{qcnn}, especially in the second case, where no simple local order parameter is known.

A possible future research direction could be to better understand the intricate connections between specific properties of the input quantum data, the structure of the \gls{qcnn} ansatz, and its trainability.
However, recent results highlight both the ability of classical machine learning to efficiently learn ground state properties of quantum systems~\cite{Huang2022classicalML, Lewis2024classicalML} and the fact that specifically \glspl{qcnn} might be classically simulable~\cite{Bermejo2024qcnnClassicallySimulable}.
This falls in line with a series of recent findings that trainable \gls{qml} models are classically simulable~\cite{Thanasilp2023subtletiesTrainability, Thanasilp2024exponentialConcentrationKernal, Cerezo2024barrenAbsenceClassical}, providing growing evidence for the inefficiency of \gls{qml} models, at least with classical input data.
It is unclear to what extent this affects \gls{qml} trained on genuine quantum input data, as correlations are preserved in input states that would be lost if fed into classical machine learning models.
However, in light of these results, future research should fundamentally aim to understand the context in which a genuine, practical quantum advantage could be achieved within the framework of \gls{qdl} and, more specifically, whether this is possible with \glspl{qcnn}.

%% file: ch7_conclusion.tex
\chapter{Conclusion}
\label{chap: conclusion}

In this thesis, we presented several new directions to leverage currently available, noisy quantum devices for the simulation of quantum many-body dynamics.
Quantum computing has left the era where every simulation could be reproduced and verified by classical brute-force methods.
Instead, current quantum processors can be emulated only by approximate classical methods, and they routinely operate over 100 qubits and thousands of two-qubit gates per circuit before the signal is lost.

On one hand, this enables increasingly interesting and complex experiments and demonstrations of algorithms.
On the other hand, simulations rarely extend beyond textbook problems and simplistic toy models.
Hence, it becomes more pressing to ask ``What can we simulate with available hardware that is of practical relevance and that we cannot solve with conventional computers?''
One of the main contributions of this thesis is providing a perspective on this question in the context of quantum dynamics.
First, in \Cref{sec: time evo algorithms}, we reviewed the most important quantum algorithms for quantum dynamics and highlighted their main features, including advantages and disadvantages.
Such an overview of the algorithmic landscape and respective strong and weak points is of particular importance since time evolution appears as a subroutine in many algorithms beyond quantum dynamics~\cite{Shor1994, Harrow2009HHL, Layden2023quantumMCMC}.
Second, in \Cref{chap: perspective}, we discussed a range of relevant time-dependent problems within the natural sciences, which we believe could benefit from a quantum computational treatment while, at the same time, being within reach of near-term devices.
This analysis culminated in a ranking of target applications based on a few qualitative metrics according to how close they are to achieving a practical quantum advantage.
In conclusion, \glspl{pf} are currently the most widely used methods to implement time evolution on digital quantum computers, and we expect that they will continue to be in the coming years.
Once the technology is mature enough for their implementation, asymptotically optimal approaches like Qubitization~\cite{Low2019qubitization, Miessen2023perspective} or entirely new methods might prevail.
In the meantime, however, they need to be better understood and, in particular, their resource requirements should be quantified away from the asymptotic limit.

In the context of the rapid hardware and algorithm developments, one of our main results is the proposal and end-to-end study of a novel method to benchmark both quantum hardware and error mitigation algorithms in \Cref{chap: benchmarking}.
Here, we developed a method to benchmark quantum simulations based on well-understood theoretical results from condensed matter physics -- universal scaling laws.
This benchmark provides a direct measure of simulation quality and the affordable resources -- specifically, the number of qubits and quantum gates -- before noise becomes detrimental.
It can be extended to benchmark against any other universal scaling law or application with universal behavior.
Crucially, and unlike conventional benchmarks, it can be straightforwardly scaled up to arbitrary system sizes.
Our resulting metric is intuitive to interpret and - importantly -- transfers to other applications that are related to Hamiltonian simulation.
This is because quantum circuits of the same structure are found, for example, in the Trotter dynamics of spin systems, when quantum optimization through \gls{qa} or its variational analog \gls{qaoa}, in certain quantum feature maps appearing in \gls{qml}, or in variational quantum circuits for spin systems.
In fact, we demonstrated its functioning on up to 133 qubits utilizing several combinations of error mitigation techniques, and showed coherent evolution up to a two-qubit gate depth of 28, featuring a maximum of 1396 two-qubit gates, before the signal was corrupted by noise.
Subsequently, these findings could be shown to transfer to another application -- quantum optimization.
Due to the above features, we are confident that this method will remain relevant beyond the near term, even when fault-tolerant quantum computers will be available.

In the near and intermediate term, however, hardware noise and its influence on simulations could not only be quantified, but also utilized.
To this end, another important result of this thesis is a proposed method to implement open quantum dynamics by combining error mitigation with noise engineering in \Cref{chap: pea oqs}.
We described in theory how the desired system-environment interaction can be implemented through first characterizing the device noise and, in a subsequent step, partly mitigating and amplifying it to reconfigure it to a different, desired noise model.
Our rigorous theoretical analysis of the method showed its feasibility and provided analytical error bounds that matched numerical simulations.
It is worth emphasizing that algorithms of this kind could be of central importance in coming years, as also pointed out in \Cref{sec: intro hardware,chap: perspective}.
As analog simulators remain highly efficient in simulating many-body quantum dynamics, such hybrid digital-analog schemes could help close this gap and bring near-term digital and analog simulators into closer competition.

Although our main focus was on time evolution, we highlight the importance of time-independent problems.
First, in \Cref{chap: benchmarking}, we demonstrated not only the transferability of our benchmark to another application, namely to \gls{qa} for solving combinatorial optimization problems.
But we also showed how to optimally employ \gls{qa} to solve specific instances of classical optimization problems within a noisy environment.
These results can help in practice both to reduce computational resources and to increase result quality.
In the future, these results should be made more robust by studying more varied sets of combinatorial optimization problems.

In another direction, we developed a variational algorithm in \Cref{sec: elph} to solve for ground states of realistic condensed matter systems with high practical relevance.
We successfully computed a range of ground-state quantities across different phases.
Moreover, we assessed the feasibility of implementing the algorithm on hardware and found it to be robust against noise.
Such mixed fermion-boson systems are indeed challenging to treat on a digital quantum computer due to the different encodings of the two subsystems, as also highlighted in \Cref{chap: perspective}, emphasizing the importance of devising novel techniques for their treatment.

Lastly, along a similar direction, we presented a \gls{qml}-based framework to classify phases of quantum states in \Cref{sec: qcnn}.
Here, we employed a \gls{qcnn}, i.e., a variational quantum circuit, to successfully train and test the model to distinguish different phases of both ground and time-evolved states.\
However, increasing evidence suggests fundamental limitations of at least variational \gls{qml} models to outperform classical counterparts.
In the future, research should therefore focus on thoroughly investigating the potential for quantum advantage in \gls{qml} in general and in variational models in particular.

Concluding, this thesis contributes to a better-informed overview of quantum simulation, including algorithms and applications.
With our main results, we hope to advance the usefulness of near-term quantum computers and to provide a guide for their utilization.
For example, we are convinced that our benchmarking scheme (\Cref{chap: benchmarking}) has been and will continue to be of practical relevance for researchers in the future, across generations of hardware developments and quantum computing platforms.
In fact, it has been directly employed already and utilized to benchmark digital~\cite{Visuri2025kipuDCA} and analog~\cite{McGeoch2024dwaveIBMcomparison} quantum simulation, as well as classical methods~\cite{Lerch2024efficientClassicalSimulation}, emphasizing the versatility and usefulness of the method.
It should be highlighted that none of the experiments in this thesis and other works would be possible without the fast and steady improvements of hardware and software.
This has to continue to enable what researchers have set as targets for the coming years, especially for the goal of reaching fault-tolerance.
We are confident that the impressive advances of recent years can continue to lead the field out of its current experimental state into a future where quantum computation contributes to the solution of practically relevant problems.
Even more so in light of the many interesting research directions and potential applications highlighted in this thesis.

%% file: appendix.tex
\appendix

\chapter[Benchmarking digital quantum simulations using quantum critical dynamics]{Benchmarking digital quantum\\simulations using quantum critical\\dynamics}
\vspace{20pt}

\section{Hardware properties and qubit selection}
\label{app: hardware properties}

\begin{figure}[h]
    \centering
    \includegraphics[width=\textwidth]{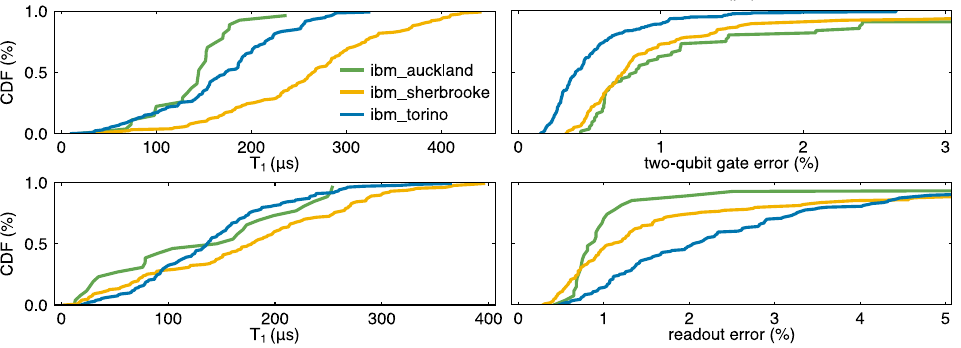}
    \mycaption{Backend Properties}{
        Cumulative distribution functions of the decoherence times $\mathrm{T}_1$ and $\mathrm{T}_2$, two-qubit gate and readout errors for the three devices used in this work, \torino{}, \sherbrooke{}, and \auckland{}.
    }
    \label{fig: hardware properties}
\end{figure}

For completeness, we report the decoherence times $\tone$ and $\ttwo$, the two-qubit gate error, and the readout error as reported for the quantum processors on which we execute the circuits, see \Cref{fig: hardware properties}.
These properties were accessed on 13$^\text{th}$ February 2024 for \torino{}, on 22$^\text{nd}$ February 2024 for \sherbrooke{}, and on 6$^\text{th}$ November 2023 for \auckland{}.
They are indicative of the device's performance when the corresponding experiments in the main text were executed, even though the data reported in the main text were gathered on several different days
The 100-qubit line in \Cref{fig: kzm device 1D} was chosen by computing the cumulative two-qubit gate error along all 100-qubit lines on the respective processor and choosing the one with the smallest error.

\section{Trotter error of benchmarking experiments}
\label{app: time step}

\begin{figure}[t]
    \centering
    \includegraphics[width=0.6\textwidth]{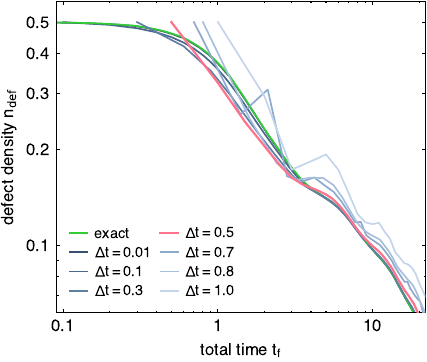}
    \mycaption{Density of defects scaling for different discretizations $\Delta t$}{
        Statevector simulations of a periodic 12-qubit chain comparing the exact, i.e., non-discrete time evolution, to Trotterized simulations with different time steps $\Delta t$.
        The time step $\Delta t = 0.5$ is highlighted in pink and corresponds to the time step chosen for the hardware experiments in \Cref{sec: benchmarking}.
    }
    \label{fig: kzm time steps}
\end{figure}

Discretization of the time evolution requires setting a time step, which in turn controls the discretization error, in our case, the Trotter error.
Ideally, the time step should be sufficiently small to reduce the Trotter error and accurately describe the continuous-time limit of the dynamics.
However, smaller time steps require more Trotter circuit layers to reach the same final time.
In \Cref{sec: benchmarking}, we adopt a practical approach and utilize a small, yet non-vanishing, time step of $\Delta t = 0.5$.

\Cref{fig: kzm time steps}\textbf{a} shows statevector simulations of a periodic 12-qubit chain with time steps varying between $\Delta t = 0.01, \ldots 1.0$ and compares the resulting density of defects to the exact, continuous-time result.
The figure shows that our choice of $\Delta = 0.5$, despite minor deviations from the continuum solution, provides a stable evolution over the entire time range we consider.
This is not the case for larger time steps, as can be seen by the increased fluctuations.
Crucially, this time step allows us to experimentally probe relevant regimes of $\tf$.
More concretely, the \gls{kz} scaling we seek to benchmark against is not present for small annealing times $\tf$, and observing the \gls{kzm} requires probing sufficiently large $\tf$.
However, this is not a limitation of the proposed method.
Rather, it is a limitation of the current hardware, but as the hardware improves and longer circuits, i.e., more Trotter steps, can be reliably executed, smaller time steps can be adopted to improve the accuracy at smaller $\tf$.

\section{Kink-kink correlator}
\label{app: kink-kink correlator}

We investigate the correlation between different defects in spin chains after digitized quantum annealing, also termed kink-kink correlation in the literature~\cite{King2022dwave2000IsingChain,Cincio2007entropy}.
Given uniform couplings $J$, the solution after annealing is ferromagnetic.
In this setting, defects are misalignments of spins on edges $i$ between lattice sites, and the correlator between defects $i$ and $i + r$ is defined as
\begin{equation}
    C_r^\mathrm{KK} = \frac{1}{N_{\mathrm{e}, r}} \sum_{i=1}^{N_{\mathrm{e}, r}} \frac{\braket{K_i K_{i + r}} - \ndef^2}{\ndef^2} \ ,
\end{equation}
where $K_i = 1 - \sigma_i^z \sigma_{i+1}^z$ measures whether there is a defect, i.e. a spin-flip, between sites $i$ and $i+1$, and $N_{\mathrm{e}, r}$ is the number of edges on the graph between edges $i$ and $i+r$.

\begin{figure}[t]
    \centering
    \includegraphics[width=0.6\textwidth]{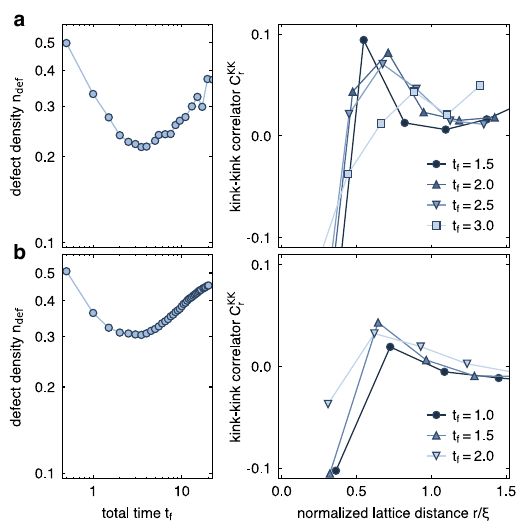}
    \mycaption{Kink-kink correlation functions for different final annealing times $\tf$}{
        \textbf{a} Density of defects (left) for a periodic 12-qubit chain on \auckland{} using REM and pulse-efficient transpilation and corresponding kink-kink correlators (right).
        \textbf{b} Density of defects (left) of an open 100-qubit chain on \sherbrooke{} using no EMS and corresponding kink-kink correlators (right).
    }
    \label{fig: kink-kink correlator}
\end{figure}

\Cref{fig: kink-kink correlator} shows the density of defects (left) of annealing a periodic 12-qubit chain on \auckland{} (\textbf{a)} using \gls{rem} and pulse-efficient transpilation and of a 100-qubit chain on \sherbrooke{} (\textbf{b}) using no \gls{ems}.
The corresponding kink-kink correlators for different $\tf$ are shown in the respective right panel as a function of the normalized lattice distance $r/\xi$ with $1/\xi = \ndef$.
The existence of a positive peak, which we observe between $r/\xi = 0.5$ and $1.0$, is expected from theoretical results\cite{Nowak2021quantum,Dziarmaga2022kinkcorrelations,Roychowdhury2021}.

\section{Quantum annealing with very large time steps}
\label{app: optimization statevector}

\begin{figure}[t]
    \centering
    \includegraphics{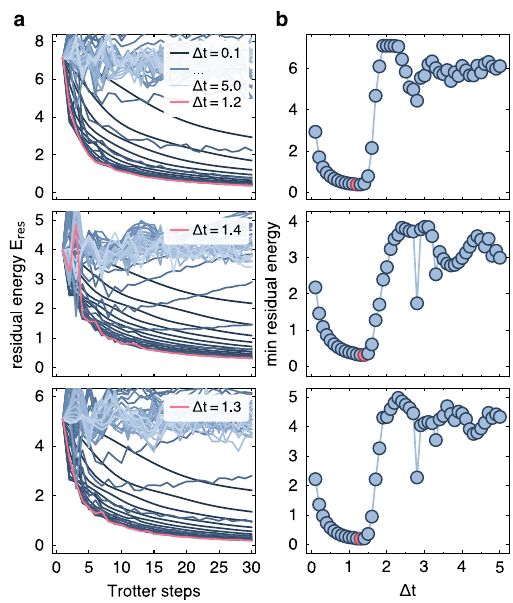}
    \mycaption{Statevector simulations of the residual energy dependence on time step and spectral gap}{
        \textbf{a} Residual energy obtained from statevector simulations of QA with fixed time steps $\Delta t \in \{ 0.1, \ldots, 5.0 \}$ as a function of circuit depth, i.e., number of time steps.
        Each row corresponds to the respective spectrum in \Cref{fig: residual energy auckland}.
        \textbf{b} Minimum residual energy from \textbf{a} as a function of the time step with fixed depth, i.e., each point corresponds to the minimum over one curve in \textbf{a}.
    }
    \label{fig: residual energy statevector}
\end{figure}

Here, we confirm that the optimal time step for digitized \gls{qa} for optimization is $\Delta t > 1$ through ideal statevector simulations.
\Cref{fig: residual energy statevector} shows the results for the same systems as in \Cref{fig: residual energy auckland} from the main text, that is, instances of disordered couplings $J_{ij} \in [-1, 1]$.
When considering a fixed number of time steps or circuit layers, the minimum residual energy obtained from up to 30 time steps decreases with growing time step size up to an optimal time step $1.2 < \Delta t < 1.4$ (depending on the system), before sharply increasing, see \Cref{fig: residual energy statevector}.
Increasing the time step even further does not yield any benefit whatsoever and, since it induces significant algorithmic errors, results in randomly fluctuating residual energies with increasing circuit depth.


\chapter{Open quantum dynamics through partial probabilistic error amplification}

\section{Evaluating expectation values at amplified noise}
\label{app: pea expectation values}

\subsubsection{Derivation of \Cref{eq: noisy expectation value}}

As mentioned in \Cref{sec: pea oqs pea}, we adopt here the notation of Ref.~\citenum{Filippov2024peaTheory}, defining left-to-right composition $\bigcirc_l^{\rightharpoonup} A_l = A_0 \circ \ldots \circ A_L$ and right-to-left composition $\bigcirc_l^{\leftharpoonup} A_l = A_L \circ \ldots \circ A_0$.
With initial state $\rho_0 = \ket{0}\bra{0}^N$ and $L$ circuit layers, the expectation value under amplified noise can be calculated to become

\bea{rCl}
\braket{\O}^{(G)}_{\rm noisy} & = &
    \tr \biggr\{
    \mathcal{O} \cdot \Bigl( \overset{\leftharpoonup}{\bigcirc}_l \U_l \circ \Lambda_l^G \Bigr) [\rho_0]
    \biggl\} \\
& \overset{(1)}{=} & 
    \tr \biggr\{
    \Bigl( \overset{\rightharpoonup}{\bigcirc}_l {\Lambda_l^G}^\dagger \circ \U_l^\dagger \Bigr) [\mathcal{O}] \cdot \rho_0
    \biggl\} \\
& \overset{(2)}{=} &
    \tr \biggr\{
    \Bigl( \overset{\rightharpoonup}{\bigcirc}_l {\Lambda_l^G}^\dagger \circ \U_l^\dagger \Bigr) [\mathcal{O}]
    \cdot \Bigl( \overset{\rightharpoonup}{\bigcirc}_l \U_l^\dagger \circ \overset{\leftharpoonup}{\bigcirc}_l \U_l \Bigr) [\rho_0]
    \biggl\} \\
& \overset{(3)}{=} & 
    \tr \biggr\{
    \overset{\leftharpoonup}{\bigcirc}_l \U_l \Bigl[ \overset{\rightharpoonup}{\bigcirc}_l {\Lambda_l^G}^\dagger \circ \U_l^\dagger [\mathcal{O}] \Bigr]
    \cdot \overset{\leftharpoonup}{\bigcirc}_l \U_l [\rho_0]
    \biggl\} \ .
\eea
In (1), we changed from Schrödinger to Heisenberg picture, $\O_\mathrm{H} = U^\dagger \O_\mathrm{S} U = \U[\O_\mathrm{S}]$.
The Schrödinger picture prescribes time evolution to the state, $\rho = \U[\rho_0] = U \rho_0 U^\dagger$, while operators remain constant in time, whereas it is the other way around in the Heisenberg picture.
The change from Schrödinger to Heisenberg picture is enabled through the trace property $\tr(ABC) = \tr(CAB) = \tr(BCA)$.
Therefore, we can write $\braket{\O_\mathrm{S}} = \tr\{ \O_\mathrm{S}  \U[\rho]\} = \tr\{ \U^\dagger[\O_\mathrm{S}] \rho \}$.
In (2), we added an identity.
In (3), we again converted the entire first factor into the Heisenberg picture, evolving under ${\bigcirc}_l^\rightharpoonup \U_l^\dagger$.

Now, the second factor resembles the noise-free evolution of the initial state,
\begin{equation}
    \Tilde{\rho}_\mathrm{ideal} := \overset{\leftharpoonup}{\bigcirc}_l \U_l [\rho_0] \ .
\end{equation}
The first factor can be rewritten as
\bea{rCl}
\overset{\leftharpoonup}{\bigcirc}_l \U_l \Bigl[ \overset{\rightharpoonup}{\bigcirc}_l {\Lambda_l^G}^\dagger \circ \U_l^\dagger [\mathcal{O}] \Bigr]
& = &
    \bigl( \U_L \circ \ldots \U_0 \bigr)
    \circ \bigl( {\Lambda_0^G}^\dagger \circ \U_0 \circ \ldots {\Lambda_L^G}^\dagger \circ \U_L \bigr) [\O] \\
& \overset{(1)}{=} &
    \bigl( \U_L \circ \ldots \U_0 \bigr)
    \circ \bigl( {\Lambda_0^G}^\dagger \circ \U_0 \circ \bigl\{ \bigl( \U_1^\dagger \circ \ldots \U_L^\dagger \bigr) \bigl( \U_L \circ \ldots \U_1 \bigr) \bigr\} \\
    && \circ {\Lambda_1^G}^\dagger \circ \U_1 \circ \ldots \circ 
    \bigl\{ \U_L^\dagger \circ \U_L\bigr\} \circ {\Lambda_L^G}^\dagger \circ \U_L \bigr) [\O] \\
& \overset{(2)}{=} &
    \overset{\rightharpoonup}{\bigcirc}_l
    \Bigl( \overset{\leftharpoonup}{\bigcirc}_{m \geq l} \U_m \Bigr)
    \circ {\Lambda_l^G}^\dagger \circ
    \Bigl( \overset{\rightharpoonup}{\bigcirc}_{m \geq l} \U_m^\dagger \Bigr)
    [\mathcal{O}] \\
& =: &
    \overset{\rightharpoonup}{\bigcirc}_l
    \U_{\geq l} \circ {\Lambda_l^G}^\dagger \circ \U_{\geq l}^\dagger [\mathcal{O}] \\
& =: &
    \overset{\rightharpoonup}{\bigcirc}_l \Tilde{\Lambda}_l^{G\dagger} [\mathcal{O}]
    \ .
\eea
In (1), we inserted identities $\overset{\rightharpoonup}{\bigcirc}_{m \geq l} \U_m^\dagger \circ \overset{\leftharpoonup}{\bigcirc}_{m \geq l} \U_m$ before every $\Lambda_{l \neq 0}^{G\dagger}$.
In (2) and the following equalities, we re-grouped the new channels stemming from inserting the identities.

\subsubsection{Derivation of \Cref{eq: noisy expectation value clifford}}

The reason is that Clifford operations transform one Pauli string into another, $\U_{\geq l}^\dagger [P_\beta] = P_{\beta(l)}$, with $P_{\beta(l)}$ being the Pauli resulting from evolution through Clifford layers $l \ldots L$.
Evolution through all noisy layers therefore results in
\begin{equation}
    \overset{\rightharpoonup}{\bigcirc}_l \Tilde{\Lambda}_l^{G\dagger} [P_\beta] = \prod_l f_{l \beta(l)}^G P_{\beta}
\end{equation}
since ($\U_{\geq l}$ undoes the action of $\U_{\geq l}^\dagger$)
\bea{rCl}
    \Tilde{\Lambda}_l^{G\dagger} [P_\beta]
    & = & \U_{\geq l} \circ \Lambda_l^{G\dagger} \circ \U_{\geq l}^\dagger [P_\beta] \\
    & = &  \U_{\geq l} \circ \Lambda_l^{G\dagger} [P_{\beta(l)}] \\
    & = & f_{l \beta(l)}^G \U_{\geq l} [P_{\beta(l)}]\\
    & = & f_{l \beta(l)}^G P_{\beta} \ .
\eea
This results in a single exponential functional form for the expectation value,
\bea{rCl}
    \braket{\O}^{(G)}_{\rm noisy} & = &
    \tr \Bigl\{
    \prod_l f_{l \beta(l)}^G P_{\beta} \cdot \Tilde{\rho}_\mathrm{ideal}
    \Bigl\} \\
    & = &
    \Bigl( \prod_l f_{l \beta(l)}^G \Bigr)
    \tr \bigl\{ P_{\beta} \cdot \Tilde{\rho}_\mathrm{ideal} \bigl\} \\ 
    & = &
    \Bigl( \prod_l f_{l \beta(l)} \Bigr)^G \braket{\O}_{\rm ideal} \ .
\eea

\section{Evaluating expectation values at locally amplified noise}
\label{app: pea oqs cases functional form}

\subsubsection{Evaluating \Cref{eq: pea oqs expectation value general} for all cases in \Cref{sec: pea different noise model}}

Here, we derive and list the functional forms of expectation values under locally amplified Pauli noise and -- importantly -- for Clifford circuits for all cases discussed in \Cref{sec: pea different noise model}.
We use $\eta = G - 1$.

\begin{enumerate}
    \item[1.] $P_\alpha \in \mathrm{SPL}(\Lambda_l), P_\alpha \in \mathrm{SPL}(\tilde{\Lambda}_l)$
    \begin{enumerate}
        \item[a)] $\tilde{\lambda}_\alpha < \lambda_\alpha \ \rightarrow \ \delta_\alpha = 1 - \frac{\tilde{\lambda}_\alpha}{\lambda_\alpha} \in (0, 1)$\\
    		This case is discussed in the main text \Cref{sec: pea different noise model}.
    		The exponential in \Cref{eq: pea oqs expectation value general} can be expanded as
    		\bea{rCl}
    			\prod_{ \{ l ; \alpha ; \beta(l) \} } e^{-2 \lambda_\alpha (\delta_\alpha \eta + 1)}
    			& = & \prod_{ \{ l ; \alpha ; \beta(l) \} } e^{-2 ( \lambda_\alpha - \tilde{\lambda}_\alpha) G + \tilde{\lambda}_\alpha} \nonumber = \biggl( \frac{f_{l\beta(l)}}{\Tilde{f}_{l\beta(l)}} \biggr)^G \Tilde{f}_{l\beta(l)}
    		\eea
		\item[b)] $\tilde{\lambda}_\alpha = \lambda_\alpha \ \rightarrow \ \delta_\alpha = 0$\\
    		In this case, $-2 \lambda_\alpha (\delta_\alpha \eta + 1) = -2 \lambda_\alpha$, and we obtain a constant pre-factor,
            \begin{equation}
                \braket{\O}^{(G)}_{\rm noisy}
                =
                \prod_l \prod_{ \{ l ; \alpha ; \beta(l) \} } e^{-2 \lambda_\alpha} \braket{\O}_{\rm ideal} 
                = \Bigl( \prod_l f_{l\beta(l)} \Bigr) \braket{\O}_{\rm ideal} \ .
            \end{equation}
		\item[c)] $\tilde{\lambda}_\alpha > \lambda_\alpha \ \rightarrow \ \delta_\alpha = \frac{1}{\eta} \bigl( \frac{\tilde{\lambda}_\alpha}{\lambda_\alpha} - 1 \bigr)$\\
			Similarly, in this case, the exponent simplifies to $-2 \lambda_\alpha (\delta_\alpha \eta + 1) = -2 \tilde{\lambda}_\alpha$, resulting in a single pre-factor of
			\begin{equation}
                \braket{\O}^{(G)}_{\rm noisy}
                =
                \prod_l \prod_{ \{ l ; \alpha ; \beta(l) \} } e^{-2 \tilde{\lambda}_\alpha} \braket{\O}_{\rm ideal} 
                = \Bigl( \prod_l \tilde{f}_{l\beta(l)} \Bigr) \braket{\O}_{\rm ideal} \ .
            \end{equation}
    \end{enumerate}
    \item[2.] $P_\alpha \notin \mathrm{SPL}(\Lambda_l), P_\alpha \in \mathrm{SPL}(\tilde{\Lambda}_l)$ \\
    	This case is very similar to (1.c), with the difference that $\lambda_\alpha = 0$, which means we sample only $\mathrm{SPL}(\tilde{\Lambda}_l)$ into the circuit.
    	This means, the noise map in the circuit is that of \Cref{eq: pea unamplified noise map} but with $\tilde{\lambda}_\alpha$, again resulting in the same expectation value as in case (1.c).
    \item[3.] $P_\alpha \in \mathrm{SPL}(\Lambda_l), P_\alpha \notin \mathrm{SPL}(\tilde{\Lambda}_l) \rightarrow \delta = 1$ \\
    This case reduces to regular \gls{pea} and the noisy expectation value is given by \Cref{eq: noisy expectation value clifford}.
\end{enumerate}

\section{Minimizing the random error in PEA with locally amplified noise}
\label{app: pea oqs random error}

\subsubsection{Derivation of \Cref{eq: pea oqs optimial shots}}

We can write the random error \Cref{eq: pea oqs error F} more concisely as
\begin{equation}
	\Delta F(0) = C \Biggl( \sum_j \frac{A_j^2}{S_j} \Biggr)^{1/2} \ ,
\end{equation}
where we collected the pre-factors into $C = 1 / (R \sum_i G_i^2 - (\sum_i G_i)^2)$, and defined $A_j = \sum_i G_i (G_i - G_j) \bigl( \frac{\Kt}{K} \bigr)^{G_j} \frac{1}{\Kt}$.
We employ the method of Lagrange multipliers to enforce the constraint $\sum_i S_i = M$.
The Lagrangian we want to minimize is then
\begin{equation}
    \mathcal{L} = C \sum_j \frac{A_j^2}{S_j} - \lambda \bigl( \sum_i S_i - M \bigr) \ .
\end{equation}
Note that we omitted the square root around the sum in the first term since it is a monotonic function and it thus suffices to minimize its argument.
The sets of resulting equations $\partial_{S_i} \mathcal{L} = 0$ together with the constraint equation $\partial_\lambda \mathcal{L} = 0$ then results in
\begin{equation}
    S_j^* = \frac{M |A_j|}{\sum_l |A_l|} \ .
\end{equation}

\subsubsection{Derivation of \Cref{eq: pea oqs optimal gain}}

Here, we detail the minimization over $\{ G_i \}$ with fixed $\{ S_i \} = \{ S_i^* \}$.
In the case of regular \gls{pea}, Ref.~\citenum{Filippov2024peaTheory} mentions numerical results providing the same minimum for $R=3, 4$ extrapolation points (noise gain factors) as for the analytically solvable $R=2$ case.
We adopt this here.
With $R=2$, the random error becomes
\bea{rCl}
    \Delta F(0, S_j^*) & = & \frac{1}{\sqrt{M}} \frac{ \bigl| G_1 (G_1 - G_2) \bigr| \bigl( \frac{\Kt}{K} \bigr)^{G_2} + \bigl| G_2 (G_2 - G_1) \bigr| \bigl( \frac{\Kt}{K} \bigr)^{G_1} }{  (G_1 - G_2)^2 } \nonumber \\
    & = & \frac{1}{\sqrt{M}} \frac{ G_1 \bigl( \frac{\Kt}{K} \bigr)^{G_2} + G_2 \bigl( \frac{\Kt}{K} \bigr)^{G_1} }{  G_2 - G_1 } \ ,
    \label{eq: min error S*}
\eea
where one factor $G_2-G_1$ is canceled in the last step.
With $G_2 > G_1 \geq 1$, we can fix the first noise gain at $G_1^*=1$.
We can then readily minimize the remaining expression $f(g) = \frac{ \kappa^{g} + g \kappa }{  g - 1 }$, where we have abbreviated $\Kt/K = \kappa, G_2=g$.
We need to solve
\begin{equation}
    \partial_g f(g) = \frac{ (g - 1)\kappa^{g} \ln \kappa - \kappa^g - \kappa }{ (g - 1)^2 } = 0 \ ,
\end{equation}
which becomes
\bea{rCl}
    && (g - 1)\kappa^{g} \ln \kappa - \kappa - \kappa^g = 0 \nonumber \\
    & \Leftrightarrow & \ \ln \kappa^{g-1} - \kappa^{1-g} - 1 = 0 \nonumber \\ 
    & \Leftrightarrow & \ \kappa^{1-g} \exp (\kappa^{1-g}) = \frac{1}{e} \ .
\eea
This expression can be solved~\cite{Filippov2024peaTheory} substituting $w = \kappa^{1-g}$, resulting in $we^w = \frac{1}{e}$.
The solution of this equation is the principal branch of the Lambert $W$ function, $w = W(\frac{1}{e})$.
Solving for the optimal $g^*$, we have
\begin{equation}
    w = \kappa^{1-g^*} \Leftrightarrow\frac{1}{1-g^*} = \frac{\ln \kappa}{\ln w} \ .
\end{equation}
Combining all of the above, we obtain the optimal noise gain values for $R=2$, 
\begin{equation}
    G_1^*=1 \ , \quad
    G_2^* = 1 + \frac{W \bigl(\frac{1}{e}\bigr) + 1}{\ln \frac{\Kt}{K}} \ .
\end{equation}

\subsubsection{Derivation of \Cref{eq: pea oqs error F min}}

Lastly, we substitute both $\{S_i^*\}$ and $\{G_i^*\}$ into $\Delta F(0)$, \Cref{eq: pea oqs error F}.
The resulting expression can be simplified as follows (using $G_2^* := 1 + B = G_1^* + B$)
\bea{rCl}
    \Delta F(0, S_j^*, G_j^*)
    & = & \frac{1}{\sqrt{M} B} \Bigl( \kappa^{1+B} + (1+B) \kappa \Bigr) \nonumber \\
    & = & \frac{\kappa}{\sqrt{M} B} \Bigl( \kappa^{B} + (1+B) \Bigr) \nonumber \\
    & = & \frac{\kappa \ln \kappa}{\sqrt{M} (1 + W)} \Bigl( \frac{1}{W} + 1 + \frac{1 + W}{\ln \kappa} \Bigr) \ . \nonumber
\eea
In the last line, we used $\kappa^B = \exp(B \ln \kappa) = \exp(1 + W)$ and, per definition (see above), $We^W = 1/e \Leftrightarrow e^{1+W} = 1/W$, therefore $\kappa^B = 1/W$.
The resulting minimum random error in the extrapolated result then becomes (analogous to the result of Ref.~\citenum{Filippov2024peaTheory})
\begin{equation}
    \Delta F(0, S_j^*, G_j^*) = \frac{\Kt}{K \sqrt{M}} \biggl( 1 + \frac{\ln(\Kt / K)}{W(\frac{1}{e})} \biggr) \ .
\end{equation}


\chapter{State preparation and phase characterization beyond dynamics}

\section{QCNN circuit structure}
\label{app: qcnn circuit structure}

The general structure of the \gls{qcnn} circuit is the same for all classification tasks presented in \Cref{sec: qcnn} and Ref.~\citenum{Nagano2023datalearning}.
In general, the \gls{qcnn} circuit is composed of alternating layers of the convolution (CL) and pooling (PL) unitaries, each constructed with repeated blocks of gates, followed by a fully-connected layer (FCL) before measuring the output (see \Cref{fig: qcnn scheme} in the main text).
In our work, we neglect the FCL since the output is measured directly after the last pooling layer.
Moreover, the gates and parameterizations to implement each CL and PL block are constructed depending on the concrete model studied, which we will detail below.

\subsection{Convolutional and pooling gates}

\begin{figure}[htbp]
    \centering
    \includegraphics[width=\textwidth]{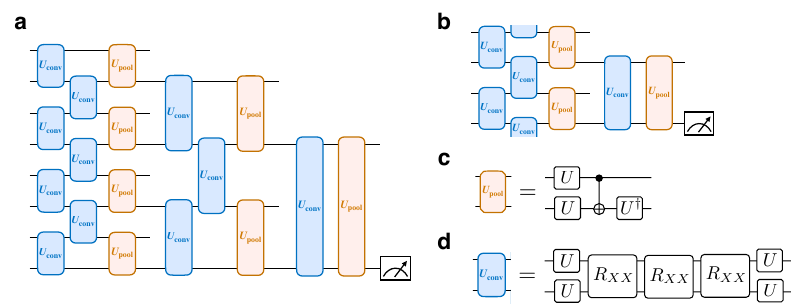}
    \mycaption{QCNN circuit details}{
    \textbf{a} \Gls{qcnn} circuit used for the phase recognition in the Schwinger model (\Cref{sec: qcnn schwinger}).
    \textbf{b} \Gls{qcnn} circuit used for the phase recognition in the $\mathbb{Z}_{2}$ model (\Cref{sec: qcnn z2}).
    The truncated $U_{\rm conv}$ gate at the first and last CL indicates periodic boundary conditions.
    \textbf{c} Pooling gates are general SU(4)-like gates parameterized with six gate angles -- three for each general single-qubit rotational gate $U(\theta_1, \theta_2, \theta_3)$ where the last gate $U^\dagger$ shares the same parameters as the first $U$-gate on the control qubit.
    \textbf{d} Convolutional gates are general SU(4)-like gates parameterized with 15 gate angles -- three for each general single-qubit rotational gate $U(\theta_1, \theta_2, \theta_3)$ and one for each two-qubit gate, e.g., $R_{XX}(\theta)$.
    }
    \label{fig: qcnn circuit details}
\end{figure}

An individual block in the CL consists of a generic SU(4)-like gate with 15 independent parameters, composed of single- and two-qubit gates with rotation angles as parameters, as seen in \Cref{fig: qcnn circuit details}\textbf{c}. 
For the Schwinger and $\mathbb{Z}_2$ models in \Cref{sec: qcnn}, the $U$ in gate is implemented with a generic single-qubit gate 
\begin{equation}
    U(\theta_1,\theta_2,\theta_3)= 
    \begin{pmatrix}
    \cos(\theta_1/2) & -e^{i\theta_3}\sin(\theta_1/2) \\
    e^{i\theta_2}\sin(\theta_1/2) & e^{i(\theta_2+\theta_3)}\cos(\theta_1/2) \\
    \end{pmatrix} \ .
\label{eq: qcnn single qubit gates}
\end{equation}
Moreover, the convolutional gates reflect the boundary conditions of the model, which is indicated by the truncated $U_{\rm conv}$ gates in the case of periodic boundary conditions in \Cref{fig: qcnn circuit details}\textbf{b}.

For the PL, each two-qubit block consists of a set of single-qubit rotation gates, \Cref{eq: qcnn single qubit gates}, followed by a CNOT gate and the adjoint of the single-qubit gate on the target qubit that is the adjoint of and sharing the same parameters as the first single-qubit gate acting on the control qubit, as seen in \Cref{fig: qcnn circuit details}\textbf{c}.

\subsection{Model-dependent QCNN circuit}

The \gls{qcnn} circuits used for the Schwinger and $\mathbb{Z}_2$ models are shown in \Cref{fig: qcnn circuit details}\textbf{a} and \textbf{b}, respectively.
For the Schwinger model ($N_{\rm step}=8$) we employ three convolutional and pooling layers each, while for the $\mathbb{Z}_2$ model, we use two layers each.
In all cases, we measure the last aubit.
The boundary conditions of the model transfer to the structure of the \gls{qcnn} circuit.
We used open boundary conditions for the Schwinger model and periodic boundary conditions for the $\mathbb{Z}_2$ model.
This is indicated by the truncated $U_{\rm conv}$ gates in \Cref{fig: qcnn circuit details}\textbf{b}.

Typically, the \gls{qcnn} circuit is translationally invariant and parameterized with only one variational parameter per convolutional layer and pooling layer.
This is the case for the \gls{qcnn} employed for the Schwinger model, \Cref{fig: qcnn circuit details}\textbf{a}.
In the case of the $\mathbb{Z}_2$ model, this translational invariant is lifted, and convolutional and pooling layers are parameterized with one parameter per individual convolutional, respectively, pooling gate.


\section{State preparation in strongly correlated electron-phonon systems}
\label{app: elph}

\subsection{NGS-VQE method and effective Hamiltonian}
\label{app: elph effective hamiltonian}

As mentioned in the main text, the variational electron-phonon wavefunction in the NGS-VQE method is given as 
\begin{equation}
    \big| \Psi \big\rangle  = U_{\rm NGS}(\{f_q\}) |\psi_{\rm ph}\rangle \otimes | \psi_{\rm e}\rangle \ ,
\label{app-eq: elph full ansatz}
\end{equation}
with the non-Gaussian transformation $U_{\rm NGS}=e^{i\mathcal{S}}$.
As benchmarked in Refs.~\citenum{Wang2020hubbardHolstein, Wang2021timeDepNonGaussian}, it is sufficient to truncate the $\mathcal{S}$ operator to the lowest-order terms
\begin{equation}
    \mathcal{S}(\{f_q\}) = -\frac{1}{\sqrt{L}} \sum_{q i \sigma} f_q  e^{iq x_i} (a_q - a_{-q}^\dagger) n_{i,\sigma} \ ,
    \label{app-eq: elph S operator}
\end{equation}
where we use the momentum-space electron density $\rho_{q} = \sum_{i, \sigma} n_{i,\sigma}e^{-i q x_i}$, and the phonon momentum operator $p_{q} = i \sum_{i}( a_{i}^{\dagger} - a_{i})  e^{-i q x_i} / \sqrt{L}$.

The goal of the NGS-VQE solver is to minimize the total energy in \Cref{eq: elph energy expectation value} in the variational parameter space spanned by $\{f_q\}$, $|\psi_{\rm ph}\rangle$, and $|\psi_{\rm e}\rangle$.
Without considering anharmonicity, the phonon state to the right of $U_{\rm NGS}$ should be weakly entangled and can be efficiently captured by variational Gaussian states
\begin{equation}
	|\psi_{\rm ph}\rangle = e^{-\frac12 R_0^T \sigma_y \Delta_R} e^{-i\frac14 \sum_\qbf \Rqd \xi_\qbf  \Rq} |0\rangle = \Ugs |0\rangle \ .
\label{app-eq: elph phonon ansatz}
\end{equation}
Here, $\Delta_R$, $\xi_\qbf$ are variational parameters and $\Rq = (x_q,p_q)^{\mathrm{T}}$ denotes the bosonic quadrature notation with canonical position $x_q$ and momentum $p_q$, where we adopt the reciprocal representation for the phonon displacement $x_{q} = \sum_{i}  (a_{j} + a_{j}^\dagger)e^{-i q r_j} / \sqrt{L}$.
For convenience, we parameterize the phonon state using the linearization of $\Ugs$ named $S_\qbf$, which satisfies $\Ugs^\dagger ( x_\qbf, p_\qbf )^T\Ugs = S_\qbf ( x_\qbf, p_\qbf )^T$. The NGS-VQE method minimizes the total energy by updating $|\psi_{\rm e}\rangle$ and  $|\psi_{\rm ph}\rangle$ iteratively.

With fixed $U_{\rm NGS}$ and $|\psi_{\rm ph}\rangle$, the electronic problem that we propose to solve using a quantum computer is the ground state of an effective Hamiltonian
\bea{rCl}
\begin{IEEEeqnarraybox}[][c]{rCl}
    \mathcal{H}_{\mathrm{eff}} & = &-\tilde{t} \sum_{\langle i,j \rangle, \sigma} \left(c_{i,\sigma}^\dagger c_{j,\sigma}^{} + \mathrm{h.c.}\right) + U \sum_i n_{i,\uparrow} n_{i,\downarrow} \\
    & & + \sum_{i,j} \sum_{\sigma,\sigma'} \tilde{V}_{ij} n_{i,\sigma} n_{j,\sigma'} + \tilde{E}_{\rm ph} \ .
\end{IEEEeqnarraybox}
\label{app-eq: elph effective hamiltonian}
\eea
Here, $\tilde{E}_{\rm ph} = \frac14 \omega \sum_\qbf \left( \mathrm{Tr} [ S_\qbf S_\qbf^\dagger ] - 2 \right)$.
The phonon-dressed nearest-neighbor hopping becomes
\begin{equation}
    \tilde{t} = t e^{-\sum_\qbf f_\qbf^2 (1 - \cos q) e_2^T S_\qbf S_\qbf^\dagger e_2/L} \ ,
\label{eq:effHopping}
\end{equation}
and the effective phonon-mediated interaction is
\begin{equation}
    \tilde{V}_{ij} = \frac1{L}\sum_{\qbf} \left[2 \omega_0f_\qbf^2-4g_\qbf f_\qbf \right]e^{iq(r_i-r_j)} \ .
\label{eq:effInteraction}
\end{equation}
In the case of Holstein couplings, Gaussian states are an efficient representation of the phonon wavefunction $|\psi_{\rm ph}\rangle$~\cite{Shi2018nonGaussianStates, Shi2020finiteTemperatureNGS, Wang2020hubbardHolstein} as they allow to represent $\tilde{t}$ and $\tilde{V}_{ij}$ in closed-form (see also Ref.~\citenum{Wang2021phononLongRange}).

The \gls{vqe} solution of the effective Hamiltonian in \Cref{app-eq: elph effective hamiltonian} gives $|\psi_{\rm e}\rangle$ in Eq.~\eqref{app-eq: elph full ansatz}.
The iterative optimization of $U_{\rm NGS}$ and $|\psi_{\rm ph}\rangle$ for fixed $|\psi_{\rm e}\rangle$ follows the imaginary time evolution in Ref.~\citenum{Wang2020hubbardHolstein}.
It is worth noting that the charge density correlation functions $\langle \rho_{q} \rho_{-q} \rangle \propto \sum_{ij} \sum_{\sigma \sigma'} n_{i,\sigma} n_{j,\sigma'}$, necessary for the imaginary time evolution of $|\psi_{\rm ph}\rangle$ appear in $\mathcal{H}_\mathrm{eff}$ as well.
They are therefore already measured with the energy expectation value during VQE and result in no additional computational cost.

\subsection{Quantum circuit and ansatz}
\label{app: elph circuit gates}

\begin{figure}[t]
    \centering
    \includegraphics[width=0.4\textwidth]{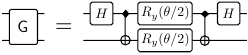}
    \mycaption{Givens rotation gate implementation}{
        Alternative implementation of the Givens rotations \textsf{G} used in the variational circuit in \Cref{fig: elph circuit vqe}.
        Here, $H$ is a Hadamard gate (not to be confused with the Hopping gates \textsf{H} in \Cref{fig: elph circuit vqe}).
    }
    \label{app-fig: givens rotation}
\end{figure}

As outlined in the main text, the employed quantum circuit is based on the \gls{hva}~\cite{Wecker2015hva, Jiang2018quantumAlgorithmsFermions}.
To encode the non-interacting $(U = 0, V_{ij} = 0)$ electronic model, a sequence of Givens rotations $\textsf{G}(\theta)$, parametrized by $\theta$, is applied to adjacent qubits. 
In the basis $|00\rangle, |01\rangle, |10\rangle, |11\rangle$, the gate is defined as 
\begin{equation}
\textsf{G}(\theta) = 
    \begin{pmatrix}
        1 & 0& 0 & 0 \\
         0 & \cos\frac{\theta}{2}& -\sin\frac{\theta}{2} & 0 \\
          0 & \sin\frac{\theta}{2}& \cos\frac{\theta}{2} & 0 \\
           0 & 0& 0 & 1 \\
    \end{pmatrix} \ ,
\end{equation}
and can be implemented using the alternative gate sequence given in \Cref{app-fig: givens rotation}.
The ground state of the full effective model is then obtained by an adiabatic evolution with the Hubbard-like Hamiltonian~\cite{Cade2020vqeFermiHubbard, Stanisic2022vqeFermiHubbard}. It can be decomposed into kinetic hopping terms,
\begin{equation}
    \textsf{H}(\theta) = e^{-i(c_i^{\dagger}c_{i+1}^{}+c_{i+1}^{\dagger}c_i^{})\theta } = e^{-i(X_i X_{i+1}+Y_i Y_{i+1})\theta/2} \ ,
\end{equation}
and on-site interactions
\begin{equation}
   \textsf{P}(\theta) = e^{-i n_i n_{j}\theta } = e^{-i | 11 \rangle\langle1 1 |_{ij}\theta} \ .
\end{equation}
The alternating sequence of phase gates (\textsf{P}) and hopping gates (\textsf{H}) is then repeated for a number of repetitions $n$, controlling the expressibility of the variational ansatz.

%% file: biblio.tex
\cleardoublepage
\phantomsection

\setcounter{biburlnumpenalty}{9000}
\emergencystretch=1em

\addcontentsline{toc}{chapter}{Bibliography}
\printbibliography